%% file: dm-dd-f1v-Bayesian-arxiv.tex
\documentclass[12pt,fleqn,a4paper]{article} % A4 size
\usepackage{epsfig,graphics,graphicx}
\usepackage{clshan-math,clshan-dm-dd}
\usepackage{multirow}
\def \lsim {\:\raisebox{-0.7ex}{$\stackrel{\textstyle<}{\sim}$}\:}
\def \gsim {\:\raisebox{-0.7ex}{$\stackrel{\textstyle>}{\sim}$}\:}
\begin{document}
\input{include-plots-arxiv}
\thispagestyle{empty}
\begin{flushright}
 March 2014
\end{flushright}
\begin{center}
{\Large\bf
 Bayesian Reconstruction of the Velocity Distribution of \\
 Weakly Interacting Massive Particles                    \\ \vspace{0.2cm}
 from Direct Dark Matter Detection Data}                 \\
\vspace*{0.7cm}
 {\sc Chung-Lin Shan} \\
\vspace*{0.5cm}
 {\it Physics Division,
      National Center for Theoretical Sciences           \\
      No.~101, Sec.~2, Kuang-Fu Road,
      Hsinchu City 30013, Taiwan, R.O.C.}                \\~\\
 {\it Department of Physics,
      Hangzhou Normal University                         \\
      No.~16, Xuelin Street, Xiasha Higher Education Zone,
      Hangzhou 310036, Zhejiang, China}                  \\~\\
 {\it Kavli Institute for Theoretical Physics China,
      Chinese Academy of Sciences                        \\
      No.~55, Zhong Guan Cun East Street,
      Beijing 100190, China}                             \\~\\
\vspace{0.05cm}
 {\it E-mail:} {\tt clshan@phys.nthu.edu.tw}             \\
\end{center}
\vspace{1cm}
\begin{abstract}
 In this paper,
 we extended our earlier work on
 the reconstruction of the (time--averaged) one--dimensional
 velocity distribution of Galactic Weakly Interacting Massive Particles (WIMPs)
 and introduce the Bayesian fitting procedure
 to the theoretically predicted velocity distribution functions.
 In this reconstruction process,
 the (rough) velocity distribution
 reconstructed by using raw data
 from direct Dark Matter detection experiments directly,
 i.e.~measured recoil energies,
 with one or more different target materials,
 has been used as ``reconstructed--input'' information.
 By assuming a fitting velocity distribution function
 and scanning the parameter space
 based on the Bayesian analysis,
 the astronomical characteristic parameters,
 e.g.~the Solar and Earth's Galactic velocities,
 will be pinned down as the output results.

 Our Monte--Carlo simulations show that
 this Bayesian scanning procedure
 could reconstruct
 the true (input) WIMP velocity distribution function
 pretty precisely
 with negligible systematic deviations
 of the reconstructed characteristic Solar and Earth's velocities
 and 1$\sigma$ statistical uncertainties
 of \mbox{$\lsim~20$ km/s}.
 Moreover,
 for the use of an improper fitting velocity distribution function,
 our reconstruction process
 could still offer useful information
 about the shape of the velocity distribution.
 In addition,
 by comparing these estimates
 to theoretical predictions,
 one could distinguish different (basic) functional forms
 of the theoretically predicted
 one--dimensional WIMP velocity distribution function
 with 2$\sigma$ to 4$\sigma$ confidence levels.
\end{abstract}
\clearpage
\section{Introduction}
 Currently,
 direct Dark Matter detection experiments
 searching for Weakly Interacting Massive Particles (WIMPs)
 are one of the promising methods
 for understanding the nature of Dark Matter (DM)
 and identifying them among new particles produced at colliders
 as well as
 studying the (sub)structure of our Galactic halo
 \cite{SUSYDM96,
       Drees12, Saab12, Baudis12c}.

 In our earlier work
 \cite{DMDDf1v},
 we developed methods
 for reconstructing the (moments of the)
 time--averaged one--dimensional velocity distribution of halo WIMPs
 by using
 the measured recoil energies directly.
 This analysis requires
 no prior knowledge about
 the WIMP density near the Earth
 nor about their scattering cross section on nucleus,
 the unique required information
 is the mass of incident WIMPs.
 We therefore turned to develop the method for
 determining the WIMP mass model--independently
 by combining two experimental data sets
 with two different target nuclei
 \cite{DMDDmchi}.
 By combining these methods
 and using two or three experimental data sets
 with different detector materials,
 one could reconstruct the one--dimensional velocity distribution
 of Galactic WIMPs directly.
 However,
 as presented in Ref.~\cite{DMDDf1v},
 with a few hundreds or even thousands recorded WIMP events,
 only estimates of the reconstructed velocity distribution
 with pretty large statistical uncertainties
 at a few ($<$ 10) points
 could be obtained.

 Therefore,
 in order to offer more detailed information
 about the Galactic WIMP velocity distribution,
 we introduce in this paper the Bayesian analysis
 into our model--independent reconstruction procedure
 developed in Ref.~\cite{DMDDf1v}
 to be able to determine,
 e.g.~the position of the peak of
 the one--dimensional velocity distribution function
 and the concrete values of
 the characteristic Solar and Earth's Galactic velocities.

 The remainder of this paper is organized as follows.
 In Sec.~2,
 we first review the model--independent method
 for reconstructing the time--averaged
 one--dimensional velocity distribution of halo WIMPs
 by using data from direct DM detection experiments directly.
 Then
 we introduce the Bayesian analysis
 and give the basic formulae
 needed in the extended reconstruction process.
 In Sec.~3,
 we present numerical results of
 the reconstructed WIMP velocity distribution functions
 based on Monte--Carlo simulations
 for different generating
 and fitting velocity distributions.
 Different input WIMP masses
 as well as
 impure (pseudo--)data sets
 mixed with (artificially added) unrejected background events
 will also be considered.
 We conclude in Sec.~4.
 Some technical details for our analysis
 will be given in Appendix.
\section{Formalism}
 In this section,
 we develop the formulae
 needed for our Bayesian reconstruction of
 the one--dimensional velocity distribution function of halo WIMPs $f_1(v)$
 by using direct Dark Matter detection data directly.

 We first review the model--independent method
 for reconstructing the time--averaged
 WIMP velocity distribution
 by using experimental data,
 i.e.~measured recoil energies,
 directly from direct detection experiments.
 These ``reconstructed data''
 (with estimated statistical uncertainties)
 will be used as input information
 for the further Bayesian analysis.
 Then,
 in the second part of this section,
 we review the basic concept of the Bayesian analysis
 and give the formulae
 needed in our extended reconstruction procedure.

\subsection{Model--independent reconstruction of
            one--dimensional WIMP velocity distribution}
 In this subsection,
 we review briefly the method for reconstructing
 the one--dimensional WIMP velocity distribution
 from experimental data directly.
 Detailed derivations and discussions
 can be found in Ref.~\cite{DMDDf1v}.

\subsubsection{From the recoil spectrum}
 The basic expression for the differential event rate
 for elastic WIMP--nucleus scattering is given by \cite{SUSYDM96}:
\beq
   \dRdQ
 = \calA \FQ \int_{\vmin}^{\vmax} \bfrac{f_1(v)}{v} dv
\~.
\label{eqn:dRdQ}
\eeq
 Here $R$ is the direct detection event rate,
 i.e.~the number of events
 per unit time and unit mass of detector material,
 $Q$ is the energy deposited in the detector,
 $F(Q)$ is the elastic nuclear form factor,
 $f_1(v)$ is the one--dimensional velocity distribution function
 of the WIMPs impinging on the detector,
 $v$ is the absolute value of the WIMP velocity
 in the laboratory frame.
 The constant coefficient $\calA$ is defined as
\beq
        \calA
 \equiv \frac{\rho_0 \sigma_0}{2 \mchi \mrN^2}
\~,
\label{eqn:calA}
\eeq
 where $\rho_0$ is the WIMP density near the Earth
 and $\sigma_0$ is the total cross section
 ignoring the form factor suppression.
 The reduced mass $\mrN$ is defined by
\beq
        \mrN
 \equiv \frac{\mchi \mN}{\mchi + \mN}
\~,
\label{eqn:mrN}
\eeq
 where $\mchi$ is the WIMP mass and
 $\mN$ that of the target nucleus.
 Finally,
 $\vmin$ is the minimal incoming velocity of incident WIMPs
 that can deposit the energy $Q$ in the detector:
\beq
   \vmin
 = \alpha \sqrt{Q}
\label{eqn:vmin}
\eeq
 with the transformation constant
\beq
        \alpha
 \equiv \sfrac{\mN}{2 \mrN^2}
\~,
\label{eqn:alpha}
\eeq
 and $\vmax$ is the maximal WIMP velocity
 in the Earth's reference frame,
 which is related to
 the escape velocity from our Galaxy
 at the position of the Solar system,
 $\vesc~\gsim~600$ km/s.

 In our earlier work \cite{DMDDf1v},
 it was found that,
 by using a time--averaged recoil spectrum $dR / dQ$
 and assuming that no directional information exists,
 the normalized one--dimensional velocity distribution function
 of incident WIMPs, $f_1(v)$, can be solved
 from Eq.~(\ref{eqn:dRdQ}) directly as
\beq
   f_1(v)
 = \calN
   \cbrac{ -2 Q \cdot \dd{Q} \bbrac{ \frac{1}{\FQ} \aDd{R}{Q} } }\Qva
\~,
\label{eqn:f1v_dRdQ}
\eeq 
 where the normalization constant $\calN$ is given by
\beq
   \calN
 = \frac{2}{\alpha}
   \cbrac{\intz \frac{1}{\sqrt{Q}}
                \bbrac{ \frac{1}{\FQ} \aDd{R}{Q} } dQ}^{-1}
\~.
\label{eqn:calN_int}
\eeq
 Here the integral
 goes over the entire physically allowed range of recoil energies:
 starting at $Q = 0$,
 and the upper limit of the integral has been written as $\infty$.
 Note that,
 because $f_1(v)$ in Eq.~(\ref{eqn:f1v_dRdQ})
 is the normalized velocity distribution,
 the normalization constant $\cal N$ here is independent of
 the constant coefficient $\cal A$
 defined in Eq.~(\ref{eqn:calA}).
 Hence,
 as the most important consequence,
 the velocity distribution function of halo WIMPs
 reconstructed by Eq.~(\ref{eqn:f1v_dRdQ})
 is independent of the local WIMP density $\rho_0$
 as well as
 of the WIMP--nucleus cross section $\sigma_0$.
 However,
 not only the overall normalization constant $\calN$
 given in Eq.~(\ref{eqn:calN_int}),
 but also the shape of the velocity distribution,
 through the transformation $Q = v^2 / \alpha^2$
 in Eq.~(\ref{eqn:f1v_dRdQ}),
 depends on the WIMP mass $\mchi$
 (involved in the coefficient $\alpha$
  defined in Eq.~(\ref{eqn:alpha})).

\subsubsection{From experimental data directly}
 In order to use the expressions
 (\ref{eqn:f1v_dRdQ}) and (\ref{eqn:calN_int})
 for reconstructing $f_1(v)$,
 one needs a functional form for the recoil spectrum $dR / dQ$.
 In practice
 this requires usually a fit to experimental data.
 However,
 data fitting will re--introduce some model dependence
 and make the error analysis more complicated.
 Hence,
 expressions that allow to reconstruct $f_1(v)$
 directly from data
 (i.e.~measured recoil energies)
 have also been developed \cite{DMDDf1v}.
 We started by considering experimental data described by
\beq
     {\T Q_n - \frac{b_n}{2}}
 \le \Qni
 \le {\T Q_n + \frac{b_n}{2}}
\~,
     ~~~~~~~~~~~~ %12
     i
 =   1,~2,~\cdots,~N_n,~
     n
 =   1,~2,~\cdots,~B.
\label{eqn:Qni}
\eeq
 Here the entire experimental possible
 energy range between the minimal and maximal cut--offs
 $\Qmin$ and $\Qmax$
 has been divided into $B$ bins
 with central points $Q_n$ and widths $b_n$.
 In each bin,
 $N_n$ events will be recorded.

 As argued in Ref.~\cite{DMDDf1v},
 the statistical uncertainty on
 the ``slope of the recoil spectrum'',
 $\bbrac{d/dQ \~ (dR / dQ)}_{Q = Q_n}$,
 appearing in the expression (\ref{eqn:f1v_dRdQ}),
 scales like the bin width to the power $-1.5$.
 In addition,
 the wider the bin width,
 the more the recorded events in this bin,
 and thus the smaller the statistical uncertainty
 on the estimator of $\bbrac{d/dQ \~ (dR / dQ)}_{Q = Q_n}$.
 Hence,
 since the recoil spectrum $dR / dQ$ is expected
 to be approximately exponential \cite{DMDDf1v},
 in order to approximate the spectrum
 in a rather wider range,
 instead of the conventional standard linear approximation,
 the following exponential ansatz
 for the measured recoil spectrum
 (before normalized by the exposure $\calE$)
 in the $n$th bin has been introduced \cite{DMDDf1v}:
\beq
        \adRdQ_{{\rm expt}, \~ n}
 \equiv \adRdQ_{{\rm expt}, \~ Q \simeq Q_n}
 \equiv \rn  \~ e^{k_n (Q - Q_{s, n})}
\~.
\label{eqn:dRdQn}
\eeq
 Here $r_n$ is the standard estimator
 for $(dR / dQ)_{\rm expt}$ at $Q = Q_n$:
\beq
   r_n
 = \frac{N_n}{b_n}
\~,
\label{eqn:rn}
\eeq
 $k_n$ is the logarithmic slope of
 the recoil spectrum in the $n$th $Q-$bin,
 which can be computed numerically
 from the average value of the measured recoil energies
 in this bin:
\beq
   \bQn
 = \afrac{b_n}{2} \coth\afrac{k_n b_n}{2}-\frac{1}{k_n}
\~,
\label{eqn:bQn}
\eeq
 where
\beq
        \bQxn{\lambda}
 \equiv \frac{1}{N_n} \sumiNn \abrac{\Qni - Q_n}^{\lambda}
\~.
\label{eqn:bQn_lambda}
\eeq
 Then the shifted point $Q_{s, n}$
 in the ansatz (\ref{eqn:dRdQn}),
 at which the leading systematic error
 due to the ansatz
 is minimal \cite{DMDDf1v},
 can be estimated by
\beq
   Q_{s, n}
 = Q_n + \frac{1}{k_n} \ln\bfrac{\sinh(k_n b_n/2)}{k_n b_n/2}
\~.
\label{eqn:Qsn}
\eeq
 Note that $Q_{s, n}$ differs from the central point of the $n$th bin, $Q_n$.

 Now,
 substituting the ansatz (\ref{eqn:dRdQn})
 into Eq.~(\ref{eqn:f1v_dRdQ})
 and then letting $Q = Q_{s, n}$,
 we can obtain that
\beq
   f_{1, {\rm rec}}(v_{s, n})
 = \calN
   \bBigg{\frac{2 Q_{s, n} r_n}{F^2(Q_{s, n})}}
   \bbrac{\dd{Q} \ln \FQ \bigg|_{Q = Q_{s, n}} - k_n}
\~.
\label{eqn:f1v_Qsn}
\eeq
 Here
\beq
   v_{s, n}
 = \alpha \sqrt{Q_{s, n}}
\~,
\label{eqn:vsn}
\eeq
 and the normalization constant $\calN$
 given in Eq.~(\ref{eqn:calN_int})
 can be estimated directly from the data by
\beq
   \calN
 = \frac{2}{\alpha}
   \bbrac{\sum_{a} \frac{1}{\sqrt{Q_a} \~ F^2(Q_a)}}^{-1}
\~,
\label{eqn:calN_sum}
\eeq
 where the sum runs over all events in the sample.

\subsubsection{Windowing the data set}
 As mentioned above,
 the statistical uncertainty on
 the slope of the recoil spectrum
 around the central point $Q_n$,
 $\bbrac{d/dQ \~ (dR / dQ)}_{Q \simeq Q_n}$,
 is approximately proportional to $b_n^{-1.5}$.
 Thus,
 in order to reduce the statistical uncertainty
 on the velocity distribution
 reconstructed by Eq.~(\ref{eqn:f1v_Qsn}),
 it seems to be better
 to use large bin width.
 However,
 neither the conventional linear approximation:
\beq
   \adRdQ_{{\rm expt}, \~ Q = Q_n}
 = \frac{N_n}{b_n}
%\label{eqn:}
\eeq
 nor the exponential ansatz
 given in Eq.~(\ref{eqn:dRdQn})
 can describe the real (but as yet unknown)
 recoil spectrum exactly.
 The neglected terms of higher powers of $Q - Q_n$
 could therefore induce some uncontrolled systematic errors
 which increase with increasing bin width.
 Moreover,
 since the number of bins scales inversely with their size,
 by using larger bins we would be able to estimate $f_1(v)$
 only at a smaller number of velocities.
 Additionally,
 once a quite large bin width is used,
 it would correspondingly lead to a quite large value
 of the first reconstructible point of $f_1(v)$,
 i.e.~$f_{1, {\rm rec}}(v_{s, 1})$,
 since the central point $Q_1$
 as well as
 the shifted point $Q_{s, 1}$
 of the first bin would be quite large.
 Finally,
 choosing a fixed bin size,
 as one conventionally does,
 would let errors on the estimated logarithmic slopes,
 and hence also on the estimates of $f_1(v)$,
 increase quickly with increasing $Q$ or $v$.
 This is due to the essentially exponential form
 of the expected recoil spectrum,
 which would lead to a quickly falling number of events
 in equal--sized bins.
 By some trial--and--error analyses
 it was found that
 the errors are roughly equal in all bins
 if the bin widths increase linearly \cite{DMDDf1v}.

 Therefore,
 it has been introduced in Ref.~\cite{DMDDf1v} that
 one can first collect experimental data
 in relatively small bins
 and then combining varying numbers of bins
 into overlapping ``windows''.
 In particular,
 the first window would be identical with the first bin.
 One starts by binning the data,
 as in Eq.~(\ref{eqn:Qni}),
 where the bin widths satisfy
\beq
   b_n
 = b_1 + (n - 1) \delta
\~,
\label{eqn:bn_delta}
\eeq
 i.e.
\beq
   Q_n
 = \Qmin + \abrac{n - \frac{1}{2}} b_1 + \bfrac{(n - 1)^2}{2} \delta
\~.
\label{eqn:Qn_delta}
\eeq
 Here the increment $\delta$ satisfies
\beq
   \delta
 = \frac{2}{B (B - 1)} \aBig{\Qmax - \Qmin - B b_1}
\~,
\label{eqn:delta_B}
\eeq
 $B$ being the total number of bins,
 and $Q_{\rm (min, max)}$ are
 the experimental minimal and maximal cut--off energies.
 Assume up to $n_W$ bins are collected into a window,
 with smaller windows at the borders of the range of $Q$.

 In order to distinguish the numbers of bins and windows,
 hereafter Latin indices $n,~m,~\cdots$ are used to label bins,
 and Greek indices $\mu,~\nu,~\cdots$ to label windows.
 For $1 \leq \mu \leq n_W$,
 the $\mu$th window simply consists of the first $\mu$ bins;
 for $n_W \leq \mu \leq B$,
 the $\mu$th window consists of bins
 $\mu-n_W + 1,~\mu-n_W + 2,~\cdots,~\mu$;
 and for $B \leq \mu \leq B+n_W-1$,
 the $\mu$th window consists of the last $n_W - (\mu - B)$ bins.
 This can also be described by introducing
 the indices $n_{\mu-}$ and $n_{\mu+}$
 which label the first and last bins
 contributing to the $\mu$th window,
 with
\cheqna
\beq
\renewcommand{\arraystretch}{1.3}
   n_{\mu-}
 = \cleft{\begin{array}{l c l}
           1,         & ~~~~~~ & {\rm for}~\mu \leq n_W, \\
           \mu-n_W+1, &        & {\rm for}~\mu \geq n_W,
          \end{array}}
\label{eqn:n_mu_minus}
\eeq
 and
\cheqnb
\beq
\renewcommand{\arraystretch}{1.3}
   n_{\mu+}
 = \cleft{\begin{array}{l c l}
           \mu, & ~~~~~~ & {\rm for}~\mu \leq B, \\
           B,   &        & {\rm for}~\mu \geq B.
          \end{array}}
\label{eqn:n_mu_plus}
\eeq
\cheqn
 The total number of windows
 defined through Eqs.~(\ref{eqn:n_mu_minus}) and (\ref{eqn:n_mu_plus})
 is evidently $W = B + n_W - 1$,
 i.e.~$1 \leq \mu \leq B + n_W - 1$.

 As shown above,
 the basic observables needed
 for the reconstruction of $f_1(v)$ by Eq.~(\ref{eqn:f1v_Qsn})
 are the number of events in the $n$th $Q-$bin, $N_n$,
 as well as
 the average value of the measured recoil energies
 in this bin, $\bQn$.
 For a ``windowed'' data set,
 one can easily calculate
 the number of events per window as
\beq
   N_{\mu}
 = \sum_{n = n_{\mu-}}^{n_{\mu+}} N_n
\~,
\label{eqn:N_mu}
\eeq
 as well as
 the average value of the measured recoil energies
\beq
   \Bar{Q - Q_{\mu}}|_{\mu}
 = \frac{1}{N_{\mu}}
   \abrac{\sum_{n = n_{\mu-}}^{n_{\mu+}} N_n \Bar{Q}|_{n}} - Q_{\mu}
\~,
\label{eqn:wQ_mu}
\eeq
 where $Q_{\mu}$ is the central point of the $\mu$th window.
 The exponential ansatz in Eq.~(\ref{eqn:dRdQn})
 is now assumed to hold over an entire window.
 We can then estimate the prefactor as
\beq
   r_{\mu}
 = \frac{N_{\mu}}{w_{\mu}}
\~,
\label{eqn:r_mu}
\eeq
 $w_{\mu}$ being the width of the $\mu$th window.
 The logarithmic slope of the recoil spectrum
 in the $\mu$th window, $k_{\mu}$,
 as well as
 the shifted point $Q_{s, \mu}$
 (from the central point of each ``window'', $Q_{\mu}$)
 can be calculated as in Eqs.~(\ref{eqn:bQn}) and (\ref{eqn:Qsn})
 with ``bin'' quantities replaced by ``window'' quantities.
 Finally,
 note that,
 due to the combination of bins
 into overlapping windows,
 these quantities are all correlated (for $n_W \neq 1$).
 The covariance matrix of
 the estimates of $f_1(v)$
 at adjacent values of $v_{s, \mu} = \alpha \sqrt{Q_{s, \mu}}$
 is given by%
\footnote{
 Note that
 Eq.~(\ref{eqn:cov_f1v_Qs_mu}) should in principle
 also include contributions involving
 the statistical error on the estimator for $\calN$
 in Eq.~(\ref{eqn:calN_sum}).
 However,
 this error and its correlations
 with the errors on the $r_{\mu}$ and $k_{\mu}$
 have been found to be negligible
 compared to the errors included in Eq.~(\ref{eqn:cov_f1v_Qs_mu})
 \cite{DMDDf1v}.
}
\beqn
 \conti {\rm cov}\aBig{f_{1, {\rm rec}}(v_{s, \mu}), f_{1, {\rm rec}}(v_{s, \nu})}
        \non\\
 \=     \bfrac{f_{1, {\rm rec}}(v_{s, \mu}) f_{1, {\rm rec}}(v_{s, \nu})}
              {r_{\mu} r_{\nu}}
        {\rm cov}\abrac{r_{\mu}, r_{\nu}}
       +\abrac{2 \calN}^2
        \bfrac{Q_{s, \mu} Q_{s, \nu} r_{\mu} r_{\nu}}{F^2(Q_{s, \mu}) F^2(Q_{s, \nu})}
        {\rm cov}\abrac{k_{\mu}, k_{\nu}}
        \non\\
\conti ~~~~~~~~~~~~ %12
       -\calN
        \cbrac{ \bfrac{f_{1, {\rm rec}}(v_{s, \mu})}{r_{\mu}}
                \bfrac{2 Q_{s, \nu} r_{\nu} }{F^2(Q_{s, \nu})}
                {\rm cov}\abrac{r_{\mu}, k_{\nu}}
               +\aBig{\mu \lgetsto \nu}}
\~.
\label{eqn:cov_f1v_Qs_mu}
\eeqn
\subsection{Bayesian analysis}
 In this subsection,
 we review the basic concept of the Bayesian analysis
 \cite{Barlow-Statistics}
 (for its applications in physics,
  see e.g.~Ref.~\cite{DAgostini95},
  and for its recent applications
  in direct DM detection phenomenology,
  see e.g.~Refs.~\cite{Akrami,
                       Pato,
                       Arina,
                       Strege12,
                       Cerdeno13,
                       Kavanagh13})
 and extend this method to the use of
 the ``reconstructed data''
 offered by our model--independent reconstruction process
 described in the previous subsection.

\subsubsection{Baye's theorem}
 We start with the {\em Baye's theorem}:
 the probability of $A$ given that $B$ is true
 multiplies the probability of that $B$ is true
 is equal to
 the probability of that
 $A$ and $B$ happen simultaneously,
 which
 is also equal to
 the probability of $B$ given that $A$ is true
 multiplies the probability of that $A$ is true.
 This can simply be expressed as
\beq
    {\rm p}(A | B) \~ {\rm p}(B)
 =  {\rm p}(A \cap B)
 =  {\rm p}(B | A) \~ {\rm p}(A)
\~,
%\label{eqn:Bayes_theorem}
\eeq
 where ${\rm p}(A | B)$ is called
 the ``conditional probability'' of $A$
 given that $B$ is true.
 As long as ${\rm p}(B) \neq 0$,
 the above equation can be rewritten as
\beqn
     {\rm p}(A | B)
 \=  \frac{{\rm p}(B | A) \~ {\rm p}(A)}{{\rm p}(B)}
     \non\\
 \=  \frac{{\rm p}(B | A) \~ {\rm p}(A)}
          {{\rm p}(B | A) \~ {\rm p}(A) + {\rm p}(B | \Bar{A}) \~ {\rm p}(\Bar{A})}
\~,
\label{eqn:Bayes_theorem}
\eeqn
 where
\(
     {\rm p}(\Bar{A})
 =   1 - {\rm p}(A)
\)
 is the probability of the complement of $A$,
 i.e.~the probability of that
 $A$ is not happen.

\subsubsection{Bayesian statistics}
 Applying the Baye's theorem
 described
 above,
 one can directly have that
\beq
     {\rm p}({\rm theory} | {\rm result})
  =  \frac{{\rm p}({\rm result} | {\rm theory})}
          {{\rm p}({\rm result})} \cdot {\rm p}({\rm theory})
\~.
\label{eqn:Bayesian_statistics}
\eeq
 This means that,
 given the observed result,
 the probability of that
 a specified theory is true
 is proportional to
 the probability of the observed result
 in the specified theory
 multiplies
 the probability of the specified theory.

 This statement can be understood as follows.
 If the observed result is predicted by a specified theory
 to be highly unlikely or even impossible/forbidden,
 then this observation
 makes the ``degree of belief'' of this specified theory small
 or even disproves this theory.
 In contrast,
 the observation of a prediction
 by a specified theory with a high probability
 will strengthen one's belief in this theory.

\subsubsection{Bayesian analysis}
 By extending Eq.~(\ref{eqn:Bayesian_statistics}),
 we can obtain that
\beq
     {\rm p}(\Theta | {\rm data})
  =  \frac{{\rm p}({\rm data} | \Theta)}
          {{\rm p}({\rm data})} \cdot {\rm p}(\Theta)
\~.
\label{eqn:Bayesian_analysis}
\eeq
 Here
 $\Theta = \cbig{a_1, a_2, \cdots, a_{N_{\rm Bayesian}}}$
 denotes a specified (combination of the) value(s)
 of the fitting parameter(s);
 ${\rm p}(\Theta)$,
 called ``prior probability'',
 represents our degree of belief about
 $\Theta$ being the true value(s) of fitting parameter(s),
 which is often given
 in form of the (multiplication of the)
 probability distribution(s) of the fitting parameter(s).
 ${\rm p}({\rm data})$,
 called ``evidence'',
 is the total probability of obtaining the particular set of data,
 which is in practice
 irrespective of the actual value(s) of the
 parameter(s) and
 can be treated as a normalization constant;
 it will not be of interest in our further discussion.
 ${\rm p}({\rm data} | \Theta)$
 denotes the probability of the observed result,
 once the specified (combination of the) value(s)
 of the fitting parameter(s)
 happens,
 which can usually be described
 by the {\em likelihood} function of $\Theta$,
 ${\cal L}(\Theta)$.
 Finally,
 ${\rm p}(\Theta | {\rm data})$,
 called the ``posterior probability density function''
 for $\Theta$,
 representes
 the probability of that
 the specified (combination of the) value(s)
 of the fitting parameter(s)
 happens,
 given the observed result.

\subsubsection{Bayesian reconstruction of \boldmath$f_1(v)$}
 Now,
 we can describe
 the procedure of our Bayesian reconstruction of
 the one--dimensional WIMP velocity distribution function
 in detail.

 First,
 by using Eqs.~(\ref{eqn:f1v_Qsn})
 to (\ref{eqn:calN_sum}),
 one can obtain $W = B + n_W - 1$ reconstructed data points:
 $\aBig{v_{s, \mu}, f_{1, {\rm rec}}(v_{s, \mu}) \pm \sigma_{f_1, s, \mu}}$,
 for $\mu = 1,~2,~\cdots,~W$,
 where
\beq
        \sigma_{f_1, s, \mu}
 \equiv \sqrt{{\rm cov}\aBig{f_{1, {\rm rec}}(v_{s, \mu}), f_{1, {\rm rec}}(v_{s, \mu})}}
%\~.
\label{eqn:sigma_f1v_Qs_mu}
\eeq
 denote the square roots of
 the diagonal entries of the covariance matrix
 given in Eq.~(\ref{eqn:cov_f1v_Qs_mu}).
 Choosing a theoretical prediction of
 the one--dimensional velocity distribution
 of halo WIMPs:
 $f_{1, \rm th}(v; a_1, a_2, \cdots, a_{N_{\rm Bayesian}})$,
 where $\abig{a_1, a_2, \cdots, a_{N_{\rm Bayesian}}}$
 are the $N_{\rm Bayesian}$ fitting parameters,
 and
 assuming that
 the reconstructed data points
 is {\em Gaussian}--distributed
 around the theoretical predictions
 $f_{1, \rm th}(v_{s, \mu}; a_1, a_2, \cdots, a_{N_{\rm Bayesian}})$,
 the likelihood function for ${\rm p}({\rm data} | \Theta)$
 can then be defined by
\beqn
 \conti
     {\cal L}\aBig{f_{1, {\rm rec}}(v_{s, \mu}),~\mu = 1,~2,~\cdots,~W;~
                   a_i,~i = 1,~2,~\cdots,~N_{\rm Bayesian}}
     \non\\
 \eqnequiv
     \prod_{\mu = 1}^{W}
     {\rm Gau}\aBig{v_{s, \mu},
                    f_{1, {\rm rec}}(v_{s, \mu}),
                    \sigma_{f_1, s, \mu};
                    a_1, a_2, \cdots, a_{N_{\rm Bayesian}}}
\~,
\label{eqn:calL}
\eeqn
 where
\beqn
 \conti
     {\rm Gau}\aBig{v_{s, \mu},
                    f_{1, {\rm rec}}(v_{s, \mu}),
                    \sigma_{f_1, s, \mu};
                    a_1, a_2, \cdots, a_{N_{\rm Bayesian}}}
     \non\\
 \eqnequiv
     \frac{1}{\sqrt{2 \pi} \~ \sigma_{f_1, s, \mu}} \~
     e^{-\bbig{  f_{1, {\rm rec}}(v_{s, \mu})
               - f_{1, \rm th}(v_{s, \mu}; a_1, a_2, \cdots, a_{N_{\rm Bayesian}})}^2 \big/
         2 \sigma_{f_1, s, \mu}^2}
\~.
\label{eqn:Bayesian_DF_Gau}
\eeqn
 Or,
 equivalently,
 we can use the {\em logarithmic} likelihood function
 given by
\beqn
 \conti
    \ln{\cal L}\aBig{f_{1, {\rm rec}}(v_{s, \mu}),~\mu = 1,~2,~\cdots,~W;~
                     a_i,~i = 1,~2,~\cdots,~N_{\rm Bayesian}}
     \non\\
 \=- \frac{1}{2}
     \sum_{\mu = 1}^{W}
     \frac{\bBig{  f_{1, {\rm rec}}(v_{s, \mu})
                 - f_{1, \rm th}(v_{s, \mu}; a_1, a_2, \cdots, a_{N_{\rm Bayesian}})}^2}
          {\sigma_{f_1, s, \mu}^2}
   - \sum_{\mu = 1}^{W}
     \ln \abrac{\sqrt{2 \pi} \~ \sigma_{f_1, s, \mu}}
\~.
\label{eqn:ln_calL}
\eeqn
 Note that
 in practical use
 the second term in Eq.~(\ref{eqn:ln_calL}) can be neglected,
 since it is just a constant
 for all scanned (combinations of) values of
 the fitting parameter(s)
 $\abig{a_1, a_2, \cdots, a_{N_{\rm Bayesian}}}$.

 Finally,
 choosing the probability distribution function
 for each fitting parameter $a_i$,
 ${\rm p}_i(a_i)$,
 the posterior probability density
 on the left--hand side of Eq.~(\ref{eqn:Bayesian_analysis})
 can then be given by
\beqn
 \conti
     {\rm p}\aBig{a_i,~i = 1,~2,~\cdots,~N_{\rm Bayesian}~\Big|~
                  f_{1, {\rm rec}}(v_{s, \mu}),~\mu = 1,~2,~\cdots,~W}
     \non\\
 \eqnpropto
     {\cal L}\aBig{f_{1, {\rm rec}}(v_{s, \mu}),~\mu = 1,~2,~\cdots,~W;~
                   a_i,~i = 1,~2,~\cdots,~N_{\rm Bayesian}}
     \prod_{i = 1}^{N_{\rm Bayesian}} {\rm p}_i(a_i)
\~.
\label{eqn:P_Bayesian}
\eeqn
\section{Numerical results}
 In this section,
 we present numerical results
 of our Bayesian reconstruction of
 the one--dimensional velocity distribution function
 of halo WIMPs
 based on Monte--Carlo simulations.

 In order to test
 whether we can reconstruct $f_1(v)$ correctly
 with even an improper adopted model
 and/or incorrect expectation value(s) of the fitting parameter(s),
 different choices of the WIMP velocity distribution function
 for generating signal events
 as well as
 different assumptions of $f_1(v)$ and/or
 (slightly) different expectation value(s) of the fitting parameter(s)
 from the commonly adopted values
 for the Bayesian fitting process
 will be condisered in our simulations.
 In Tables \ref{tab:setup_Gau},
 \ref{tab:setup_sh} and \ref{tab:setup_Gau_k},
 we list the {\em input} setup
 for the chosen {\em input} velocity distribution functions
 used for {\em generating} WIMP signals
 as well as
 the scanning ranges,
 the expectation values of
 and 1$\sigma$ uncertainties on
 the fitting parameters
 used for different {\em fitting} velocity distributions.
 Additionally,
 a common maximal cut--off
 on the one--dimensional WIMP velocity distribution
 has been set as \mbox{$\vmax = 700$ km/s}.

 The WIMP mass $\mchi$
 involved in the coefficient $\alpha$
 in Eqs.~(\ref{eqn:vsn}) and (\ref{eqn:calN_sum})
 for estimating the reconstructed points $v_{s, \mu}$
 as well as
 the normalization constant $\calN$
 has been assumed
 either to be known precisely with a negligible uncertainty
 from other (e.g.~collider) experiments
 or determined from
 direct detection experiments
 with {\em different} data sets.
 As in Ref.~\cite{DMDDf1v},
 a $\rmXA{Ge}{76}$ nucleus has been chosen
 as our detector material for reconstructing $f_1(v)$,
 whereas a $\rmXA{Si}{28}$ target
 and a {\em second} $\rmXA{Ge}{76}$ target
 have been used for determining $\mchi$
 \cite{DMDDmchi}.

 As in Refs.~\cite{DMDDf1v, DMDDbg-f1v},
 the WIMP--nucleus cross section in Eq.~(\ref{eqn:calA})
 has been assumed
 to be only spin--independent (SI),
 \mbox{$\sigmapSI = 10^{-9}$ pb},
 and
 the commonly used analytic form
 for the elastic nuclear form factor:
\beq
   F_{\rm SI}^2(Q)
 = \bfrac{3 j_1(q R_1)}{q R_1}^2 e^{-(q s)^2}
\label{eqn:FQ_WS}
\eeq
 has been adopted.
 Here $Q$ is the recoil energy
 transferred from the incident WIMP to the target nucleus,
 $j_1(x)$ is a spherical Bessel function,
\(
   q
 = \sqrt{2 m_{\rm N} Q}
%\label{eqn:qq}
\)
 is the transferred 3-momentum,
 for the effective nuclear radius we use
\(
   R_1
 = \sqrt{R_A^2 - 5 s^2}
%\label{eqn:R1}
\)
 with
\(
        R_A
 \simeq 1.2 \~ A^{1/3}~{\rm fm}
%\~,
%\label{eqn:RA}
\)
 and a nuclear skin thickness
\(
        s
 \simeq 1~{\rm fm}
%\~.
%\label{eqn:ss}
\).

 In our simulations,
 the experimental threshold energies
 have been assumed to be negligible ($\Qmin = 0$)%
\footnote{
 Note that,
 once the experimental threshold energy
 of the data set
 used for reconstructing $f_1(v)$
 is non--negligible,
 the estimate (\ref{eqn:calN_sum})
 of the normalization constant $\calN$
 would need to be modified properly,
 especially for light WIMPs.
}
 and the maximal cut--off energies
 are set as \mbox{$\Qmax = 100$ keV}
 for all target nuclei%
\footnote{
 Note that,
 due to the maximal cut--off
 on the one--dimensional WIMP velocity distribution,
 $\vmax$,
 a kinematic maximal cut--off energy
\beq
   Q_{\rm max, kin}
 = \frac{\vmax^2}{\alpha^2}
%\~.
\label{eqn:Qmax_kin}
\eeq
 has also been taken into account.
};
 the widths of the first energy bin in Eq.~(\ref{eqn:bn_delta})
 have also been set commonly as \mbox{$b_1 = 10$ keV}.
 Additionally,
 we assumed that
 all experimental systematic uncertainties
 as well as
 the uncertainty on the measurement of the recoil energy
 could be ignored.
 The energy resolution of most existing and next--generation detectors
 should be good enough so that
 the very small measurement uncertainties can be neglected
 compared to the statistical uncertainties
 on our reconstructed results
 with only few events.
 Energy range between $\Qmin$ and $\Qmax$
 have been divided into five bins
 and up to three bins
 have been combined to a window.
 (\mbox{3 $\times$}) 5,000 experiments with 500 total events on average
 in one experiment have been simulated.
 In Secs.~3.1 to 3.3,
 the input WIMP mass has been fixed as \mbox{$\mchi = 100$ GeV}.
 In Sec.~3.4,
 we consider the cases with
 a light WIMP mass of \mbox{$\mchi = 25$ GeV}
 and a heavy one of \mbox{$\mchi = 250$ GeV}.
 In addition,
 in Sec.~3.5,
 we consider also briefly
 the effects of unrejected background events
 for different input WIMP masses
 \cite{DMDDbg-mchi, DMDDbg-f1v}.

 Note that,
 for our numerical simulations
 presented in this section,
 the actual numbers of generated signal and background events
 in each simulated experiment
 are Poisson--distributed around their expectation values
 {\em independently}.
 This means that,
 for example,
 for simulations shown in Figs.~\ref{fig:dRdQ-bg-ex-const-000-100-20-100}
 we generate 400 (100) events {\em on average} for WIMP--signals (backgrounds)
 and the total event number recorded in one experiment
 is then the sum of these two numbers.

 Regarding our degree of belief about
 each fitting parameter $a_i$,
 i.e.~${\rm p}_i(a_i)$ in Eq.~(\ref{eqn:P_Bayesian}),
 two probability distribution functions
 have been considered.
 The simplest one is the flat--distribution:
\beq
    {\rm p}_i(a_i)
 =  1
\~,
    ~~~~~~~~~~~~~~~~ % 16
    {\rm for~} a_{i, \rm min} \le a_i \le a_{i, \rm max},
\label{eqn:Bayesian_DF_a_flat}
\eeq
 where $a_{i, \rm (min, max)}$ denote
 the minimal and maximal bounds of the scanning interval of
 the fitting parameter $a_i$.
 On the other hand,
 for the case that
 we have already prior knowledge about
 one fitting parameter,
 a Gaussian--distribution:
\beq
    {\rm p}_i(a_i; \mu_{a, i}, \sigma_{a, i})
 =  \frac{1}{\sqrt{2 \pi} \~ \sigma_{a, i}} \~
    e^{-(a_i - \mu_{a, i})^2 / 2 \sigma_{a, i}^2}
%\~,
\label{eqn:Bayesian_DF_a_Gau}
\eeq
 with the expectation value $\mu_{a, i}$ of and
 the 1$\sigma$ uncertainty $\sigma_{a, i}$ on
 the fitting parameter $a_i$
 is used.

 Note that,
 in one simulated experiment,
 we scan the parameter space
 $(a_1, a_2, \cdots, a_{N_{\rm Bayesian}})$
 in the volume
 $a_i \in [a_{i, {\rm min}}, a_{i, {\rm max}}]$,
 $i = 1, 2, \cdots, N_{\rm Bayesian}$
 to find a particular point
 $(a_1^{\ast}, a_2^{\ast}, \cdots, a_{N_{\rm Bayesian}}^{\ast})$,
 which maximizes the (numerator of the)
 posterior probability density: \\
 ${\rm p}\aBig{a_i,~i = 1,~2,~\cdots,~N_{\rm Bayesian}~\Big|~
               f_{1, {\rm rec}}(v_{s, \mu}),~\mu = 1,~2,~\cdots,~W}$.
 After that
 all simulations have been done,
 we determine the median value of
 the (1$\sigma$ lower and upper bounds of the) velocity distribution
 reconstructed by Eq.~(\ref{eqn:f1v_Qsn}) from all experiments,
 denoted as
 $f_{1, {\rm median}}\abrac{\alpha_{\rm (median)} \sqrt{Q_{s, \mu, {\rm Bayesian}}}}$,
 for $\mu = 1,~2,~\cdots,~W$,
 and shown as solid black crosses in the (top--)left frame(s) of,
 e.g.~Figs.~\ref{fig:f1v-Ge-100-0500-Gau-Gau-flat}.
 And
 we define further
\beqn
 \conti    {\rm p}_{\rm median}\aBig{a_i,~i = 1,~2,~\cdots,~N_{\rm Bayesian}}
           \non\\
 \eqnequiv {\rm p}\aBig{a_i,~i = 1,~2,~\cdots,~N_{\rm Bayesian}~\Big|~
                        f_{1, {\rm median}}\abrac{\alpha_{\rm (median)} \sqrt{Q_{s, \mu, {\rm Bayesian}}}},~%
                        \mu = 1,~2,~\cdots,~W}
\~,
           \non\\
\label{eqn:P_Bayesian_median}
\eeqn
 and check the points
 $(a_1^{\ast}, a_2^{\ast}, \cdots, a_{N_{\rm Bayesian}}^{\ast})$
 obtained from all simulated experiments
 to find the special (``best--fit'') point
 $(a_{1, {\rm max}}, a_{2, {\rm max}}, \cdots, a_{N_{\rm Bayesian}, {\rm max}})$,
 which maximizes ${\rm p}_{\rm median}(a_i,~i = 1,~2,~\cdots,~N_{\rm Bayesian})$.

\subsection{Simple Maxwellian velocity distribution}
\begin{table}[t!]
\InsertSetupTable
{simple Maxwellian velocity distribution $f_{1, \Gau}(v)$}
{Simple
 & $v_0$      [km/s] &        220 & [160,    300] &   230 & 20   \\
}
{The input setup for
 the simple Maxwellian velocity distribution $f_{1, \Gau}(v)$
 used for generating WIMP events
 as well as
 the scanning range,
 the expectation value of
 and the 1$\sigma$ uncertainty on
 the unique fitting parameter $v_0$.
}
{tab:setup_Gau}

 We consider first
 the simplest isothermal spherical Galactic halo model
 for generating WIMP events.
 The normalized one--dimensional
 simple Maxwellian velocity distribution function
 can be expressed as
 \cite{SUSYDM96, DMDDf1v}:
\beq
   f_{1, \Gau}(v)
 = \frac{4}{\sqrt{\pi}}
   \afrac{v^2}{v_0^3} e^{-v^2 / v_0^2}
\~,
\label{eqn:f1v_Gau}
\eeq
 where $v_0 \approx 220$ km/s
 is the Solar orbital velocity in the Galactic frame.

 In Table \ref{tab:setup_Gau},
 we list the input setup for
 the simple Maxwellian velocity distribution $f_{1, \Gau}(v)$
 used for generating WIMP events
 as well as
 the scanning range,
 the expectation value of
 and the 1$\sigma$ uncertainty on
 the unique fitting parameter $v_0$.
 Note that,
 for generating WIMP signals,
 \mbox{$v_0 = 220$ km/s} has been used,
 whereas for Bayesian fitting of this unique parameter
 we used a slightly different value of \mbox{$v_0 = 230$ km/s}
 and set a 1$\sigma$ uncertainty
 of \mbox{$\sigma(v_0) = 20$ km/s}.

\subsubsection{Simple Maxwellian velocity distribution}

 As the simplest Galactic halo model,
 we consider first the use of
 the simple Maxwellian velocity distribution $f_{1, \Gau}(v)$
 given in Eq.~(\ref{eqn:f1v_Gau})
 with an unique fitting parameter $v_0$
 to fit the reconstructed--input data
 given by Eqs.~(\ref{eqn:f1v_Qsn}) to (\ref{eqn:calN_sum})
 and (\ref{eqn:cov_f1v_Qs_mu}).

 In Fig.~\ref{fig:f1v-Ge-100-0500-Gau-Gau-flat}(a),
 we show the reconstructed
 simple Maxwellian velocity distribution function
 for an input WIMP mass of \mbox{$\mchi = 100$ GeV}
 with a $\rmXA{Ge}{76}$ target.
 Here the flat distribution
 given by Eq.~(\ref{eqn:Bayesian_DF_a_flat})
 for the fitting parameter $v_0$
 has been used.
 The black crosses are the velocity distribution
 reconstructed by Eqs.~(\ref{eqn:vsn}) and (\ref{eqn:f1v_Qsn}):
 the vertical error bars show
 the square roots of the diagonal entries of the covariance matrix
 given in Eq.~(\ref{eqn:cov_f1v_Qs_mu})
 (i.e.~$\sigma_{f_1, s, \mu}$
  given in Eq.~(\ref{eqn:sigma_f1v_Qs_mu}))
 and
 the horizontal bars indicate
 the sizes of the windows used
 for estimating $f_{1, {\rm rec}}(v_{s, \mu})$,
 respectively.
 The solid red curve
 is the {\em generating} simple Maxwellian velocity distribution
 with an input value of \mbox{$v_0 = 220$ km/s}.
 While
 the dashed green curve indicates
 the {\em reconstructed} simple Maxwellian velocity distribution
 with the fitting parameter $v_0$
 given by the {\em median} value of all simulated experiments,
 the dash--dotted blue curve indicates
 the {\em reconstructed} simple Maxwellian velocity distribution
 with
 $v_0$
 which maximizes ${\rm p}_{\rm median}\abrac{a_i,~i = 1,~2,~\cdots,~N_{\rm Bayesian}}$
 defined in Eq.~(\ref{eqn:P_Bayesian_median}).

\plotGeGauGauflat
\plotGeGauGauGau

 Meanwhile,
 the light--green (light--blue) area shown here
 indicate the $1\~(2)\~\sigma$ statistical uncertainty bands
 of the Bayesian reconstructed velocity distribution function,
 which has been determined as follows.
 After scanning
 the reconstructed fitting parameter $v_0$
 obtained from all simulated experiments
 and ordering
 according to their ${\rm p}_{\rm median}$ values
 defined in Eq.~(\ref{eqn:P_Bayesian_median})
 {\em descendingly},
 we can not only
 determine the point
 which maximizes ${\rm p}_{\rm median}$
 (labeled with the subscript ``max'' in our plots%
\footnote{
 Note that
 the subscript ``input'' in Figs.~\ref{fig:f1v-Ge-100-0500-Gau-Gau-flat}
 and \ref{fig:f1v-Ge-100-0500-Gau-Gau-Gau}
 indicate that
 the WIMP mass needed
 in Eqs.~(\ref{eqn:vsn}) and (\ref{eqn:f1v_Qsn})
 has been given as the input WIMP mass;
 whereas,
 the subscript ``algo'' in Figs.~\ref{fig:f1v-Ge-SiGe-100-0500-Gau-Gau-flat}
 and \ref{fig:f1v-Ge-SiGe-100-0500-Gau-Gau-Gau}
 indicate that
 the WIMP mass
 is reconstructed by the algorithmic procedure
 developed in Ref.~\cite{DMDDmchi}.
}),
 but also
 the smallest and largest values of
 the first 68.27\% (95.45\%) of all these reconstructed $v_0$'s.
 We then use the smallest (largest) value of
 the first 68.27\% (95.45\%) reconstructed $v_0$'s
 to give the $1\~(2)\~\sigma$ lower (upper) boundaries
 of the Bayesian reconstructed velocity distribution function.
 This means that
 all of the velocity distributions with $v_0$'s
 which give the largest 68.27\% (95.45\%) ${\rm p}_{\rm median}$ values
 should be in the $1\~(2)\~\sigma$ light--green (light--blue) areas.

\begin{table}[t!]
\InsertResultsTable
{simple Maxwellian velocity distribution $f_{1, \Gau}(v)$}
{simple Maxwellian velocity distribution $f_{1, \Gau}(v)$}
{
 \multirow{4}{*}{$v_0$ [km/s]} &
 \multirow{2}{*}{Input} &
 Flat &
 218.8 & $218.8\~^{+12.6}_{-11.2}\~(^{+33.6}_{-22.4})$ & [207.6, 231.4] & [196.4, 252.4] \\
 \cline{3-7}
 & & Gaussian &
 221.6 & $221.6 \pm  9.8\~(^{+23.8}_{-18.2})$ & [211.8, 231.4] & [203.4, 245.4] \\
 \cline{2-7}
 & \multirow{2}{*}{Reconst.} &
 Flat &
 220.2 & $218.8\~^{+21.0}_{-16.8}\~(^{+49.0}_{-33.6})$ & [202.0, 239.8] & [185.2, 267.8] \\
 \cline{3-7}
 & & Gaussian &
 220.2 & $221.6\~^{+15.4}_{-12.6}\~(^{+33.6}_{-26.6})$ & [209.0, 237.0] & [195.0, 255.2] \\
}
{The reconstructed results of $v_0$
 for all four considered cases
 with the simple Maxwellian velocity distribution $f_{1, \Gau}(v)$
 as well as
 the $1\~(2)\~\sigma$ uncertainty ranges
 of the median values.
}
{tab:results_Gau_Gau}

 On the other hand,
 Fig.~\ref{fig:f1v-Ge-100-0500-Gau-Gau-flat}(b) shows
 the distribution of the Bayesian reconstructed fitting parameter $v_0$
 in all simulated experiments.
 The red vertical line indicates the true (input) value of $v_0$,
 which has been labeled with the subscript ``gen''.
 The green vertical line indicates
 the median value of the simulated results,
 whereas
 the blue one
 indicates the value which maximizes
 ${\rm p}_{\rm median}$.
 In addition,
 the horizontal thick (thin) green bars show
 the $1\~(2)\~\sigma$ ranges of the reconstructed results%
\footnote{
 Note that
 the $1\~(2)\~\sigma$ ranges given here mean that,
 according to the order of the reconstructed values of $v_0$ {\em along}
 and centered at their {\em median} value $v_{0, {\rm median}}$,
 68.27\% (95.45\%) of the reconstructed values in the simulated experiments
 are in this range.
}.
 Note that
 the bins at \mbox{$v_0 = 160$ km/s} and \mbox{$v_0 = 300$ km/s}
 are ``overflow'' bins,
 which contain also the experiments
 with the best--fit $v_0$ value of
 either \mbox{$v_0 < 160$ km/s} or \mbox{$v_0 > 300$ km/s}.

 In Figs.~\ref{fig:f1v-Ge-100-0500-Gau-Gau-flat},
 it can be seen clearly that,
 {\em without} a prior knowledge about
 the Solar Galactic velocity,
 one could in principle pin down the parameter $v_0$ very precisely
 with $1\~(2)\~\sigma$ statistical uncertainties of only
 \mbox{$^{+12.6}_{-11.2}\~(^{+33.6}_{-22.4})$ km/s}
 (see Table \ref{tab:results_Gau_Gau}).
 Moreover,
 by using the Bayesian reconstruction of
 the one--dimensional velocity distribution function,
 the large (1$\sigma$) statistical uncertainty
 given by Eq.~(\ref{eqn:sigma_f1v_Qs_mu})
 can be reduced significantly:
 the band of the 2$\sigma$ statistical uncertainty
 would be approximately equal to or even smaller than
 the (solid black) vertical 1$\sigma$ uncertainty bars!

 Furthermore,
 in Figs.~\ref{fig:f1v-Ge-100-0500-Gau-Gau-Gau}
 we consider the case
 with a rough prior knowledge about the Solar Galactic velocity $v_0$.
 It has been found that,
 firstly,
 by using a Gaussian probability distribution
 for $v_0$ with a 1$\sigma$ uncertainty of \mbox{20 km/s},
 one could reduce the $1\~(2)\~\sigma$ statistical uncertainties
 on the Bayesian reconstructed parameter $v_0$
 to \mbox{$\pm  9.8\~(^{+23.8}_{-18.2})$ km/s}
 (see Table \ref{tab:results_Gau_Gau}).
 Secondly and more importantly,
 although an expectation value of \mbox{$v_0 = 230$ km/s},
 which differs (slightly) from the true (input) one,
 is used,
 the value of the fitting parameter $v_0$
 could still be pinned down precisely
 with a tiny systematic deviation
 ($<$ 2 km/s).

\plotGeSiGeGauGauflat
\plotGeSiGeGauGauGau

 In Figs.~\ref{fig:f1v-Ge-SiGe-100-0500-Gau-Gau-flat}
 and \ref{fig:f1v-Ge-SiGe-100-0500-Gau-Gau-Gau},
 we consider the case that
 the WIMP mass $\mchi$
 needed in Eqs.~(\ref{eqn:vsn}) and (\ref{eqn:calN_sum})
 is reconstructed by the algorithmic procedure
 developed in Ref.~\cite{DMDDmchi}
 with a $\rmXA{Si}{28}$ target
 and a second $\rmXA{Ge}{76}$ target.
 Note that,
 while the vertical bars show
 the 1$\sigma$ statistical uncertainties
 estimated by Eq.~(\ref{eqn:sigma_f1v_Qs_mu})
 taking into account an extra contribution from
 the 1$\sigma$ statistical uncertainty
 on the reconstructed WIMP mass,
 the horizontal bars shown here
 indicate the 1$\sigma$ statistical uncertainties
 on the estimates of $v_{s, \mu}$
 given in Eq.~(\ref{eqn:vsn})
 due to the uncertainty on the reconstructed WIMP mass;
 the statistical and systematic uncertainties
 due to estimating of $Q_{s, \mu}$
 have been neglected here.

 It can be seen that,
 due to the extra {\em statistical fluctuation}
 on the reconstructed WIMP mass
 \cite{DMDDmchi}
 and in turn
 the uncertainty on the estimated $v_{s, \mu}$
 (the horizontal bars in the left frames),
 the statistical uncertainties on the Bayesian reconstructed $v_0$
 with both of the flat and the Gaussian probability distributions
 become \mbox{$\sim 30$\%} to \mbox{$\sim 60$\%} larger.
 However,
 same as the case with an input WIMP mass,
 the use of the Gaussian probability distribution
 can not only reduce
 the statistical uncertainty on $v_0$ significantly
 and therefore improve
 the Bayesian reconstructed velocity distribution,
 but also alleviate
 the ``imprecisely'' expected value of $v_0$.

 In Table \ref{tab:results_Gau_Gau},
 we list the reconstructed results of $v_0$
 for all four considered cases
 with the simple Maxwellian velocity distribution $f_{1, \Gau}(v)$
 as well as
 the $1\~(2)\~\sigma$ uncertainty ranges
 of the {\em median} values of $v_0$.
 It would be worth to emphasize that,
 the statistical uncertainties shown
 in Figs.~\ref{fig:f1v-Ge-100-0500-Gau-Gau-Gau}
 and \ref{fig:f1v-Ge-SiGe-100-0500-Gau-Gau-Gau}
 are (much) smaller than the {\em input} uncertainty
 on the expectation value of $v_0$ of \mbox{20 km/s}.

\subsection{Shifted Maxwellian velocity distribution}
\begin{table}[t!]
\InsertSetupTable
{shifted Maxwellian velocity distribution $f_{1, \sh}(v)$}
{Simple
 & $v_0$      [km/s] & $\sim 295$ & [160,    400] &   280 & 40   \\
 \hline
 1--para. shifted
 & $v_0$      [km/s] &        220 & [160,    300] &   230 & 20   \\
 \hline
 \multirow{2}{*}{Shifted}
 & $v_0$      [km/s] &        220 & [160,    300] &   230 & 20   \\
 & $\ve$      [km/s] &        231 & [160,    300] &   245 & 20   \\
 \hline
 \multirow{2}{*}{Variated shifted}
 & $v_0$      [km/s] &        220 & [160,    300] &   230 & 20   \\
 & $\Delta v$ [km/s] &         11 & [$-50$,   80] &    15 & 20   \\
}
{The input setup for
 the shifted Maxwellian velocity distribution $f_{1, \sh}(v)$
 used for generating WIMP signals
 as well as
 the theoretically estimated values,
 the scanning ranges,
 the expectation values of
 and the 1$\sigma$ uncertainties on
 the fitting parameters
 used for different fitting velocity distribution functions.
}
{tab:setup_sh}

 By taking into account
 the orbital motion of the Solar system around our Galaxy
 as well as
 that of the Earth around the Sun,
 a more realistic
 shifted Maxwellian velocity distribution of halo WIMPs
 has been given by
 \cite{SUSYDM96, DMDDf1v}:
\beq
   f_{1, \sh}(v)
 = \frac{1}{\sqrt{\pi}} \afrac{v}{v_0 \ve}
   \bBig{  e^{-(v - \ve)^2 / v_0^2}
         - e^{-(v + \ve)^2 / v_0^2}  }
\~.
\label{eqn:f1v_sh}
\eeq
 Here
 $\ve$ is the time-–dependent
 Earth's velocity in the Galactic frame
 \cite{Freese,
       SUSYDM96}:
\beq
   \ve(t)
 = v_0 \bbrac{1.05 + 0.07 \cos\afrac{2 \pi (t - t_{\rm p})}{1~{\rm yr}}}
\~,
\label{eqn:ve}
\eeq
 with $t_{\rm p} \simeq$ June 2nd is the date
 on which the velocity of the Earth relative to the WIMP halo is maximal%
\footnote{
 In our simulations,
 the time dependence of the Earth's velocity in the Galactic frame,
 the second term of $\ve(t)$,
 will be ignored,
 i.e.~$\ve = 1.05 \~ v_0$
 is used.
}.

 In Table \ref{tab:setup_sh},
 we list the input setup for
 the shifted Maxwellian velocity distribution $f_{1, \sh}(v)$
 used for generating WIMP signals
 as well as
 the theoretically estimated values,
 the scanning ranges,
 the expectation values of
 and the 1$\sigma$ uncertainties on
 the fitting parameters 
 used for different fitting velocity distribution functions.

\subsubsection{Simple Maxwellian velocity distribution}
\plotGeSiGeshGauflat
\plotGeSiGeshGauGau
\begin{table}[b!]
\InsertResultsTable
{shifted Maxwellian velocity distribution $f_{1, \sh}(v)$}
{simple  Maxwellian velocity distribution $f_{1, \Gau}(v)$}
{
 \multirow{4}{*}{$v_0$ [km/s]} &
 \multirow{2}{*}{Input} &
 Flat &
 296.8 & $296.8\~^{+21.6}_{-19.2}\~(^{+45.6}_{-36.0})$ & [277.6, 318.4] & [260.8, 342.4] \\
 \cline{3-7}
 & & Gaussian &
 294.4 & $292.0\~^{+16.8}_{-12.0}\~(^{+31.8}_{-26.4})$ & [280.0, 308.8] & [265.6, 323.8] \\
 \cline{2-7}
 & \multirow{2}{*}{Reconst.} &
 Flat &
 299.2 & $296.8\~^{+40.8}_{-31.2}\~(^{+86.4}_{-60.0})$ & [265.6, 337.6] & [236.8, 383.2] \\
 \cline{3-7}
 & & Gaussian &
 294.4 & $294.4 \pm 26.4\~(\pm 52.8)$ & [268.0, 320.8] & [241.6, 347.2] \\
}
{The reconstructed results of $v_0$
 for all four considered cases
 with the simple Maxwellian velocity distribution $f_{1, \Gau}(v)$
 as well as
 the $1\~(2)\~\sigma$ uncertainty ranges
 of the median values.
}
{tab:results_sh_Gau}

 We consider first the simplest case of
 the simple Maxwellian velocity distribution function $f_{1, \Gau}(v)$
 with the unique fitting parameter $v_0$
 to fit the reconstructed--input data points
 given by Eqs.~(\ref{eqn:f1v_Qsn}) and (\ref{eqn:cov_f1v_Qs_mu}).

 As in Sec.~3.1.1,
 in Figs.~%
 \ref{fig:f1v-Ge-SiGe-100-0500-sh-Gau-flat}
 we use first the flat probability distribution
 for the fitting parameter $v_0$
 with either the precisely known (input) (upper)
 or the reconstructed (lower) WIMP mass,
 respectively.
 Figs.~\ref{fig:f1v-Ge-SiGe-100-0500-sh-Gau-flat}(a)
 and
 (c)
 show clearly that,
 although an ``improper'' choice for
 the fitting velocity distribution function
 and a simple flat probability distribution for $v_0$
 (i.e.~without any prior knowledge about $v_0$)
 have been used,
 the 1$\sigma$ statistical uncertainty bands
 of the reconstructed WIMP velocity distribution function
 could in principle still cover the true (input) distribution.
 More precisely and quantitatively,
 the deviations of the peaks of
 the reconstructed velocity distributions
 from that of the true (input) one
 are only \mbox{$\sim 10$ km/s}.
 Note that
 the 1$\sigma$ statistical uncertainty
 on the reconstructed fitting parameter $v_0$
 is \mbox{$\sim 20$ km/s} or even \mbox{$\sim 40$ km/s}
 (see Table \ref{tab:results_sh_Gau}).

 However,
 our simulations show also that,
 with an ``improper'' assumption about
 the fitting velocity distribution function,
 one would obtain an ``unexpected'' result
 for the fitting parameter $v_0$:
 2.5$\sigma$
 (with the reconstructed WIMP mass,
  Fig.~\ref{fig:f1v-Ge-SiGe-100-0500-sh-Gau-flat}(d))
 to 4$\sigma$
 (with the input WIMP mass,
  Fig.~\ref{fig:f1v-Ge-SiGe-100-0500-sh-Gau-flat}(b))
 deviations of the
 reconstructed Solar Galactic velocity
 from the theoretical estimate of
 \mbox{$v_0 \approx 220$ km/s}.
 Such observation would indicate clearly that
 our initial assumption about
 the fitting velocity distribution function
 would be incorrect 
 or at least need to be modified.

 Moreover,
 in Figs.~%
 \ref{fig:f1v-Ge-SiGe-100-0500-sh-Gau-Gau}
 we assume that
 a rough prior knowledge about
 the Solar Galactic velocity $v_0$ exists
 and use the Gaussian probability distribution
 for $v_0$
 with an expectation value of \mbox{$v_0 = {\it 280}$ km/s}
 and a 1$\sigma$ uncertainty of \mbox{{\em 40} km/s}.
 As observed in Sec.~3.1.1,
 with a prior knowledge about the fitting parameter $v_0$,
 one could reconstruct the velocity distribution function better:
 the $1\~(2)\~\sigma$ statistical uncertainties on $v_0$
 could be reduced to \mbox{$\sim 60$\%}.
 Additionally,
 the reconstructed 1$\sigma$ statistical uncertainties on $v_0$
 with both of the input and the reconstructed WIMP masses
 are much smaller than
 the input 1$\sigma$ value of \mbox{40 km/s}.

 In Table \ref{tab:results_sh_Gau},
 we list the reconstructed values of $v_0$
 for all four considered cases
 with the simple Maxwellian velocity distribution $f_{1, \Gau}(v)$
 as well as
 the $1\~(2)\~\sigma$ uncertainty ranges
 of the median values of $v_0$.

\subsubsection{One--parameter shifted Maxwellian velocity distribution}
\plotGeSiGeshshvflat
\plotGeSiGeshshvGau

 In the previous Sec.~3.2.1,
 we have found that,
 by assuming (improperly)
 the simple Maxwellian velocity distribution $f_{1, \Gau}(v)$,
 in both cases with and without an expectation value of
 the fitting parameter $v_0$,
 one would obtain a {\em much higher} reconstruction result:
 \mbox{$v_{\rm 0, rec} \simeq 295$ km/s},
 which is 2$\sigma$ to 4$\sigma$ apart from
 the theoretical estimate of
 \mbox{$v_0 \approx 220$ km/s}.
 This observation implies the need of
 a more suitable fitting velocity distribution function.
 Hence,
 as the second trial,
 we consider now the use of
 the {\em shifted} Maxwellian velocity distribution $f_{1, \sh}(v)$
 given in Eq.~(\ref{eqn:f1v_sh})
 with {\em only one} fitting parameter,
 i.e.~the Solar Galactic velocity $v_0$.
 In this case,
 we fix simply that
\beq
   \ve
 = 1.05\~v_0
\~,
\label{eqn:ve_ave}
\eeq
 and neglect the time--dependence of $\ve(t)$.%
\footnote{
 Note that
 hereafter we use $f_{1, \sh, v_0}(v)$
 to denote the ``one--parameter''
 shifted Maxwellian velocity distribution function,
 in order to distinguish this
 from the ``original'' one
 given in Eq.~(\ref{eqn:f1v_sh})
 with $v_0$ and $\ve$
 as two independent fitting parameters.
}

 In Figs.~%
 \ref{fig:f1v-Ge-SiGe-100-0500-sh-sh_v0-flat},
 we consider first the flat probability distribution
 for the fitting parameter $v_0$
 with either the precisely known (input) (upper)
 or the reconstructed (lower) WIMP mass,
 respectively.
 It can be seen clearly that,
 with a more suitable assumption about
 the fitting function,
 one could indeed reconstruct the WIMP velocity distribution
 much closer to the true (input) one.
 Although
 no prior knowledge about $v_0$
 is used,
 this most important characteristic parameter
 could in principle be pinned down very precisely:
 the difference between the {\em median} values of
 the reconstructed $v_0$
 and the true (input) one
 would be \mbox{$\lsim\~{\it 5}$ km/s}
 (see also Table \ref{tab:results_sh_sh_v0}).

 In addition,
 with the input WIMP mass,
 the $1\~(2)\~\sigma$ statistical uncertainties
 on the reconstructed $v_0$
 are only \mbox{$^{+14.0}_{-11.2}\~(^{+29.4}_{-23.8})$ km/s}.
 Even with the reconstructed WIMP mass,
 the $1\~(2)\~\sigma$ statistical uncertainties
 could still be limited as small as only
 \mbox{$^{+25.2}_{-22.4}\~(^{+56.0}_{-44.8})$ km/s}.
 It would be worth to emphasize that,
 compared to the uncertainty
 on the astronomical measurement of $v_0$
 of \mbox{$\sim 20$ km/s} (or probably larger),
 the result offered by our Bayesian reconstruction method
 would be a pretty precise estimate
 and could help us to confirm
 the astronomical measurement of $v_0$.

 Moreover,
 in Figs.~%
 \ref{fig:f1v-Ge-SiGe-100-0500-sh-sh_v0-Gau}
 we give the reconstruction results
 with the Gaussian probability distribution
 for $v_0$
 with an expectation value of \mbox{$v_0 = 230$ km/s}
 and a 1$\sigma$ uncertainty of \mbox{20 km/s}.
 As summarized in Table \ref{tab:results_sh_sh_v0},
 with a prior knowledge of the parameter $v_0$,
 the $1\~(2)\~\sigma$ statistical uncertainties
 could be reduced significantly
 to be \mbox{$\lsim$ 70\%}.
 Remind here that
 the expectation value of the Gaussian probability distribution
 of $v_0$
 has been set as \mbox{$v_0 = 230$ km/s},
 a bit different from the input value.
 Nevertheless,
 our simulations show that
 this ``artificial'' (systematic) error
 could be corrected in our reconstruction process for $v_0$.

 In Table \ref{tab:results_sh_sh_v0},
 we list the reconstructed results of $v_0$
 for all four considered cases
 with the one--parameter shifted Maxwellian velocity distribution
 $f_{1, \sh, v_0}(v)$
 as well as
 the $1\~(2)\~\sigma$ uncertainty ranges
 of the median values of $v_0$.

\begin{table}[b!]
\InsertResultsTable
{shifted Maxwellian velocity distribution $f_{1, \sh}(v)$}
{one--parameter shifted Maxwellian velocity distribution $f_{1, \sh, v_0}(v)$}
{
 \multirow{4}{*}{$v_0$ [km/s]} &
 \multirow{2}{*}{Input} &
 Flat &
 217.4 & $217.4\~^{+14.0}_{-11.2}\~(^{+29.4}_{-23.8})$ & [206.2, 231.4] & [193.6, 246.8] \\
 \cline{3-7}
 & & Gaussian &
 221.6 & $221.6\~^{+ 9.8}_{- 8.4}\~(\pm 18.2)$ & [213.2, 231.4] & [203.4, 239.8] \\
 \cline{2-7}
 & \multirow{2}{*}{Reconst.} &
 Flat &
 218.8 & $218.8\~^{+25.2}_{-22.4}\~(^{+56.0}_{-44.8})$ & [196.4, 244.0] & [174.0, 274.8] \\
 \cline{3-7}
 & & Gaussian &
 221.6 & $221.6\~^{+16.8}_{-15.4}\~(\pm 33.6)$ & [206.2, 238.4] & [188.0, 255.2] \\
}
{The reconstructed results of $v_0$
 for all four considered cases
 with the one--parameter shifted Maxwellian velocity distribution
 $f_{1, \sh, v_0}(v)$
 as well as
 the $1\~(2)\~\sigma$ uncertainty ranges
 of the median values.
}
{tab:results_sh_sh_v0}
\subsubsection{Shifted Maxwellian velocity distribution}
\plotGeshshGau
\plotGeSiGeshshGau

 Now we release the fixed relation between $v_0$ and $\ve$
 given in Eq.~(\ref{eqn:ve_ave})
 and consider the reconstruction of these two parameters
 {\em simultaneously} and {\em independently}.
 In addition,
 we assume here that,
 from the (naive) trials with the simple and one--parameter shifted
 Maxwellian velocity distributions
 done previously,
 one could already obtain a rough idea about
 the shape of the velocity distribution of halo WIMPs.
 This information would in turn give us
 the prior knowledge about
 the expectation values of the Solar and Earth's Galactic velocities
 $v_0$ and $\ve$.
 Hence,
 we consider here only the Gaussian probability distribution
 for both fitting parameters
 with expectation values of \mbox{$v_0 = 230$ km/s}
 and \mbox{$\ve = 245$ km/s}
 and a common 1$\sigma$ uncertainty of \mbox{20 km/s}
 (see Table \ref{tab:setup_sh})%
\footnote{
 Remind that
 both of the expectation values of the fitting parameters $v_0$ and $\ve$
 differ slightly from the true (input) values.
 Note also that
 the time--dependence of the Earth's Galactic velocity
 is ignored here
 and $\ve$ is thus treated as
 a {\em time--independent} fitting parameter.
}.

 In Fig.~\ref{fig:f1v-Ge-100-0500-sh-sh-Gau}(a),
 we show the reconstructed
 shifted Maxwellian velocity distribution function
 as well as
 the $1\~(2)\~\sigma$ statistical uncertainty bands.
 Here the true (input) WIMP mass has been used.
 Comparing to Fig.~\ref{fig:f1v-Ge-SiGe-100-0500-sh-sh_v0-Gau}(a),
 it can be seen clearly that,
 with a prior knowledge about
 the Solar and Earth's Galactic velocities $v_0$ and $\ve$,
 one could in principle reconstruct the velocity distribution function
 with two fitting parameters
 more precisely:
 the $1\~(2)\~\sigma$ statistical uncertainty bands
 are much thinner and
 the deviations of $v_0$ and $\ve$ are {\em only a few} km/s
 (see also Table \ref{tab:results_sh_sh})%
\footnote{
 Note however that,
 for using velocity distributions
 with two or more fitting parameters
 {\em without} constraints on these parameters
 (i.e.~the use of the ``flat'' probability distribution),
 the distribution of the reconstructed results
 by our Bayesian analysis
 would be pretty wide,
 a part of them would even be
 on the boundary of the scanning ranges of these parameters.
}.

 Fig.~\ref{fig:f1v-Ge-100-0500-sh-sh-Gau}(b) shows
 the distribution of
 the Bayesian reconstructed fitting parameter $v_0$ and $\ve$
 in all simulated experiments
 on the $v_0 - \ve$ plane.
 The light--green (light--blue, gold) points
 indicate the $1\~(2)\~(> 2)\~\sigma$ areas
 of the reconstructed combination of $v_0$ and $\ve$.
 Note here that
 these $1\~(2)\~(> 2)\~\sigma$ areas are determined
 according to the {\em descending} order of the ${\rm p}_{\rm median}$ values
 of the reconstructed combination of $v_0$ and $\ve$.
 This means that
 the light--green (light--blue, gold) areas
 are the reconstructed combinations of $v_0$ and $\ve$
 which give the largest 68.27\% (95.45\%) ${\rm p}_{\rm median}$ values
 in all of the simulated experiments.

 Moreover,
 the red upward--triangle indicates
 the input values of $v_0$ and $\ve$,
 which has been labeled with the subscript ``gen''.
 The green disk shows
 the median values of the simulated results,
 whereas
 and blue downward--triangle
 the point which maximizes
 ${\rm p}_{\rm median}$.
 In addition,
 the thick (thin) green crosses show
 the $1\~(2)\~\sigma$ (68.27\% (95.45\%)) ranges of the reconstructed results
 according to the order of
 the reconstructed values of $v_0$ or $\ve$ {\em along}
 (centered at their median values
  $v_{0, {\rm median}}$ or $v_{{\rm e, median}}$).

 Additionally,
 in Figs.~\ref{fig:f1v-Ge-100-0500-sh-sh-Gau}(c) and (d),
 we give
 the distributions of
 the Bayesian reconstructed fitting parameters $v_0$ and $\ve$
 as well as
 the $1\~(2)\~\sigma$ statistical uncertainty ranges
 on the $v_0-$ and $\ve-$ axes
 {\em separately}.
 Note here that
 we project all reconstructed combinations of $\abig{v_0, \ve}$
 on the $v_0-$ or $\ve-$axis.
 Comparing Fig.~\ref{fig:f1v-Ge-100-0500-sh-sh-Gau}(c)
 to Fig.~\ref{fig:f1v-Ge-SiGe-100-0500-sh-sh_v0-Gau}(b),
 it can be found that,
 while
 the deviations of $v_0$ are a little bit larger
 then the results
 with the ``one--parameter'' fitting velocity distribution,
 the $1\~(2)\~\sigma$ statistical uncertainties
 are reduced to \mbox{$\sim$ 70\%}
 (see Table \ref{tab:results_sh_sh} and
  Table \ref{tab:results_sh_sh_v0}).

\begin{table}[t!]
\InsertResultsTable
{shifted Maxwellian velocity distribution $f_{1, \sh}(v)$}
{shifted Maxwellian velocity distribution $f_{1, \sh}(v)$}
{
 \multirow{2}{*}{$v_0$ [km/s]} &
 Input &
 Gaussian &
 223.0 & $223.0 \pm  7.0\~(\pm 14.0)$ & [216.0, 230.0] & [209.0, 237.0] \\
 \cline{2-7}
 & Reconst. &
 Gaussian &
 223.0 & $223.0 \pm 12.6\~(^{+26.6}_{-23.8})$ & [210.4, 235.6] & [199.2, 249.6] \\
\hline
 \multirow{2}{*}{$\ve$ [km/s]} &
 Input &
 Gaussian &
 238.4 & $238.4\~^{+ 7.0}_{- 8.4}\~(^{+15.4}_{-16.9})$ & [230.0, 245.4] & [221.6, 253.8] \\
 \cline{2-7}
 & Reconst. &
 Gaussian &
 238.4 & $238.4 \pm 12.6\~(^{+25.2}_{-26.6})$ & [225.8, 251.0] & [211.8, 263.6] \\
}
{The reconstructed results of $v_0$ and $\ve$
 with the shifted Maxwellian velocity distribution $f_{1, \sh}(v)$
 as well as
 the $1\~(2)\~\sigma$ uncertainty ranges
 of the median values.
}
{tab:results_sh_sh}

 As a comparison,
 in Figs.~\ref{fig:f1v-Ge-SiGe-100-0500-sh-sh-Gau}
 we use the reconstructed WIMP mass.
 Fig.~\ref{fig:f1v-Ge-SiGe-100-0500-sh-sh-Gau}(b)
 shows that,
 as expected,
 the distribution of $v_0$ and $\ve$ on the $v_0 - \ve$ plane
 becomes wider and
 extends toward the directions of
 higher (lower)--$v_0$ and higher (lower)--$\ve$.
 Nevertheless,
 the ``best--fit'' combination of them
 which maximizes ${\rm p}_{\rm median}$
 as well as
 the median values of $v_0$ and $\ve$
 consistant very well
 with the true (input) values.

 In Table \ref{tab:results_sh_sh_v0},
 we list the reconstructed results of $v_0$ and $\ve$
 with the shifted Maxwellian velocity distribution $f_{1, \sh}(v)$
 as well as
 the $1\~(2)\~\sigma$ uncertainty ranges
 of the median values.

\subsubsection{Variated shifted Maxwellian velocity distribution}
\plotGeshshDvGau
\plotGeSiGeshshDvGau

 In the previous Sec.~3.2.3,
 it has been found that,
 by using the shifted Maxwellian velocity distribution
 given in Eq.~(\ref{eqn:f1v_sh})
 with two fitting parameters:
 $v_0$ and $\ve$,
 one could reconstruct
 the ($1\~(2)\~\sigma$ statistical uncertainty bands of the)
 velocity distribution function
 as well as
 pin down the Solar Galactic velocity $v_0$
 pretty precisely.
 However,
 as shown in Figs.~\ref{fig:f1v-Ge-100-0500-sh-sh-Gau}(d)
 and \ref{fig:f1v-Ge-SiGe-100-0500-sh-sh-Gau}(d),
 the deviations of the reconstructed Earth's Galactic velocity $\ve$
 from the true (input) value
 seem to be (much) larger than
 the deviations of the reconstructed $v_0$.
 This might be caused by
 the strong (anti--)correlation
 between $v_0$ and $\ve$.
 Hence,
 we consider now a variation of
 the shifted Maxwellian distribution function,
 in the hope that
 this pretty large systematic deviation of $\ve$
 could be reduced.

 We rewrite the shifted Maxwellian velocity distribution
 given in Eq.~(\ref{eqn:f1v_sh})
 to the following ``variated'' form:
\beq
   f_{1, \sh, \Delta v}(v)
 = \frac{1}{\sqrt{\pi}} \bfrac{v}{v_0 \abrac{v_0 + \Delta v}}
   \cbigg{  e^{-\bbrac{v - \abrac{v_0 + \Delta v}}^2 / v_0^2}
          - e^{-\bbrac{v + \abrac{v_0 + \Delta v}}^2 / v_0^2}  }
\~,
\label{eqn:f1v_sh_Dv}
\eeq
 where
\beq
        \Delta v
 \equiv \ve - v_0
%\~,
\label{eqn:Delta_v}
\eeq
 is the difference between $v_0$ and $\ve(t)$.%
\footnote{
 As in the previous Sec.~3.2.3,
 the time--dependence of the Earth's Galactic velocity
 is ignored here
 and $\Delta v$ is thus treated as
 a {\em time--independent} fitting parameter.
}
\begin{table}[b!]
\InsertResultsTable
{shifted Maxwellian velocity distribution $f_{1, \sh}(v)$}
{variated shifted Maxwellian velocity distribution $f_{1, \sh, \Delta v}(v)$}
{
 \multirow{2}{*}{$v_0$ [km/s]} &
 Input &
 Gaussian &
 221.6 & $221.6 \pm  8.4\~(\pm 15.4)$ & [213.2, 230.0] & [206.2, 237.0] \\
 \cline{2-7}
 & Reconst. &
 Gaussian &
 221.6 & $221.6 \pm 14.0\~(\pm 29.4)$ & [207.6, 235.6] & [192.2, 251.0] \\
\hline
 \multirow{2}{*}{$\Delta v$ [km/s]} &
 Input &
 Gaussian &
  11.1 & $ 11.1 \pm  5.2\~(^{+10.4}_{-11.7})$ & [5.9,  16.3] & [$- 0.6$,  21.5] \\
 \cline{2-7}
 & Reconst. &
 Gaussian &
  11.1 & $ 11.1 \pm  7.8\~(^{+15.6}_{-14.3})$ & [3.3,  18.9] & [$- 3.2$,  26.7] \\
}
{The reconstructed results of $v_0$ and $\Delta v$
 with the variated shifted Maxwellian velocity distribution
 $f_{1, \sh, \Delta v}(v)$
 as well as
 the $1\~(2)\~\sigma$ uncertainty ranges
 of the median values.
}
{tab:results_sh_sh_Dv}

 As in the previous Sec.~3.2.3,
 we assume here that,
 we have already a rough idea about
 the shape of the velocity distribution of halo WIMPs,
 and thus
 the prior knowledge about
 the expectation values of the (difference between the)
 Solar and Earth's Galactic velocities
 $v_0$ and $\Delta v$.
 Hence,
 we consider here only the Gaussian probability distribution
 for both of the fitting parameters
 with expectation values of \mbox{$v_0 = 230$ km/s}
 and \mbox{$\Delta v = 15$ km/s}
 and a common 1$\sigma$ uncertainty of \mbox{20 km/s}
 (see Table \ref{tab:setup_sh}).
 Two cases with both of
 the true (input) and the reconstructed WIMP masses
 have been considered.

 By comparing Figs.~\ref{fig:f1v-Ge-100-0500-sh-sh_Dv-Gau}
 to Figs.~\ref{fig:f1v-Ge-100-0500-sh-sh-Gau}
 and Figs.~\ref{fig:f1v-Ge-SiGe-100-0500-sh-sh_Dv-Gau}
 to Figs.~\ref{fig:f1v-Ge-SiGe-100-0500-sh-sh-Gau},
 it can be seen obviously that
 the reconstructed velocity distribution function
 could indeed match the true (input) one much precisely,
 with however
 (slightly) wider $1\~(2)\~\sigma$ statistical uncertainty bands.
 The systematic deviations of
 both fitting parameters
 from the true (input) values
 would also be much smaller than
 those given in the $v_0 - \ve$ case.
 This implies importantly that,
 the use of our variated shifted velocity distribution function
 given in Eq.~(\ref{eqn:f1v_sh_Dv})
 could indeed offer an estimate of
 the Earth's Galactic velocity $\ve = v_0 + \Delta v$
 with a much higher precision.
 Such a trick would be helpful to improve
 the estimation(s) of the Earth's Galactic velocity $\ve$
 (and/or other fitting parameter(s)).

 Note however that,
 as shown in Tables \ref{tab:results_sh_sh}
 and \ref{tab:results_sh_sh_Dv},
 by using the variated shifted Maxwellian velocity distribution,
 the $1\~(2)\~\sigma$ statistical uncertainties
 on the reconstructed $v_0$
 would be \mbox{$\sim$ 10\%} larger.
 Meanwhile,
 from Eq.~(\ref{eqn:Delta_v})
 the statistical uncertainty on $\ve$
 can be estimated by%
\footnote{
 Since,
 according to Eq.~(\ref{eqn:Delta_v}),
 for a fixed value of $\ve$
 two fitting parameters $v_0$ and $\Delta v$
 should be ``anti--correlated''.
}
\beqn
      \sigma\abrac{\ve}
 \=   \sqrt{  \sigma^2\abrac{v_0}
            + \sigma^2\abrac{\Delta v}
            + 2 \~ {\rm cov}\abrac{v_0, \Delta v} \sigma\abrac{v_0} \sigma\abrac{\Delta v}}
      \non\\
 \eqnle  \sqrt{  \sigma^2\abrac{v_0}
            + \sigma^2\abrac{\Delta v}}
\~.
\label{eqn:sigma_ve}
\eeqn
 Then,
 according to the results
 given in Table \ref{tab:results_sh_sh_Dv},
 the $1\~(2)\~\sigma$ statistical uncertainties on $\ve$
 would be maximal \mbox{$\lsim~15\%$} enlarged,
 whereas
 the systematic deviation of $\ve$
 is much smaller.

 In Table \ref{tab:results_sh_sh_Dv},
 we list the reconstructed results of $v_0$ and $\Delta v$
 with the variated shifted Maxwellian velocity distribution
 $f_{1, \sh, \Delta v}(v)$
 as well as
 the $1\~(2)\~\sigma$ uncertainty ranges
 of the median values.

\subsection{Modified Maxwellian velocity distribution}

 In this subsection,
 we consider another recently often used theoretical
 one--dimensional WIMP velocity distribution function.
 In Refs.~\cite{Lisanti10, Pato},
 a modification of the simple Maxwellian velocity distribution
 in Eq.~(\ref{eqn:f1v_Gau})
 has been suggested:
\beq
   f_{1, \Gau, k}(v)
 = \left\{\renewcommand{\arraystretch}{1.5}
          \begin{array}{l c c}
          \frac{v^2}{N_{f, k}}
          \abrac{  e^{-v^2     / k v_0^2}
                 - e^{-\vmax^2 / k v_0^2}  }^k,
%\~,
          & ~~~~~~~~~~~~ & % 12
          ({\rm for}~v \le \vmax), \\
          0,
%\~,
          & ~~~~~~~~~~~~ & % 12
          ({\rm for}~v >   \vmax). \\
          \end{array}
   \right.
\label{eqn:f1v_Gau_k}
\eeq
 Here
 $\vmax$ is the maximal cut-–off
 velocity of $f_{1, \Gau, k}(v)$
 and $N_{f, k}$ is the normalization constant
 depending on the value of the power index $k$.

\plotfvGauk
\begin{table}[b!]
\InsertSetupTable
{modified Maxwellian velocity distribution $f_{1, \Gau, k}(v)$}
{Simple
 & $v_0$      [km/s] & $\sim 220$ & [160,    300] &   230 & 20   \\
\hline
 1--para. shifted
 & $v_0$      [km/s] & $\sim 175$ & [100,    300] &   200 & 40   \\
\hline
 \multirow{2}{*}{Shifted}
 & $v_0$      [km/s] & $\sim 175$ & [100,    300] &   200 & 40   \\
 & $\ve$      [km/s] & $\sim 185$ & [50,     300] &   200 & 40   \\
\hline
 \multirow{2}{*}{Variated shifted}
 & $v_0$      [km/s] & $\sim 175$ & [100,    300] &   200 & 40   \\
 & $\Delta v$ [km/s] & $\sim  10$ & [$-120$,  80] & $-20$ & 40   \\
\hline
 \multirow{2}{*}{Modified}
 & $v_0$      [km/s] &        220 & [160,    300] &   230 & 20   \\
 & $k$               &          2 & [0.5,    3.5] &     1 &  0.5 \\
}
{The input setup for
 the modified Maxwellian velocity distribution $f_{1, \Gau, k}(v)$
 used for generating WIMP signals
 as well as
 the theoretically estimated values,
 the scanning ranges,
 the expectation values of
 and the 1$\sigma$ uncertainties on
 the fitting parameters
 used for different fitting velocity distribution functions.
}
{tab:setup_Gau_k}

 In Fig.~\ref{fig:f1v_Gau_k},
 we give
 the {\em normalized}
 modified Maxwellian velocity distribution function $f_{1, \Gau, k}(v)$
 with a common value of the Solar Galactic velocity
 \mbox{$v_0 = 220$ km/s}
 and different power indices $k$:
 $k = 1$ (dashed light--green),
 $k = 2$ (solid red),
 $k = 3$ (dash--dotted blue) and
 $k = 4$ (dotted magenta).
 As a comparison,
 the simple Maxwellian velocity distribution $f_{1, \Gau}(v)$
 with \mbox{$v_0 = 220$ km/s}
 is also given
 as the short--dashed black curve.

 In Table \ref{tab:setup_Gau_k},
 we list the input setup for
 the modified Maxwellian velocity distribution $f_{1, \Gau, k}(v)$
 used for generating WIMP signals
 as well as
 the theoretically estimated values,
 the scanning ranges,
 the expectation values of
 and the 1$\sigma$ uncertainties on
 the fitting parameters
 used for different fitting velocity distribution functions.

\subsubsection{Simple Maxwellian velocity distribution}
\plotGeSiGeGaukGauflat
\plotGeSiGeGaukGauGau

 As in Sec.~3.2,
 we start to fit to the reconstructed--input data points
 given by Eqs.~(\ref{eqn:f1v_Qsn}) and (\ref{eqn:cov_f1v_Qs_mu})
 with the simple Maxwellian velocity distribution function $f_{1, \Gau}(v)$.

 In Figs.~%
 \ref{fig:f1v-Ge-SiGe-100-0500-Gau_k-Gau-flat},
 we use first the flat probability distribution
 for the fitting parameter $v_0$
 with either the precisely known (input) (upper)
 or the reconstructed (lower) WIMP mass,
 respectively.
 As shown in Figs.~\ref{fig:f1v-Ge-100-0500-Gau-Gau-flat}
 and \ref{fig:f1v-Ge-SiGe-100-0500-Gau-Gau-flat},
 although
 prior knowledge about the Solar Galactic velocity is {\em absent},
 one could in principle pin down the fitting parameter $v_0$ very precisely
 with systematic deviations of {\em only a few} km/s and
 $1\~(2)\~\sigma$ statistical uncertainties of only
 \mbox{$^{+11.2}_{-12.6}\~(^{+36.4}_{-22.4})$ km/s} and
 \mbox{$^{+21.0}_{-18.2}\~(^{+53.2}_{-32.2})$ km/s}.

\begin{table}[b!]
\InsertResultsTable
{modified Maxwellian velocity distribution $f_{1, \Gau, k = 2}(v)$}
{simple   Maxwellian velocity distribution $f_{1, \Gau}(v)$}
{
 \multirow{4}{*}{$v_0$ [km/s]} &
 \multirow{2}{*}{Input} &
 Flat &
 217.4 & $217.4\~^{+11.2}_{-12.6}\~(^{+36.4}_{-22.4})$ & [204.8, 228.6] & [195.0, 253.8] \\
 \cline{3-7}
 & & Gaussian &
 218.8 & $220.2 \pm  9.8\~(^{+26.6}_{-18.2})$ & [210.4, 230.0] & [202.0, 246.8] \\
 \cline{2-7}
 & \multirow{2}{*}{Reconst.} &
 Flat &
 218.8 & $217.4\~^{+21.0}_{-18.2}\~(^{+53.2}_{-32.2})$ & [199.2, 238.4] & [185.2, 270.6] \\
 \cline{3-7}
 & & Gaussian &
 220.2 & $220.2\~^{+15.4}_{-14.0}\~(^{+36.4}_{-26.6})$ & [206.2, 235.6] & [193.6, 256.6] \\
}
{The reconstructed results of $v_0$
 for all four considered cases
 with the simple Maxwellian velocity distribution $f_{1, \Gau}(v)$
 as well as
 the $1\~(2)\~\sigma$ uncertainty ranges
 of the median values.
}
{tab:results_Gau_k_Gau}

 In Figs.~%
 \ref{fig:f1v-Ge-SiGe-100-0500-Gau_k-Gau-Gau},
 we use the Gaussian probability distribution
 for $v_0$
 with an expectation value of \mbox{$v_0 = 230$ km/s}
 and a 1$\sigma$ uncertainty of \mbox{20 km/s}.
 As summarized in Table \ref{tab:results_Gau_k_Gau},
 the statistical uncertainties on the reconstructed $v_0$
 could now in principle be reduced by
 \mbox{$\sim 10$\%} to \mbox{$\sim 30$\%}.
 In addition,
 even with the reconstructed WIMP mass
 and thus a larger statistical fluctuation,
 the 1$\sigma$ statistical uncertainties shown
 in Figs.~%
 \ref{fig:f1v-Ge-SiGe-100-0500-Gau_k-Gau-Gau}
 are (much) smaller than the given uncertainty
 on the expectation value of $v_0$ (\mbox{20 km/s}).
 Moreover,
 although an expectation value of \mbox{$v_0 = 230$ km/s},
 which differs (slightly) from the true (input) one,
 is used,
 the value of the fitting parameter $v_0$
 could still be pinned down precisely
 with a negligible systematic deviation.

 Furthermore,
 remind here that,
 either,
 as shown in Figs.~\ref{fig:f1v-Ge-SiGe-100-0500-Gau_k-Gau-flat},
 two reconstructed
 (dashed green and dash-dotted blue)
 velocity distribution functions
 match the true (input) (solid red) one
 very well,
 but the reconstructed Solar Galactic velocity $v_0$
 would shift slightly away from the true (input) value,
 or,
 as shown in Figs.~\ref{fig:f1v-Ge-SiGe-100-0500-Gau_k-Gau-Gau},
 $v_0$ could be determined very precisely,
 but the reconstructed velocity distribution functions
 would differ slightly from the true (input) one.
 This is because of the fact that
 we have {\em artificially} used an input value of $k = 2$
 and,
 as shown in Fig.~\ref{fig:f1v_Gau_k},
 the modified Maxwellian velocity distribution function
 with $k = 2$
 is slightly different from the simple Maxwellian distribution.

 In Table \ref{tab:results_Gau_k_Gau},
 we list the reconstructed results of $v_0$
 for all four considered cases
 with the simple Maxwellian velocity distribution $f_{1, \Gau}(v)$
 as well as
 the $1\~(2)\~\sigma$ uncertainty ranges
 of the median values of $v_0$.

\subsubsection{One--parameter shifted Maxwellian velocity distribution}
\plotGeSiGeGaukshvflat
\plotGeSiGeGaukshvGau
\begin{table}[b!]
\InsertResultsTable
{modified Maxwellian velocity distribution $f_{1, \Gau, k = 2}(v)$}
{one--parameter shifted Maxwellian velocity distribution $f_{1, \sh, v_0}(v)$}
{
 \multirow{4}{*}{$v_0$ [km/s]} &
 \multirow{2}{*}{Input} &
 Flat &
 162.0 & $160.0\~^{+10.0}_{- 8.0}\~(^{+28.0}_{-14.0})$ & [152.0, 170.0] & [146.0, 188.0] \\
 \cline{3-7}
 & & Gaussian &
 162.0 & $162.0\~^{+10.0}_{- 8.0}\~(^{+26.0}_{-14.0})$ & [154.0, 172.0] & [148.0, 188.0] \\
 \cline{2-7}
 & \multirow{2}{*}{Reconst.} &
 Flat &
 162.0 & $162.0 \pm 14.0\~(^{+36.0}_{-24.0})$ & [148.0, 176.0] & [138.0, 198.0] \\
 \cline{3-7}
 & & Gaussian &
 164.0 & $162.0\~^{+16.0}_{-12.0}\~(^{+38.0}_{-24.0})$ & [150.0, 178.0] & [138.0, 200.0] \\
}
{The reconstructed results of $v_0$
 for all four considered cases
 with the one--parameter shifted Maxwellian velocity distribution
 $f_{1, \sh, v_0}(v)$
 as well as
 the $1\~(2)\~\sigma$ uncertainty ranges
 of the median values.
}
{tab:results_Gau_k_sh_v0}

 Now,
 as a comparison of Sec.~3.3.1,
 we consider as the next trail the reconstruction with
 the one--parameter shifted Maxwellian velocity distribution function
 to fit the modified simple Maxwellian velocity distribution.

 As usual,
 in Figs.~%
 \ref{fig:f1v-Ge-SiGe-100-0500-Gau_k-sh_v0-flat}
 we use first the flat probability distribution
 for the fitting parameter $v_0$
 with either the precisely known (input) (upper)
 or the reconstructed (lower) WIMP mass,
 respectively.
 It can be seen that,
 firstly,
 although the $1\~(2)\~\sigma$ statistical uncertainty bands
 could still cover the true (input) velocity distribution
 (somehow well),
 the systematic deviations of the peaks of
 the reconstructed velocity distributions
 from that of the true (input) one
 would be \mbox{$\sim 15$ km/s}.
 The comparisons between reconstructed results shown in
 Figs.~\ref{fig:f1v-Ge-SiGe-100-0500-Gau_k-sh_v0-flat}(a) and (c)
 to Figs.~\ref{fig:f1v-Ge-SiGe-100-0500-Gau_k-Gau-flat}(a) and (c)
 could indicate a high probability of the improper assumption of
 the fitting
 (one--parameter) shifted Maxwellian velocity distribution.

 Moreover,
 Figs.~\ref{fig:f1v-Ge-SiGe-100-0500-Gau_k-sh_v0-flat}(b) and (d)
 (see also Table \ref{tab:results_Gau_k_sh_v0})
 show clearly 4$\sigma$ to even 6$\sigma$ deviations of
 the reconstructed $v_0$
 from the true (input) value of \mbox{$v_0 = 220$ km/s}.
 As discussed in Sec.~3.2.1,
 this observation implies also
 an improper use of
 the (one--parameter) shifted Maxwellian velocity distribution
 as our fitting function.

 In Figs.~%
 \ref{fig:f1v-Ge-SiGe-100-0500-Gau_k-sh_v0-Gau},
 we use the Gaussian probability distribution
 for $v_0$
 with a {\em slightly smaller} expectation value of \mbox{$v_0 = {\it 200}$ km/s}
 and a 1$\sigma$ uncertainty of \mbox{{\em 40} km/s}.
 In contrast to our observations presented before,
 for this case
 the use of the Gaussian probability distribution
 would {\em not} reduce the $1\~(2)\~\sigma$ statistical uncertainties
 for both cases with the true (input) and
 the reconstructed WIMP masses
 (see Table \ref{tab:results_Gau_k_sh_v0}).
 This would be caused by
 the large difference between
 the given expectation value
 and the (theoretically estimated) most suitable one of the parameter $v_0$
 (\mbox{175 km/s},
  see Table \ref{tab:setup_Gau_k}).
 This would indicate that,
 in practical use of our Bayesian reconstruction method,
 some trial--and--error tests
 for determining a more suitable expectation value of $v_0$
 (and the other fitting parameters)
 would be necessary and
 could then improve the fitting results.

 In Table \ref{tab:results_Gau_k_sh_v0},
 we list the reconstructed results of $v_0$
 for all four considered cases
 with the one--parameter shifted Maxwellian velocity distribution
 $f_{1, \sh, v_0}(v)$
 as well as
 the $1\~(2)\~\sigma$ uncertainty ranges
 of the median values of $v_0$.

\subsubsection{Shifted Maxwellian velocity distribution}
\plotGeGaukshGau
\plotGeSiGeGaukshGau

 As in Sec.~3.2.3,
 we release now the fixed relation between $v_0$ and $\ve$
 given in Eq.~(\ref{eqn:ve_ave})
 and consider the reconstruction of these two parameters
 simultaneously and independently.
 In addition,
 we also assume here that,
 from the (naive) trials with the simple and one--parameter shifted
 Maxwellian velocity distributions
 done previously,
 one could already obtain a rough idea about
 the shape of the velocity distribution of halo WIMPs
 as well as
 expectation values of the Solar and Earth's Galactic velocities
 $v_0$ and $\ve$.
 Hence,
 we consider here only the Gaussian probability distribution
 for both fitting parameters
 with a common expectation value of \mbox{$v_0 = \ve = {\it 200}$ km/s}
 and a common 1$\sigma$ uncertainty of \mbox{{\em 40} km/s}
 (see Table \ref{tab:setup_Gau_k}).
 Note also that,
 after some trial--and--error tests,
 we set the scanning ranges of $v_0$ and $\ve$
 as \mbox{$v_0 \in [100, 300]$ km/s} and
 \mbox{$\ve \in [50, 300]$ km/s}.

 In Figs.~\ref{fig:f1v-Ge-100-0500-Gau_k-sh-Gau},
 we consider first the case with the true (input) WIMP mass.
 As observed in Sec.~3.3.2,
 although the (low--velocity parts of the)
 $1\~(2)\~\sigma$ statistical uncertainty bands
 could cover the true (input) velocity distribution,
 the systematic deviations of the peaks of
 the reconstructed velocity distributions
 from that of the true (input) one
 would be \mbox{$\sim 10$ km/s}
 (a little bit better then results shown
  in Figs.~\ref{fig:f1v-Ge-SiGe-100-0500-Gau_k-sh_v0-flat}(a) and (c)).
 Moreover,
 Figs.~\ref{fig:f1v-Ge-100-0500-Gau_k-sh-Gau}(c) and (d)
 show \mbox{$\gsim 4\sigma$} deviations of
 the reconstructed values of $v_0$ and $\ve$
 from the true (input, estimated) values of
 \mbox{$v_0 = 220$ km/s} and
 \mbox{$\ve = 231$ km/s}.
 Additionally,
 the best--fit value of $\ve$ is now even
 (\mbox{$\sim 10$ km/s}) {\em smaller} than
 the best--fit value of $v_0$.
 Hence,
 as discussed in Sec.~3.3.2,
 this observation
 (combined with the results given in Sec.~3.3.2)
 would indicate a high probability of
 the improper assumption of the fitting
 shifted Maxwellian velocity distribution.
 Moreover,
 in Figs.~\ref{fig:f1v-Ge-SiGe-100-0500-Gau_k-sh-Gau}
 we show our simulations with the reconstructed WIMP mass.
 The $1\~(2)\~\sigma$ statistical uncertainties on $v_0$ and $\ve$
 would be \mbox{$\sim$ 10\%} to \mbox{$\sim$ 50\%} enlarged.

\begin{table}[b!]
\InsertResultsTable
{modified Maxwellian velocity distribution $f_{1, \Gau, k = 2}(v)$}
{shifted  Maxwellian velocity distribution $f_{1, \sh}(v)$}
{
 \multirow{2}{*}{$v_0$ [km/s]} &
 Input &
 Gaussian &
 172.0 & $170.0 \pm 12.0\~(^{+38.0}_{-24.0})$ & [158.0, 182.0] & [146.0, 208.0] \\
 \cline{2-7}
 & Reconst. &
 Gaussian &
 172.0 & $170.0\~^{+18.0}_{-14.0}\~(^{+44.0}_{-30.0})$ & [156.0, 188.0] & [140.0, 214.0] \\
\hline
 \multirow{2}{*}{$\ve$ [km/s]} &
 Input &
 Gaussian &
 162.5 & $162.5\~^{+17.5}_{-15.0}\~(^{+42.5}_{-27.5})$ & [147.5, 180.0] & [135.0, 205.0] \\
 \cline{2-7}
 & Reconst. &
 Gaussian &
 162.5 & $162.5\~^{+20.0}_{-15.0}\~(^{+48.1}_{-30.0})$ & [147.5, 182.5] & [132.5, 210.6] \\
}
{The reconstructed results of $v_0$ and $\ve$
 with the shifted Maxwellian velocity distribution $f_{1, \sh}(v)$
 as well as
 the $1\~(2)\~\sigma$ uncertainty ranges
 of the median values.
}
{tab:results_Gau_k_sh}

 Figs.~\ref{fig:f1v-Ge-100-0500-Gau_k-sh-Gau} and
 \ref{fig:f1v-Ge-SiGe-100-0500-Gau_k-sh-Gau}
 as well as
 Table \ref{tab:results_Gau_k_sh}
 show that
 one could fit an improper model
 to experimental data (somehow) well
 with combinations of special, probably unusual values of
 the fitting parameters.
 On one hand,
 the reconstructed results could offer us (rough) information
 about the (shape of the) velocity distribution of Galactic WIMPs.
 On the other hand,
 however,
 the observation that
 the reconstructed values of $v_0$ and $\ve$
 are $2\sigma$ to even $> 4\sigma$ different
 from our (theoretically) expected values
 would indicate evidently the improper assumption about
 the fitting velocity distribution function.

 In Table \ref{tab:results_Gau_k_sh},
 we list the reconstructed results of $v_0$ and $\ve$
 with the shifted Maxwellian velocity distribution $f_{1, \sh}(v)$
 as well as
 the $1\~(2)\~\sigma$ uncertainty ranges
 of the median values.

\subsubsection{Variated shifted Maxwellian velocity distribution}
\plotGeGaukshDvGau
\plotGeSiGeGaukshDvGau

 As in Sec.~3.2.4,
 in order to correct
 the systematic deviations of the results given
 with the shifted Maxwellian velocity distribution function
 in Eq.~(\ref{eqn:f1v_sh}),
 we consider here the use of
 its variation
 given in Eq.~(\ref{eqn:f1v_sh_Dv}).
 As previously,
 we consider here only the Gaussian probability distribution
 for both fitting parameters $v_0$ and $\Delta v$,
 with a common 1$\sigma$ uncertainty of \mbox{{\em 40} km/s}.
 Moreover,
 according to the fitting results given in Sec.~3.3.3
 and some trial--and--error tests,
 we set \mbox{$v_0 = {\it 200}$ km/s} and \mbox{$\Delta v = {\it -20}$ km/s}
 as the expectation values
 as well as
 \mbox{$v_0 \in [100, 300]$ km/s} and
 \mbox{$\Delta v \in [-120, 80]$ km/s}
 as the scanning ranges.

 Comparing Figs.~\ref{fig:f1v-Ge-100-0500-Gau_k-sh_Dv-Gau}
 to Figs.~\ref{fig:f1v-Ge-100-0500-Gau_k-sh-Gau},
 it can be seen clearly that
 the use of the variated shifted Maxwellian velocity distribution
 could indeed offer a more reasonable and preciser reconstruction:
 not only that
 the ($1\~(2)\~\sigma$ statistical uncertainty bands of the)
 reconstructed velocity distribution functions
 can match the true (input) one more closer,
 the $1\~(2)\~\sigma$ statistical uncertainties on $v_0$
 is also only \mbox{$\lsim$ 70\%} of the uncertainties
 shown in Fig.~\ref{fig:f1v-Ge-100-0500-Gau_k-sh-Gau}(c).
 Additionally,
 by using Eq.~(\ref{eqn:sigma_ve}),
 the upper bound of the $1\~(2)\~\sigma$ statistical uncertainties
 on $\ve$
 can be (approximately) given as:
 \mbox{$^{+16.12}_{-12.80}\~(^{+41.23}_{-24.41})$ km/s},
 which is also maximal (approximately) equal to
 or even smaller than the uncertainties
 given in Table \ref{tab:results_Gau_k_sh}.
 Hence,
 we would like to conclude that
 our introduction of the variated
 shifted Maxwellian velocity distribution
 (and/or probably some other variations)
 could indeed be a useful trick
 for the practical use of our Bayesian reconstruction procedure.

\begin{table}[b!]
\InsertResultsTable
{modified Maxwellian velocity distribution $f_{1, \Gau, k = 2}(v)$}
{variated shifted Maxwellian velocity distribution $f_{1, \sh, \Delta v}(v)$}
{
 \multirow{2}{*}{$v_0$ [km/s]} &
 Input &
 Gaussian &
 180.0 & $178.0 \pm  8.0\~(^{+26.0}_{-14.0})$ & [170.0, 186.0] & [164.0, 204.0] \\
 \cline{2-7}
 & Reconst. &
 Gaussian &
 180.0 & $178.0\~^{+14.0}_{-12.0}\~(^{+34.5}_{-22.0})$ & [166.0, 192.0] & [156.0, 212.5] \\
\hline
 \multirow{2}{*}{$\Delta v$ [km/s]} &
 Input &
 Gaussian &
 $-34.0$ & $-32.0\~^{+14.0}_{-10.0}\~(^{+32.0}_{-20.0})$ & [$-42.0$, $-18.0$] & [$-52.0$,   0.0] \\
 \cline{2-7}
 & Reconst. &
 Gaussian &
 $-32.0$ & $-32.0\~^{+14.0}_{-10.0}\~(^{+30.5}_{-20.0})$ & [$-42.0$, $-18.0$] & [$-52.0$,  $- 1.5$] \\
}
{The reconstructed results of $v_0$ and $\Delta v$
 with the variated shifted Maxwellian velocity distribution
 $f_{1, \sh, \Delta v}(v)$
 as well as
 the $1\~(2)\~\sigma$ uncertainty ranges
 of the median values.
}
{tab:results_Gau_k_sh_Dv}

 On the other hand,
 in Figs.~\ref{fig:f1v-Ge-SiGe-100-0500-Gau_k-sh_Dv-Gau}
 we use the reconstructed WIMP mass.
 Comparing Figs.~\ref{fig:f1v-Ge-SiGe-100-0500-Gau_k-sh_Dv-Gau}
 to Figs.~\ref{fig:f1v-Ge-100-0500-Gau_k-sh_Dv-Gau}
 (see also Table \ref{tab:results_Gau_k_sh_Dv}),
 although
 the width of the $1\~(2)\~\sigma$ statistical uncertainty bands
 of the reconstructed velocity distribution functions
 as well as
 the $1\~(2)\~\sigma$ statistical uncertainties on $v_0$
 would be clearly larger,
 the $1\~(2)\~\sigma$ statistical uncertainties on $\Delta v$
 are approximately equal to results
 with the true (input) WIMP mass.
 Meanwhile,
 even for this case with the reconstructed WIMP mass,
 the $1\~(2)\~\sigma$ statistical uncertainties on $v_0$ and $\Delta v$
 are (much) smaller than the given 1$\sigma$ uncertainty
 of their Gaussian probability distributions
 (\mbox{40 km/s}).

 Remind that,
 our simulations of the use of
 the variated shifted Maxwellian velocity distribution function
 for fitting data generated by (modified) simple Maxwellian one
 shown here
 indicates again that,
 by using an ``improper'' assumption about
 the fitting velocity distribution function
 with prior knowledge about
 the fitting parameters ($v_0$, $\ve$ or $\Delta v$),
 one would still reconstruct
 an approximate shape of the WIMP velocity distribution;
 the deviations of the peaks of
 the reconstructed velocity distributions
 from that of the true (input) one
 could even be \mbox{$\lsim {\it 10}$ km/s},
 much smaller than the commonly used 1$\sigma$ uncertainty
 on the Solar Galactic velocity of \mbox{$\sim 20$ km/s}.

 However,
 our simulations show also that,
 with an ``improper'' assumption about
 the fitting velocity distribution,
 one would obtain an ``unexpected'' result
 about each single fitting parameter.
 E.g.~here we get 3$\sigma$ to 5$\sigma$ deviations
 of the reconstructed Solar Galactic velocity
 from the theoretical estimate
 (see
  Table
  \ref{tab:results_Gau_k_sh_Dv})
 and large ``negative'' values for
 the difference between the Solar and Earth's Galactic velocities.
 This observation indicates clearly that
 our initial assumption about
 the fitting velocity distribution function
 would be incorrect 
 or at least need to be modified.

 In Table \ref{tab:results_Gau_k_sh_Dv},
 we list the reconstructed results of $v_0$ and $\Delta v$
 with the variated shifted Maxwellian velocity distribution
 $f_{1, \sh, \Delta v}(v)$
 as well as
 the $1\~(2)\~\sigma$ uncertainty ranges
 of the median values.

\subsubsection{Modified Maxwellian velocity distribution}
\plotGeGaukGaukGauflat
\plotGeSiGeGaukGaukGauflat

 As the last tested fitting velocity distribution function
 with data generated by the modified Maxwellian velocity distribution
 given by Eq.~(\ref{eqn:f1v_Gau_k}),
 we consider now the reconstruction of
 the modified Maxwellian velocity distribution itself
 with two fitting parameters:
 the Solar Galactic velocity $v_0$ and
 the power index $k$.

 In Figs.~\ref{fig:f1v-Ge-100-0500-Gau_k-Gau_k-Gau-flat}
 and \ref{fig:f1v-Ge-SiGe-100-0500-Gau_k-Gau_k-Gau-flat},
 we use the Gaussian probability distribution
 for the fitting parameter $v_0$
 with an expectation value of \mbox{$v_0 = 230$ km/s}
 and a 1$\sigma$ uncertainty of \mbox{20 km/s}
 but
 the flat distribution
 for the parameter $k$,
 with either the precisely known (input)
 or the reconstructed WIMP mass,
 respectively.
 Remind that,
 in Figs.~\ref{fig:f1v-Ge-100-0500-Gau_k-Gau_k-Gau-flat}(d)
 and \ref{fig:f1v-Ge-SiGe-100-0500-Gau_k-Gau_k-Gau-flat}(d)
 the bins at $k = 0.5$ and $k = 3.5$ are ``overflow'' bins.
 This means that
 they contain also the experiments
 whose best--fit values of $k$ would be
 either $k < 0.5$ or $k > 3.5$.

 First,
 as shown in Sec.~3.3.1,
 the Solar Galactic velocity $v_0$
 could be pinned down very precisely:
 a small systematic deviation of \mbox{$<$ 10 km/s}
 and 1$\sigma$ statistical uncertainty of
 \mbox{$\lsim\~10$ km/s}
 could be achieved.
 However,
 Figs.~\ref{fig:f1v-Ge-100-0500-Gau_k-Gau_k-Gau-flat}(b) and (d)
 as well as
 Figs.~\ref{fig:f1v-Ge-SiGe-100-0500-Gau_k-Gau_k-Gau-flat}(b) and (d)
 show that,
 due to the small difference between
 the modified Maxwellian velocity distribution
 with different power indices $k$
 (see Fig.~\ref{fig:f1v_Gau_k})
 and the large statistical fluctuation and
 1$\sigma$ reconstruction uncertainty
 with only 500 WIMP events (on average),
 our Bayesian reconstruction of the velocity distribution
 would be {\em totally non--sensitive}
 on the second fitting parameter (power index) $k$.

 Nevertheless,
 the wide spread of the reconstructed power index $k$
 (in particular,
  the high cumulative numbers in the (overflow) bin $k = 3.5$)
 and,
 in contrast,
 the narrow widths of
 the $1\~(2)\~\sigma$ statistical uncertainty bands of
 the reconstructed velocity distribution functions
 imply that,
 for reconstructing the rough information about
 the one--dimensional WIMP velocity distribution,
 a precise value of the power index $k$
 would {\em not be crucial},
 and
 the {\em simple} Maxwellian velocity distribution
 $f_{1, \Gau}(v)$ given in Eq.~(\ref{eqn:f1v_Gau})
 would already be a good approximation%
\footnote{
 As described in Ref.~\cite{Lisanti10},
 the modification of the simple Maxwellian velocity distribution function
 $f_{1, \Gau, k}(v)$ given in Eq.~(\ref{eqn:f1v_Gau_k})
 has significant difference from
 the simple Maxwellian one given in Eq.~(\ref{eqn:f1v_Gau})
 in the high--velocity tail.
 Moreover,
 in Refs.~\cite{YYMao, Kuhlen13},
 another empirical modification of
 the simple Maxwellian velocity distribution
 with also a significant difference
 in the high--velocity tail
 has been introduced.
 Unfortunately,
 our simulations presented here
 show that
 it would be {\em impossible}
 to distinguish these subtle modifications
 by our Bayesian reconstruction method
 (with only a few hundreds of recorded WIMP events).

 On the other hand,
 simulations with other (well--motivated) halo models
 and different fitting velocity distribution functions
 can be tested on the {\tt AMIDAS} website
 \cite{AMIDAS-web, % AMIDAS-web-TiResearch,
       AMIDAS-II}.
}.

 In Table \ref{tab:results_Gau_k_Gau_k},
 we give
 the reconstructed results of $v_0$ and $k$
 with the modified Maxwellian velocity distribution
 $f_{1, \Gau, k}(v)$
 as well as
 the $1\~(2)\~\sigma$ uncertainty ranges
 of the median values.
 Note that,
 since the 1$\sigma$ lower and upper bounds of the median values of $k$
 are already beyond our scanning range,
 the 2$\sigma$ bounds are meanless to give here.

\begin{table}[b!]
\InsertResultsTable
{modified Maxwellian velocity distribution $f_{1, \Gau, k = 2}(v)$}
{modified Maxwellian velocity distribution $f_{1, \Gau, k}(v)$}
{
 \multirow{2}{*}{$v_0$ [km/s]} &
 Input &
 Gaussian &
 228.6 & $227.2\~^{+ 7.0}_{- 8.4}\~(^{+19.6}_{-18.2})$ & [218.8, 234.2] & [209.0, 246.8] \\
 \cline{2-7}
 & Reconst. &
 Gaussian &
 227.2 & $227.2\~^{+ 9.8}_{-12.6}\~(^{+29.4}_{-28.0})$ & [214.6, 237.0] & [199.2, 256.6] \\
\hline
 \multirow{2}{*}{$k$} &
 Input &
 Flat &
 3.47 & $3.38                  $ & [0.5, 3.5] & $\times$ \\
 \cline{2-7}
 & Reconst. &
 Flat &
 2.72 & $3.29                  $ & [0.5, 3.5] & $\times$ \\
}
{The reconstructed results of $v_0$ and $k$
 with the modified Maxwellian velocity distribution
 $f_{1, \Gau, k}(v)$
 as well as
 the $1\~(2)\~\sigma$ uncertainty ranges
 of the median values.
 Note that,
 firstly,
 we use here the Gaussian probability distribution for $v_0$
 but the flat one for $k$.
 Secondly,
 since the 1$\sigma$ lower and upper bounds of the median values of $k$
 are already beyond our scanning range,
 the 2$\sigma$ bounds are meanless to give here.
}
{tab:results_Gau_k_Gau_k}
\subsection{For different WIMP masses}
\plotGeSiGeshGauflatL
\plotGeSiGeshshGauL
\plotGeSiGeshshDvGauL

 In the previous Secs.~3.1 to 3.3,
 we fixed the input WIMP mass as \mbox{$\mchi = 100$ GeV}.
 As a further test of our Bayesian reconstruction method
 for the one--dimensional WIMP velocity distribution,
 in this subsection,
 we consider the cases for
 either a light WIMP mass of \mbox{$\mchi = 25$ GeV}
 or a heavy WIMP mass of \mbox{$\mchi = 250$ GeV}.

 Here we consider
 only the {\em shifted} Maxwellian velocity distribution
 given in Eq.~(\ref{eqn:f1v_sh})
 for generating WIMP events;
 three fitting functions:
 simple,
 shifted
 and variated shifted
 Maxwellian velocity distributions
 will be tested.
 All input setup and fitting parameters
 are the same as in Sec.~3.2
 (see Table \ref{tab:setup_sh}).
 Additionally,
 only the {\em reconstructed} WIMP mass is used.

\subsubsection{For a light WIMP mass}

 We consider first
 a rather light WIMP mass of \mbox{$\mchi = 25$ GeV}.
 Note that,
 firstly,
 since for our tested targets $\rmXA{Si}{28}$ and $\rmXA{Ge}{76}$,
 the kinematic maximal cut--off energies
 given in Eq.~(\ref{eqn:Qmax_kin})
 are only \mbox{$Q_{\rm max, kin, Si} = 68.12$ keV}
 and \mbox{$Q_{\rm max, kin, Ge} = 52.65$ keV},
 respectively,
 the maximal experimental cut--off energy for both targets
 in our simulations demonstrated here
 has been reduced to only \mbox{$\Qmax = 50$ keV}.
 Secondly,
 since the lighter the WIMP mass,
 the shaper the expected recoil energy spectrum,
 the width of the first energy bin in Eq.~(\ref{eqn:bn_delta})
 has been set as \mbox{$b_1 = 5$ keV}
 and the total number of bins
 has been reduced to only {\em four} bins
 between $\Qmin$ and $\Qmax$ ($B = 4$);
 up to {\em two} bins have been combined to a window
 and thus {\em four} windows ($W = 4$)
 will be reconstructed%
\footnote{
 The last window is neglected automatically
 in the \amidas\ code,
 due to a very few expected event number
 in the last bin (window).
}$^{,~}$%
\footnote{
 It has been found that,
 by reducing the total number of the energy bins
 (and in turn that of the windows)
 and thus collecting more events in one bin (window),
 the Bayesian reconstructed velocity distribution
 could be improved (significantly).
}.
\paragraph{Simple Maxwellian velocity distribution \\}
\begin{table}[t!]
\InsertResultsTable
{shifted Maxwellian velocity distribution $f_{1, \sh}(v)$}
{simple  Maxwellian velocity distribution $f_{1, \Gau}(v)$}
{
 \multirow{4}{*}{$v_0$ [km/s]} &
 \multirow{2}{*}{Input} &
 Flat &
 296.8 & $294.4\~^{+14.4}_{-12.0}\~(^{+31.2}_{-21.6})$ & [282.4, 308.8] & [272.8, 325.6] \\
 \cline{3-7}
 & & Gaussian &
 294.4 & $294.4\~^{+ 9.6}_{-12.0}\~(^{+24.0}_{-21.6})$ & [282.4, 304.0] & [272.8, 318.4] \\
 \cline{2-7}
 & \multirow{2}{*}{Reconst.} &
 Flat &
 292.0 & $289.6\~^{+28.8}_{-24.0}\~(^{+62.4}_{-43.2})$ & [265.6, 318.4] & [246.4, 352.0] \\
 \cline{3-7}
 & & Gaussian &
 292.0 & $289.6 \pm 21.6\~(^{+48.0}_{-40.8})$ & [268.0, 311.2] & [248.8, 337.6] \\
\hline
\hline
 \multicolumn{7}{|| c ||}
 {\bf\boldmath
  Reconstruction:
  one--parameter shifted Maxwellian velocity distribution $f_{1, \sh, v_0}(v)$} \\
\hline
 \multirow{4}{*}{$v_0$ [km/s]} &
 \multirow{2}{*}{Input} &
 Flat &
 220.2 & $218.8\~^{+ 9.8}_{- 7.0}\~(^{+19.6}_{-14.0})$ & [211.8, 228.6] & [204.8, 238.4] \\
 \cline{3-7}
 & & Gaussian &
 221.6 & $221.6 \pm  7.0\~(^{+15.4}_{-14.0})$ & [214.6, 228.6] & [207.6, 237.0] \\
 \cline{2-7}
 & \multirow{2}{*}{Reconst.} &
 Flat &
 217.4 & $214.6\~^{+21.0}_{-16.8}\~(^{+44.8}_{-30.8})$ & [197.8, 235.6] & [183.8, 259.4] \\
 \cline{3-7}
 & & Gaussian &
 218.8 & $218.8 \pm 15.4\~(^{+32.2}_{-29.4})$ & [203.4, 234.2] & [189.4, 251.0] \\
\hline
\hline
 \multicolumn{7}{|| c ||}
 {\bf\boldmath
  Reconstruction:
  shifted Maxwellian velocity distribution $f_{1, \sh}(v)$} \\
\hline
 \multirow{2}{*}{$v_0$ [km/s]} &
 Input &
 Gaussian &
 221.6 & $221.6 \pm  7.0\~(^{+15.4}_{-12.9})$ & [214.6, 228.6] & [208.7, 237.0] \\
 \cline{2-7}
 & Reconst. &
 Gaussian &
 218.8 & $220.2\~^{+12.6}_{-14.0}\~(^{+28.0}_{-26.6})$ & [206.2, 232.8] & [193.6, 248.2] \\
\hline
 \multirow{2}{*}{$\ve$ [km/s]} &
 Input &
 Gaussian &
 237.0 & $237.0\~^{+ 7.0}_{- 8.4}\~(^{+14.0}_{-16.8})$ & [228.6, 244.0] & [220.2, 251.0] \\
 \cline{2-7}
 & Reconst. &
 Gaussian &
 234.2 & $234.2 \pm 12.6(\pm 25.2)$ & [221.6, 246.8] & [209.0, 259.4] \\
\hline
\hline
 \multicolumn{7}{|| c ||}
 {\bf\boldmath
  Reconstruction:
  variated shifted Maxwellian velocity distribution $f_{1, \sh, \Delta v}(v)$} \\
\hline
 \multirow{2}{*}{$v_0$ [km/s]} &
 Input &
 Gaussian &
 221.6 & $221.6 \pm  5.6\~(^{+12.6}_{-11.2})$ & [216.0, 227.2] & [210.4, 234.2] \\
 \cline{2-7}
 & Reconst. &
 Gaussian &
 218.8 & $218.8\~^{+14.0}_{-12.6}\~(^{+29.4}_{-23.8})$ & [206.2, 232.8] & [195.0, 248.2] \\
\hline
 \multirow{2}{*}{$\Delta v$ [km/s]} &
 Input &
 Gaussian &
   9.8 & $ 11.1\~^{+ 3.9}_{- 5.2}\~(^{+ 9.1}_{-11.7})$ & [5.9,  15.0] & [$- 0.6$,  20.2] \\
 \cline{2-7}
 & Reconst. &
 Gaussian &
   9.8 & $  9.8 \pm  6.5\~(^{+14.3}_{-13.0})$ & [3.3,  16.3] & [$- 3.2$,  24.1] \\
}
{The reconstructed results
 with four fitting velocity distribution functions
 for the input WIMP mass of \mbox{$\mchi = 25$ GeV}.
}
{tab:results_sh_L}

 In Figs.~\ref{fig:f1v-Ge-SiGe-025-0500-sh-Gau-flat}
 (cf.~Figs.~\ref{fig:f1v-Ge-SiGe-100-0500-sh-Gau-flat}(c) and (d)),
 we use first the (improper)
 simple Maxwellian velocity distribution function
 with one parameter $v_0$
 to fit the reconstructed--input data.
 As the first trial of the reconstruction of
 the one--dimensional WIMP velocity distribution
 {\em without} prior knowledge about the Solar Galactic velocity,
 the flat probability distribution
 has been used here
 (results with the Gaussian probability distribution
  are given in Table \ref{tab:results_sh_L}).

 Although we use the improper assumption about
 the fitting velocity distribution
 and have only {\em four} available data points
 (solid black vertical bars),
 the $1\~(2)\~\sigma$ statistical uncertainty bands
 could still cover the true (input) velocity distribution
 with a systematic deviation of the peak of
 the reconstructed velocity distribution
 of \mbox{$\lsim$ 15 km/s}
 from that of the true (input) one.
 However,
 the best--fit values of the Solar Galactic velocity
 are now \mbox{$\simeq 294$ km/s}
 and \mbox{$\sim 3\sigma$} apart from the theoretically expected value.

 Moreover,
 our further simulations
 with the Gaussian probability distribution
 for the fitting parameter $v_0$
 with an expectation value of \mbox{$v_0 = {\it 280}$ km/s}
 and a 1$\sigma$ uncertainty of \mbox{{\em 40} km/s}
 show that,
 the $1\~(2)\~\sigma$ statistical uncertainties
 for such a light WIMP mass
 could be reduced to \mbox{$\sim 60$\%} to \mbox{$\sim 80$\%}
 (compare Table \ref{tab:results_sh_L} to Table \ref{tab:results_sh_Gau}).

\paragraph{Shifted Maxwellian velocity distribution \\}

 In Figs.~\ref{fig:f1v-Ge-SiGe-025-0500-sh-sh-Gau}
 (cf.~Figs.~\ref{fig:f1v-Ge-SiGe-100-0500-sh-sh-Gau}),
 the shifted Maxwellian velocity distribution
 with two fitting parameters $v_0$ and $\ve$
 has been tested to fit to the reconstructed--input data.
 Only the Gaussian probability distribution
 for both fitting parameters
 with expectation values of \mbox{$v_0 = 230$ km/s}
 and \mbox{$\ve = 245$ km/s}
 and a common 1$\sigma$ uncertainty of \mbox{20 km/s}
 is considered here.

 Astonishingly,
 with {\em only four} available data points
 the reconstructed velocity distribution functions
 could match the true (input) one very precisely:
 the systematic deviation of $v_0$ is negligible
 and that of $\ve$ is {\em only a few} km/s,
 the $1\~(2)\~\sigma$ statistical uncertainties
 on two fitting parameters are also only
 \mbox{$^{+12.6}_{-14.0}\~(^{+28.0}_{-26.6})$} and
 \mbox{$\pm 12.6(\pm 25.2)$},
 respectively.

\paragraph{Variated shifted Maxwellian velocity distribution \\}

 In Figs.~\ref{fig:f1v-Ge-SiGe-025-0500-sh-sh_Dv-Gau}
 (cf.~Figs.~\ref{fig:f1v-Ge-SiGe-100-0500-sh-sh_Dv-Gau}),
 the variated shifted Maxwellian velocity distribution
 with two parameters $v_0$ and $\Delta v$
 has been tested to fit to the reconstructed--input data.
 Only the Gaussian probability distribution
 for both fitting parameters
 with expectation values of \mbox{$v_0 = 230$ km/s}
 and \mbox{$\Delta v = 15$ km/s}
 and a common 1$\sigma$ uncertainty of \mbox{20 km/s}
 is considered here.

 It can be seen that,
 although
 the $1\~(2)\~\sigma$ statistical uncertainty bands
 are a bit wider than
 those given with the shifted Maxwellian distribution,
 with only four available data points
 the reconstructed velocity distribution function
 could also match the true (input) one very well.
 Moreover,
 as shown in Secs.~3.2.4 and 3.3.4,
 the second fitting parameter $\Delta v$
 could also be pinned down very precisely
 with a negligible systematic deviation
 (see Table \ref{tab:results_sh_L}).

 ~

 In Table \ref{tab:results_sh_L},
 we give the reconstructed results
 with all four fitting velocity distribution functions
 for the input WIMP mass of \mbox{$\mchi = 25$ GeV}.
 Both cases with
 the true (input) and the reconstructed WIMP masses
 have been simulated and summarized.

\subsubsection{For a heavy WIMP mass}
\plotGeSiGeshGauflatH
\plotGeSiGeshshGauH
\plotGeSiGeshshDvGauH

 We consider now
 a rather heavy WIMP mass of \mbox{$\mchi = 250$ GeV}.
 The maximal experimental cut--off energy for both targets
 in our simulations demonstrated here
 has been set again as \mbox{$\Qmax = 100$ keV}.
 And,
 as usual,
 the width of the first energy bin in Eq.~(\ref{eqn:bn_delta})
 has been set as \mbox{$b_1 = 10$ keV},
 {\em five} bins between $\Qmin$ and $\Qmax$ are used ($B = 5$)
 and up to {\em three} bins have been combined to a window ($W = 6$).

\paragraph{Simple Maxwellian velocity distribution \\}

 For a heavy WIMP mass,
 due to the (large) statistical fluctuation
 discussed in Ref.~\cite{DMDDmchi},
 our reconstructed velocity distribution functions
 given in Fig.~\ref{fig:f1v-Ge-SiGe-250-0500-sh-Gau-flat}(a)
 (cf.~Figs.~\ref{fig:f1v-Ge-SiGe-100-0500-sh-Gau-flat}(c)
  and \ref{fig:f1v-Ge-SiGe-025-0500-sh-Gau-flat}(a))
 have clearly
 a (much) wider $1\~(2)\~\sigma$ statistical uncertainty bands;
 the $1\~(2)\~\sigma$ statistical uncertainties
 on the reconstructed Solar Galactic velocity
 are also (much) larger as
 \mbox{$^{+33.6}_{-31.2}\~(^{+64.8}_{-60.0})$ km/s}.
 And,
 as shown in Fig.~\ref{fig:f1v-Ge-SiGe-250-0500-sh-Gau-flat}(b),
 a considerable fraction of the reconstructed $v_0$
 would excess our scanning upper bound of \mbox{400 km/s}.
 Remind that
 the bin at \mbox{$v_0 = 400$ km/s}
 is an ``overflow'' bin,
 which contains also the experiments
 with the best--fit $v_0$ value of \mbox{$> 400$ km/s}.

 Nevertheless,
 comparing to the much larger 1$\sigma$ statistical uncertainty
 on the reconstructed--input data (solid black vertical bars),
 our Bayesian reconstruction with
 an in fact improper fitting velocity distribution
 could still offer an approximation
 with only \mbox{$\lsim\~15$ km/s}
 systematic deviation of the peak of
 the reconstructed velocity distribution
 from that of the true (input) one.

\paragraph{Shifted Maxwellian velocity distribution \\}

 In Figs.~\ref{fig:f1v-Ge-SiGe-250-0500-sh-sh-Gau}
 (cf.~Figs.~\ref{fig:f1v-Ge-SiGe-100-0500-sh-sh-Gau}
  and \ref{fig:f1v-Ge-SiGe-025-0500-sh-sh-Gau}),
 the shifted Maxwellian velocity distribution
 with two fitting parameters $v_0$ and $\ve$
 has been tested to fit to the reconstructed--input data.
 Only the Gaussian probability distribution
 for both fitting parameters
 with expectation values of \mbox{$v_0 = 230$ km/s}
 and \mbox{$\ve = 245$ km/s}
 and a common 1$\sigma$ uncertainty of \mbox{20 km/s}
 is considered here.

 With a proper fitting velocity distribution
 and {\em one more} fitting parameter,
 the reconstructed velocity distribution
 could now match the true (input) one very precisely
 with much narrower $1\~(2)\~\sigma$ statistical uncertainty bands.
 Additionally,
 the $1\~(2)\~\sigma$ statistical uncertainties
 on the fitting parameters $v_0$ and $\ve$
 can be significantly reduced to only
 \mbox{$\pm 9.8\~(^{+22.4}_{-18.2})$} and
 \mbox{$^{+14.0}_{-11.2}\~(^{+26.6}_{-21.0})$},
 respectively.
 Note that,
 as shown in Table \ref{tab:results_sh_H},
 the use of the {\em one--parameter}
 shifted Maxwellian velocity distribution
 with {\em only one} fitting parameter $v_0$
 and the fixed relation between $v_0$ and $\ve$
 would give
 a (much) wider $1\~(2)\~\sigma$ statistical uncertainty bands
 of the reconstructed velocity distribution
 as well as
 a (much) larger $1\~(2)\~\sigma$ statistical uncertainties
 on the reconstructed Solar Galactic velocity:
 \mbox{$\sim 50$\%} to a factor of \mbox{$\sim 2$} larger.

\paragraph{Variated shifted Maxwellian velocity distribution \\}
\begin{table}[t!]
\InsertResultsTable
{shifted Maxwellian velocity distribution $f_{1, \sh}(v)$}
{simple  Maxwellian velocity distribution $f_{1, \Gau}(v)$}
{
 \multirow{4}{*}{$v_0$ [km/s]} &
 \multirow{2}{*}{Input} &
 Flat &
 287.2 & $287.2\~^{+28.8}_{-21.6}\~(^{+69.6}_{-40.8})$ & [265.6, 316.0] & [246.4, 356.8] \\
 \cline{3-7}
 & & Gaussian &
 284.8 & $284.8\~^{+19.2}_{-16.8}\~(^{+36.0}_{-31.2})$ & [268.0, 304.0] & [253.6, 320.8] \\
 \cline{2-7}
 & \multirow{2}{*}{Reconst.} &
 Flat &
 289.6 & $289.6\~^{+55.2}_{-43.2}\~(^{+110.4}_{-74.4})$ & [246.4, 344.8] & [215.2, 400.0] \\
 \cline{3-7}
 & & Gaussian &
 289.6 & $287.2\~^{+33.6}_{-31.2}\~(^{+64.8}_{-60.0})$ & [256.0, 320.8] & [227.2, 352.0] \\
\hline
\hline
 \multicolumn{7}{|| c ||}
 {\bf\boldmath
  Reconstruction:
  one--parameter shifted Maxwellian velocity distribution $f_{1, \sh, v_0}(v)$} \\
\hline
 \multirow{4}{*}{$v_0$ [km/s]} &
 \multirow{2}{*}{Input} &
 Flat &
 209.0 & $209.0\~^{+18.2}_{-15.4}\~(^{+43.4}_{-26.6})$ & [193.6, 227.2] & [182.4, 252.4] \\
 \cline{3-7}
 & & Gaussian &
 218.8 & $218.8 \pm  9.8\~(\pm 19.6)$ & [209.0, 228.6] & [199.2, 238.4] \\
 \cline{2-7}
 & \multirow{2}{*}{Reconst.} &
 Flat &
 211.8 & $210.4\~^{+37.8}_{-29.4}\~(^{+89.6}_{-50.4})$ & [181.0, 248.2] & [160.0, 300.0] \\
 \cline{3-7}
 & & Gaussian &
 218.8 & $220.2 \pm 18.2\~(^{+36.4}_{-35.0})$ & [202.0, 238.4] & [185.2, 256.6] \\
\hline
\hline
 \multicolumn{7}{|| c ||}
 {\bf\boldmath
  Reconstruction:
  shifted Maxwellian velocity distribution $f_{1, \sh}(v)$} \\
\hline
 \multirow{2}{*}{$v_0$ [km/s]} &
 Input &
 Gaussian &
 223.0 & $223.0\~^{+ 5.6}_{- 7.0}\~(^{+12.6}_{-14.0})$ & [216.0, 228.6] & [209.0, 235.6] \\
 \cline{2-7}
 & Reconst. &
 Gaussian &
 223.0 & $224.4 \pm  9.8\~(^{+22.4}_{-18.2})$ & [214.6, 234.2] & [206.2, 246.8] \\
\hline
 \multirow{2}{*}{$\ve$ [km/s]} &
 Input &
 Gaussian &
 237.0 & $237.0 \pm  7.0\~(^{+14.0}_{-15.4})$ & [230.0, 244.0] & [221.6, 251.0] \\
 \cline{2-7}
 & Reconst. &
 Gaussian &
 237.0 & $237.0\~^{+14.0}_{-11.2}\~(^{+26.6}_{-21.0})$ & [225.8, 251.0] & [216.0, 263.6] \\
\hline
\hline
 \multicolumn{7}{|| c ||}
 {\bf\boldmath
  Reconstruction:
  variated shifted Maxwellian velocity distribution $f_{1, \sh, \Delta v}(v)$} \\
\hline
 \multirow{2}{*}{$v_0$ [km/s]} &
 Input &
 Gaussian &
 218.8 & $218.8\~^{+ 9.8}_{- 8.4}\~(\pm 18.2)$ & [210.4, 228.6] & [200.6, 237.0] \\
 \cline{2-7}
 & Reconst. &
 Gaussian &
 220.2 & $220.2\~^{+15.4}_{-16.8}\~(^{+32.2}_{-30.8})$ & [203.4, 235.6] & [189.4, 252.4] \\
\hline
 \multirow{2}{*}{$\Delta v$ [km/s]} &
 Input &
 Gaussian &
   9.8 & $  9.8 \pm  5.2\~(^{+ 9.1}_{-10.4})$ & [4.6,  15.0] & [$- 0.6$,  18.9] \\
 \cline{2-7}
 & Reconst. &
 Gaussian &
   9.8 & $  9.8\~^{+ 9.1}_{- 7.8}\~(^{+18.2}_{-15.6})$ & [2.0,  18.9] & [$- 5.8$,  28.0] \\
}
{The reconstructed results
 with four fitting velocity distribution functions
 for the input WIMP mass of \mbox{$\mchi = 250$ GeV}.
}
{tab:results_sh_H}

 Finally,
 in Figs.~\ref{fig:f1v-Ge-SiGe-250-0500-sh-sh_Dv-Gau}
 (cf.~Figs.~\ref{fig:f1v-Ge-SiGe-100-0500-sh-sh_Dv-Gau}
  and \ref{fig:f1v-Ge-SiGe-025-0500-sh-sh_Dv-Gau}),
 we test the possibility of
 improving the reconstruction precision
 by the use of
 the variated shifted Maxwellian velocity distribution
 with two fitting parameters $v_0$ and $\Delta v$.
 Only the Gaussian probability distribution
 for both fitting parameters
 with expectation values of \mbox{$v_0 = 230$ km/s}
 and \mbox{$\Delta v = 15$ km/s}
 and a common 1$\sigma$ uncertainty of \mbox{20 km/s}
 is considered here.

 While
 the reconstructed velocity distribution function
 could still match the true (input) one very precisely
 with however
 wider $1\~(2)\~\sigma$ statistical uncertainty bands,
 the ``best--fit'' values of
 both parameters $v_0$ and $\Delta v$
 (and in turn $\ve$)
 could again be very precisely determined
 with negligible systematic deviations
 (see Table \ref{tab:results_sh_H}).

 ~

 Note that,
 since the heavier the WIMP mass,
 the smaller the transformation constant $\alpha$
 defined in Eq.~(\ref{eqn:alpha}),
 for an experimental maximal cut--off energy
 \mbox{$\Qmax \approx 100$ GeV},
 the reconstructible velocity range
 of our model--independent data analysis method
 would be much smaller than our maximal cut--off velocity $\vmax$
 (e.g.~\mbox{$\sim 285$ km/s}
  for \mbox{$\mchi = 250$ GeV}
  and the $\rmXA{Ge}{76}$ target).
 Therefore,
 our simulations shown in this subsection
 demonstrate meaningfully that,
 our Bayesian reconstruction of
 the one--dimensional WIMP velocity distribution
 would be an important improvement
 for offering more and preciser information
 about the Galactic halo,
 e.g.~the position of the peak of the WIMP velocity distribution,
 for the WIMP mass between \mbox{$\cal O$(20) GeV}
 and even \mbox{$\cal O$(500) GeV}.

 In Table \ref{tab:results_sh_H},
 we give the reconstructed results
 with all four fitting velocity distribution functions
 for the input WIMP mass of \mbox{$\mchi = 250$ GeV}.
 Both cases with the true (input) and the reconstructed WIMP masses
 have been simulated and summarized.

\subsection{Background effects}
\plotdRdQbg

 In this last part of
 our presentation of the numerical simulations
 of the Bayesian reconstruction of
 the WIMP velocity distribution function,
 we consider the effects of {\em unrejected} background events.
 Similar to our earlier works
 in Refs.~\cite{DMDDbg-mchi, DMDDbg-f1v},
 we take into account a small fraction of
 {\em artificially} generated background events
 in the fake experimental data sets
 and want to study
 how well the WIMP velocity distribution
 as well as
 the fitting parameters
 could be reconstructed.

 In all simulations demonstrated in this subsection,
 a combination of
 the {\em target--dependent exponential} form
 of the residue background spectrum
 introduced in Ref.~\cite{DMDDbg-mchi}
 with a small {\em constant} component
 has been used:
\beq
    \aDd{R}{Q}_{\rm bg}
 =  \aDd{R}{Q}_{\rm bg, ex}
  + r_{\rm const} \aDd{R}{Q}_{\rm bg, const}
\~,
\label{eqn:dRdQ_bg}
\eeq
 where
\beq
   \aDd{R}{Q}_{\rm bg, ex}
 = \exp\abrac{-\frac{Q /{\rm keV}}{A^{0.6}}}
\~,
\label{eqn:dRdQ_bg_ex}
\eeq
 and
\beq
   \aDd{R}{Q}_{\rm bg, const}
 = 1
\~.
\label{eqn:dRdQ_bg_const}
\eeq
 Here $Q$ is the recoil energy,
 $A$ is the atomic mass number of the target nucleus.
 The power index of $A$, 0.6, is an empirical constant,
 which has been chosen so that
 the exponential background spectrum is
 somehow similar to,
 but still different from
 the expected recoil spectrum of the target nuclei;
 otherwise,
 there is in practice no difference between
 the WIMP scattering and background spectra%
\footnote{
 Note that,
 among different possible choices,
 we use in our simulations the atomic mass number $A$
 as the simplest, unique characteristic parameter
 in the general analytic form (\ref{eqn:dRdQ_bg_ex})
 for defining the artificial residue background spectra
 for different target nuclei.
 However,
 it does not mean that
 the (superposition of the real) background spectra
 would depend simply/primarily on $A$ or
 on the mass of the target nucleus, $\mN$.
 In other words,
 it is practically equivalent to
 use the expression (\ref{eqn:dRdQ_bg_ex})
 or $(dR / dQ)_{\rm bg, ex} = e^{-Q / 13.5~{\rm keV}}$ directly
 for a $\rmXA{Ge}{76}$ target
 (cf.~\cite{Green-mchi08}).
}.
 Additionally,
 $r_{\rm const}$ is the ratio
 between the exponential and constant components
 in the total {\em background} spectrum,
 which has been fixed as $r_{\rm const} = 0.05$
 for all simulations.

 Note that,
 firstly,
 as argued in Ref.~\cite{DMDDbg-mchi},
 the exponential form of background spectrum
 is rather naive;
 but,
 since we consider here
 only {\em a few tens residue} background events
 induced by {\em several different} sources,
 pass all discrimination criteria,
 and then mix with other WIMP--induced events
 in our data sets of a few hundreds {\em total} events,
 an exact form of background spectrum
 for each target nucleus
 would not be crucial and
 the exponential + constant form of background spectrum
 in Eq.~(\ref{eqn:dRdQ_bg})
 should practically not be unrealistic.
 Secondly,
 as demonstrated in Refs.~\cite{DMDDf1v, DMDDmchi}
 and in the previous subsections,
 our Bayesian reconstruction of
 the one--dimensional WIMP velocity distribution
 requires only measured recoil energies
 and occasionally prior knowledge about
 the Solar and Earth's Galactic velocities.
 Hence,
 for applying this method to future real experimental data,
 prior knowledge about (different) background source(s)
 is {\em not required at all}.

 In Figs.~\ref{fig:dRdQ-bg-ex-const-000-100-20-100},
 we show
 the measured recoil energy spectrum (solid red histogram)
 for a $\rmXA{Ge}{76}$ (a) and a $\rmXA{Si}{28}$ (b) targets
 with an input WIMP mass of \mbox{$\mchi = 100$ GeV}.
 The dotted blue curve is
 the elastic WIMP--nucleus scattering spectrum
 for {\em generating} signal events,
 whereas
 the dashed green curve shows
 the {\em artificial} background spectrum:
 the exponential background spectrum
 given in Eq.~(\ref{eqn:dRdQ_bg_ex})
 accompanied with an extra constant component,
 normalized to fit to the background ratio of 20\%.

\plotGeSiGeshshvGaubg
\plotGeSiGeshshGaubg
\subsubsection{For a moderate WIMP mass}

 We consider first a moderate input WIMP mass of
 \mbox{$\mchi = 100$ GeV}.
 The {\em shifted} Maxwellian velocity distribution
 given in Eq.~(\ref{eqn:f1v_sh})
 is used for generating WIMP signals.
 All input setup and fitting parameters
 are the same as in Sec.~3.2
 (see Table \ref{tab:setup_sh})
 and
 a fraction of 20\% background events
 has been taken into account.
 Additionally,
 as in Sec.~3.4,
 we consider only the use of
 the Gaussian probability distribution
 for the fitting parameters:
 $v_0$ and $\ve$ or $\Delta v$
 as well as
 the use of the reconstructed WIMP mass.

\begin{table}[t!]
\InsertResultsTable
{shifted Maxwellian velocity distribution $f_{1, \sh}(v)$}
{simple  Maxwellian velocity distribution $f_{1, \Gau}(v)$}
{
 \multirow{4}{*}{$v_0$ [km/s]} &
 \multirow{2}{*}{Input} &
 Flat &
 284.8 & $284.8\~^{+24.0}_{-19.2}\~(^{+52.8}_{-40.8})$ & [265.6, 308.8] & [244.0, 337.6] \\
 \cline{3-7}
 & & Gaussian &
 284.8 & $284.8 \pm 16.8\~(^{+36.0}_{-33.6})$ & [268.0, 301.6] & [251.2, 320.8] \\
 \cline{2-7}
 & \multirow{2}{*}{Reconst.} &
 Flat &
 253.6 & $253.6\~^{+36.0}_{-31.2}\~(^{+79.2}_{-60.0})$ & [222.4, 289.6] & [193.6, 332.8] \\
 \cline{3-7}
 & & Gaussian &
 258.4 & $258.4\~^{+28.8}_{-26.4}\~(^{+57.6}_{-55.2})$ & [232.0, 287.2] & [203.2, 316.0] \\
\hline
\hline
 \multicolumn{7}{|| c ||}
 {\bf\boldmath
  Reconstruction:
  one--parameter shifted Maxwellian velocity distribution $f_{1, \sh, v_0}(v)$} \\
\hline
 \multirow{4}{*}{$v_0$ [km/s]} &
 \multirow{2}{*}{Input} &
 Flat &
 211.8 & $211.8\~^{+16.8}_{-15.4}\~(^{+36.4}_{-30.8})$ & [196.4, 228.6] & [181.0, 248.2] \\
 \cline{3-7}
 & & Gaussian &
 217.4 & $218.8\~^{+ 9.8}_{-11.2}\~(^{+21.0}_{-23.8})$ & [207.6, 228.6] & [195.0, 239.8] \\
 \cline{2-7}
 & \multirow{2}{*}{Reconst.} &
 Flat &
 188.0 & $188.0\~^{+26.6}_{-22.4}\~(^{+56.0}_{-28.0})$ & [165.6, 214.6] & [160.0, 244.0] \\
 \cline{3-7}
 & & Gaussian &
 202.0 & $202.0\~^{+18.2}_{-16.8}\~(^{+36.4}_{-35.0})$ & [185.2, 220.2] & [167.0, 238.4] \\
\hline
\hline
 \multicolumn{7}{|| c ||}
 {\bf\boldmath
  Reconstruction:
  shifted Maxwellian velocity distribution $f_{1, \sh}(v)$} \\
\hline
 \multirow{2}{*}{$v_0$ [km/s]} &
 Input &
 Gaussian &
 225.8 & $225.8\~^{+ 7.0}_{- 8.4}\~(\pm 16.8)$ & [217.4, 232.8] & [189.4, 239.8] \\
 \cline{2-7}
 & Reconst. &
 Gaussian &
 214.6 & $214.6\~^{+12.2}_{-12.6}\~(\pm 25.2)$ & [202.0, 226.8] & [189.4, 239.8] \\
\hline
 \multirow{2}{*}{$\ve$ [km/s]} &
 Input &
 Gaussian &
 231.4 & $231.4 \pm 8.4\~(^{+16.8}_{-18.2})$ & [223.0, 239.8] & [213.2, 248.2] \\
 \cline{2-7}
 & Reconst. &
 Gaussian &
 220.2 & $220.2\~^{+12.6}_{-11.2}\~(^{+25.2}_{-23.8})$ & [209.0, 232.8] & [196.4, 245.4] \\
\hline
\hline
 \multicolumn{7}{|| c ||}
 {\bf\boldmath
  Reconstruction:
  variated shifted Maxwellian velocity distribution $f_{1, \sh, \Delta v}(v)$} \\
\hline
 \multirow{2}{*}{$v_0$ [km/s]} &
 Input &
 Gaussian &
 220.2 & $220.2\~^{+ 8.4}_{- 9.8}\~(\pm 19.6)$ & [210.4, 228.6] & [200.6, 239.8] \\
 \cline{2-7}
 & Reconst. &
 Gaussian &
 204.8 & $204.8\~^{+16.8}_{-15.4}\~(^{+32.2}_{-30.8})$ & [189.4, 221.6] & [174.0, 237.0] \\
\hline
 \multirow{2}{*}{$\Delta v$ [km/s]} &
 Input &
 Gaussian &
   5.9 & $  5.9\~^{+ 5.2}_{- 6.5}\~(^{+10.4}_{-13.0})$ & [$- 0.6$,  11.1] & [$- 7.1$,  16.3] \\
 \cline{2-7}
 & Reconst. &
 Gaussian &
 $- 0.6$ & $- 0.6 \pm  7.8\~(\pm 15.6)$ & [$- 8.4$,   7.2] & [$-16.2$,  15.0] \\
}
{The reconstructed results
 with four fitting velocity distribution functions
 for data sets mixed with 20\% background events
 and the input WIMP mass of \mbox{$\mchi = 100$ GeV}.
}
{tab:results_sh_bg}
\paragraph{One--parameter shifted Maxwellian velocity distribution \\}

 In Fig.~\ref{fig:f1v-Ge-SiGe-100-0500-sh-sh_v0-Gau-bg}(a),
 it can be seen first that,
 due to the extra background events
 in both of the low and high energy ranges
 (see Figs.~\ref{fig:dRdQ-bg-ex-const-000-100-20-100}),
 the reconstructed--input data (solid black vertical bars)
 would be shifted (strongly) to the low--velocity range%
\footnote{
 Note that,
 as shown in Fig.~\ref{fig:f1v-Ge-SiGe-100-0500-sh-sh_v0-Gau-bg}(a)
 and \ref{fig:f1v-Ge-SiGe-100-0500-sh-sh-Gau-bg}(a),
 the reconstructed WIMP mass
 is now {\em overestimated}:
 \mbox{$m_{\chi, {\rm rec}} \approx 136$ GeV}.
}:
 the peak of the solid black crosses is now at \mbox{$\sim 220$ km/s},
 i.e.~\mbox{$\sim 90$ km/s} smaller then
 the position of the true (input) velocity distribution.
 However,
 our simulation shown in
 Fig.~\ref{fig:f1v-Ge-SiGe-100-0500-sh-sh_v0-Gau-bg}(a)
 indicates clearly and importantly that,
 by assuming the shifted Maxwellian WIMP velocity distribution
 and the {\em time--averaged} relation
 between the Solar and Earth's Galactic velocities,
 the reconstructed WIMP velocity distributions
 could alleviate this systematic shift:
 the deviations of the peaks of
 the ($1\~(2)\~\sigma$ statistical uncertainty bands of the)
 reconstructed velocity distributions
 would only be \mbox{$\sim\~30\~^{+30}_{-20}\~(^{+60}_{-40})$ km/s}.

 In fact,
 it has also been found that,
 once an (approximately) precisely determined (true) WIMP mass
 could be used,
 the reconstructed WIMP velocity distribution
 could match the true (input) one very precisely:
 the deviation of the reconstructed $v_0$
 would be \mbox{$\lsim\~10$ km/s} (flat)
 or even negligible (Gaussian)
 (see Table \ref{tab:results_sh_bg}).

 Note that,
 although a fraction of 20\% unrejected background events
 has been mixed (artificially) into the analyzed (pseudo--)data sets,
 the (1$\sigma$ statistical uncertainty on the)
 median value of the reconstructed $v_0$'s
 (\mbox{$202.0\~^{+18.2}_{-16.8}$ km/s})
 would still cover the true (input) Solar Galactic velocity of
 \mbox{$v_0 = 220$ km/s}.
 Moreover,
 once we take into account the statistical fluctuation
 of the reconstructed--input data (the solid black vertical bars),
 the effect of 20\% residue background events
 on reconstructing information about
 the (shape of the) WIMP velocity distribution function
 would {\em not be very significant}.

\paragraph{Shifted Maxwellian velocity distribution \\}

 In Figs.~\ref{fig:f1v-Ge-SiGe-100-0500-sh-sh-Gau-bg},
 we release the fixed relation between $v_0$ and $\ve$
 and determine these two parameters
 simultaneously and independently.

 It can be seen that,
 firstly,
 the $1\~(2)\~\sigma$ statistical uncertainty bands
 are obviously narrower than those shown
 in Fig.~\ref{fig:f1v-Ge-SiGe-100-0500-sh-sh_v0-Gau-bg}(a);
 the deviations of the peaks of
 the reconstructed velocity distributions
 from that of the true (input) one
 would be reduced to only \mbox{$\lsim\~15$ km/s}.
 In addition,
 the systematic deviations and
 the (1$\sigma$ statistical uncertainties on the)
 median values of the reconstructed fitting paramaters $v_0$ and $\ve$
 shown in Figs.~\ref{fig:f1v-Ge-SiGe-100-0500-sh-sh-Gau-bg}(c) and (d)
 are also (much) smaller than that shown in
 Fig.~\ref{fig:f1v-Ge-SiGe-100-0500-sh-sh_v0-Gau-bg}(b)
 (see Table \ref{tab:results_sh_bg}).
 Note here that,
 as given in Table \ref{tab:results_sh_bg},
 once an (approximately) precisely determined (true) WIMP mass
 could be used,
 one could reconstruct
 the WIMP velocity distribution function
 very precisely
 with very small or even negligible systematic deviations
 of both two fitting parameters.
 This means that,
 our Bayesian reconstruction method
 for the WIMP velocity distribution function
 would {\em not be affected (significantly)}
 by a fraction of \mbox{$\sim 20$\%}
 unrejected background events
 mixed in the analyzed data sets
 (for a WIMP mass of ${\cal O}(100)$ GeV). 

 ~

 In Table \ref{tab:results_sh_bg},
 we give the reconstructed results
 with all four fitting velocity distribution functions
 for data sets mixed with 20\% background events
 and the input WIMP mass of \mbox{$\mchi = 100$ GeV}.
 Both cases with the true (input) and the reconstructed WIMP masses
 have been simulated and summarized.

\subsubsection{For a light WIMP mass}

 Now,
 we consider the case
 with a light input WIMP mass of \mbox{$\mchi = 25$ GeV}.
 Simulation setup is the same as in Sec.~3.4.1
 and
 a fraction of 20\% background events
 has been taken into account.

\plotGeSiGeshshvGaubgL
\plotGeSiGeshshGaubgL
\paragraph{One--parameter shifted Maxwellian velocity distribution \\}

 Since the reconstructed WIMP mass
 would be \mbox{$\sim 30$\%} {\em overestimated}
 (\mbox{$m_{\chi, {\rm rec}} \approx 31.5$ GeV})
 due to the extra background events
 and {\em only four} reconstructed--input data points are available,
 Fig.~\ref{fig:f1v-Ge-SiGe-025-0500-sh-sh_v0-Gau-bg}(a)
 shows that
 the ``best--fit'' {\em one--parameter}
 shifted Maxwellian velocity distribution functions
 would match
 not the true (input) velocity distribution (solid red curve),
 but the analyzed data points (solid black crosses)
 well.
 Nevertheless,
 at least,
 the 2$\sigma$ statistical uncertainty band
 of the reconstructed velocity distributions
 could cover the true (input) one;
 the systematic deviations of the peaks of
 the reconstructed velocity distributions
 from that of the true (input) one
 would also only be \mbox{$\sim 30$ km/s}.
 Meanwhile,
 the 2$\sigma$ statistical uncertainty on the
 median value of the reconstructed $v_0$'s
 (\mbox{$200.6^{+29.4}_{-26.6}$ km/s})
 would still cover
 the true (input) Solar Galactic velocity of
 \mbox{$v_0 = 220$ km/s}.
 This could be further improved
 by using an (approximately) precisely determined (true) WIMP mass
 to be
 \mbox{$227.2\~^{+ 9.8}_{- 8.4}$ km/s} (flat) and
 \mbox{$228.6 \pm  7.0$ km/s} (Gaussian)
 (see Table \ref{tab:results_sh_bg_L}).

\paragraph{Shifted Maxwellian velocity distribution \\}

 As the case of the \mbox{100 GeV} WIMP mass
 shown in Figs.~\ref{fig:f1v-Ge-SiGe-100-0500-sh-sh_v0-Gau-bg}(a)
 and \ref{fig:f1v-Ge-SiGe-100-0500-sh-sh-Gau-bg}(a),
 the $1\~(2)\~\sigma$ statistical uncertainty bands
 reconstructed with the shifted Maxwellian velocity distribution
 with two independent fitting parameters $v_0$ and $\ve$
 shown in Fig.~\ref{fig:f1v-Ge-SiGe-025-0500-sh-sh-Gau-bg}(a)
 would clearly be much narrower than those
 reconstructed with only one parameter $v_0$.
 Meanwhile,
 in contrast to other (presented) cases,
 our simulations with the {\em true} ({\em input}) WIMP mass
 show that
 the reconstructed--input data
 as well as
 the ($1\~(2)\~\sigma$ statistical uncertainty bands of the)
 reconstructed velocity distribution function
 would slightly shift to the {\em high}--velocity range.

 ~

 Furthermore,
 comparing results given in Table \ref{tab:results_sh_bg_L}
 to those in Table \ref{tab:results_sh_bg},
 it has been found interesting and probably importantly that,
 for an (input) WIMP mass of \mbox{$\cal O$(20) GeV},
 the use of our {\em variated} shifted Maxwellian velocity distribution
 given in Eq.~(\ref{eqn:f1v_sh_Dv})
 could offer preciser reconstruction results
 with (relatively) smaller statistical uncertainties,
 although
 fewer (four in our simulations) data points are available.

 In Table \ref{tab:results_sh_bg_L},
 we give the reconstructed results
 with all four fitting velocity distribution functions
 for data sets mixed with 20\% background events
 and the input WIMP mass of \mbox{$\mchi = 25$ GeV}.
 Both cases with the true (input) and the reconstructed WIMP masses
 have been simulated and summarized.

\begin{table}[t!]
\InsertResultsTable
{shifted Maxwellian velocity distribution $f_{1, \sh}(v)$}
{simple  Maxwellian velocity distribution $f_{1, \Gau}(v)$}
{
 \multirow{4}{*}{$v_0$ [km/s]} &
 \multirow{2}{*}{Input} &
 Flat &
 308.8 & $306.4\~^{+16.8}_{-12.0}\~(^{+36.0}_{-24.0})$ & [294.4, 323.2] & [282.4, 342.4] \\
 \cline{3-7}
 & & Gaussian &
 304.0 & $304.0 \pm 12.0\~(\pm 24.0)$ & [292.0, 316.0] & [280.0, 328.0] \\
 \cline{2-7}
 & \multirow{2}{*}{Reconst.} &
 Flat &
 260.8 & $260.8\~^{+24.0}_{-21.6}\~(^{+51.0}_{-40.8})$ & [239.2, 284.8] & [220.0, 311.8] \\
 \cline{3-7}
 & & Gaussian &
 265.6 & $263.2\~^{+21.6}_{-19.2}\~(^{+43.2}_{-38.4})$ & [244.0, 284.8] & [224.8, 306.4] \\
\hline
\hline
 \multicolumn{7}{|| c ||}
 {\bf\boldmath
  Reconstruction:
  one--parameter shifted Maxwellian velocity distribution $f_{1, \sh, v_0}(v)$} \\
\hline
 \multirow{4}{*}{$v_0$ [km/s]} &
 \multirow{2}{*}{Input} &
 Flat &
 228.6 & $227.2\~^{+ 9.8}_{- 8.4}\~(^{+22.4}_{-15.4})$ & [218.8, 237.0] & [211.8, 249.6] \\
 \cline{3-7}
 & & Gaussian &
 228.6 & $228.6 \pm  7.0\~(^{+15.4}_{-14.0})$ & [221.6, 235.6] & [214.6, 244.0] \\
 \cline{2-7}
 & \multirow{2}{*}{Reconst.} &
 Flat &
 193.6 & $193.6\~^{+16.8}_{-15.4}\~(^{+35.4}_{-29.4})$ & [178.2, 210.4] & [164.2, 229.0] \\
 \cline{3-7}
 & & Gaussian &
 200.6 & $200.6\~^{+15.4}_{-14.0}\~(^{+29.4}_{-26.6})$ & [186.6, 216.0] & [174.0, 230.0] \\
\hline
\hline
 \multicolumn{7}{|| c ||}
 {\bf\boldmath
  Reconstruction:
  shifted Maxwellian velocity distribution $f_{1, \sh}(v)$} \\
\hline
 \multirow{2}{*}{$v_0$ [km/s]} &
 Input &
 Gaussian &
 230.0 & $230.0 \pm  7.0\~(\pm 14.0)$ & [223.0, 237.0] & [216.0, 244.0] \\
 \cline{2-7}
 & Reconst. &
 Gaussian &
 207.6 & $207.6 \pm 12.6\~(^{+23.8}_{-25.2})$ & [195.0, 220.2] & [182.4, 231.4] \\
\hline
 \multirow{2}{*}{$\ve$ [km/s]} &
 Input &
 Gaussian &
 239.8 & $239.8\~^{+ 8.4}_{- 7.0}\~(\pm 15.4)$ & [232.8, 248.2] & [224.4, 255.2] \\
 \cline{2-7}
 & Reconst. &
 Gaussian &
 218.8 & $220.2\~^{+11.2}_{-12.6}\~(^{+23.8}_{-25.2})$ & [207.6, 231.4] & [195.0, 244.0] \\
\hline
\hline
 \multicolumn{7}{|| c ||}
 {\bf\boldmath
  Reconstruction:
  variated shifted Maxwellian velocity distribution $f_{1, \sh, \Delta v}(v)$} \\
\hline
 \multirow{2}{*}{$v_0$ [km/s]} &
 Input &
 Gaussian &
 228.6 & $228.6 \pm  7.0\~(^{+14.0}_{-12.6})$ & [221.6, 235.6] & [216.0, 242.6] \\
 \cline{2-7}
 & Reconst. &
 Gaussian &
 203.4 & $204.8 \pm 12.6\~(\pm 23.8)$ & [192.2, 217.4] & [181.0, 228.6] \\
\hline
 \multirow{2}{*}{$\Delta v$ [km/s]} &
 Input &
 Gaussian &
  12.4 & $ 11.1 \pm  5.2\~(\pm 10.4)$ & [5.9,  16.3] & [0.7,  21.5] \\
 \cline{2-7}
 & Reconst. &
 Gaussian &
   2.0 & $  2.0 \pm  6.5\~(\pm 13.0)$ & [$- 4.5$,   8.5] & [$-11.0$,  15.0] \\
}
{The reconstructed results
 with four fitting velocity distribution functions
 for data sets mixed with 20\% background events
 and the input WIMP mass of \mbox{$\mchi = 25$ GeV}.
}
{tab:results_sh_bg_L}
\subsubsection{For a heavy WIMP mass}

 As the last case,
 we consider here
 a heavy input WIMP mass of \mbox{$\mchi = 250$ GeV}.
 Simulation setup is the same as in Sec.~3.4.2.
 Note however that,
 since the constant component of the background spectrum
 used in our simulations
 would cause a strongly {\em overestimated} WIMP mass,
 in particular,
 once WIMPs are heavy
 (e.g.~the \mbox{250 GeV} input WIMP mass
  would now be reconstructed as \mbox{$\simeq 338$ GeV})
 \cite{DMDDbg-mchi},
 the ratio of the background events
 in the analyzed data sets
 has been set as {\em only 10\%}
 \cite{DMDDbg-f1v}.

\paragraph{One--parameter shifted Maxwellian velocity distribution \\}

 As usual,
 we consider first the one--parameter
 shifted Maxwellian velocity distribution function
 to fit the reconstructed--input data points.

\plotGeSiGeshshvGaubgH
\plotGeSiGeshshGaubgH

 In Figs.~\ref{fig:f1v-Ge-SiGe-250-0500-sh-sh_v0-Gau-bg},
 we can see unexpectedly that,
 although the input WIMP mass is pretty heavy,
 the systematic deviations of the peak positions of
 the reconstructed WIMP velocity distribution functions
 from that of the true (input) one
 would be only \mbox{$\sim 10$ km/s}
 (for 10\% background ratio!).
 This might be due to that,
 as shown in Sec.~3.4.2,
 for our used experimental maximal cut--off energy
 \mbox{$\Qmax = 100$ GeV}
 and the $\rmXA{Ge}{76}$ target,
 the reconstructible velocity range
 would only be
 \mbox{$\sim 270$ km/s}
 (shifted slightly to the {\em low}--velocity range
  due to the {\em overestimated} WIMP mass)
 and thus
 this maximal reconstructible velocity
 is theoretically smaller than
 the position of the peak of the velocity distribution function
 (see e.g.~Fig.~\ref{fig:f1v-Ge-SiGe-250-0500-sh-sh_v0-Gau-bg}(a)).
 This means that
 the approximately {\em monotonically increased} shape of
 the reconstructed--input data points (solid black vertical bars)
 with pretty large 1$\sigma$ statistical uncertainties
 would alleviate the effects of the overestimations of
 the analyzed (reconstructed--input) data points
 and the reconstructed WIMP mass
 caused by the extra background events.

\begin{table}[t!]
\InsertResultsTable
{shifted Maxwellian velocity distribution $f_{1, \sh}(v)$}
{simple  Maxwellian velocity distribution $f_{1, \Gau}(v)$}
{
 \multirow{4}{*}{$v_0$ [km/s]} &
 \multirow{2}{*}{Input} &
 Flat &
 277.6 & $277.6\~^{+26.4}_{-21.6}\~(^{+64.8}_{-38.4})$ & [256.0, 304.0] & [239.2, 342.4] \\
 \cline{3-7}
 & & Gaussian &
 277.6 & $278.8 \pm 18.0\~(^{+37.2}_{-32.4})$ & [260.8, 296.8] & [246.4, 316.0] \\
 \cline{2-7}
 & \multirow{2}{*}{Reconst.} &
 Flat &
 263.2 & $263.2\~^{+48.0}_{-38.4}\~(^{+115.2}_{-62.4})$ & [224.8, 311.2] & [200.8, 378.4] \\
 \cline{3-7}
 & & Gaussian &
 268.0 & $268.0\~^{+33.6}_{-31.2}\~(^{+67.2}_{-55.2})$ & [236.8, 301.6] & [212.8, 335.2] \\
\hline
\hline
 \multicolumn{7}{|| c ||}
 {\bf\boldmath
  Reconstruction:
  one--parameter shifted Maxwellian velocity distribution $f_{1, \sh, v_0}(v)$} \\
\hline
 \multirow{4}{*}{$v_0$ [km/s]} &
 \multirow{2}{*}{Input} &
 Flat &
 203.4 & $203.4\~^{+19.2}_{-14.0}\~(^{+44.8}_{-28.0})$ & [189.4, 222.6] & [175.4, 248.2] \\
 \cline{3-7}
 & & Gaussian &
 216.0 & $216.0\~^{+ 9.8}_{-11.2}\~(^{+19.6}_{-21.0})$ & [204.8, 225.8] & [195.0, 235.6] \\
 \cline{2-7}
 & \multirow{2}{*}{Reconst.} &
 Flat &
 195.0 & $193.6\~^{+33.6}_{-28.0}\~(^{+78.4}_{-33.6})$ & [165.6, 227.2] & [160.0, 272.0] \\
 \cline{3-7}
 & & Gaussian &
 211.8 & $210.4 \pm 18.2\~(^{+36.4}_{-35.0})$ & [192.2, 228.6] & [175.4, 246.8] \\
\hline
\hline
 \multicolumn{7}{|| c ||}
 {\bf\boldmath
  Reconstruction:
  shifted Maxwellian velocity distribution $f_{1, \sh}(v)$} \\
\hline
 \multirow{2}{*}{$v_0$ [km/s]} &
 Input &
 Gaussian &
 221.6 & $223.0\~^{+ 5.6}_{- 8.4}\~(^{+12.6}_{-15.4})$ & [214.6, 228.6] & [207.6, 235.6] \\
 \cline{2-7}
 & Reconst. &
 Gaussian &
 218.8 & $220.2\~^{+ 9.8}_{- 8.4}\~(^{+22.4}_{-18.2})$ & [211.8, 230.0] & [202.0, 242.6] \\
\hline
 \multirow{2}{*}{$\ve$ [km/s]} &
 Input &
 Gaussian &
 232.8 & $234.2\~^{+ 7.0}_{- 8.4}\~(^{+14.0}_{-16.8})$ & [225.8, 241.2] & [217.4, 248.2] \\
 \cline{2-7}
 & Reconst. &
 Gaussian &
 230.0 & $230.0\~^{+12.6}_{- 9.8}\~(^{+25.2}_{-21.0})$ & [220.2, 242.6] & [209.0, 255.2] \\
\hline
\hline
 \multicolumn{7}{|| c ||}
 {\bf\boldmath
  Reconstruction:
  variated shifted Maxwellian velocity distribution $f_{1, \sh, \Delta v}(v)$} \\
\hline
 \multirow{2}{*}{$v_0$ [km/s]} &
 Input &
 Gaussian &
 216.0 & $217.4\~^{+ 8.4}_{- 9.8}\~(^{+18.2}_{-19.6})$ & [207.6, 225.8] & [197.8, 235.6] \\
 \cline{2-7}
 & Reconst. &
 Gaussian &
 210.4 & $211.8\~^{+16.8}_{-15.4}\~(^{+33.6}_{-30.8})$ & [196.4, 228.6] & [181.0, 245.4] \\
\hline
 \multirow{2}{*}{$\Delta v$ [km/s]} &
 Input &
 Gaussian &
   7.2 & $  7.2\~^{+ 5.2}_{- 5.2}\~(^{+10.4}_{-11.7})$ & [2.0,  12.4] & [$- 4.5$,  17.6] \\
 \cline{2-7}
 & Reconst. &
 Gaussian &
   4.6 & $  4.6\~^{+ 7.8}_{- 7.8}\~(^{+16.9}_{-14.3})$ & [$- 3.2$,  12.4] & [$- 9.7$,  21.5] \\
}
{The reconstructed results
 with four fitting velocity distribution functions
 for data sets mixed with {\em 10\%} background events
 and the input WIMP mass of \mbox{$\mchi = 250$ GeV}.
}
{tab:results_sh_bg_H}
\paragraph{Shifted Maxwellian velocity distribution \\}

 Now
 we use the shifted Maxwellian velocity distribution function
 with two independent fitting parameters $v_0$ and $\ve$
 to fit the reconstructed--input data points.
 Astonishingly and unexpectedly
 (probably accidentally),
 Figs.~\ref{fig:f1v-Ge-SiGe-250-0500-sh-sh-Gau-bg}
 show that
 both of the ``best--fit'' results of the parameters $v_0$ and $\ve$
 are {\em almost exact} as the true (input) values
 and
 the 1$\sigma$ statistical uncertainties on $v_0$ and $\ve$
 are only \mbox{$\sim 10$ km/s}.

 Meanwhile,
 in contrast to our simulation results
 with the variated shifted Maxwellian velocity distribution function
 shown previously,
 for the case of the 250 GeV WIMP mass
 with 10\% background ratio,
 the Bayesian reconstructed parameter $\Delta v$
 could have a (much) larger deviations
 from the true (estimated) value
 (see Table \ref{tab:results_sh_bg_H})!

 ~

 In Table \ref{tab:results_sh_bg_H},
 we give the reconstructed results
 with all four fitting velocity distribution functions
 for data sets mixed with {\em 10\%} background events
 and the input WIMP mass of \mbox{$\mchi = 250$ GeV}.
 Both cases with the true (input) and the reconstructed WIMP masses
 have been simulated and summarized.

\section{Summary and conclusions}

 In this paper,
 we extended our earlier work on
 the development of the model--independent data analysis method
 for the reconstruction of the (time--averaged) one--dimensional
 velocity distribution of Galactic WIMPs
 and introduced the Bayesian fitting procedure
 of the theoretical velocity distribution functions.

 In this fitting procedure,
 the (rough) velocity distribution
 reconstructed by using raw experimental data,
 i.e.~measured recoil energies,
 with one or more different target nuclei
 has been used as reconstructed--input data (points).
 By assuming a fitting WIMP velocity distribution function
 and scanning the parameter space
 based on the Bayesian analysis,
 the (fitting) astronomical characteristic parameters,
 e.g.~the Solar and Earth's Galactic velocities $v_0$ and $\ve$,
 would be pinned down as the output results
 and thus
 the functional form of the one--dimensional velocity distribution
 can be reconstructed
 (instead of only a few discrete points).

 As the first test of our Bayesian reconstruction method
 for the one--dimensional WIMP velocity distribution function,
 we used the simplest isothermal spherical Galactic halo model
 for both generating WIMP--signal events
 and as the assumed velocity distribution
 with the unique fitting parameter:
 the Solar Galactic velocity $v_0$.
 Our simulations show that,
 with (only) 500 recorded events (on average)
 and without prior knowledge about the Solar Galactic velocity,
 $v_0$ could in principle be pinned down
 with a negligible deviation and
 a 1$\sigma$ statistical uncertainty of only \mbox{$\sim 12$ km/s}
 (with a precisely known WIMP mass)
 or \mbox{$\sim 20$ km/s}
 (with a reconstructed WIMP mass),
 respectively.
 Moreover,
 once (rough) information about the Solar Galactic velocity
 can be given,
 the statistical uncertainties on the reconstructed $v_0$
 could even be reduced to \mbox{$\sim 70$\%}.
 
 For more realistic consideration,
 we then took into account
 the orbital motion of the Solar system around our Galaxy
 as well as
 that of the Earth around the Sun
 and
 turned to use the shifted Maxwellian velocity distribution function
 for generating WIMP signals.
 As comparisons,
 four different fitting
 functions
 have been considered:
 the simple and
 the (one--parameter and variated) shifted Maxwellian velocity distributions.
 It has been found
 that,
 firstly,
 with an improper assumed fitting function
 (e.g.~the simple Maxwellian velocity distribution here),
 the WIMP velocity distribution
 could still be reconstructed and
 offer some important information about Galactic WIMPs,
 e.g.~the rough position of the peak of
 the one--dimensional velocity distribution function.
 The deviations of the peaks of
 the reconstructed velocity distributions
 from that of the true (input) one
 would be only \mbox{$\sim 10$ km/s}.
 However,
 the best--fit value(s) of the fitting parameter(s)
 would be unexpected/unreasonable.
 For instance,
 the reconstructed Solar Galactic velocity $v_0$
 would be
 2$\sigma$ (with the reconstructed WIMP mass)
 or even
 4$\sigma$ (with the input WIMP mass)
 apart from its theoretical estimate.
 Such an observation
 could in turn be an important criterion
 on the assumption of fitting velocity distribution function.

 Moreover,
 our simulations with
 the (one--parameter and variated)
 shifted Maxwellian velocity distributions
 show that,
 although in all of these three cases
 the reconstructed velocity distributions
 could match the true (input) one pretty precisely,
 with two fitting parameters
 the $1\~(2)\~\sigma$ statistical uncertainty bands
 of the reconstructed velocity distributions
 would be narrower then those
 with only one fitting parameter.
 In addition,
 the use of the variation of the shifted Maxwellian velocity distribution
 could (strongly) reduce the systematic deviations of
 the determinations of the characteristic
 Solar and Earth's Galactic velocities $v_0$ and $\ve$,
 with however a bit larger statistical uncertainties.

 Furthermore,
 we considered also a modification of
 the simple Maxwellian velocity distribution
 with an extra power index
 as the generating WIMP velocity distribution.
 First,
 we used the simple Maxwellian velocity distribution
 without the power index
 as the test fitting function.
 Since the difference between the modification and the original
 simple Maxwellian velocity distributions
 are very tiny,
 the reconstructed velocity distribution function
 could match the true (input) one very precisely
 and the characteristic Solar Galactic velocity
 could also be reconstructed
 with negligible systematic deviation.

 Meanwhile,
 our simulations with the
 (one--parameter and variated)
 shifted Maxwellian velocity distribution functions
 show that,
 although the positions of the peak of
 the reconstructed velocity distribution
 would be only \mbox{$\lsim\~10$ km/s} deviated
 from the true (input) one,
 a clear 2$\sigma$ to even 6$\sigma$ difference
 between the best--fit values of
 the Solar and/or the Earth's Galactic velocities
 and the true (input) ones
 could be observed.
 Such results would in turn
 indicate evidently the improper assumption of
 the shifted Maxwellian velocity distribution function.

 Moreover,
 from the simulations with
 the modified simple Maxwellian velocity distribution
 with the power index
 as the second fitting parameter,
 it has been found that
 our Bayesian reconstruction of the WIMP velocity distribution
 would be (totally) non--sensitive
 on the power index.
 This means that,
 unfortunately,
 with only a few hundreds of recorded WIMP events
 it would still be impossible
 to distinguish (evidently)
 different subtle variations of
 the (simple and shifted) WIMP velocity distribution functions.

 As comparisons,
 we considered also a light and a heavy input WIMP masses.
 For the case of light WIMPs,
 due to the sharp shapes of the recoil energy spectra
 and the small kinetic maximal cut--off energies,
 the recorded WIMP events
 would need to be separated into fewer bins/windows.
 However,
 our simulations show that,
 with only four available reconstructed--input data points,
 the true (input) velocity distribution function
 could astonishingly be reconstructed very precisely.
 On the other hand,
 once WIMPs are heavy,
 the statistical fluctuation on the reconstructed WIMP mass
 becomes pretty large
 and hence
 the Bayesian reconstructions of the velocity distribution
 as well as
 of the Solar and Earth's Galactic velocities
 would have large statistical uncertainties.
 Nevertheless,
 the reconstructed velocity distribution function
 with the best--fit characteristic Solar and Earth's Galactic velocities
 could still match the true (input) one very precisely.

 Finally,
 the effects of residue (unrejected) background events
 mixed in data sets to analyze
 have also been considered.
 Three different WIMP masses
 with background ratios of 10\% or 20\%
 have been tested.
 Although,
 due to the choice of our artificial residue background spectrum,
 the reconstructed WIMP masses
 would be overestimated
 and the (rough shape of) the reconstructed--input data points
 would thus be shifted (significantly) to lower velocities,
 the functional forms of the chosen fitting velocity distributions
 could somehow alleviate these systematic shifts
 and the $1\~(2)\~\sigma$ statistical uncertainty bands
 could still cover the true (input) velocity distribution.
 In particular,
 for heavy WIMPs,
 since the reconstructed--input data points
 should be in the velocity range smaller than
 the position of the peak of the velocity distribution function,
 its approximately monotonically increased shape
 with pretty large 1$\sigma$ statistical uncertainties
 would alleviate the effects of the overestimations of
 the analyzed (reconstructed--input) data points
 and the reconstructed WIMP mass
 caused by the extra background events.
 The reconstructed velocity distribution function
 could then match the true (input) one pretty well.

 It would be worth to emphasize that,
 first,
 comparing to the pretty large
 (1$\sigma$) statistical uncertainties
 on the reconstructed--input data points
 (offered by our model--independent method
  developed in Ref.~\cite{DMDDf1v}
  with raw experimental measured recoil energies),
 our Bayesian reconstruction of
 the WIMP velocity distribution function
 introduced here
 with only a few km/s deviation and
 \mbox{$\cal O$(10) km/s} 1$\sigma$ statistical uncertainties
 on the reconstructed Solar and Earth's Galactic velocities
 would be a remarkable improvement.

 Second,
 all our simulations show importantly that,
 even initial values different slightly from the true (input) setup
 have been used as the expectation values
 for the Gaussian probability distribution of the fitting parameters,
 these fitting parameters
 could still be pinned down (pretty) precisely.
 As long as
 a proper assumed fitting velocity distribution function
 is used,
 the best--fit values of the reconstructed parameters
 could always be less than 1$\sigma$
 apart from the true (input/theoretical) values.
 This observation indicates that
 rough, slightly incorrect prior knowledge
 about our fitting parameters
 would not affect (significantly)
 the reconstructed results
 in our Bayesian reconstruction procedure.

 Moreover,
 by rewriting the functional form of
 the (basic) fitting velocity distribution function,
 one could not only pin down the fitting parameters more precisely,
 but also occasionally reduce the statistical uncertainties
 on the reconstructed parameters.

 In summary,
 we developed in this paper
 the Bayesian reconstruction procedure
 for fitting theoretically predicted models of
 the one--dimensional WIMP velocity distribution function
 to data (points),
 which can be reconstructed directly
 from experimental measured recoil energies.
 Hopefully,
 this extension of our earlier work
 could offer more useful information about the Dark Matter halo,
 which could further be used
 in e.g.~indirect DM detection experiments.

\subsubsection*{Acknowledgments}
 The author appreciates Mei-Yu Wang
 for useful discussions
 about models of the velocity distribution of Galactic WIMPs.
 The author would also like to thank
 the Physikalisches Institut der Universit\"at T\"ubingen
 for the technical support of the computational work
 presented in this paper
 as well as
 the friendly hospitality of
 the Graduate School of Science and Engineering for Research,
 University of Toyama,
 the Institute of Modern Physics,
 Chinese Academy of Sciences,
 the Center for High Energy Physics,
 Peking University,
 and
 the Xinjiang Astronomical Observatory,
 Chinese Academy of Sciences,
 where part of this work was completed.
 This work
 was partially supported
 by the National Science Council of R.O.C.~%
 under the contracts no.~NSC-98-2811-M-006-044 and
 no.~NSC-99-2811-M-006-031
 as well as
 the CAS Fellowship for Taiwan Youth Visiting Scholars
 under the grant no.~2013TW2JA0002.
\appendix
\setcounter{equation}{0}
\setcounter{figure}{0}
\renewcommand{\theequation}{A\arabic{equation}}
\renewcommand{\thefigure}{A\arabic{figure}}
%
%
% Appendix A
%
\section{Formulae needed in Sec.~2.1}
 Here we list all formulae needed
 for the model--independent method
 for the reconstruction of
 the one--dimensional WIMP velocity distribution function
 described in Sec.~2.1.
 Detailed derivations and discussions
 can be found in Ref.~\cite{DMDDf1v}.

 First,
 by using the standard Gaussian error propagation,
 the expression for the uncertainty
 on the logarithmic slope $k_n$ can be given
 from Eq.~(\ref{eqn:bQn}) directly as
\beq
   \sigma^2(k_n)
 = k_n^4
   \cbrac{  1
          - \bfrac{k_n b_n / 2}{\sinh (k_n b_n / 2)}^2}^{-2}
            \sigma^2\abrac{\bQn}
\~,
\label{eqn:sigma_kn}
\eeq
 with
\beq
   \sigma^2\abrac{\bQn}
 = \frac{1}{N_n - 1} \bbigg{\bQQn - \bQn^2}
\~.
\label{eqn:sigma_bQn}
\eeq
 For replacing the ``bin'' quantities
 by ``window'' quantities,
 one needs the covariance matrix
 for $\Bar{Q - Q_{\mu}}|_{\mu}$,
 which follows directly from the definition (\ref{eqn:wQ_mu}):
\beqn
 \conti {\rm cov}\abrac{\Bar{Q - Q_{\mu}}|_{\mu}, \Bar{Q - Q_{\nu}}|_{\nu}}
        \non\\
 \=     \frac{1}{N_{\mu} N_{\nu}}
        \sum_{n = n_{\nu-}}^{n_{\mu+}}
        \bbigg{  N_n
                 \abrac{\Bar{Q}|_n - \Bar{Q}|_{\mu}}
                 \abrac{\Bar{Q}|_n - \Bar{Q}|_{\nu}}
               + N_n^2 \sigma^2\abrac{\bQn}}
\~.
\label{eqn:cov_wQ_mu}
\eeqn
 Note that,
 firstly,
 $\mu \leq \nu$ has been assumed here
 and the covariance matrix is, of course, symmetric.
 Secondly,
 the sum is understood to vanish
 if the two windows $\mu$, $\nu$ do not overlap,
 i.e.~if $n_{\mu+} < n_{\nu-}$.
 Moreover,
 from Eq.~(\ref{eqn:r_mu}),
 we can get
\beq
   {\rm cov}(r_{\mu}, r_{\nu})
 = \frac{1}{w_{\mu} w_{\nu}} \sum_{n = n_{\nu-}}^{n_{\mu+}} N_n
\~,
\label{eqn:cov_r_mu}
\eeq
 where $\mu \leq \nu$ has again been taken.
 And the mixed covariance matrix can be given by
\beq
   {\rm cov}\abrac{r_{\mu}, \Bar{Q - Q_{\nu}}|_{\nu}}
 = \frac{1}{w_{\mu} N_{\nu}}
   \sum_{n = n_{-}}^{n_{+}} N_n \abrac{\Bar{Q}|_n - \Bar{Q}|_{\nu}}
\~.
\label{eqn:cov_r_mu_wQ_nu}
\eeq
 Note here that
 this sub--matrix is {\em not} symmetric
 under the exchange of $\mu$ and $\nu$.
 In the definition of $n_{-}$ and $n_{+}$
 we therefore have to distinguish two cases:
\beqn
\renewcommand{\arraystretch}{1.6}
\begin{array}{l c l}
   n_{-} = n_{\nu-},~
   n_{+} = n_{\mu+},  &
   ~~~~~~~~~~~~       &
   {\rm if}~\mu \leq \nu\~; \\
   n_{-} = n_{\mu-},~
   n_{+} = n_{\nu+},  &
                      &
   {\rm if}~\mu \geq \nu\~.
\end{array}
\label{eqn:def_n_pm}
\eeqn
 As before,
 the sum in Eq.~(\ref{eqn:cov_r_mu_wQ_nu})
 is understood to vanish if $n_{-} > n_{+}$.

 Furthermore,
 the covariance matrices
 involving the estimators of the logarithmic slopes $k_{\mu}$,
 estimated by Eq.~(\ref{eqn:bQn}) with replacing $n \to \mu$,
 can be given from Eq.~(\ref{eqn:sigma_kn}) as
\beqn
        {\rm cov}\abrac{k_{\mu}, k_{\nu}}
 \=     k_{\mu}^2 k_{\nu}^2
        \cbrac{  1
               - \bfrac{k_{\mu} b_{\mu} / 2}{\sinh (k_{\mu} b_{\mu} / 2)}^2}^{-1}
        \cbrac{  1
               - \bfrac{k_{\nu} b_{\nu} / 2}{\sinh (k_{\nu} b_{\nu} / 2)}^2}^{-1}
        \non\\
        \non\\
 \conti ~~~~~~~~~~~~~~~~ \times %16
        {\rm cov}\abrac{\Bar{Q - Q_{\mu}}|_{\mu}, \Bar{Q - Q_{\nu}}|_{\nu}}
\~,
\label{eqn:cov_k_mu}
\eeqn
 and
\beq
   {\rm cov}\abrac{r_{\mu}, k_{\nu}}
 = k_{\nu}^2
   \cbrac{  1
          - \bfrac{k_{\nu} b_{\nu} / 2}{\sinh (k_{\nu} b_{\nu} / 2)}^2}^{-1}
   {\rm cov}\abrac{r_{\mu}, \Bar{Q - Q_{\nu}}|_{\nu}}
\~.
\label{eqn:cov_r_mu_k_nu}
\eeq
\end{document}

%% file: include-plots-arxiv.tex
\newcommand{\InsertSetupTable} [4] {
\begin{center}
\renewcommand{\arraystretch}{1.75}
\setlength{\tabcolsep}{3pt}
{\footnotesize
\begin{tabular}{|| c | c | c | c | c | c ||}
\hline
\hline
 \multicolumn{6}{|| c ||}
 {\bf\boldmath
  Input: #1} \\
\hline
\hline
 \makebox[2.5 cm][c]{Fitting model}           &
 \makebox[1.7 cm][c]{Parameter}               &
 \makebox[3.7 cm][c]{Input/theoretical value} &
 \makebox[2.5 cm][c]{Scanning range}          &
 \makebox[2.8 cm][c]{Expectation value}       &
 \makebox[2.4 cm][c]{1$\sigma$ uncertainty}   \\
\hline
 #2
\hline
\hline
\end{tabular}
}
\caption{
 #3
}
\label{#4}
\end{center}
\end{table}
}
\newcommand{\InsertResultsTable} [5] {
\begin{center}
\renewcommand{\arraystretch}{1.75}
\setlength{\tabcolsep}{3pt}
{\footnotesize
\begin{tabular}{|| c | c | c | c | c | c | c ||}
\hline
\hline
 \multicolumn{7}{|| c ||}
 {Input: #1} \\
\hline
\hline
 \multicolumn{7}{|| c ||}
 {\bf\boldmath
  Reconstruction: #2} \\
\hline
 \makebox[1.8 cm][c]{Parameter}                   &
 \makebox[2   cm][c]{WIMP mass}                   &
 \makebox[1.8 cm][c]{Prob.~dist.}                 &
 \makebox[2.2 cm][c]{Max.~${\rm p}_{\rm median}$} &
 \makebox[3.2 cm][c]{Median}                      &
 \makebox[2.2 cm][c]{1$\sigma$ range}             &
 \makebox[2.2 cm][c]{2$\sigma$ range}             \\
\hline
 #3
\hline
\hline
\end{tabular}
}
\caption{
 #4
}
\label{#5}
\end{center}
\end{table}
}
\newcommand{\InsertTwoPlots} [1] {
\begin{center}
 \begin{picture}(16.5, 6.5)
   \put(0  , 0.5){\framebox(8  , 6  ){#1}}
   \put(8.5, 0.5){\framebox(8  , 6  ){}}
   \put(0  , 0  ){\framebox(8  , 0.5){(a)}}
   \put(8.5, 0  ){\framebox(8  , 0.5){(b)}}
 \end{picture}
\end{center}
}
\newcommand{\InsertFourPlots} [1] {
\begin{center}
 \begin{picture}(16.5, 12.5)
   \put(0  , 7  ){\framebox(8 , 6  ){#1}}
   \put(8.5, 7  ){\framebox(8 , 6  ){}}
   \put(0  , 6.5){\framebox(8  , 0.5){(a)}}
   \put(8.5, 6.5){\framebox(8  , 0.5){(b)}}
   \put(0  , 0.5){\framebox(8 , 6  ){}}
   \put(8.5, 0.5){\framebox(8 , 6  ){}}
   \put(0  , 0  ){\framebox(8  , 0.5){(c)}}
   \put(8.5, 0  ){\framebox(8  , 0.5){(d)}}
 \end{picture}
\end{center}
}
\newcommand{\plotfvGauk}{
\begin{figure}[t!]
\begin{center}
\vspace{-0.25cm}
{
\includegraphics[width=10cm]{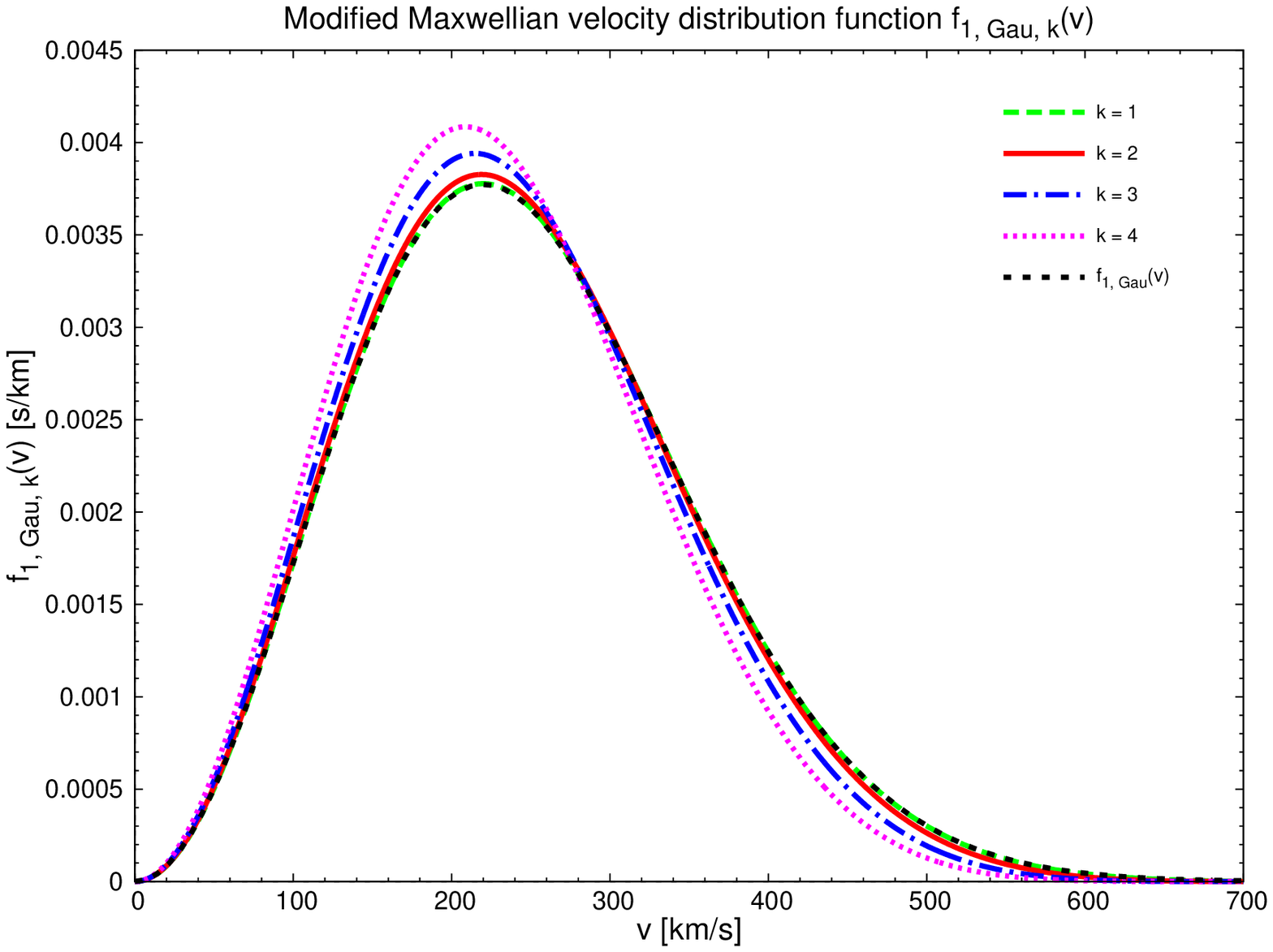}
}
\vspace{-0.35cm}
\end{center}
\caption{
 The normalized
 modified Maxwellian velocity distribution function $f_{1, \Gau, k}(v)$
 given in Eq.~(\ref{eqn:f1v_Gau_k})
 with a common value of the Solar Galactic velocity
 \mbox{$v_0 = 220$ km/s}
 and different power indices $k$:
 $k = 1$ (dashed light--green),
 $k = 2$ (solid red),
 $k = 3$ (dash--dotted blue) and
 $k = 4$ (dotted magenta).
 As a comparison,
 the simple Maxwellian velocity distribution $f_{1, \Gau}(v)$
 with \mbox{$v_0 = 220$ km/s}
 is also given
 as the short--dashed black curve.
}
\label{fig:f1v_Gau_k}
\end{figure}
}
\newcommand{\plotdRdQbg}{
\begin{figure}[t!]
\begin{center}
\vspace{-0.25cm}
{
\hspace*{-1.6cm}
\includegraphics[width=8.5cm]{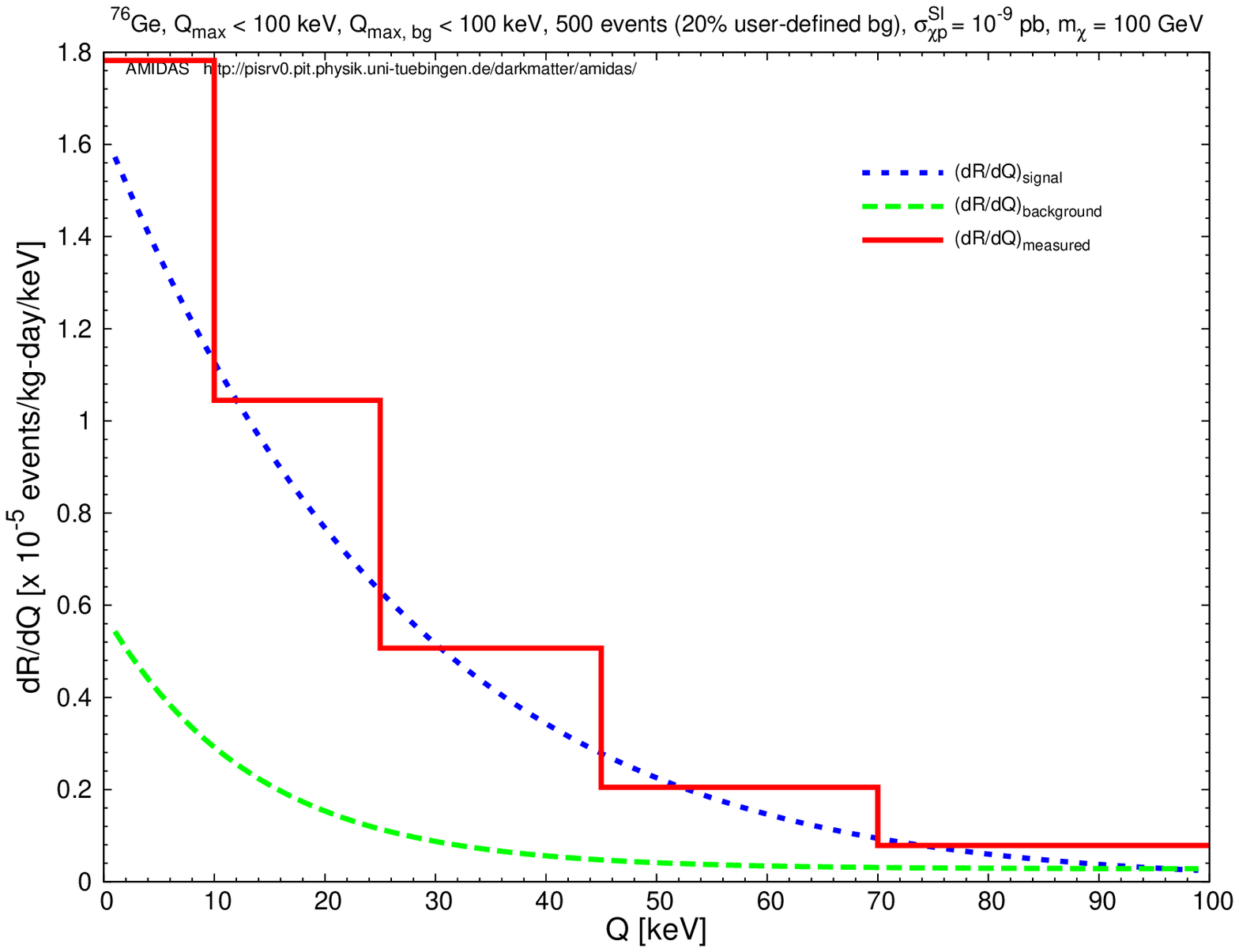}
\includegraphics[width=8.5cm]{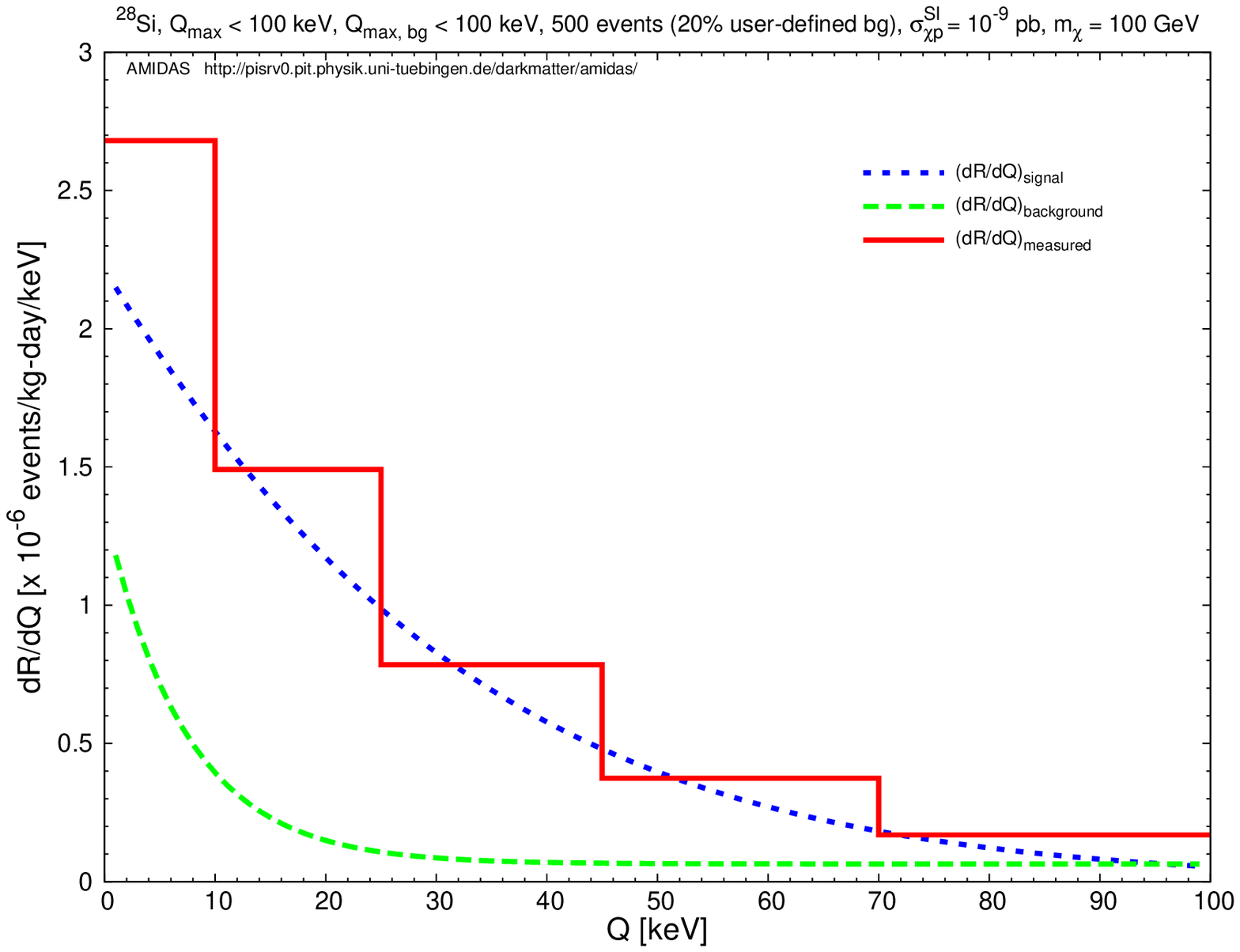} \hspace*{-1.6cm} \par
\makebox[8.5cm]{(a)}\hspace{0.325cm}\makebox[8.175cm]{(b)}%
}
\vspace{-0.35cm}
\end{center}
\caption{
 Measured recoil energy spectrum (solid red histogram)
 for a $\rmXA{Ge}{76}$ (a) and a $\rmXA{Si}{28}$ (b) targets
 with an input WIMP mass of \mbox{$\mchi = 100$ GeV}.
 The dotted blue curve is
 the elastic WIMP--nucleus scattering spectrum
 for generating signal events,
 whereas
 the dashed green curve shows
 the artificial background spectrum,
 normalized to fit to the background ratio of 20\%.
}
\label{fig:dRdQ-bg-ex-const-000-100-20-100}
\end{figure}
}
\newcommand{\plotGeGauGauflat}{
\begin{figure}[t!]
\begin{center}
\vspace{-0.25cm}
{
\hspace*{-1.6cm}
\includegraphics[width=8.5cm]{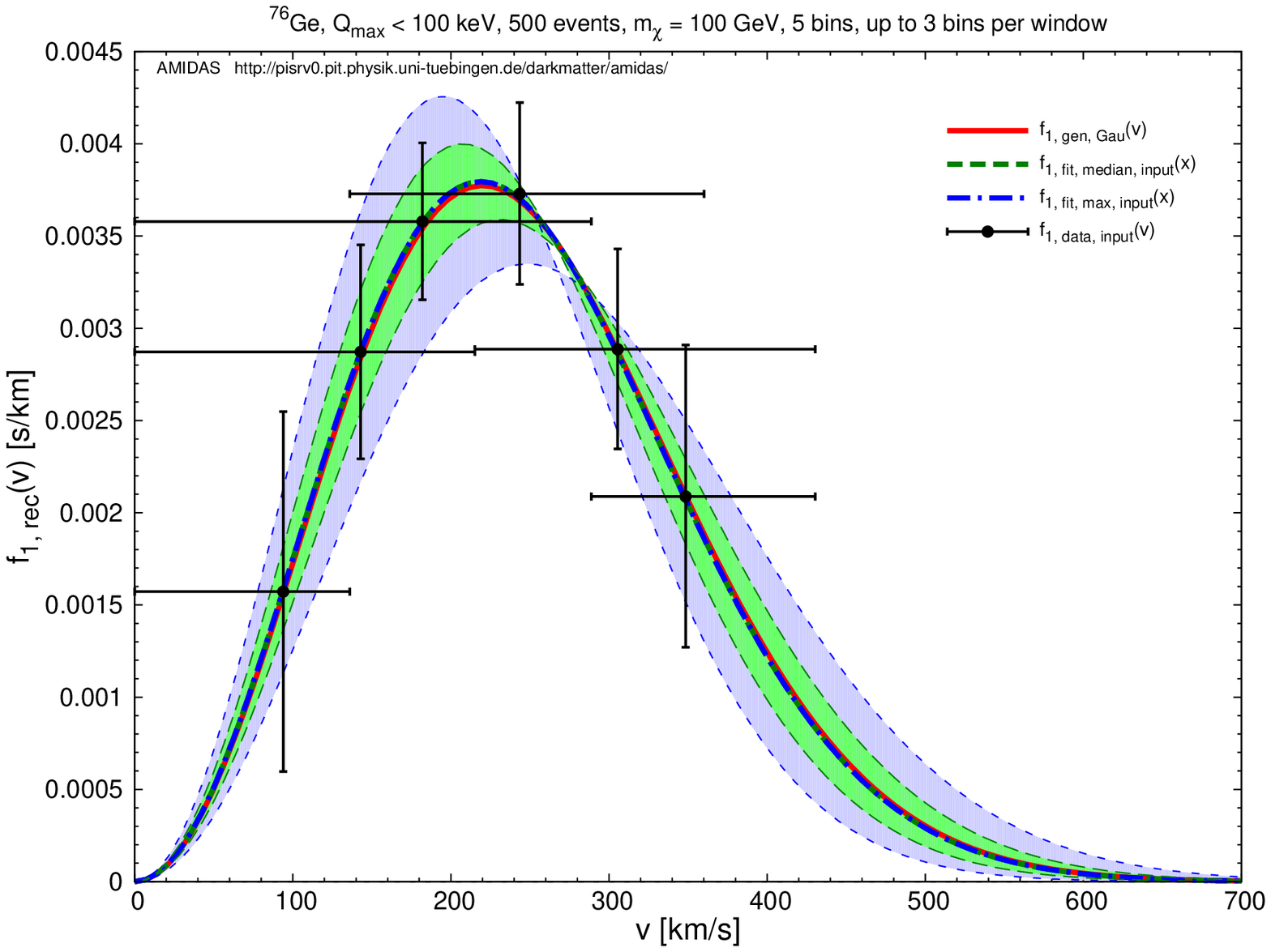}
\includegraphics[width=8.5cm]{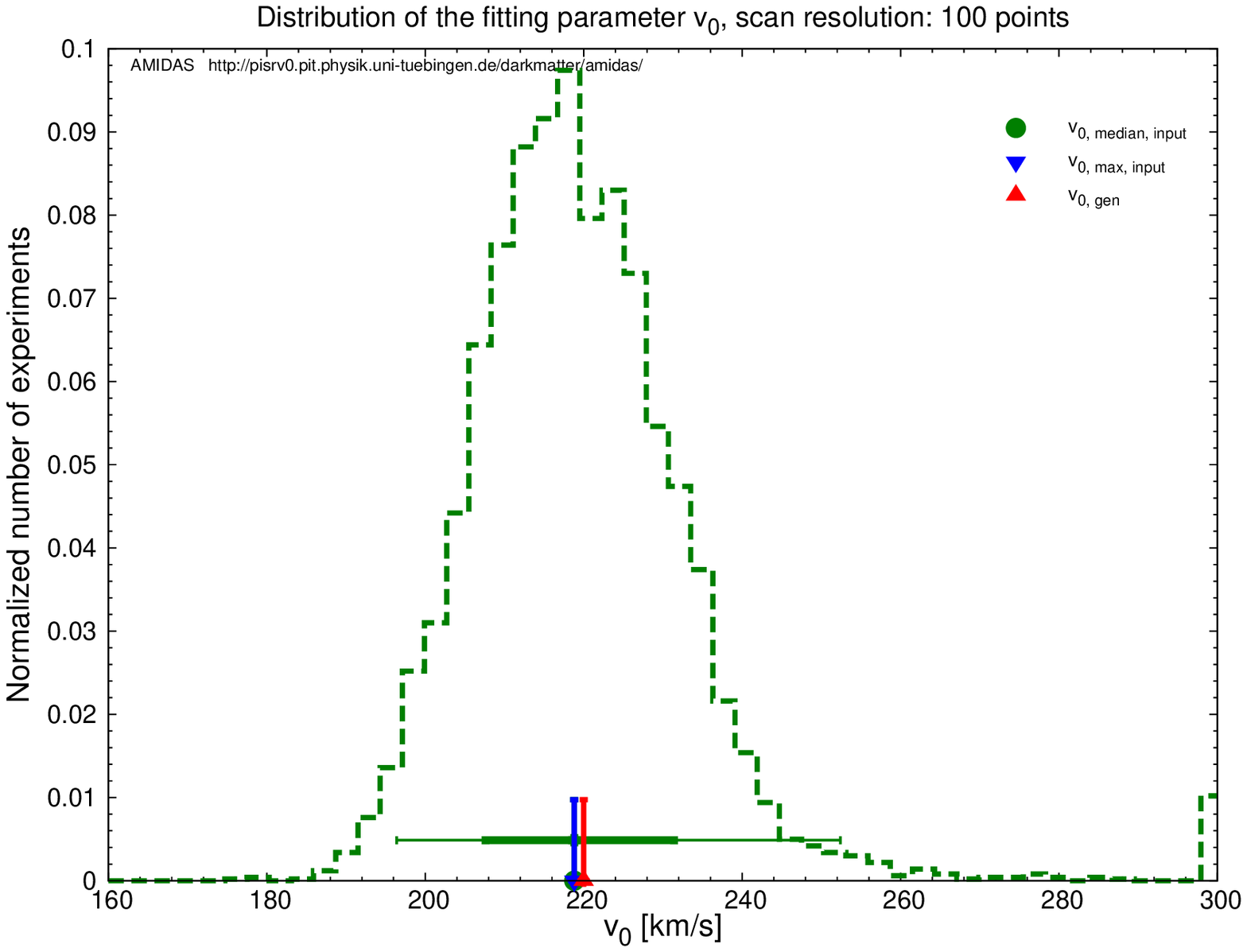} \hspace*{-1.6cm} \par
\makebox[8.5cm]{(a)}\hspace{0.325cm}\makebox[8.175cm]{(b)}%
}
\vspace{-0.5cm}
\end{center}
\caption{
 (a)
 The reconstructed
 simple Maxwellian velocity distribution function
 for an input WIMP mass of \mbox{$\mchi = 100$ GeV}
 with a $\rmXA{Ge}{76}$ target.
 The black crosses are the velocity distribution
 reconstructed by Eqs.~(\ref{eqn:vsn}) and (\ref{eqn:f1v_Qsn}):
 the vertical error bars show
 the square roots of the diagonal entries of the covariance matrix
 estimated by Eq.~(\ref{eqn:cov_f1v_Qs_mu})
 (i.e.,
  $\sigma_{f_1, s, \mu}$ given in Eq.~(\ref{eqn:sigma_f1v_Qs_mu}))
 and
 the horizontal bars indicate
 the sizes of the windows used
 for estimating $f_{1, {\rm rec}}(v_{s, \mu})$,
 respectively.
 The solid red curve
 is the {\em generating} simple Maxwellian velocity distribution
 with an input value of \mbox{$v_0 = 220$ km/s}.
 While
 the dashed green curve indicates
 the {\em reconstructed} simple Maxwellian velocity distribution
 with the fitting parameter $v_0$
 given by the {\em median} value of all simulated experiments,
 the dash--dotted blue curve indicates
 the {\em reconstructed} simple Maxwellian velocity distribution
 with $v_0$
 which maximizes ${\rm p}_{\rm median}\abrac{a_i,~i = 1,~2,~\cdots,~N_{\rm Bayesian}}$
 defined in Eq.~(\ref{eqn:P_Bayesian_median}).
 (b)
 The distribution of the Bayesian reconstructed fitting parameter $v_0$
 in all simulated experiments.
 The red vertical line indicates the true (input) value of $v_0$,
 which has been labeled with the subscript ``gen''.
 The green vertical line indicates
 the median value of the simulated results,
 whereas
 the blue one indicates the value which maximizes
 ${\rm p}_{\rm median}$.
 In addition,
 the horizontal thick (thin) green bars show
 the $1\~(2)\~\sigma$ ranges of the reconstructed results.
 Note that
 the bins at \mbox{$v_0 = 160$ km/s} and \mbox{$v_0 = 300$ km/s}
 are ``overflow'' bins,
 which contain also the experiments
 with the best--fit $v_0$ value of
 either \mbox{$v_0 < 160$ km/s} or \mbox{$v_0 > 300$ km/s}.
 See the text for further details.
}
\label{fig:f1v-Ge-100-0500-Gau-Gau-flat}
\end{figure}
}
\newcommand{\plotGeGauGauGau}{
\begin{figure}[b!]
\begin{center}
{
\hspace*{-1.6cm}
\includegraphics[width=8.5cm]{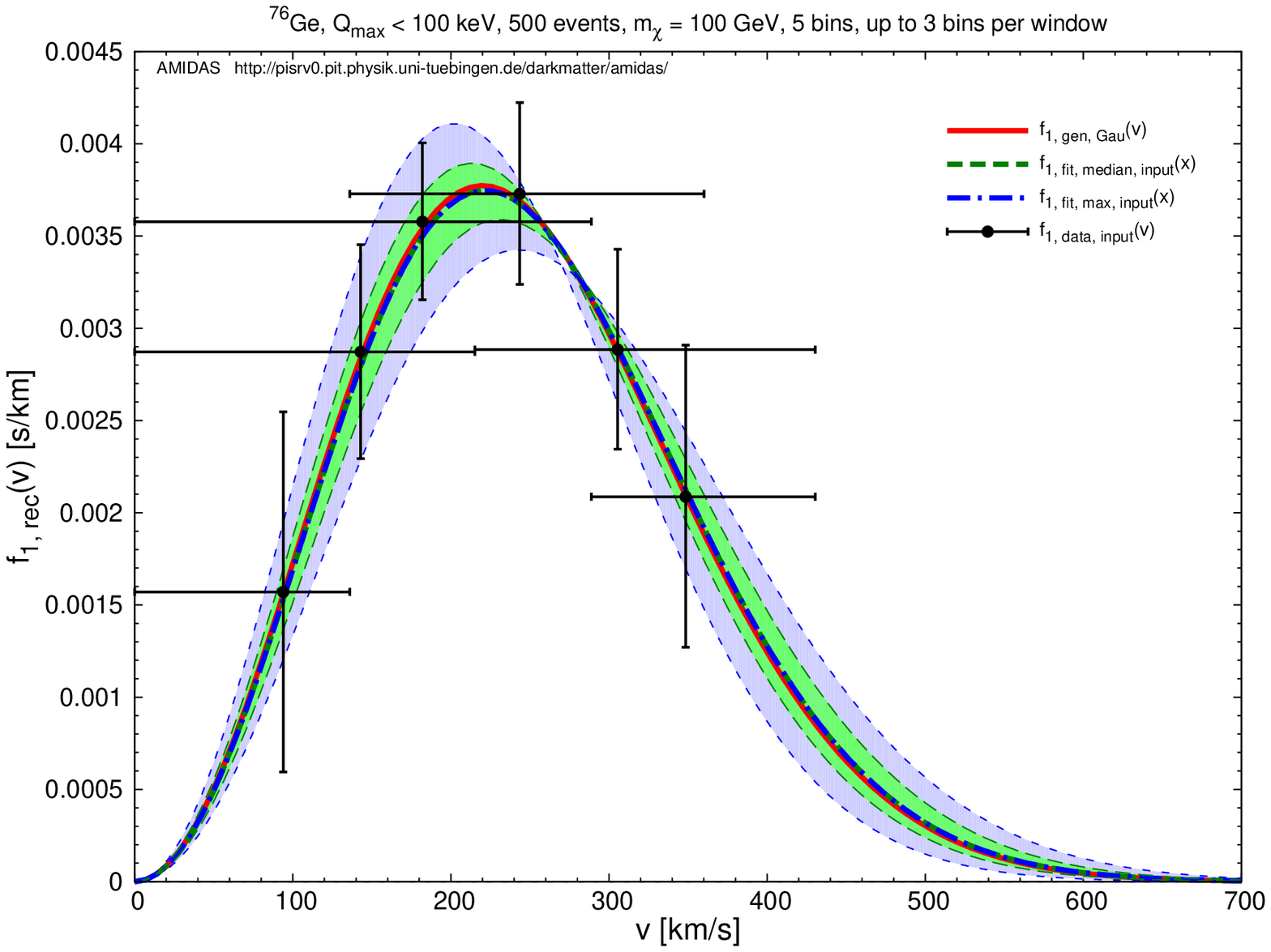}
\includegraphics[width=8.5cm]{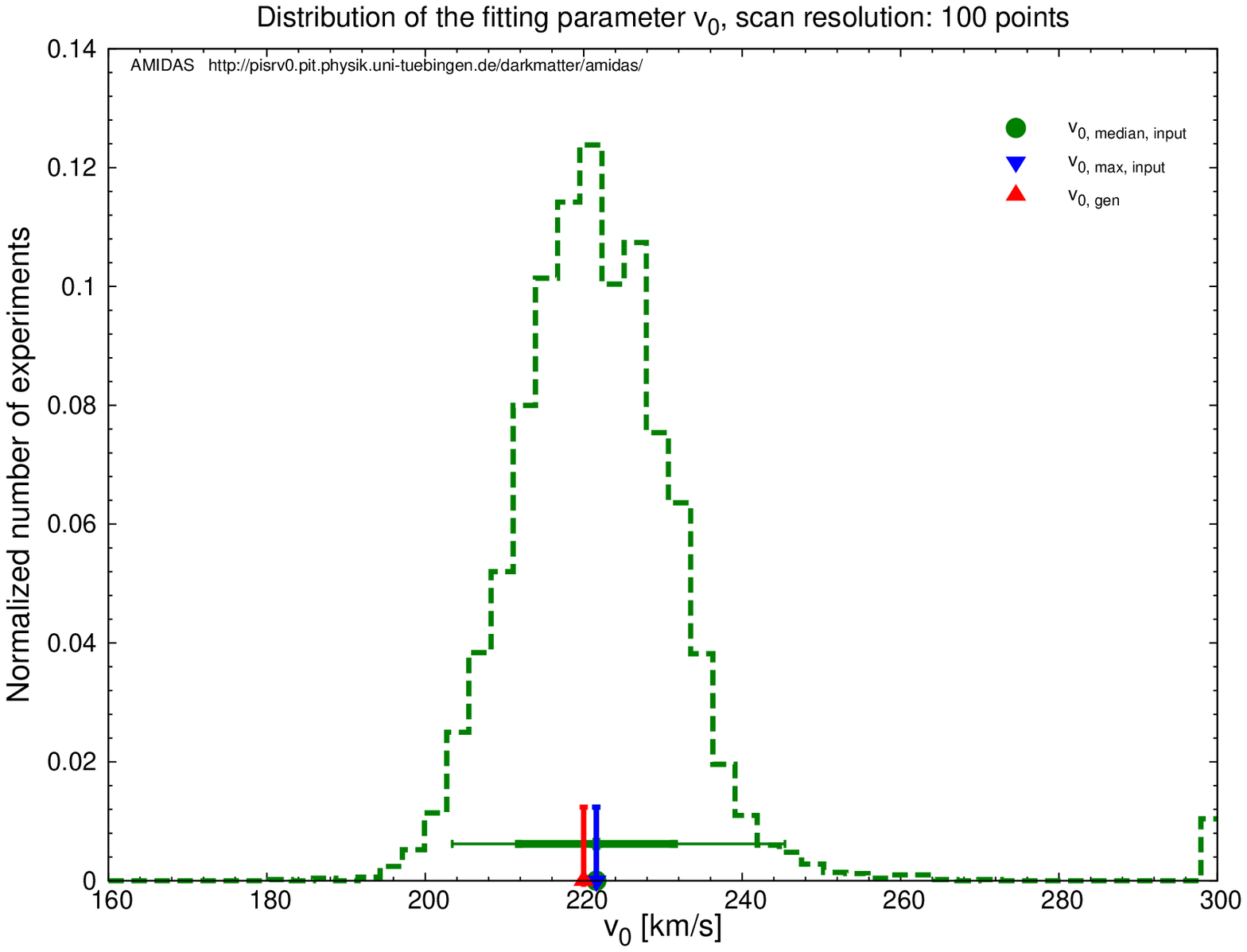} \hspace*{-1.6cm} \par
\makebox[8.5cm]{(a)}\hspace{0.325cm}\makebox[8.175cm]{(b)}%
}
\vspace{-0.5cm}
\end{center}
\caption{
 As in Figs.~\ref{fig:f1v-Ge-100-0500-Gau-Gau-flat},
 except that
 the Gaussian probability distribution
 given in Eq.~(\ref{eqn:Bayesian_DF_a_Gau})
 for $v_0$
 with an expectation value of \mbox{$v_0 = 230$ km/s}
 and a 1$\sigma$ uncertainty of \mbox{20 km/s}
 has been used.
}
\label{fig:f1v-Ge-100-0500-Gau-Gau-Gau}
\end{figure}
}
\newcommand{\plotGeSiGeGauGauflat}{
\begin{figure}[t!]
\begin{center}
\vspace{-0.25cm}
{
\hspace*{-1.6cm}
\includegraphics[width=8.5cm]{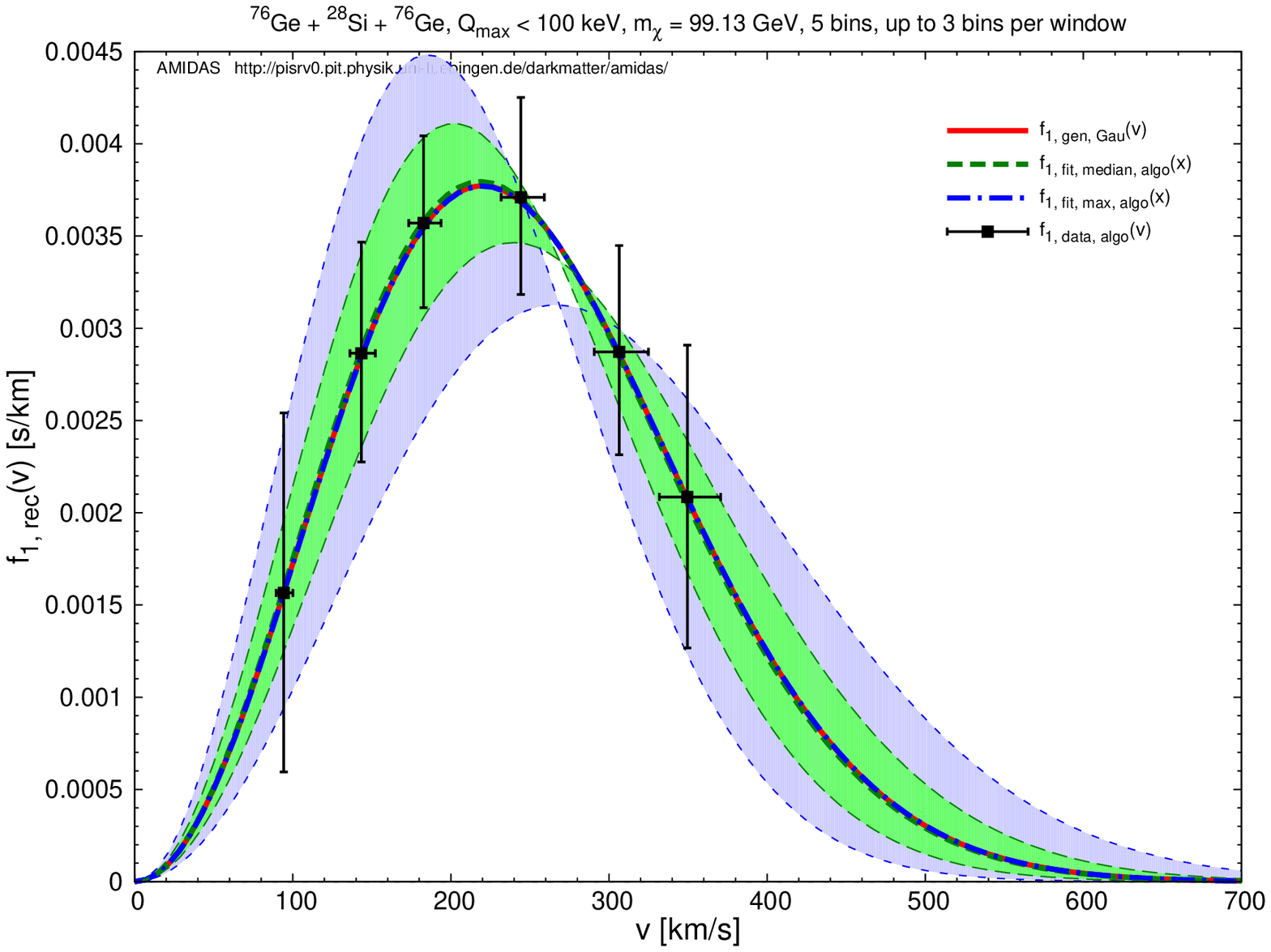}
\includegraphics[width=8.5cm]{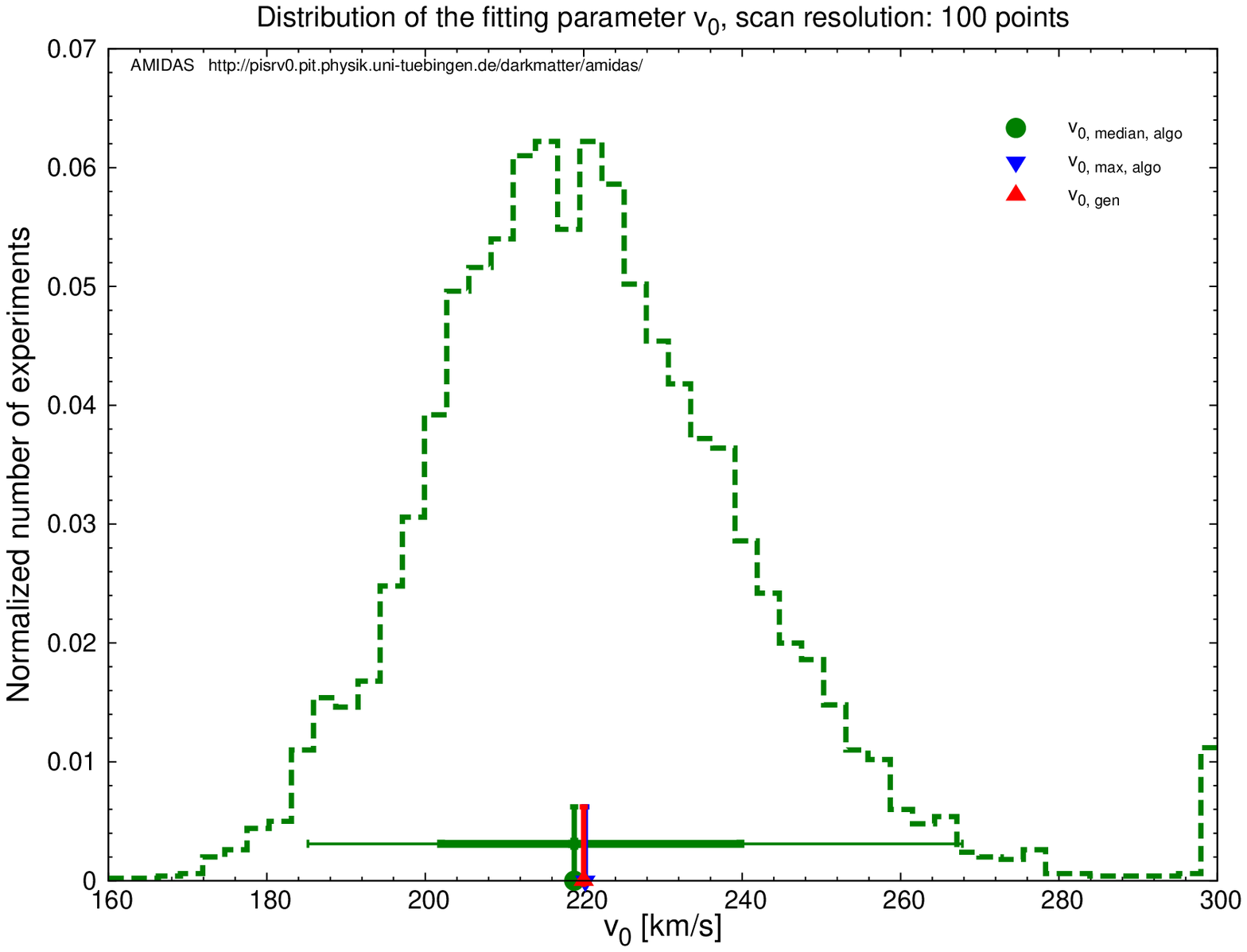} \hspace*{-1.6cm} \par
\makebox[8.5cm]{(a)}\hspace{0.325cm}\makebox[8.175cm]{(b)}%
}
\vspace{-0.35cm}
\end{center}
\caption{
 As in Figs.~\ref{fig:f1v-Ge-100-0500-Gau-Gau-flat},
 except that
 the WIMP mass $\mchi$
 needed in Eqs.~(\ref{eqn:vsn}) and (\ref{eqn:calN_sum})
 is reconstructed by the algorithmic procedure
 developed in Ref.~\cite{DMDDmchi}
 with a $\rmXA{Si}{28}$ target
 and a second $\rmXA{Ge}{76}$ target.
 While the vertical bars show
 the 1$\sigma$ statistical uncertainties
 estimated by Eq.~(\ref{eqn:sigma_f1v_Qs_mu})
 taking into account an extra contribution from
 the 1$\sigma$ statistical uncertainty
 on the reconstructed WIMP mass,
 the horizontal bars shown here
 indicate the 1$\sigma$ statistical uncertainties
 on the estimates of $v_{s, \mu}$
 given in Eq.~(\ref{eqn:vsn})
 due to the uncertainty on the reconstructed WIMP mass;
 the statistical and systematic uncertainties
 due to estimating of $Q_{s, \mu}$
 have been neglected here.
}
\label{fig:f1v-Ge-SiGe-100-0500-Gau-Gau-flat}
\end{figure}
}
\newcommand{\plotGeSiGeGauGauGau}{
\begin{figure}[b!]
\begin{center}
\vspace{0.25cm}
{
\hspace*{-1.6cm}
\includegraphics[width=8.5cm]{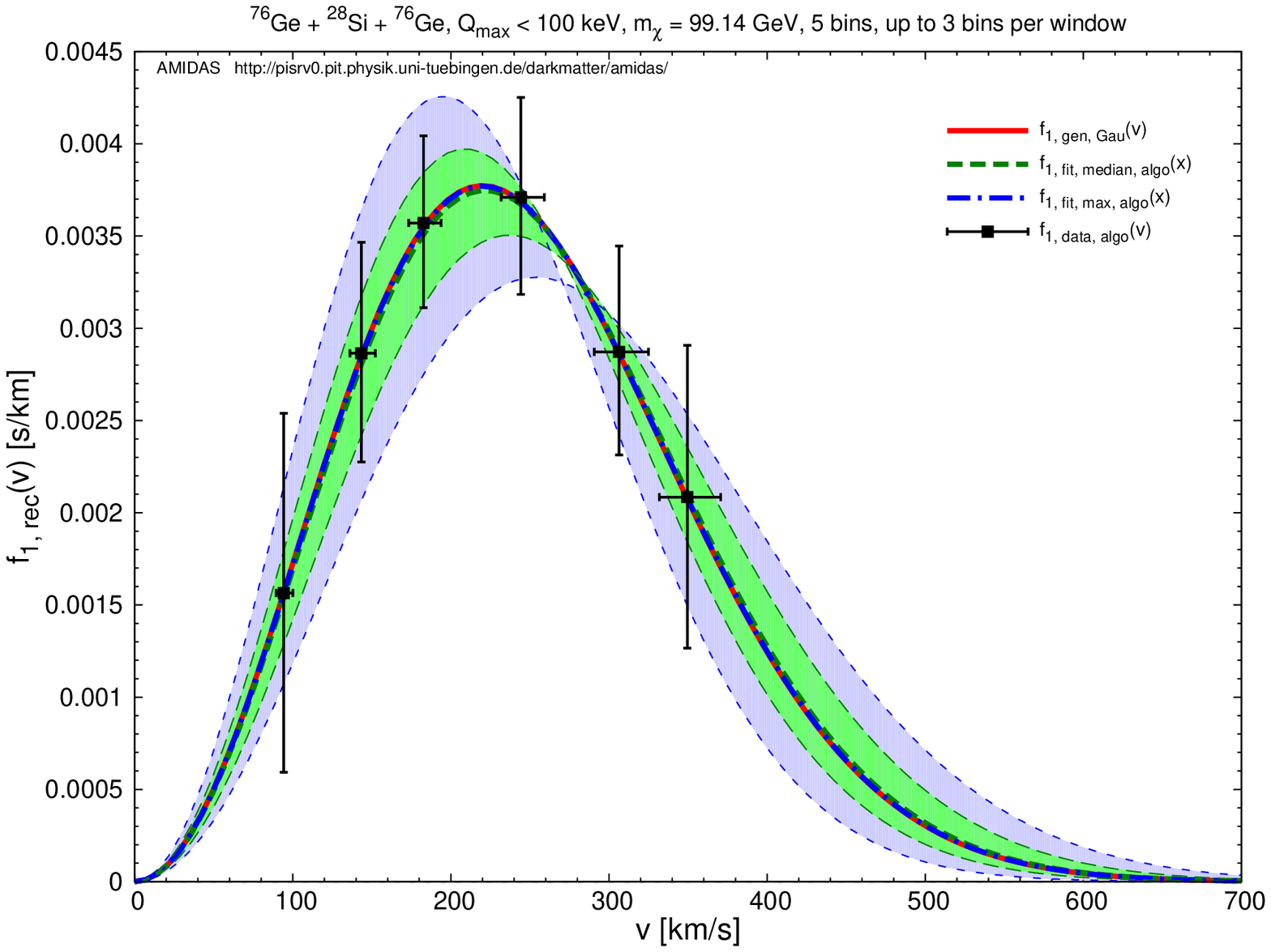}
\includegraphics[width=8.5cm]{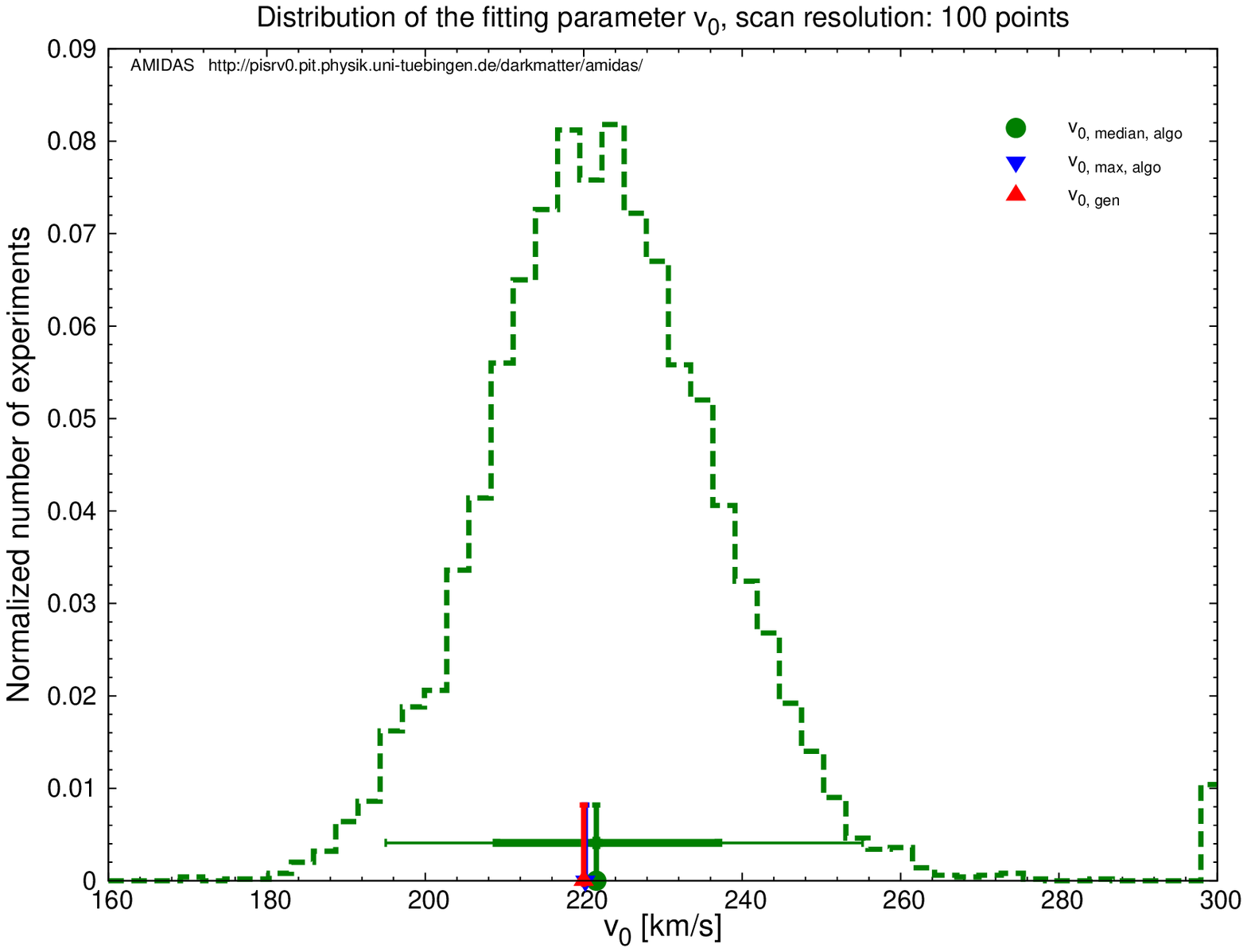} \hspace*{-1.6cm} \par
\makebox[8.5cm]{(a)}\hspace{0.325cm}\makebox[8.175cm]{(b)}%
}
\vspace{-0.35cm}
\end{center}
\caption{
 As in Figs.~\ref{fig:f1v-Ge-SiGe-100-0500-Gau-Gau-flat},
 except that
 the Gaussian probability distribution
 for $v_0$
 with an expectation value of \mbox{$v_0 = 230$ km/s}
 and a 1$\sigma$ uncertainty of \mbox{20 km/s}
 has been used.
}
\label{fig:f1v-Ge-SiGe-100-0500-Gau-Gau-Gau}
\end{figure}
}
\newcommand{\plotGeSiGeshGauflat}{
\begin{figure}[t!]
\begin{center}
\vspace{-0.25cm}
{
\hspace*{-1.6cm}
\includegraphics[width=8.5cm]{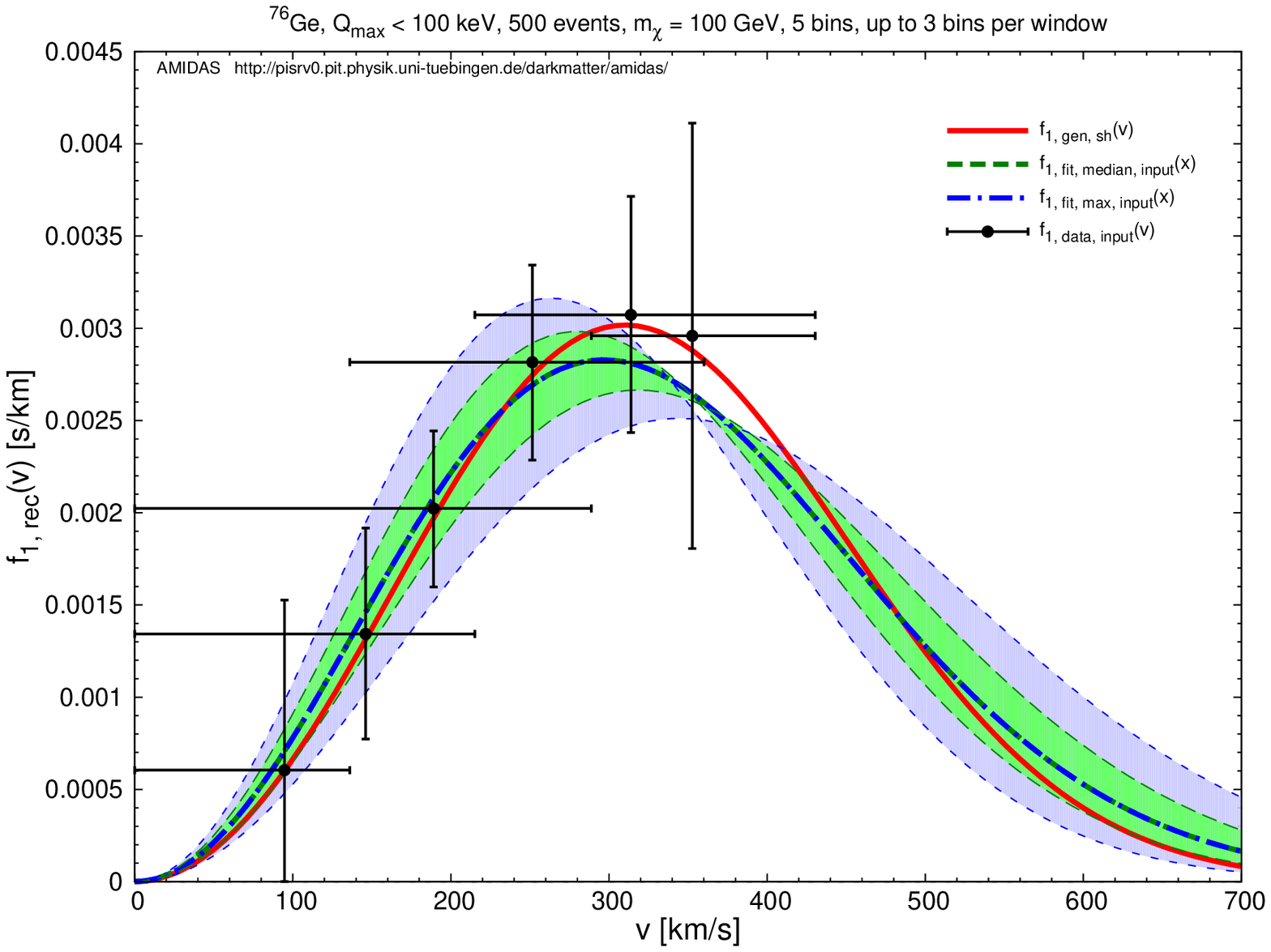}
\includegraphics[width=8.5cm]{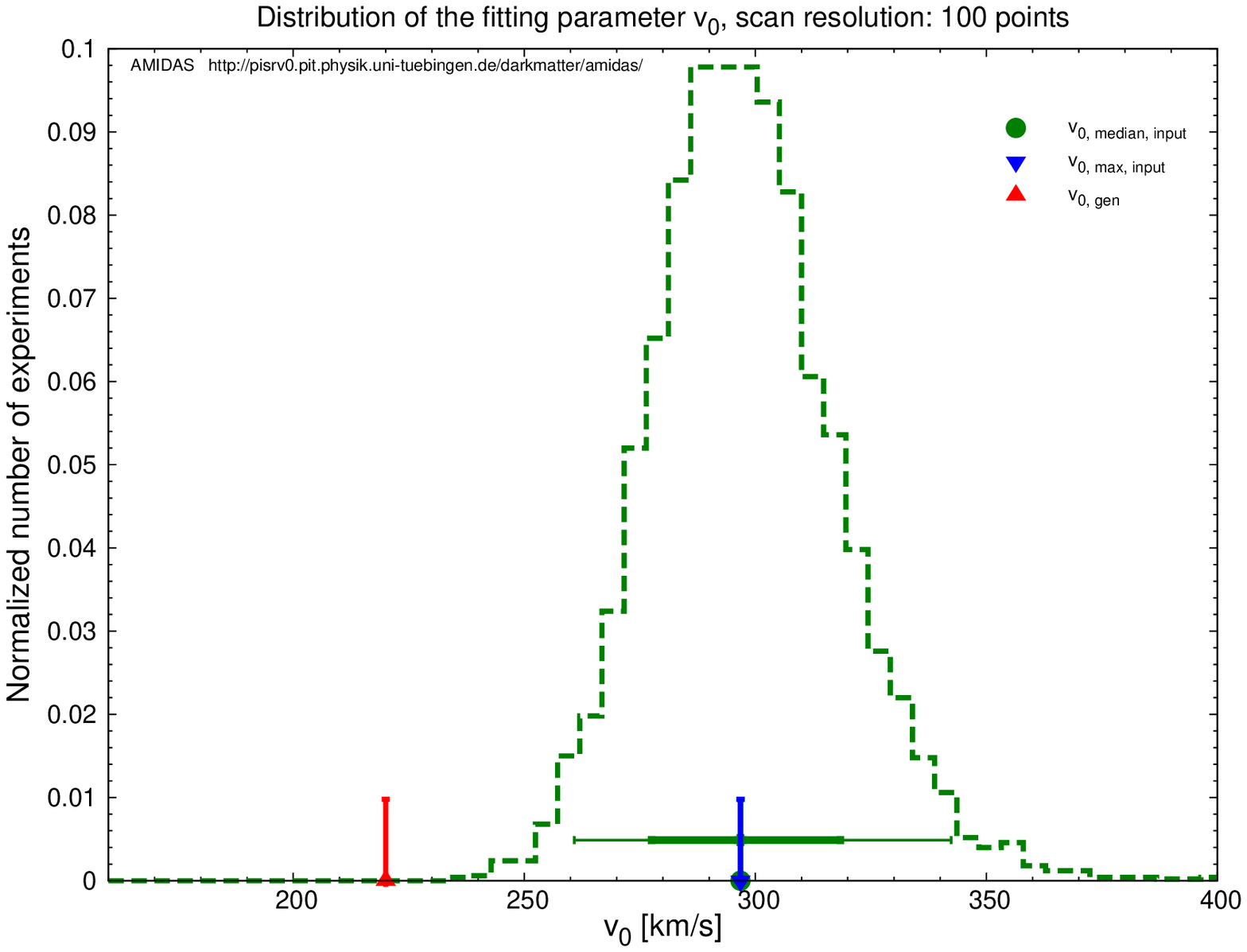}      \hspace*{-1.6cm} \par
\makebox[8.5cm]{(a)}\hspace{0.325cm}\makebox[8.175cm]{(b)}         \\ \vspace{0.5cm}
\hspace*{-1.6cm}
\includegraphics[width=8.5cm]{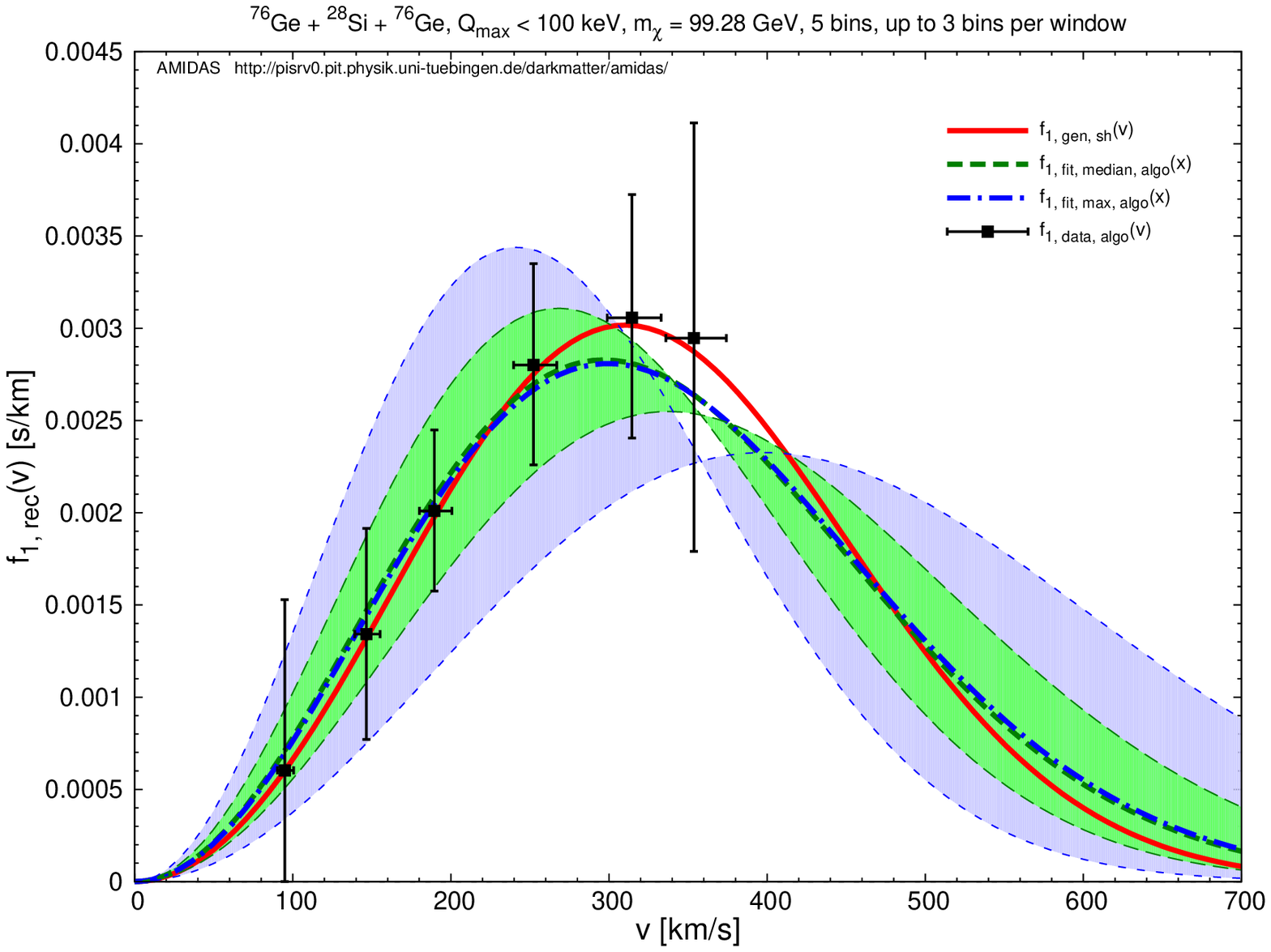}
\includegraphics[width=8.5cm]{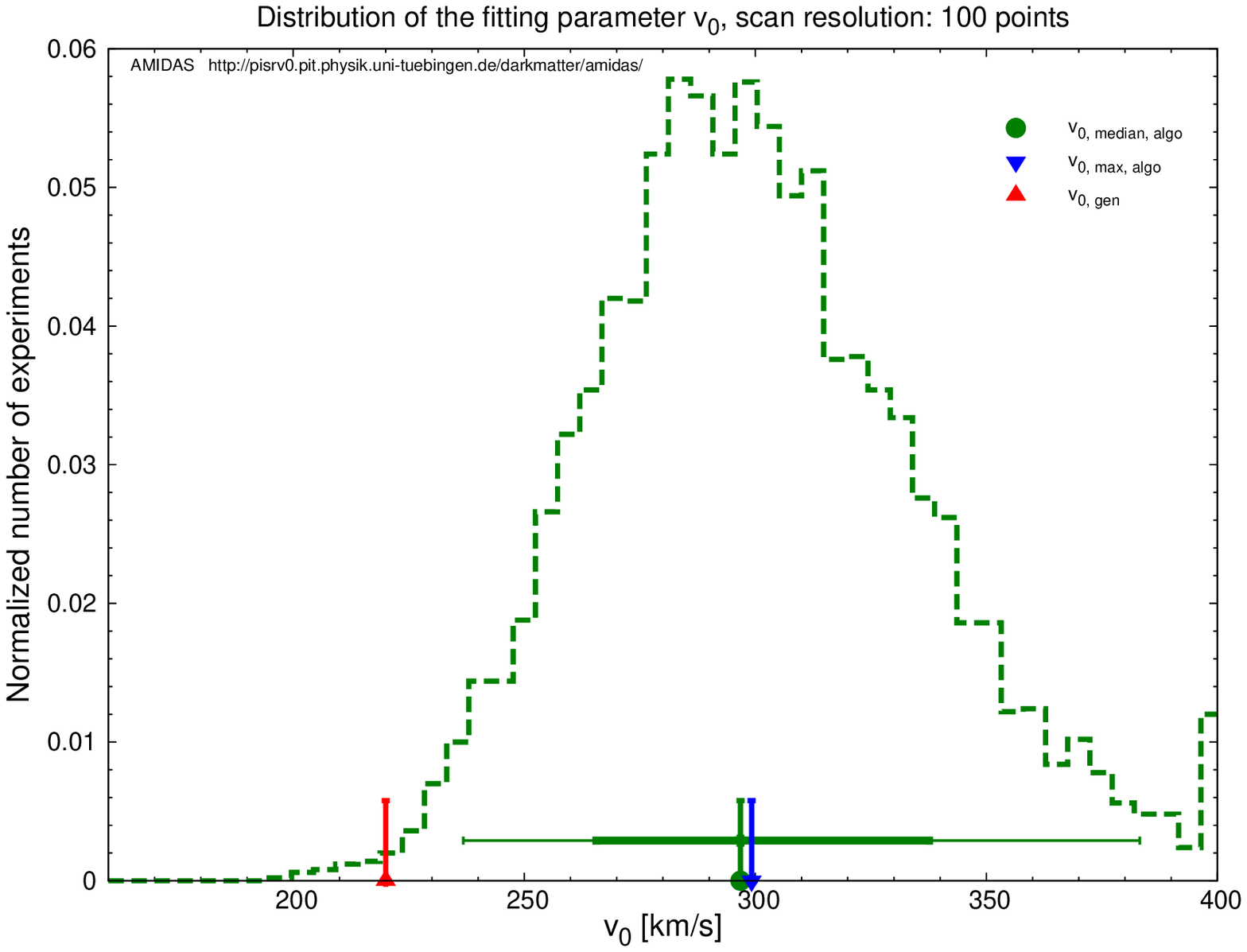} \hspace*{-1.6cm} \par
\makebox[8.5cm]{(c)}\hspace{0.325cm}\makebox[8.175cm]{(d)}%
}
\vspace{-0.35cm}
\end{center}
\caption{
 (a) (b)
 As in Figs.~\ref{fig:f1v-Ge-100-0500-Gau-Gau-flat},
 except that
 the shifted Maxwellian velocity distribution function
 given in Eq.~(\ref{eqn:f1v_sh})
 has been used
 for generating WIMP signals.
 (c) (d)
 As in Figs.~\ref{fig:f1v-Ge-SiGe-100-0500-Gau-Gau-flat}:
 the WIMP mass $\mchi$
 has been
 reconstructed
 with a $\rmXA{Si}{28}$ target
 and a second $\rmXA{Ge}{76}$ target.
}
\label{fig:f1v-Ge-SiGe-100-0500-sh-Gau-flat}
\end{figure}
}
\newcommand{\plotGeSiGeshGauGau}{
\begin{figure}[t!]
\begin{center}
\vspace{-0.25cm}
{
\hspace*{-1.6cm}
\includegraphics[width=8.5cm]{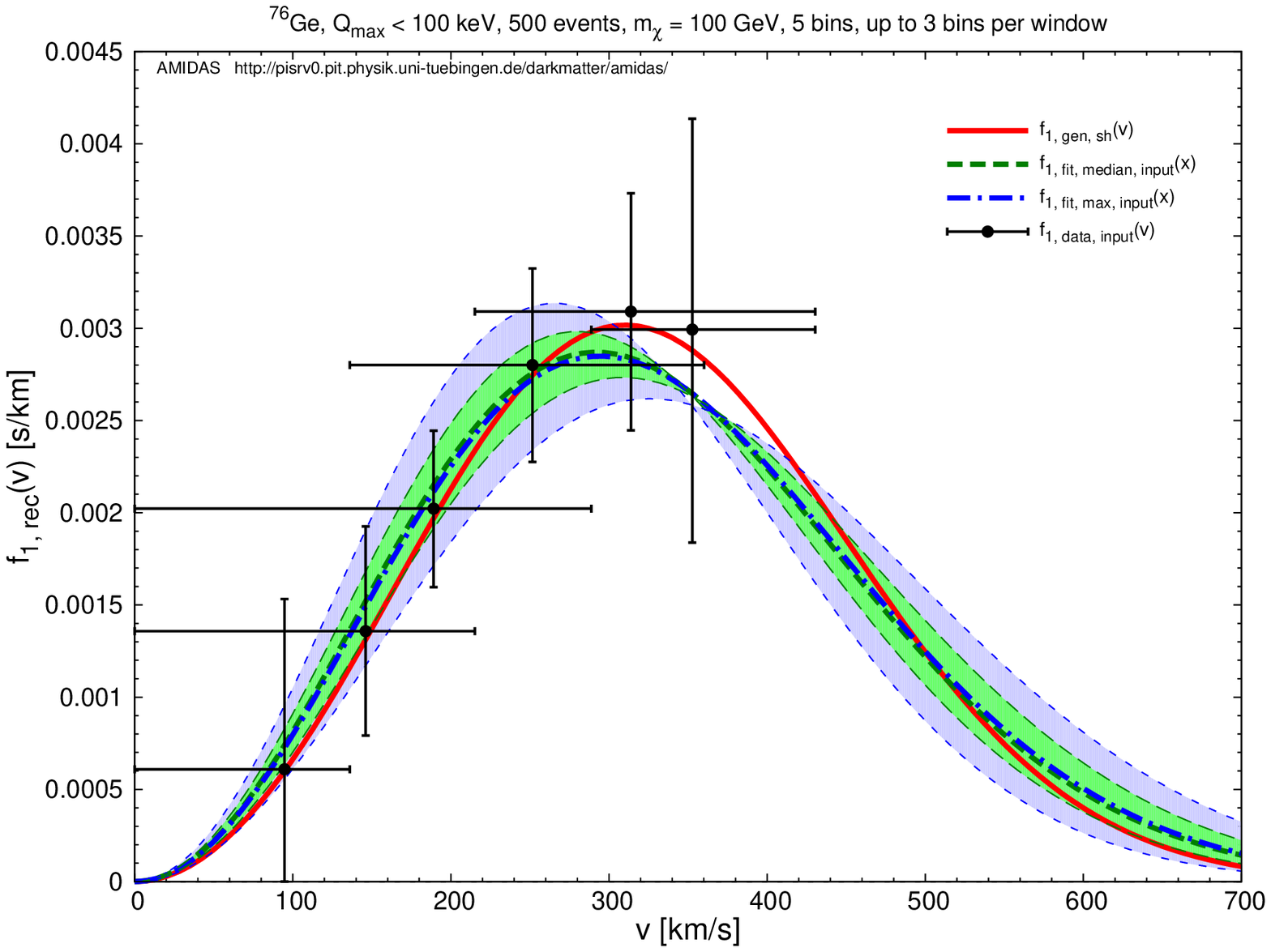}
\includegraphics[width=8.5cm]{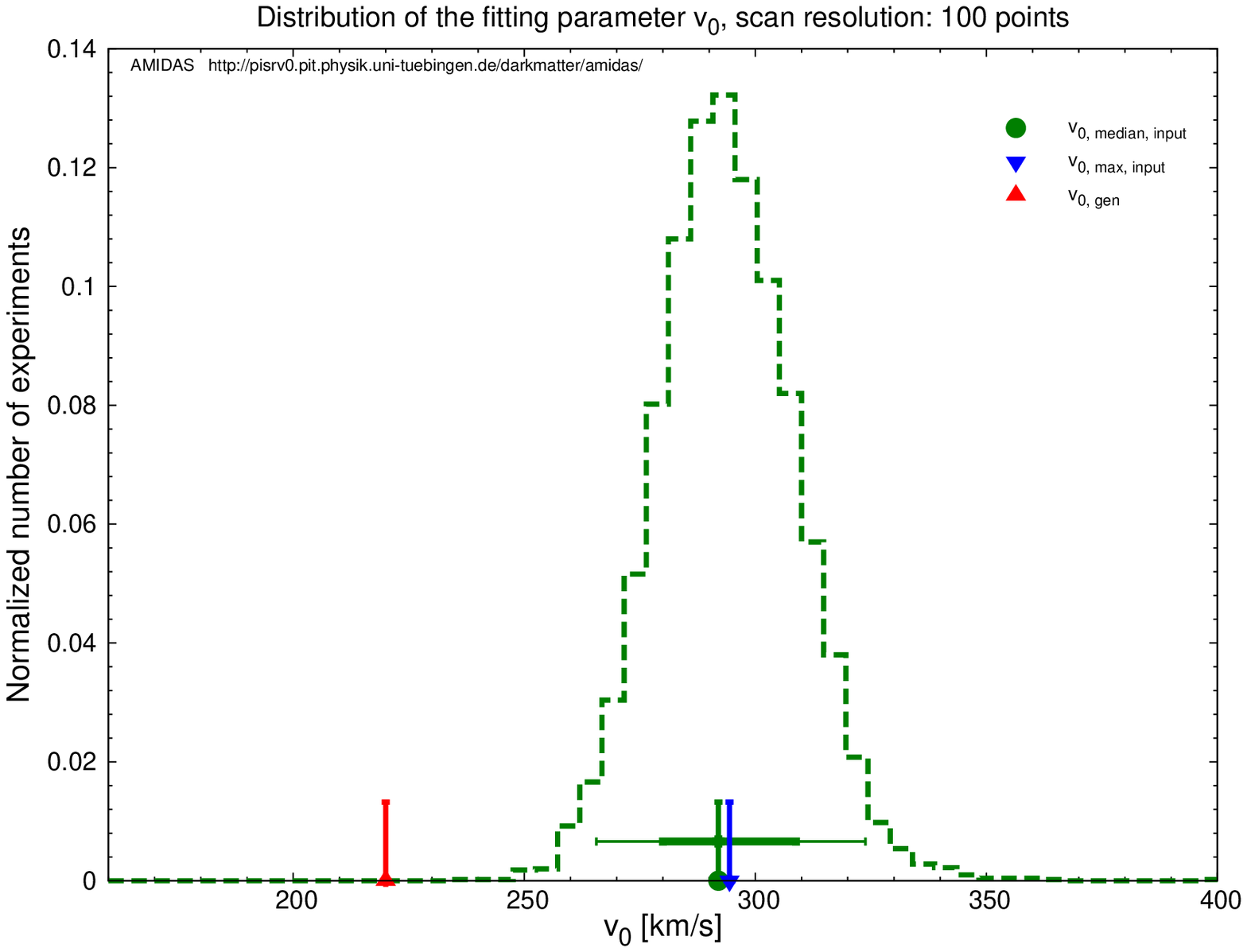}      \hspace*{-1.6cm} \par
\makebox[8.5cm]{(a)}\hspace{0.325cm}\makebox[8.175cm]{(b)}        \\ \vspace{0.5cm}
\hspace*{-1.6cm}
\includegraphics[width=8.5cm]{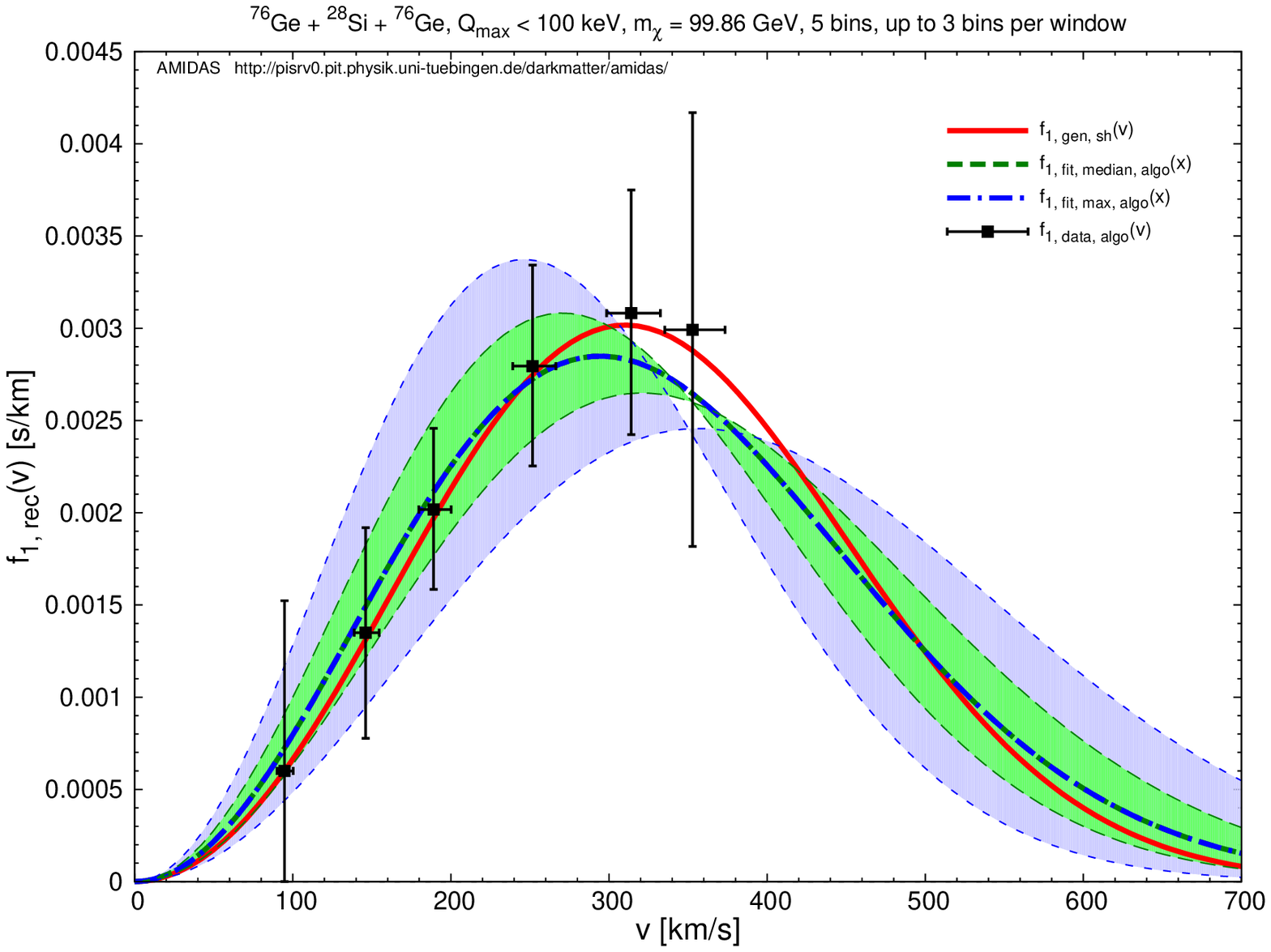}
\includegraphics[width=8.5cm]{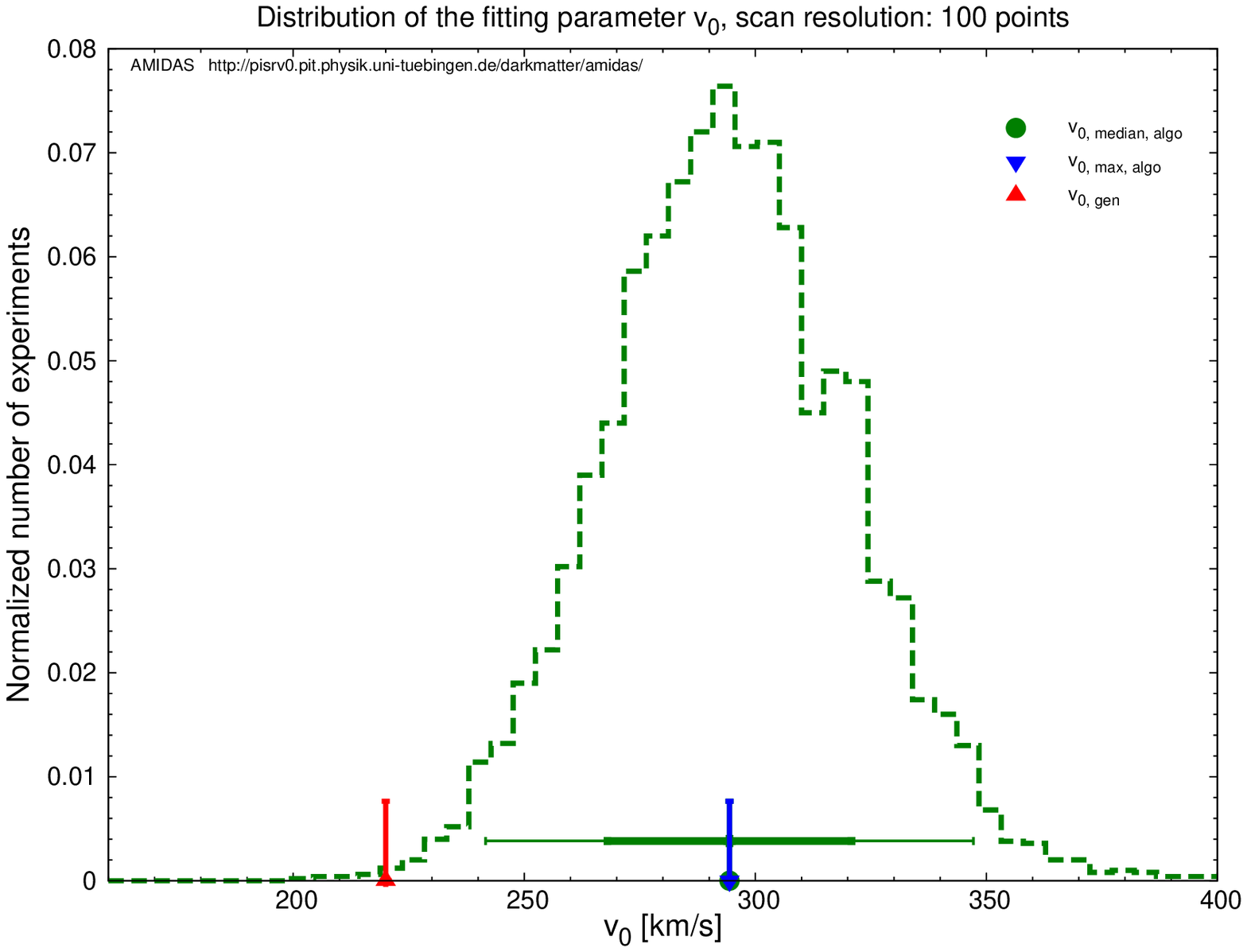} \hspace*{-1.6cm} \par
\makebox[8.5cm]{(c)}\hspace{0.325cm}\makebox[8.175cm]{(d)}%
}
\vspace{-0.35cm}
\end{center}
\caption{
 As in Figs.~\ref{fig:f1v-Ge-SiGe-100-0500-sh-Gau-flat},
 except that
 the Gaussian probability distribution
 for $v_0$
 with an expectation value of \mbox{$v_0 = 280$ km/s}
 and a 1$\sigma$ uncertainty of \mbox{40 km/s}
 has been used.
}
\label{fig:f1v-Ge-SiGe-100-0500-sh-Gau-Gau}
\end{figure}
}
\newcommand{\plotGeSiGeshshvflat}{
\begin{figure}[t!]
\begin{center}
\vspace{-0.25cm}
{
\hspace*{-1.6cm}
\includegraphics[width=8.5cm]{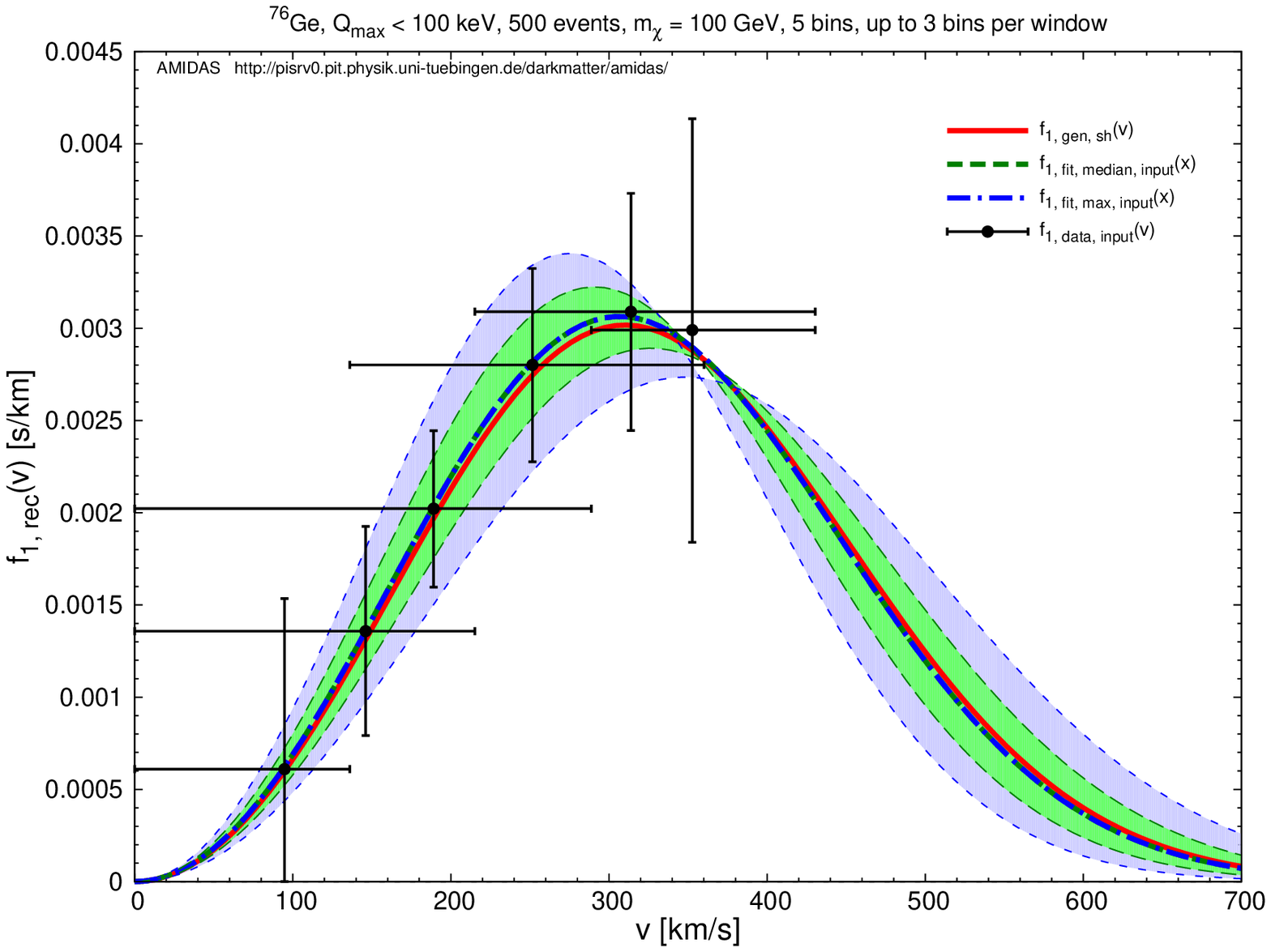}
\includegraphics[width=8.5cm]{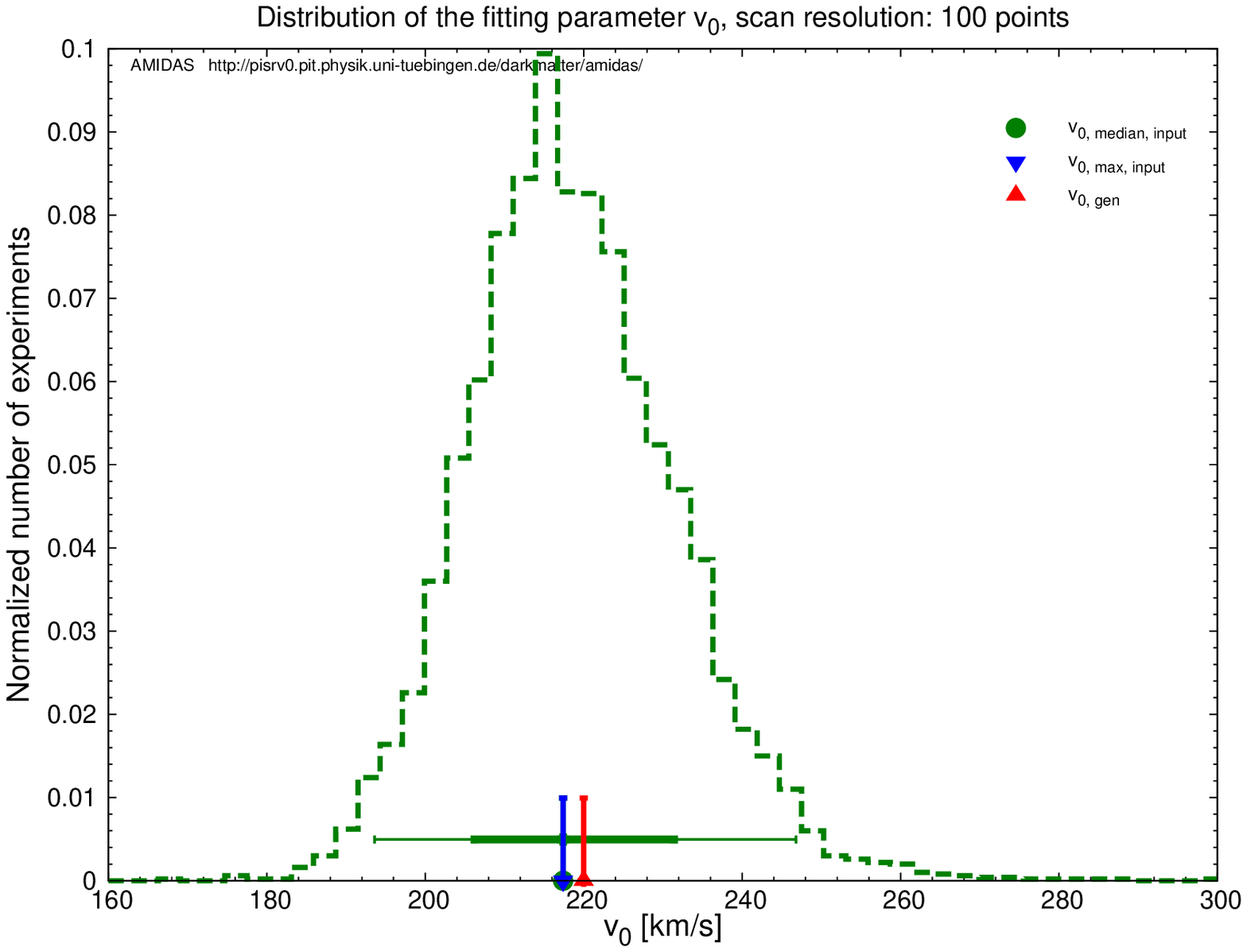}      \hspace*{-1.6cm} \par
\makebox[8.5cm]{(a)}\hspace{0.325cm}\makebox[8.175cm]{(b)}           \\ \vspace{0.5cm}
\hspace*{-1.6cm}
\includegraphics[width=8.5cm]{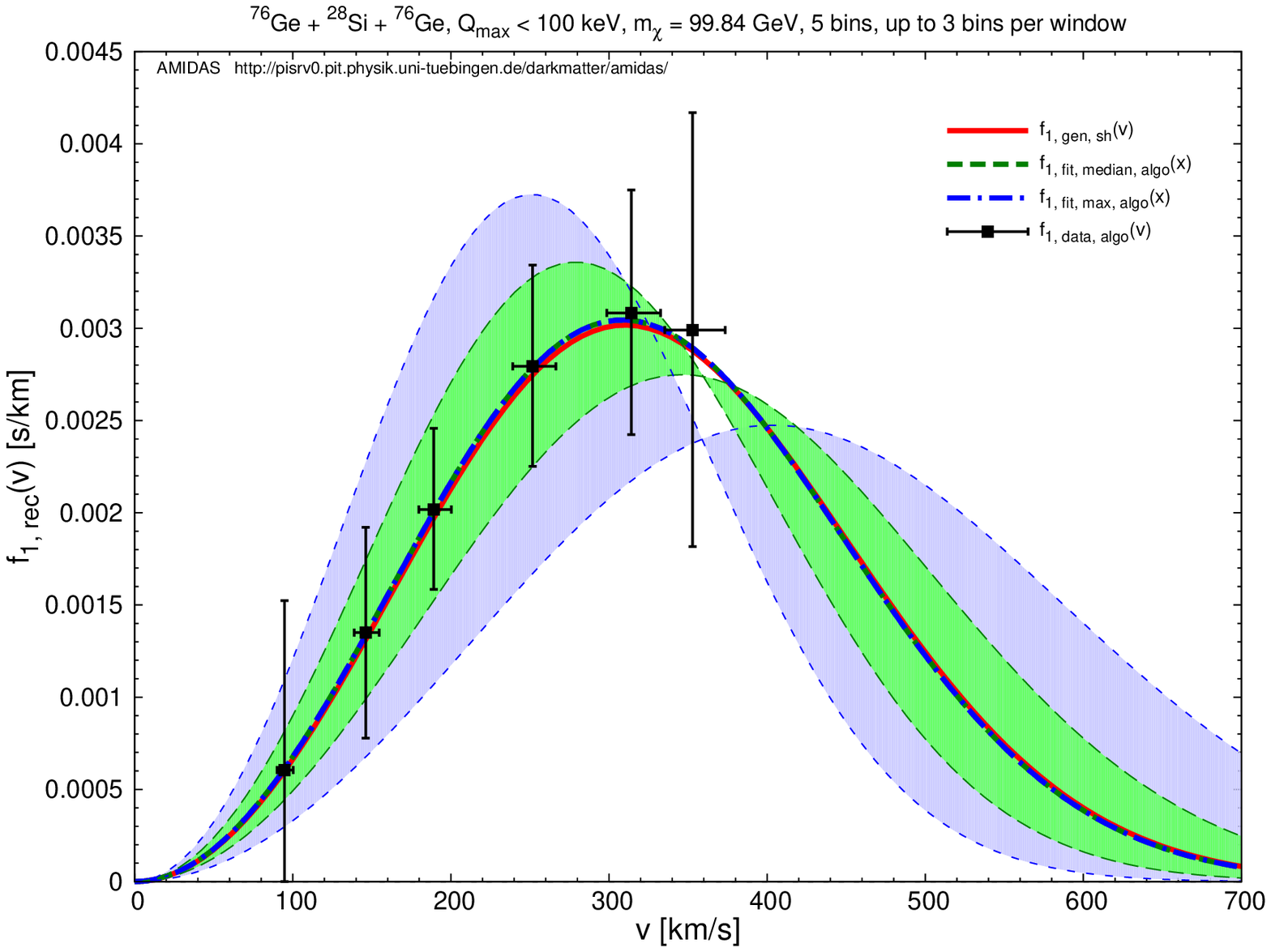}
\includegraphics[width=8.5cm]{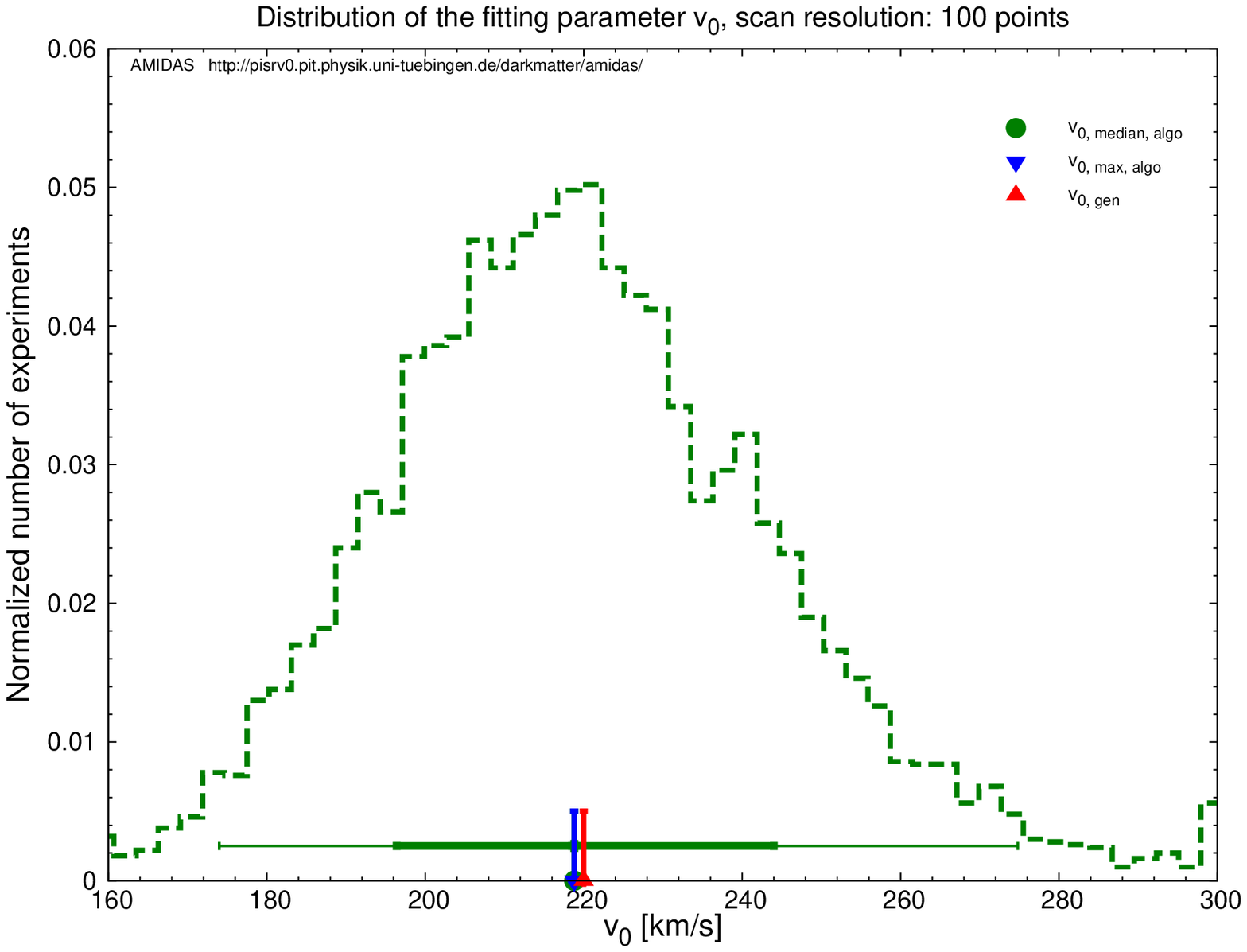} \hspace*{-1.6cm} \par
\makebox[8.5cm]{(c)}\hspace{0.325cm}\makebox[8.175cm]{(d)}%
}
\vspace{-0.35cm}
\end{center}
\caption{
 As in Figs.~\ref{fig:f1v-Ge-SiGe-100-0500-sh-Gau-flat},
 except that
 the one--parameter shifted
 Maxwellian velocity distribution function $f_{1, \sh, v_0}(v)$
 with the unique fitting parameter $v_0$
 has been used
 as the fitting velocity distribution.
}
\label{fig:f1v-Ge-SiGe-100-0500-sh-sh_v0-flat}
\end{figure}
}
\newcommand{\plotGeSiGeshshvGau}{
\begin{figure}[t!]
\begin{center}
\vspace{-0.25cm}
{
\hspace*{-1.6cm}
\includegraphics[width=8.5cm]{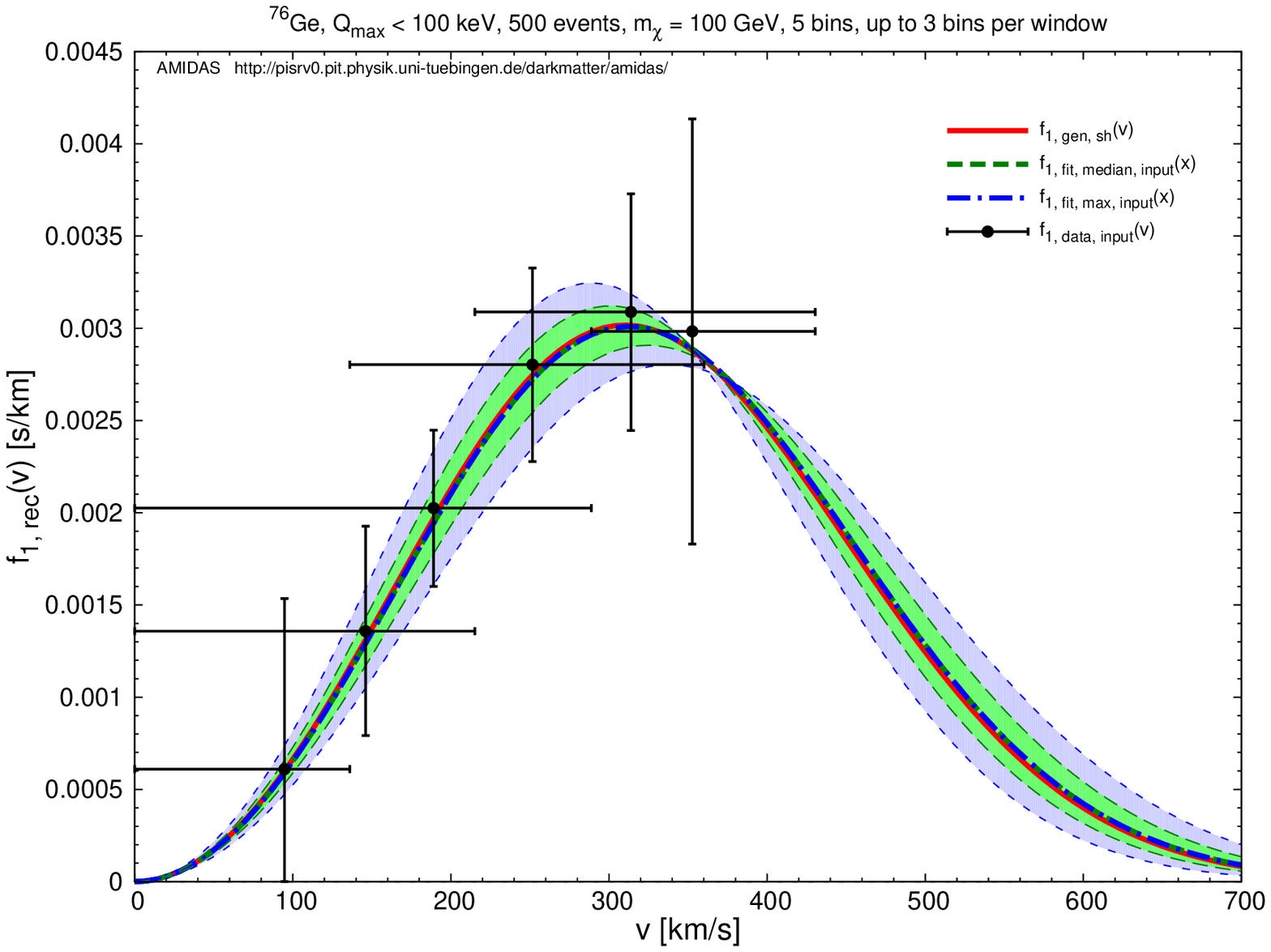}
\includegraphics[width=8.5cm]{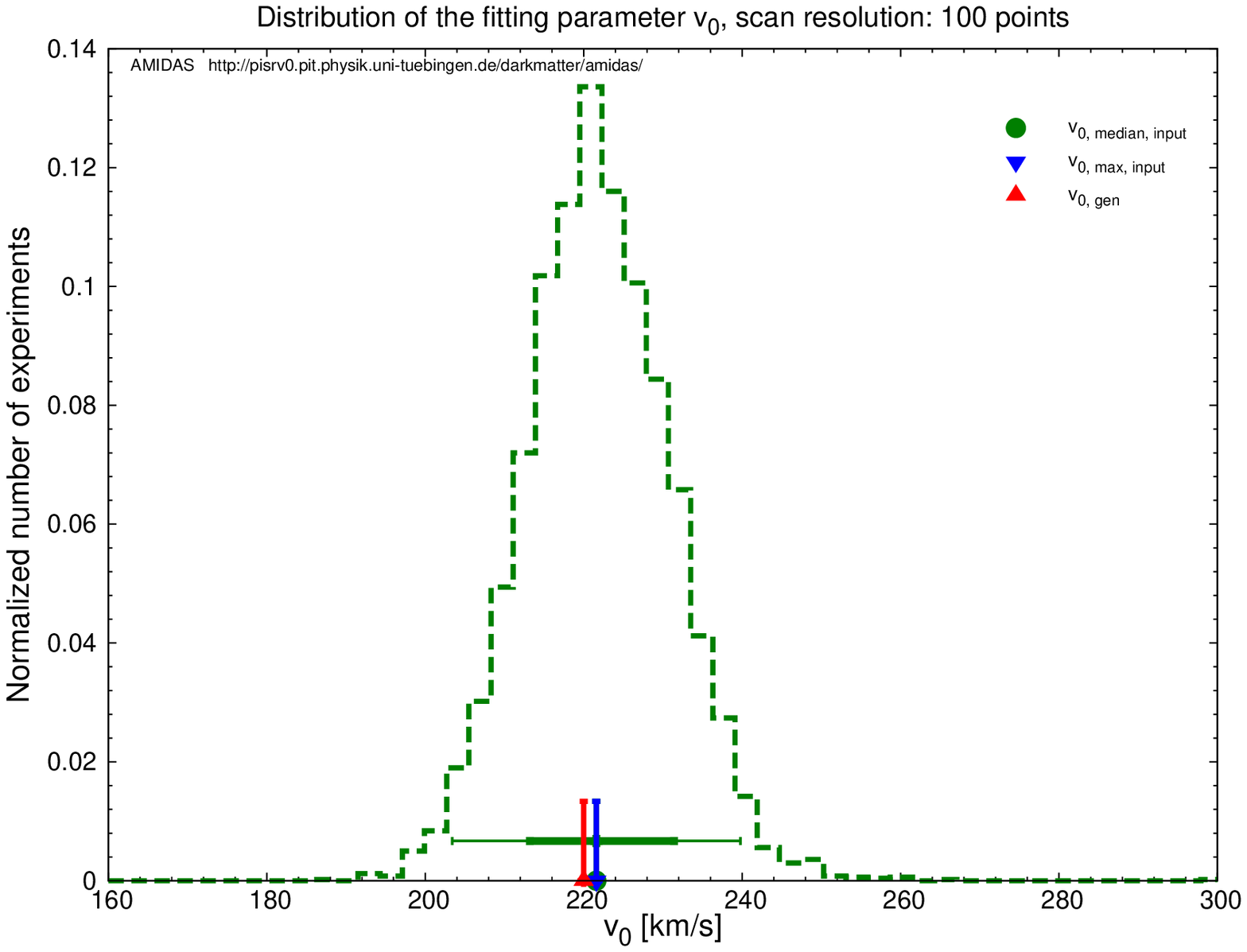}      \hspace*{-1.6cm} \par
\makebox[8.5cm]{(a)}\hspace{0.325cm}\makebox[8.175cm]{(b)}          \\ \vspace{0.5cm}
\hspace*{-1.6cm}
\includegraphics[width=8.5cm]{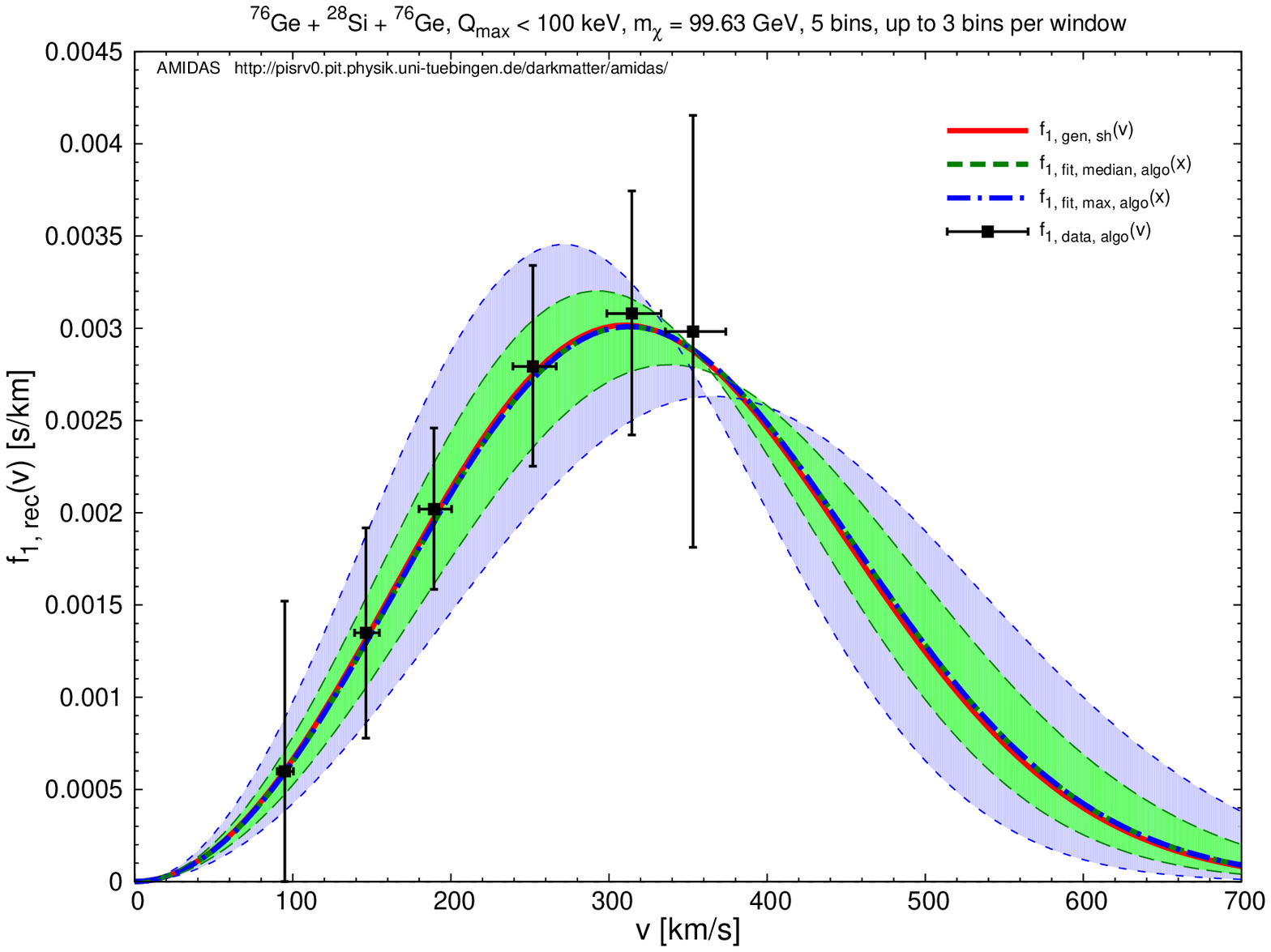}
\includegraphics[width=8.5cm]{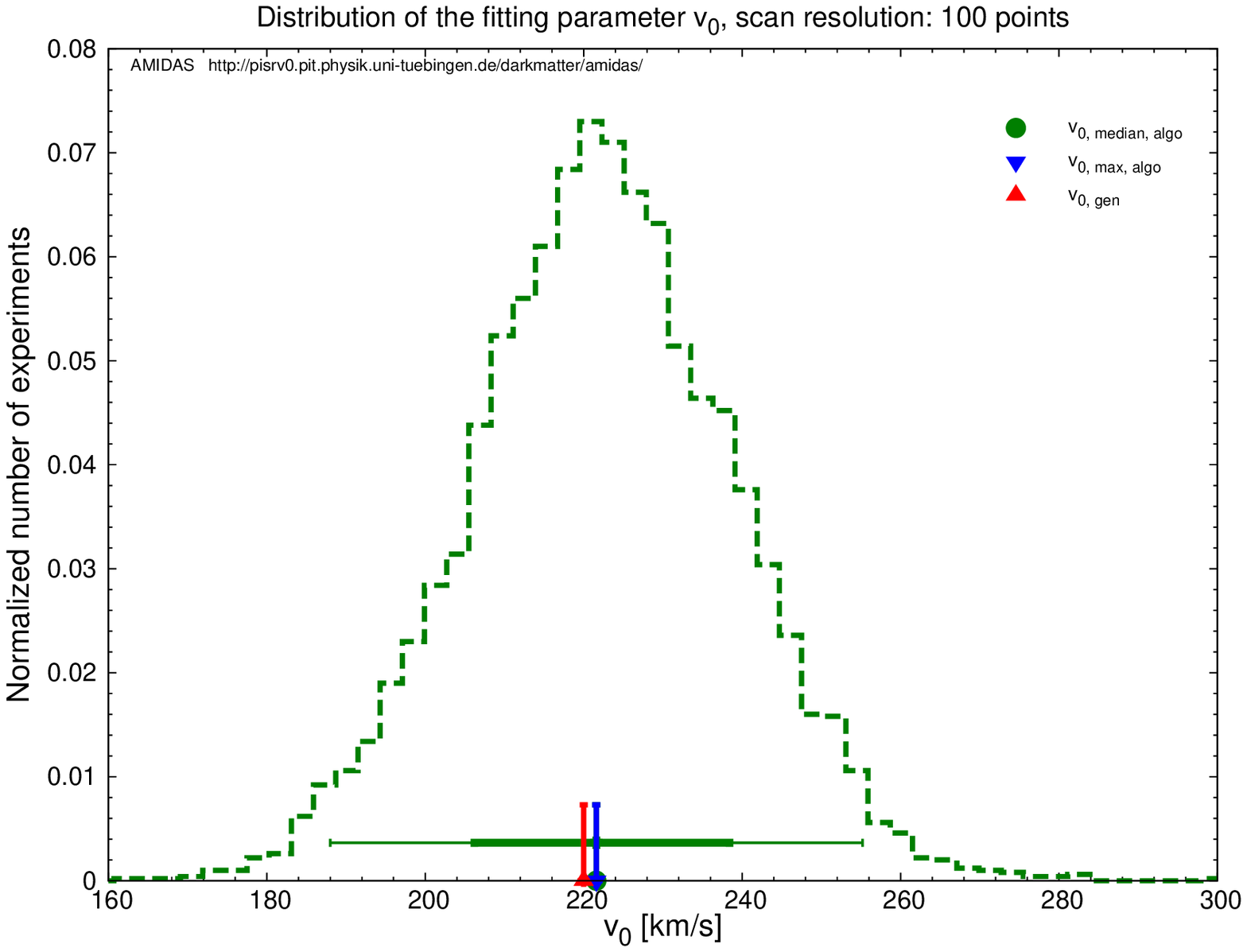} \hspace*{-1.6cm} \par
\makebox[8.5cm]{(c)}\hspace{0.325cm}\makebox[8.175cm]{(d)}%
}
\vspace{-0.35cm}
\end{center}
\caption{
 As in Figs.~\ref{fig:f1v-Ge-SiGe-100-0500-sh-sh_v0-flat},
 except that
 the Gaussian probability distribution
 for $v_0$
 with an expectation value of \mbox{$v_0 = 230$ km/s}
 and a 1$\sigma$ uncertainty of \mbox{20 km/s}
 has been used.
}
\label{fig:f1v-Ge-SiGe-100-0500-sh-sh_v0-Gau}
\end{figure}
}
\newcommand{\plotGeshshGau}{
\begin{figure}[t!]
\begin{center}
\vspace{-0.25cm}
{
\hspace*{-1.6cm}
\includegraphics[width=8.5cm]{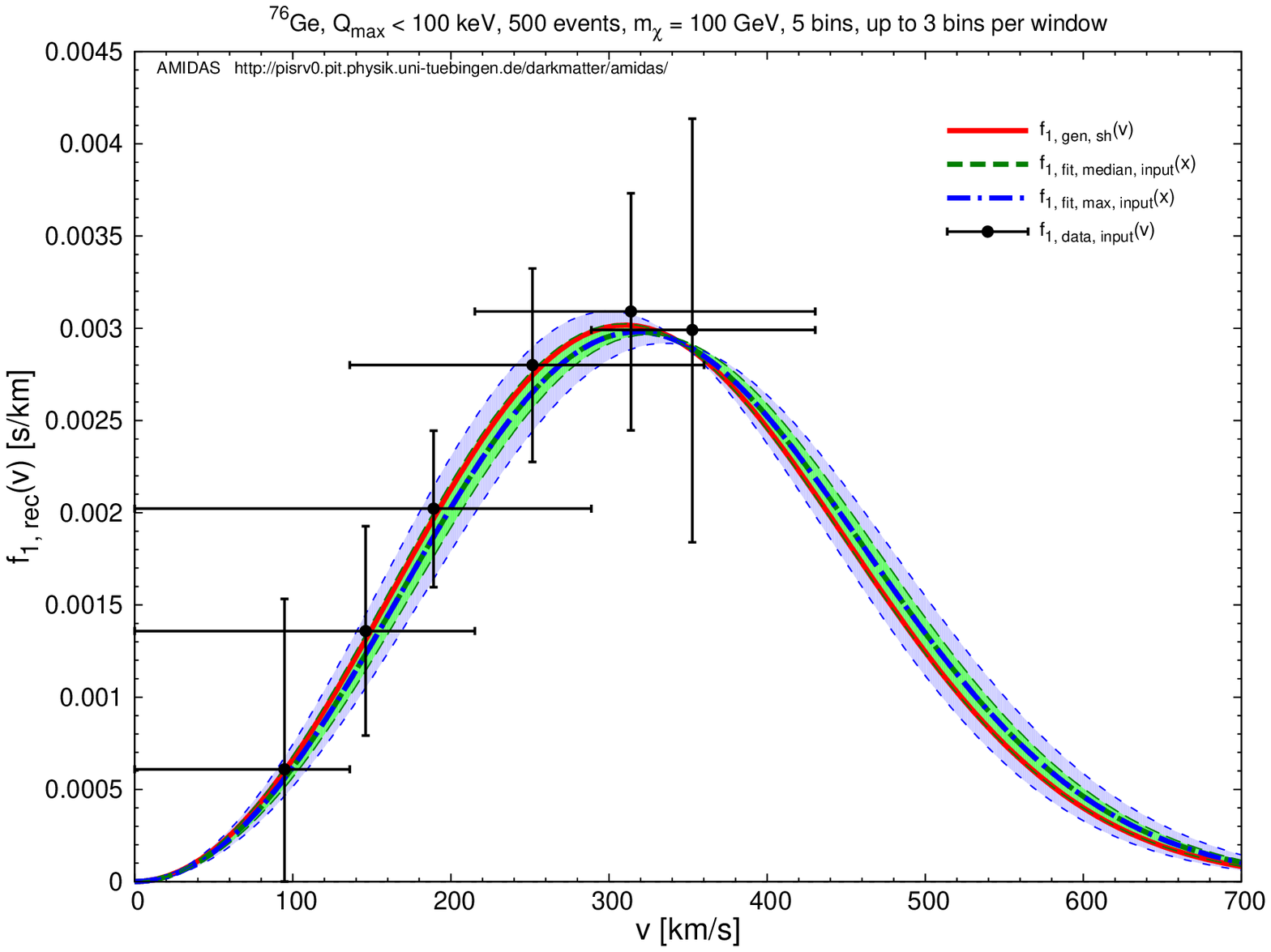}
\includegraphics[width=8.5cm]{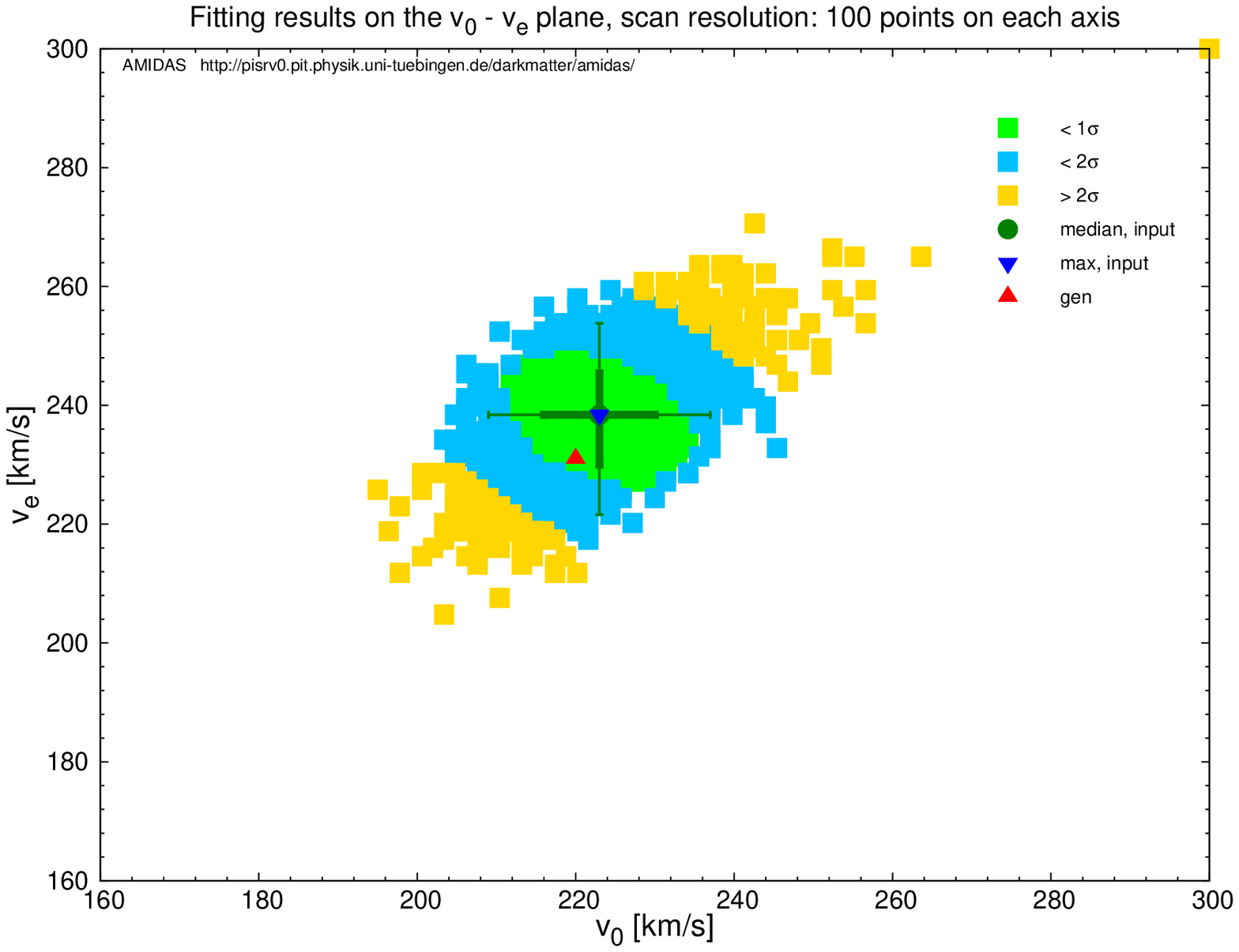} \hspace*{-1.6cm} \par
\makebox[8.5cm]{(a)}\hspace{0.325cm}\makebox[8.175cm]{(b)}     \\ \vspace{0.5cm}
\hspace*{-1.6cm}
\includegraphics[width=8.5cm]{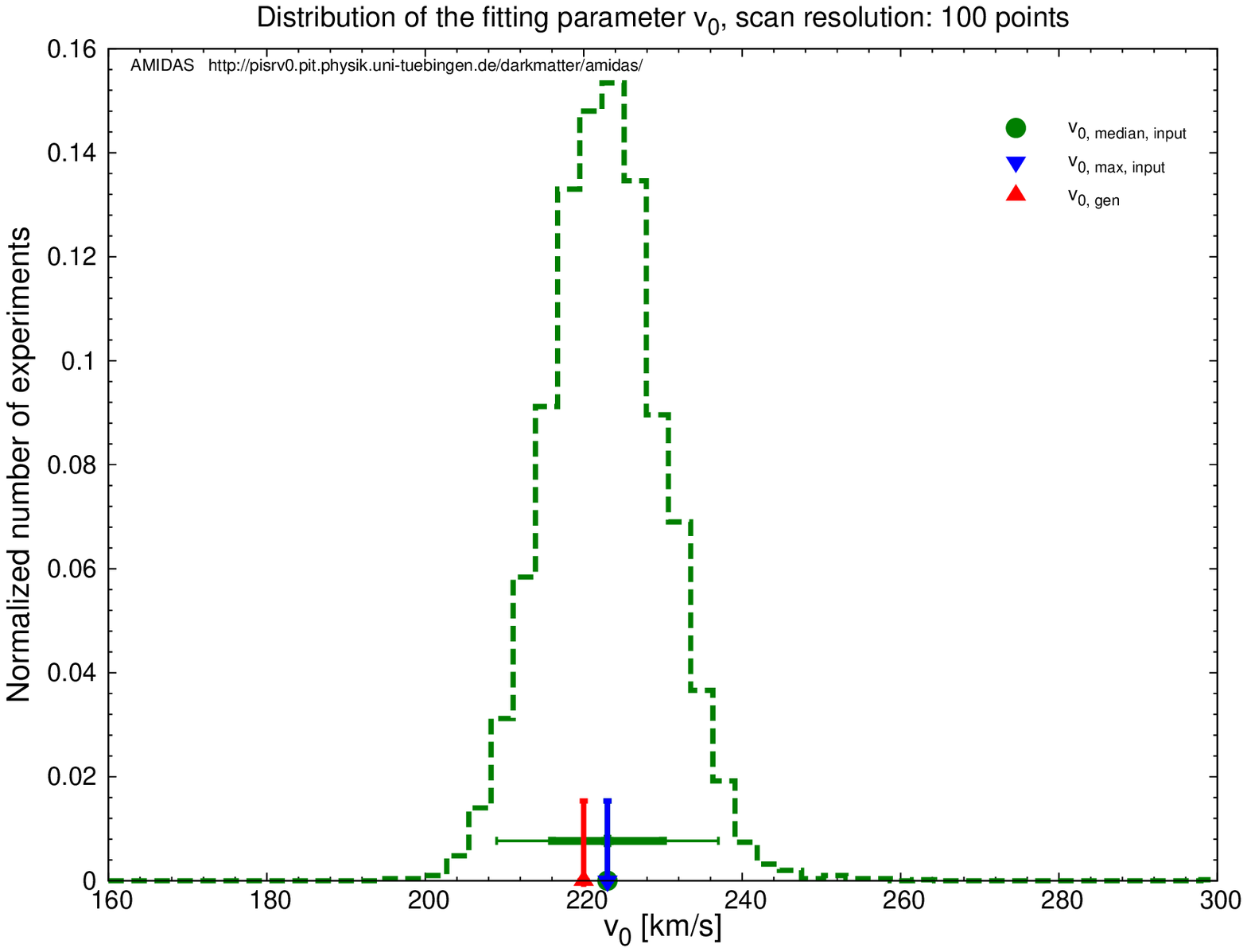}
\includegraphics[width=8.5cm]{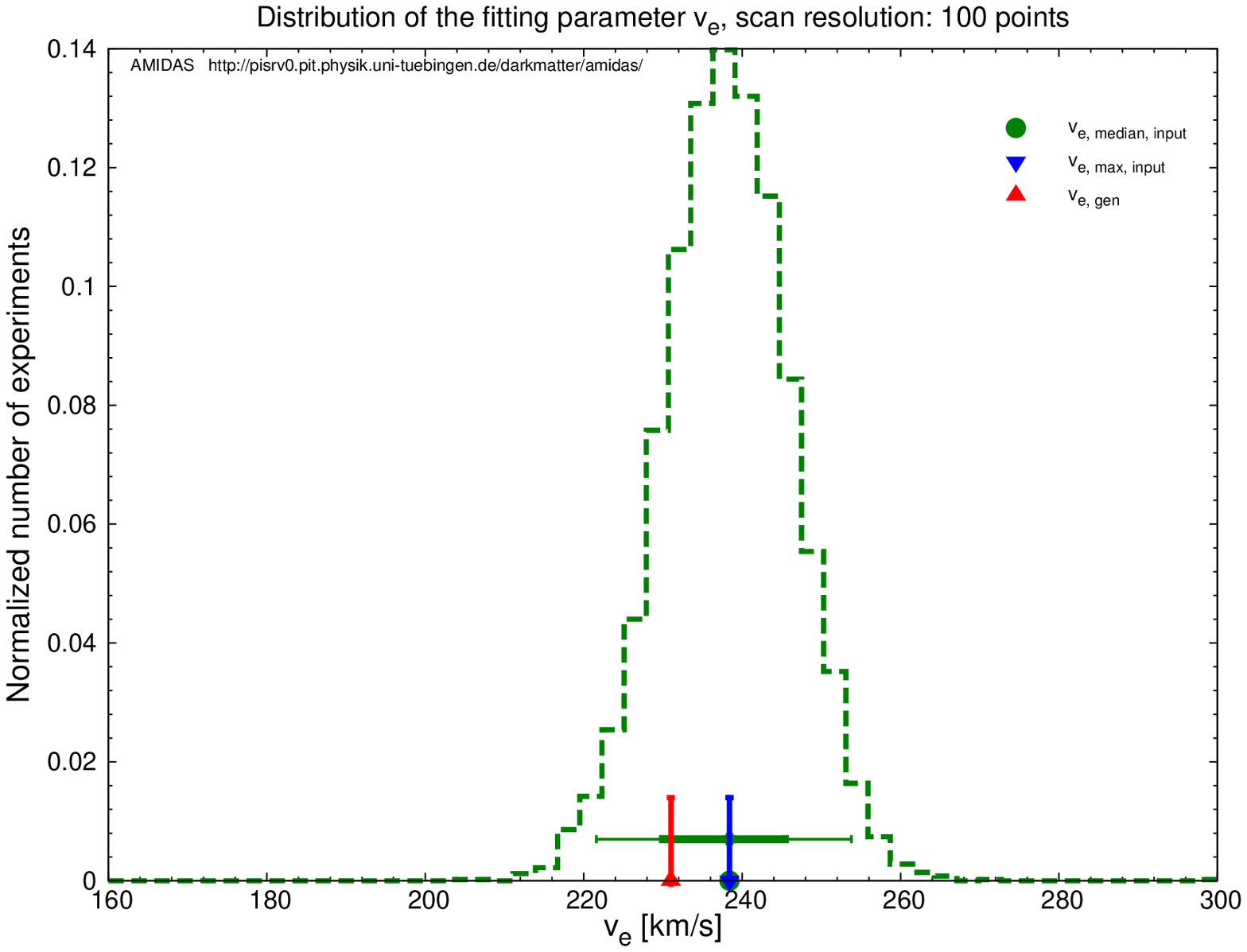}    \hspace*{-1.6cm} \par
\makebox[8.5cm]{(c)}\hspace{0.325cm}\makebox[8.175cm]{(d)}     \\
}
\vspace{-0.35cm}
\end{center}
\caption{
 (a)
 As in Fig.~\ref{fig:f1v-Ge-SiGe-100-0500-sh-sh_v0-flat}(a),
 except that
 the shifted Maxwellian velocity distribution function
 given in Eq.~(\ref{eqn:f1v_sh})
 with two fitting parameters $v_0$ and $\ve$
 is used.
 The Gaussian probability distribution
 for both fitting parameters
 with expectation values of \mbox{$v_0 = 230$ km/s}
 and \mbox{$\ve = 245$ km/s}
 and a common 1$\sigma$ uncertainty of \mbox{20 km/s}
 has been used.
 (b)
 The distribution of
 the Bayesian reconstructed fitting parameters $v_0$ and $\ve$
 in all simulated experiments
 on the $v_0 - \ve$ plane.
 The light--green (light--blue, gold) points
 indicate the $1\~(2)\~(> 2)\~\sigma$ areas
 of the reconstructed combination of $v_0$ and $\ve$.
 The red upward--triangle indicates
 the input values of $v_0$ and $\ve$,
 which has been labeled with the subscript ``gen''.
 The green disk shows
 the median values of the simulated results,
 whereas
 the blue downward--triangle
 the point which maximizes
 ${\rm p}_{\rm median}$.
 The meaning of
 the horizontal thick (thin) green bars
 are the same as
 in Fig.~\ref{fig:f1v-Ge-SiGe-100-0500-sh-sh_v0-flat}(b).
 (c)
 As in Fig.~\ref{fig:f1v-Ge-SiGe-100-0500-sh-sh_v0-flat}(b).
 (d)
 Similar to (c):
 the distribution of
 the Bayesian reconstructed second fitting parameter $\ve$.
 See the text for further details.
}
\label{fig:f1v-Ge-100-0500-sh-sh-Gau}
\end{figure}
}
\newcommand{\plotGeSiGeshshGau}{
\begin{figure}[t!]
\begin{center}
\vspace{-0.25cm}
{
\hspace*{-1.6cm}
\includegraphics[width=8.5cm]{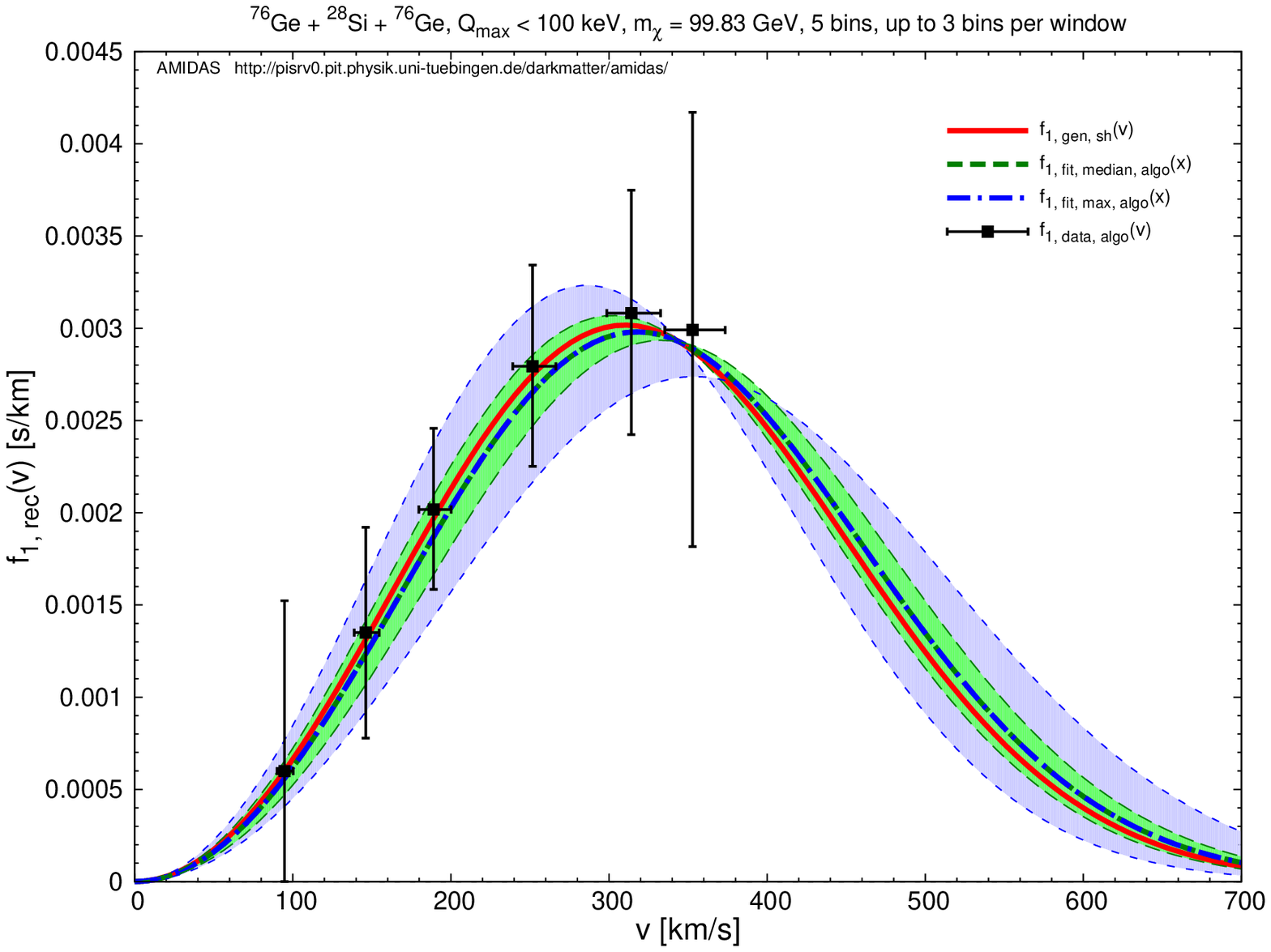}
\includegraphics[width=8.5cm]{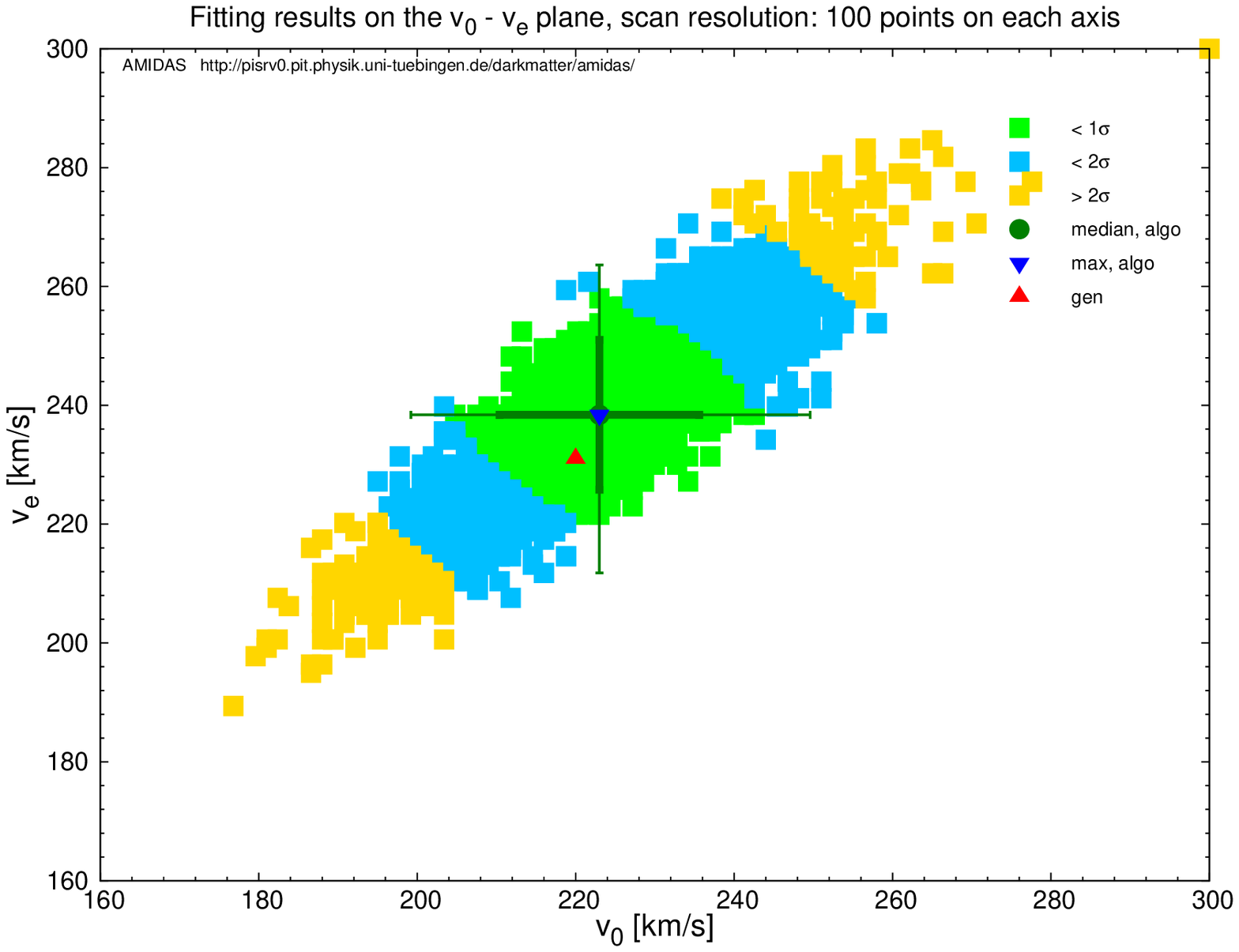} \hspace*{-1.6cm} \par
\makebox[8.5cm]{(a)}\hspace{0.325cm}\makebox[8.175cm]{(b)}          \\ \vspace{0.5cm}
\hspace*{-1.6cm}
\includegraphics[width=8.5cm]{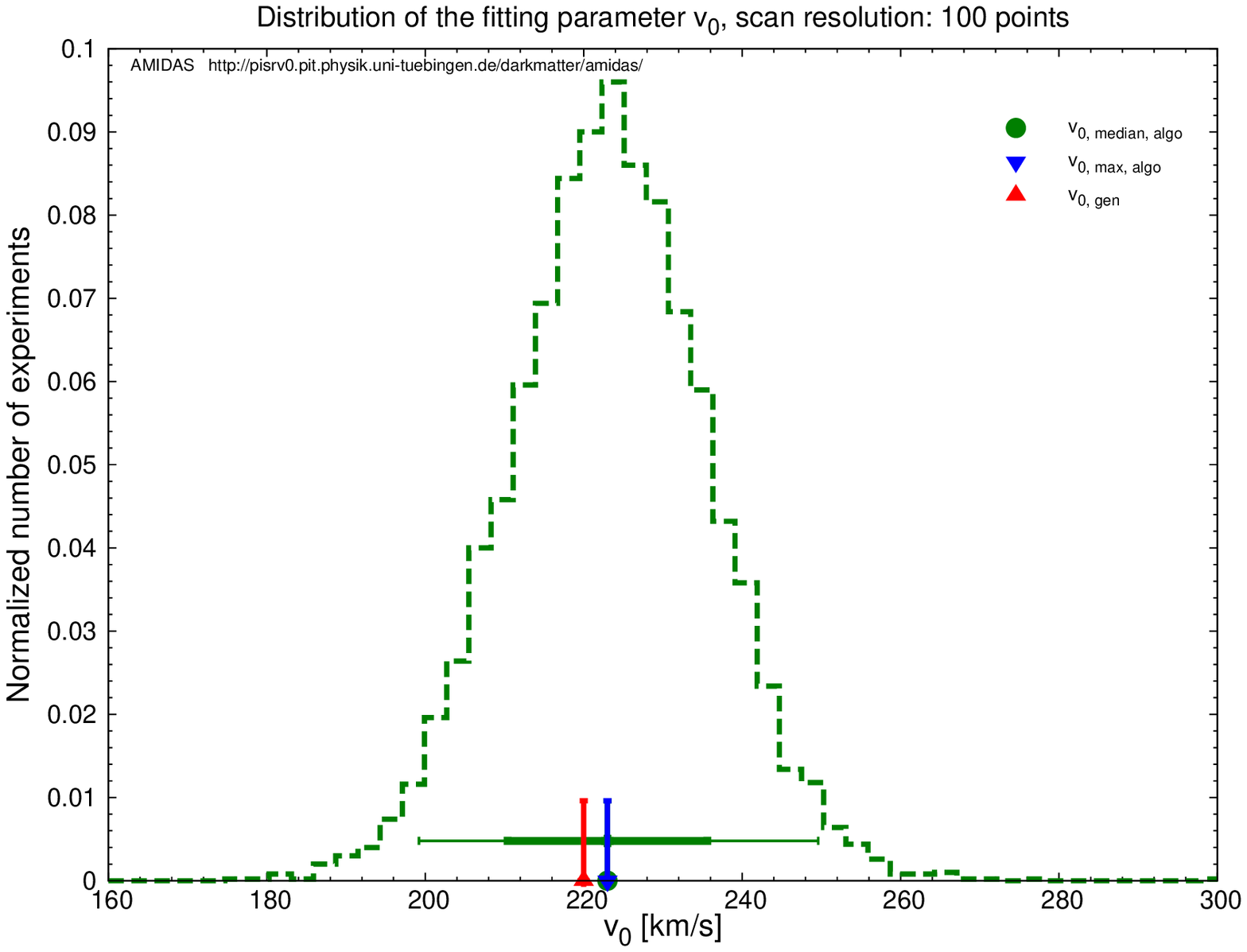}
\includegraphics[width=8.5cm]{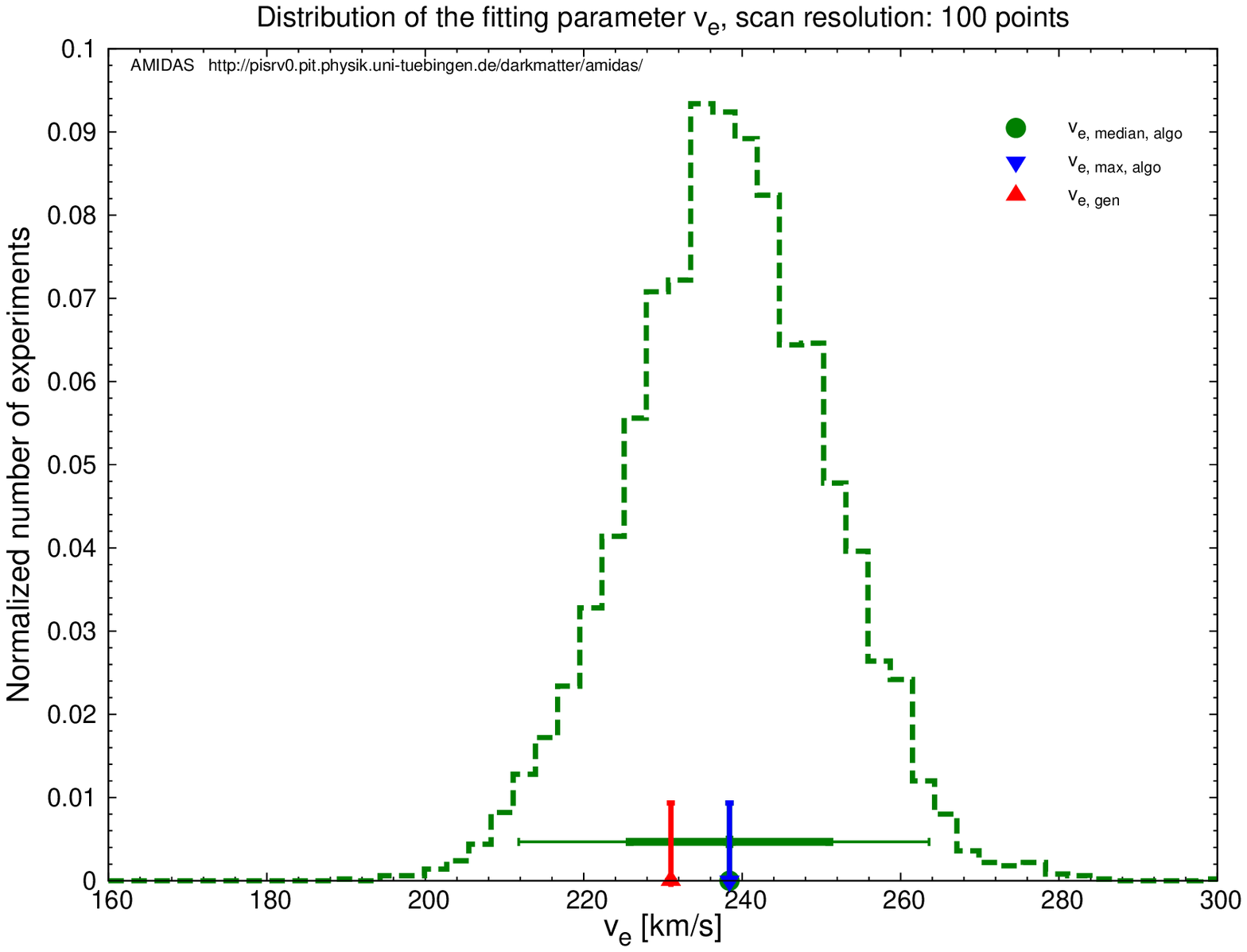}    \hspace*{-1.6cm} \par
\makebox[8.5cm]{(c)}\hspace{0.325cm}\makebox[8.175cm]{(d)}          \\
}
\vspace{-0.35cm}
\end{center}
\caption{
 As in Figs.~\ref{fig:f1v-Ge-100-0500-sh-sh-Gau},
 except that
 the reconstructed WIMP mass
 has been
 used.
}
\label{fig:f1v-Ge-SiGe-100-0500-sh-sh-Gau}
\end{figure}
}
\newcommand{\plotGeshshDvGau}{
\begin{figure}[t!]
\begin{center}
\vspace{-0.25cm}
{
\hspace*{-1.6cm}
\includegraphics[width=8.5cm]{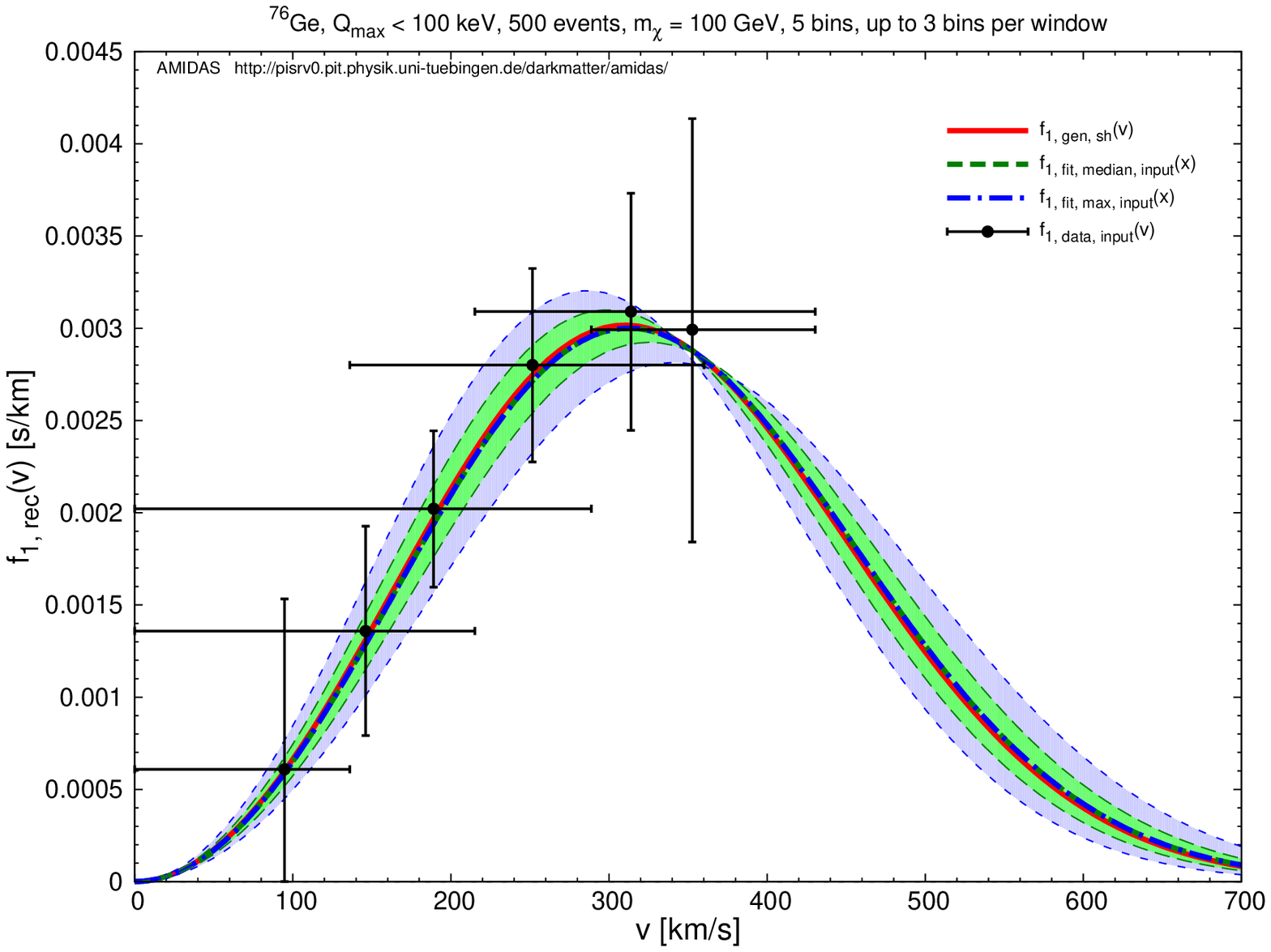}
\includegraphics[width=8.5cm]{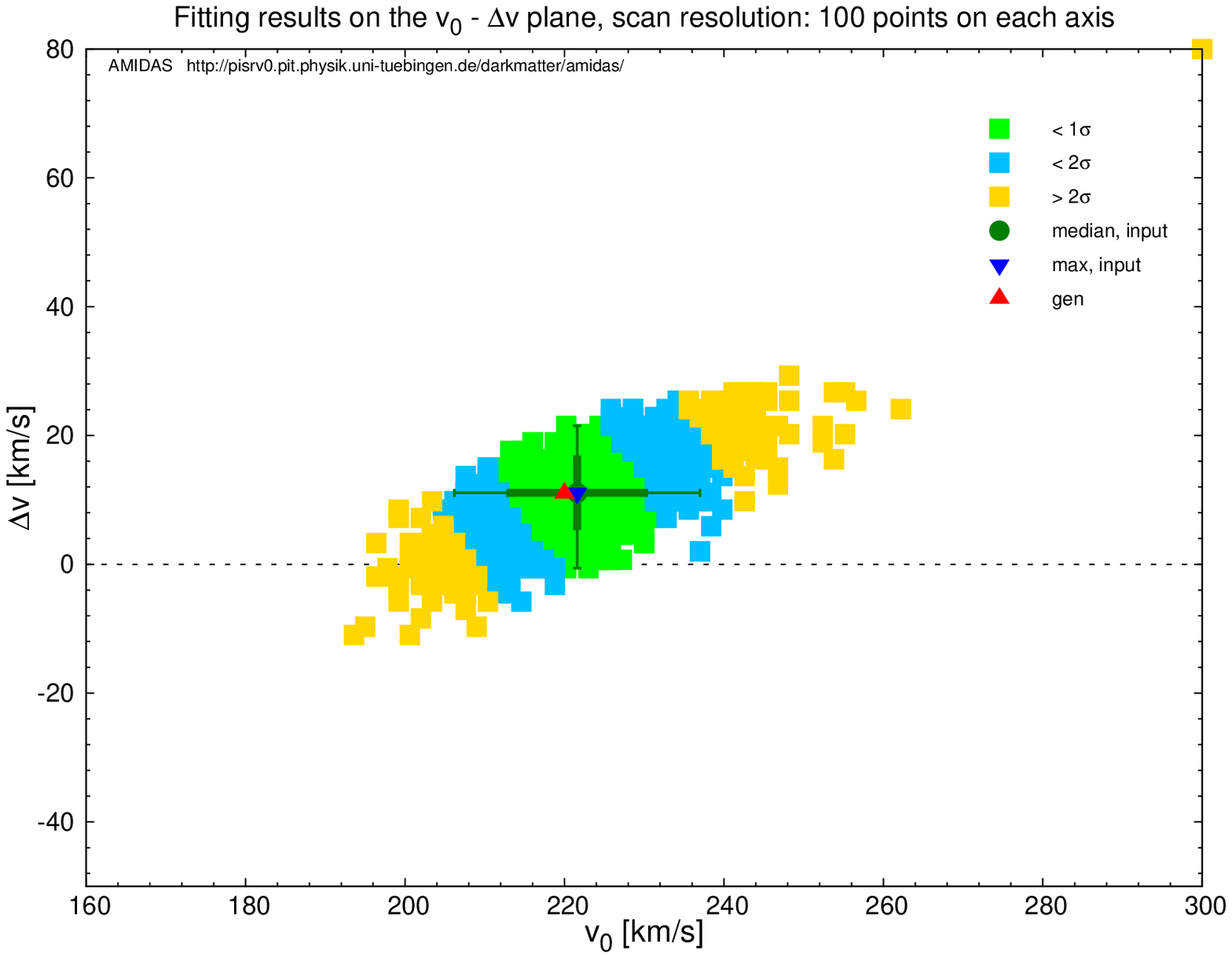} \hspace*{-1.6cm} \par
\makebox[8.5cm]{(a)}\hspace{0.325cm}\makebox[8.175cm]{(b)}        \\ \vspace{0.5cm}
\hspace*{-1.6cm}
\includegraphics[width=8.5cm]{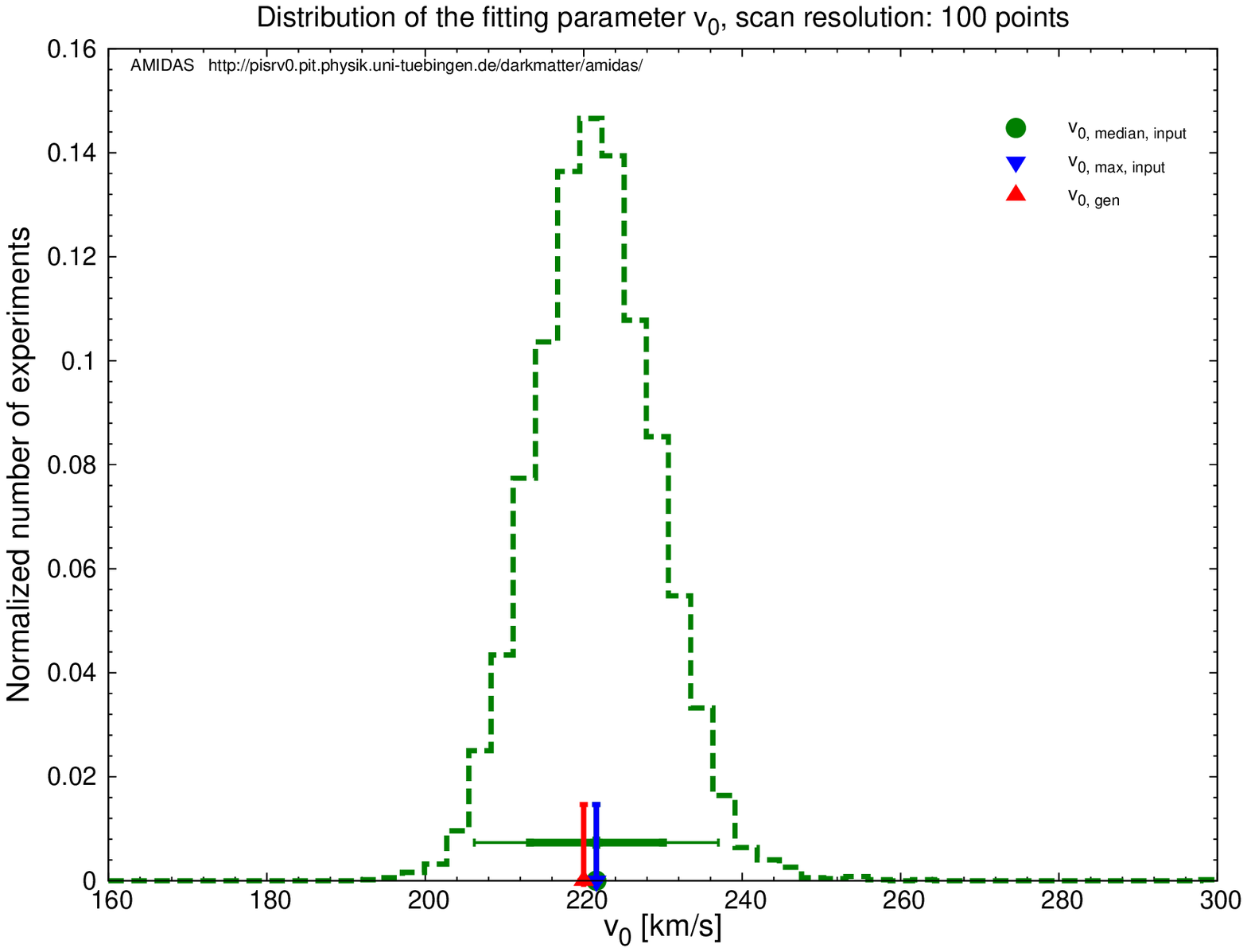}
\includegraphics[width=8.5cm]{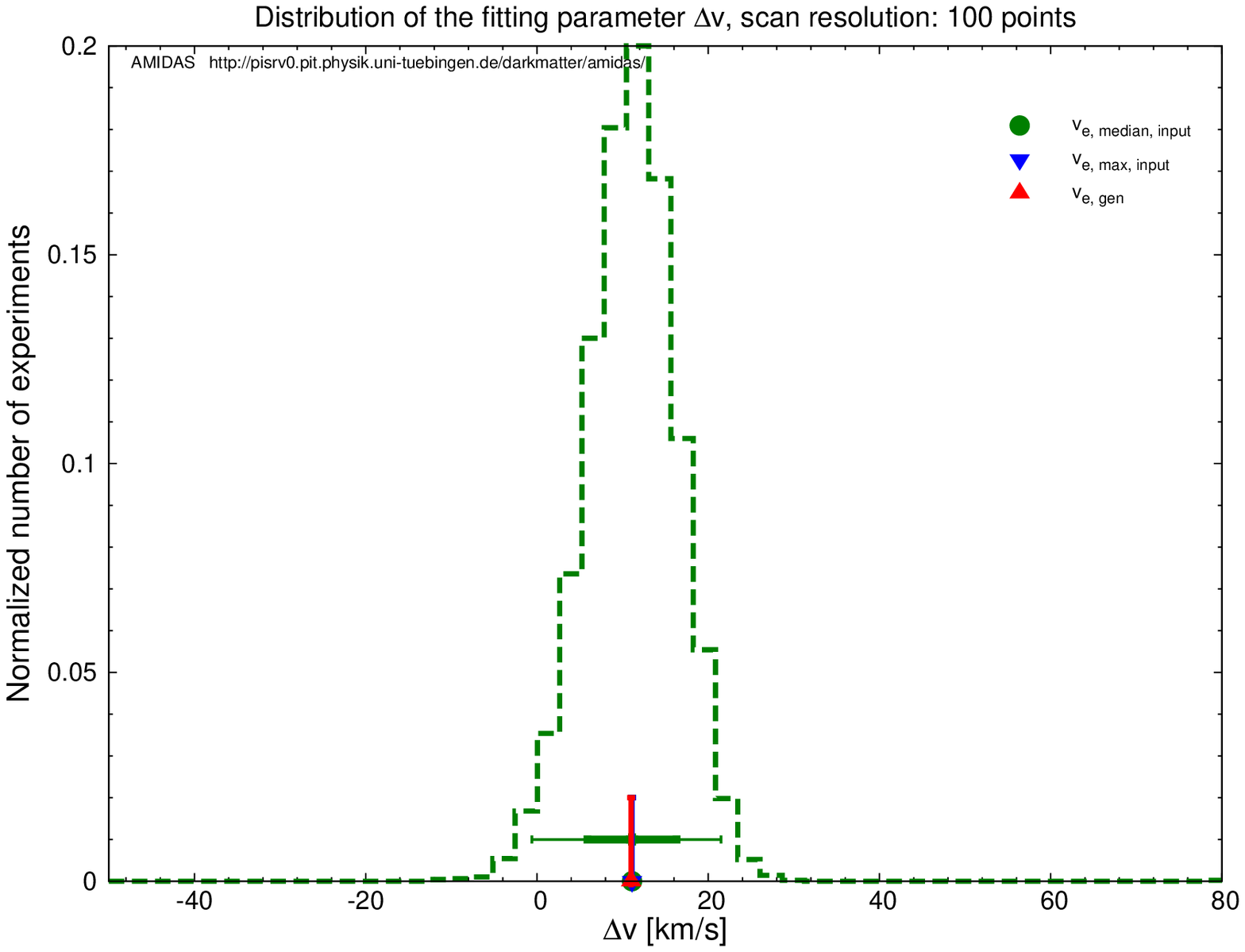}    \hspace*{-1.6cm} \par
\makebox[8.5cm]{(c)}\hspace{0.325cm}\makebox[8.175cm]{(d)}        \\
}
\vspace{-0.35cm}
\end{center}
\caption{
 As in Figs.~\ref{fig:f1v-Ge-100-0500-sh-sh-Gau},
 except that
 the variated shifted Maxwellian velocity distribution function
 given in Eq.~(\ref{eqn:f1v_sh_Dv})
 with two fitting parameters $v_0$ and $\Delta v$
 is used.
 The Gaussian probability distribution
 for both fitting parameters
 with expectation values of \mbox{$v_0 = 230$ km/s}
 and \mbox{$\Delta v = 15$ km/s}
 and a common 1$\sigma$ uncertainty of \mbox{20 km/s}
 has been used.
}
\label{fig:f1v-Ge-100-0500-sh-sh_Dv-Gau}
\end{figure}
}
\newcommand{\plotGeSiGeshshDvGau}{
\begin{figure}[t!]
\begin{center}
\vspace{-0.25cm}
{
\hspace*{-1.6cm}
\includegraphics[width=8.5cm]{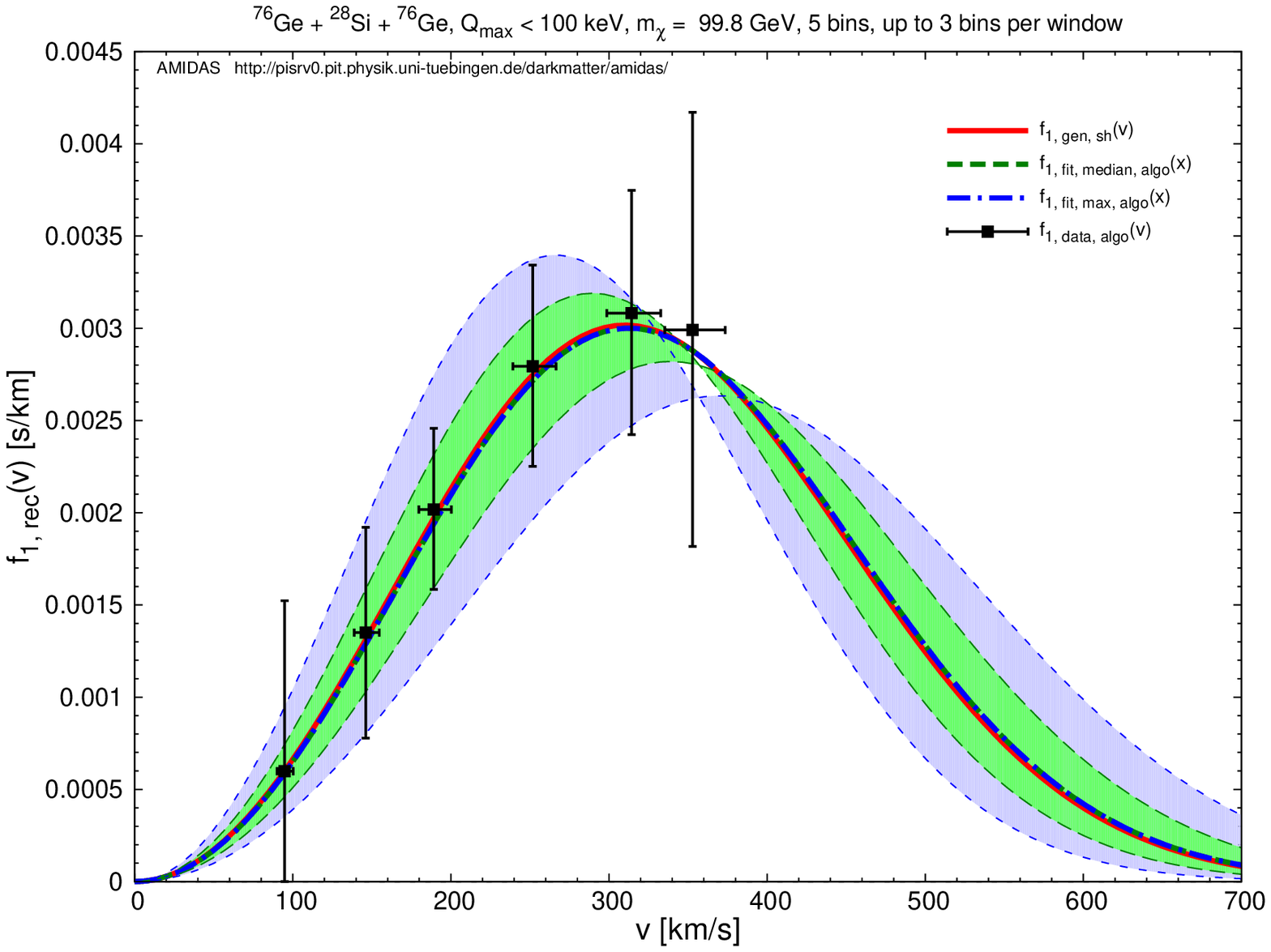}
\includegraphics[width=8.5cm]{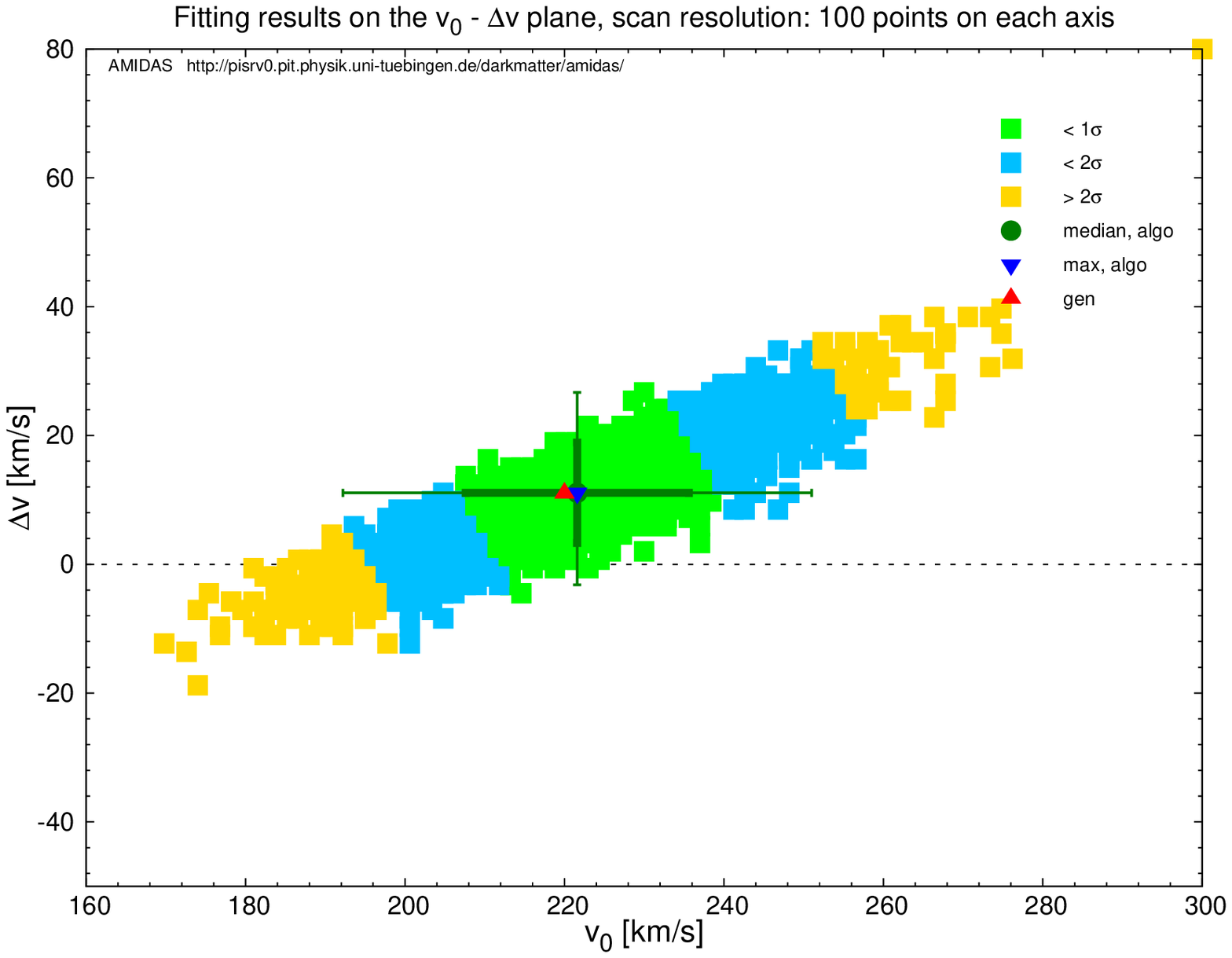} \hspace*{-1.6cm} \par
\makebox[8.5cm]{(a)}\hspace{0.325cm}\makebox[8.175cm]{(b)}             \\ \vspace{0.5cm}
\hspace*{-1.6cm}
\includegraphics[width=8.5cm]{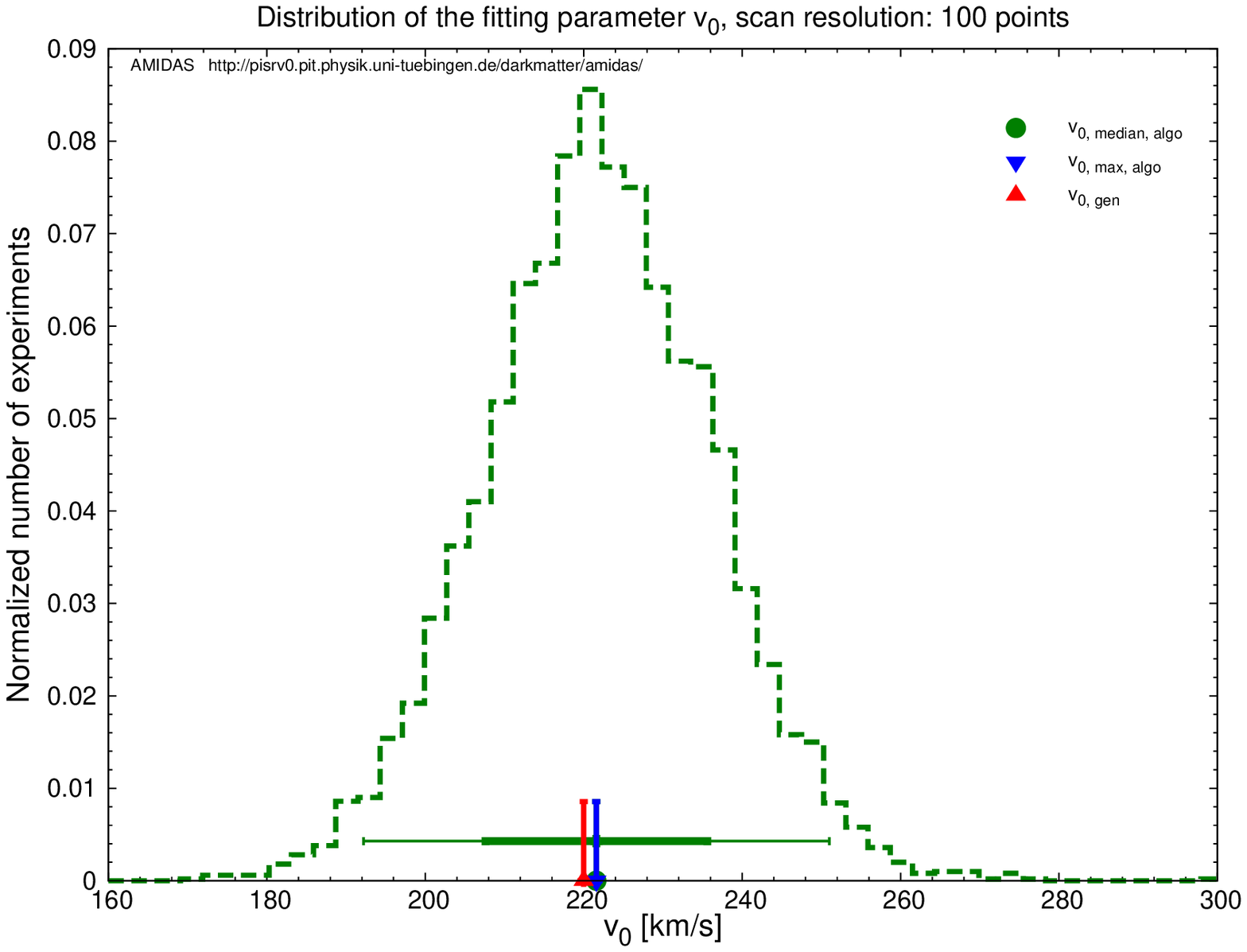}
\includegraphics[width=8.5cm]{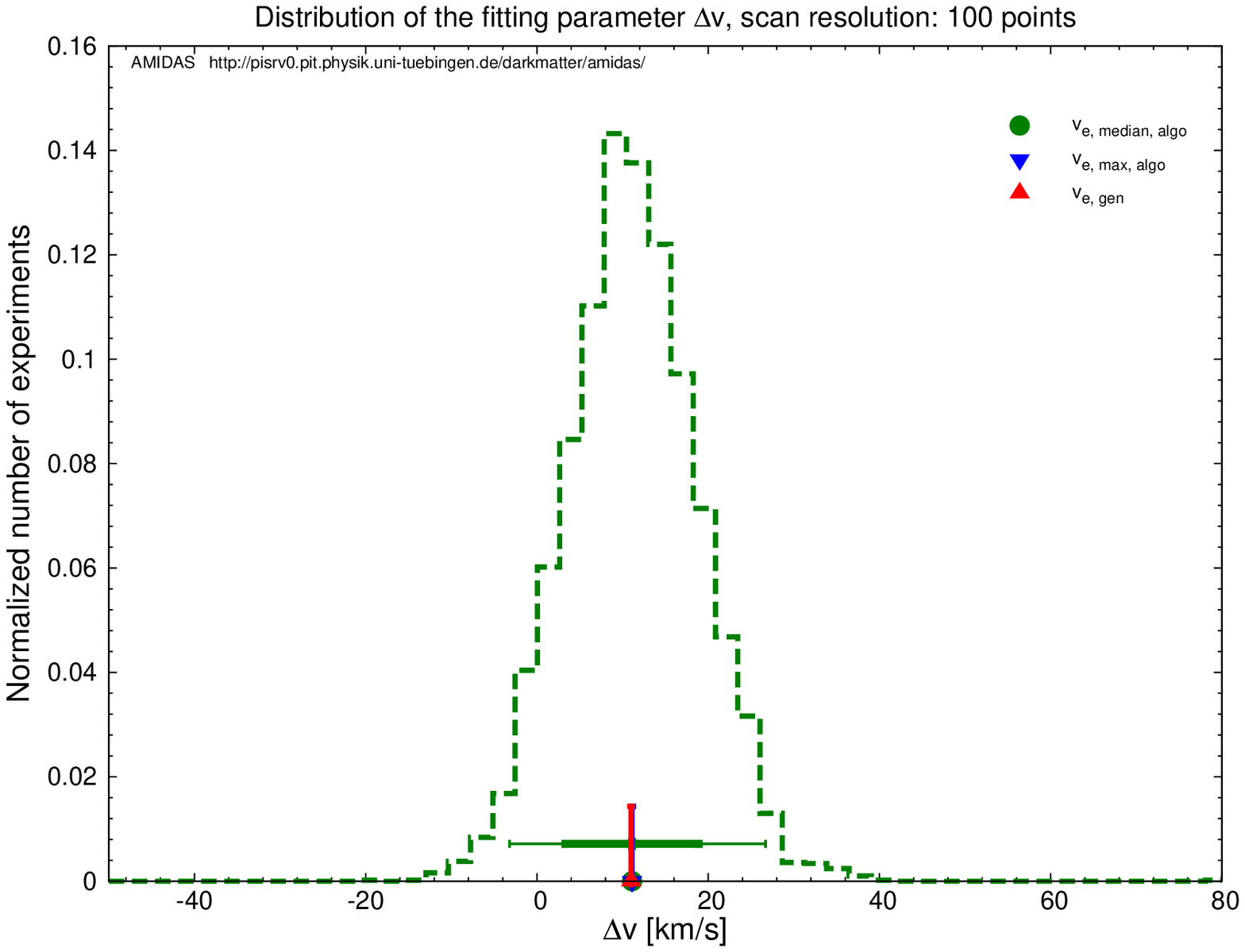}    \hspace*{-1.6cm} \par
\makebox[8.5cm]{(c)}\hspace{0.325cm}\makebox[8.175cm]{(d)}             \\
}
\vspace{-0.35cm}
\end{center}
\caption{
 As in Figs.~\ref{fig:f1v-Ge-100-0500-sh-sh_Dv-Gau},
 except that
 the reconstructed WIMP mass
 has been
 used.
}
\label{fig:f1v-Ge-SiGe-100-0500-sh-sh_Dv-Gau}
\end{figure}
}
\newcommand{\plotGeSiGeGaukGauflat}{
\begin{figure}[t!]
\begin{center}
\vspace{-0.25cm}
{
\hspace*{-1.6cm}
\includegraphics[width=8.5cm]{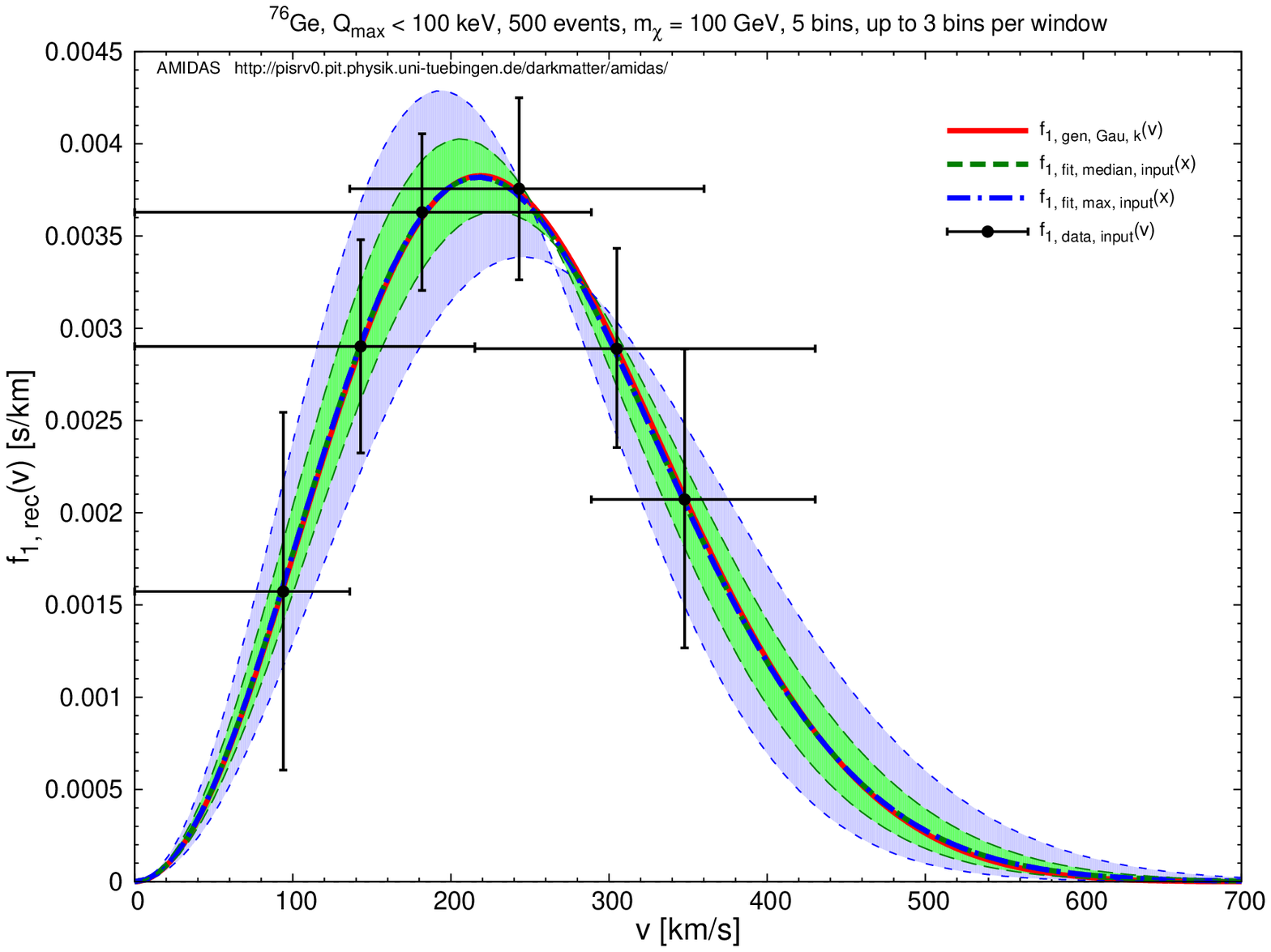}
\includegraphics[width=8.5cm]{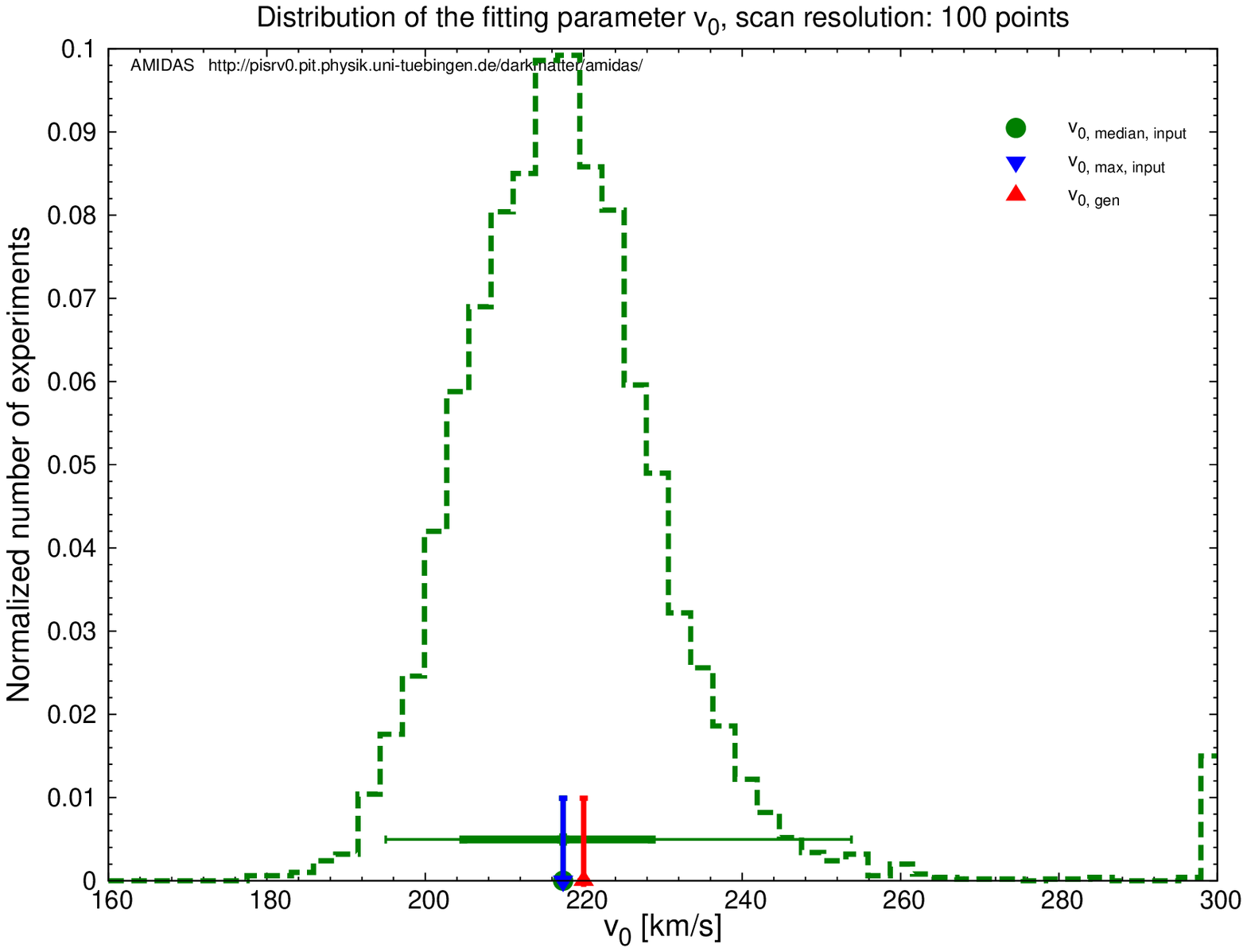}      \hspace*{-1.6cm} \par
\makebox[8.5cm]{(a)}\hspace{0.325cm}\makebox[8.175cm]{(b)}            \\ \vspace{0.5cm}
\hspace*{-1.6cm}
\includegraphics[width=8.5cm]{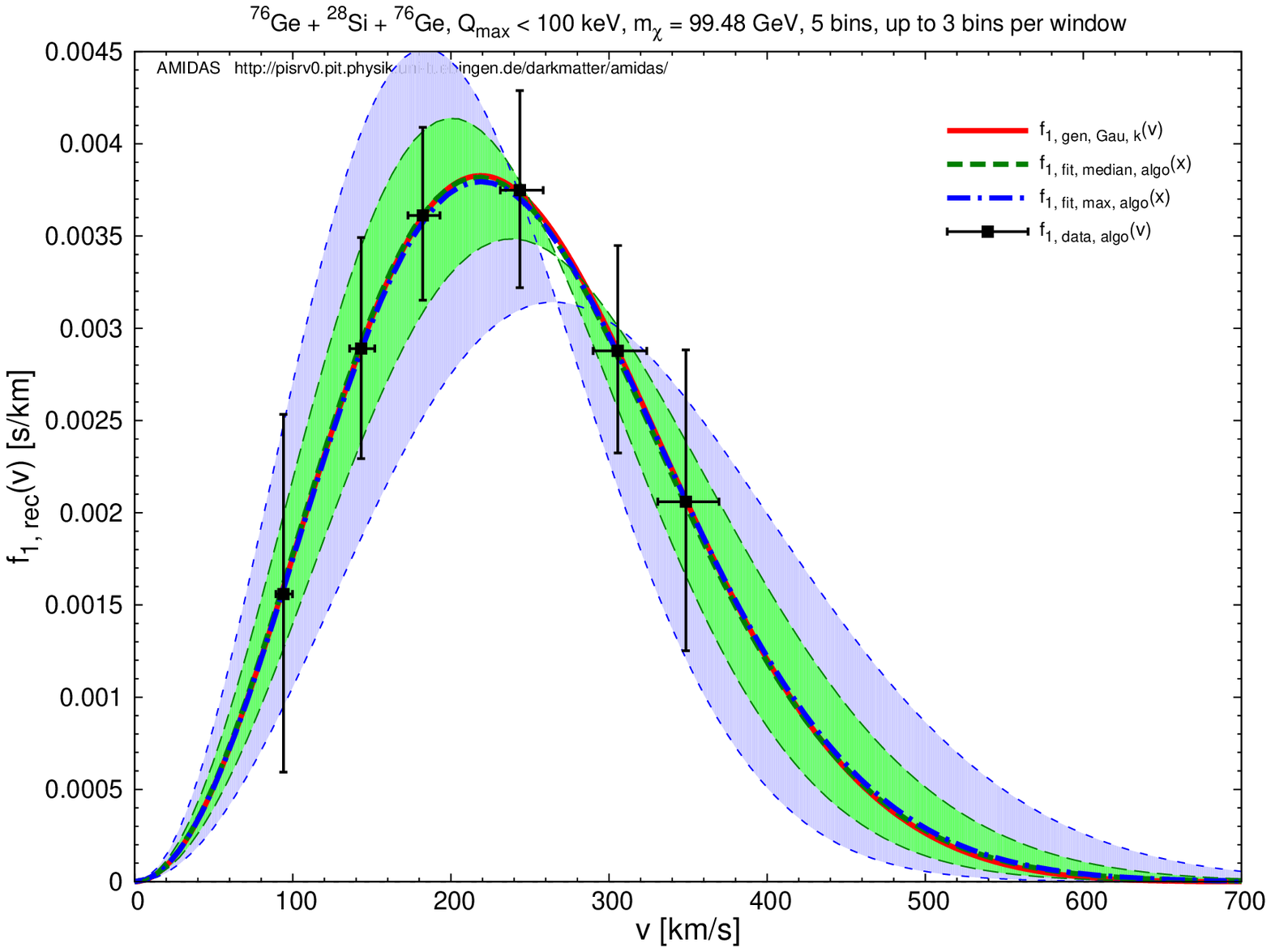}
\includegraphics[width=8.5cm]{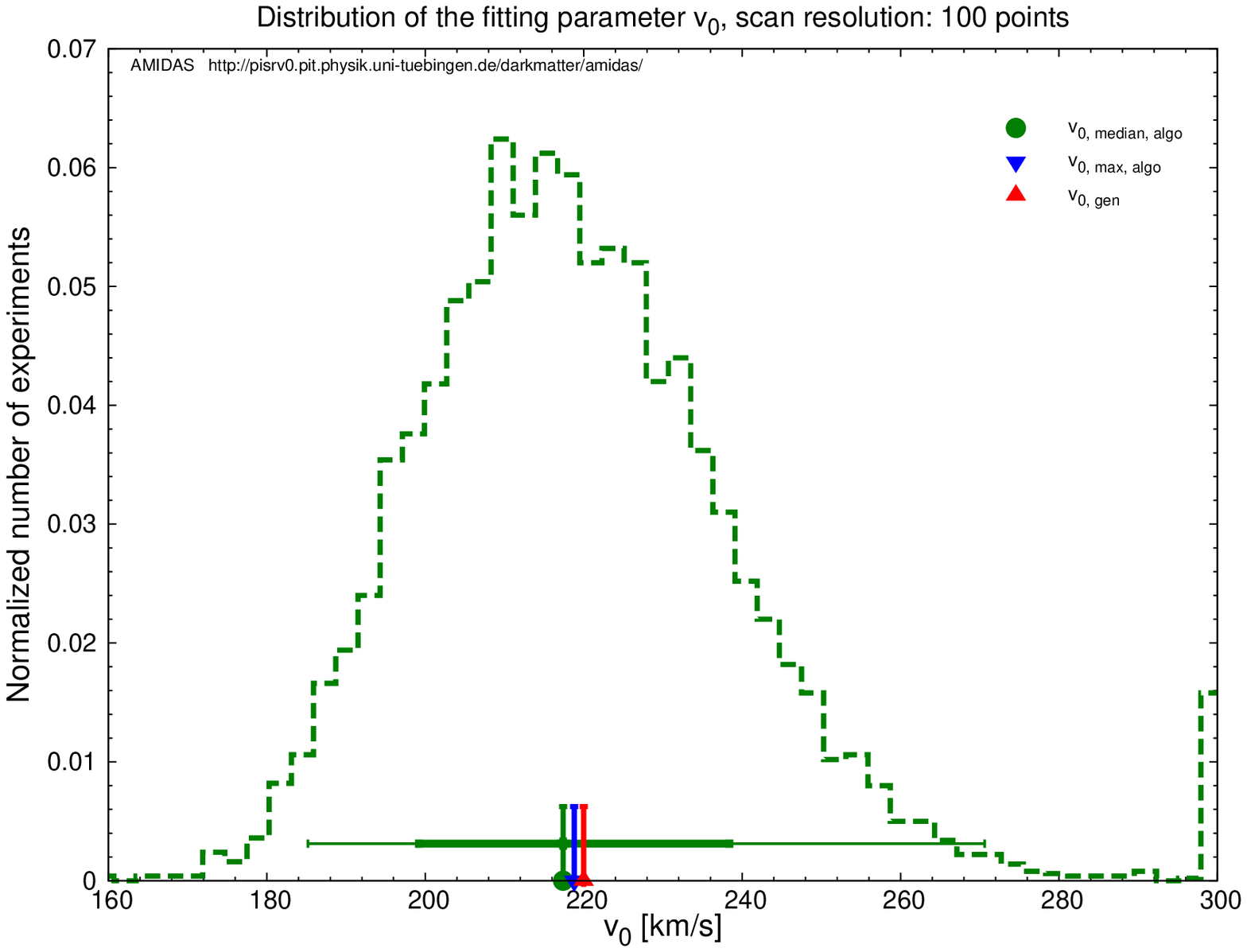} \hspace*{-1.6cm} \par
\makebox[8.5cm]{(c)}\hspace{0.325cm}\makebox[8.175cm]{(d)}%
}
\vspace{-0.35cm}
\end{center}
\caption{
 As in
 Figs.~\ref{fig:f1v-Ge-SiGe-100-0500-sh-Gau-flat},
 except that
 the modified Maxwellian velocity distribution function
 given in Eq.~(\ref{eqn:f1v_Gau_k})
 has been used
 for generating WIMP signals.
}
\label{fig:f1v-Ge-SiGe-100-0500-Gau_k-Gau-flat}
\end{figure}
}
\newcommand{\plotGeSiGeGaukGauGau}{
\begin{figure}[t!]
\begin{center}
\vspace{-0.25cm}
{
\hspace*{-1.6cm}
\includegraphics[width=8.5cm]{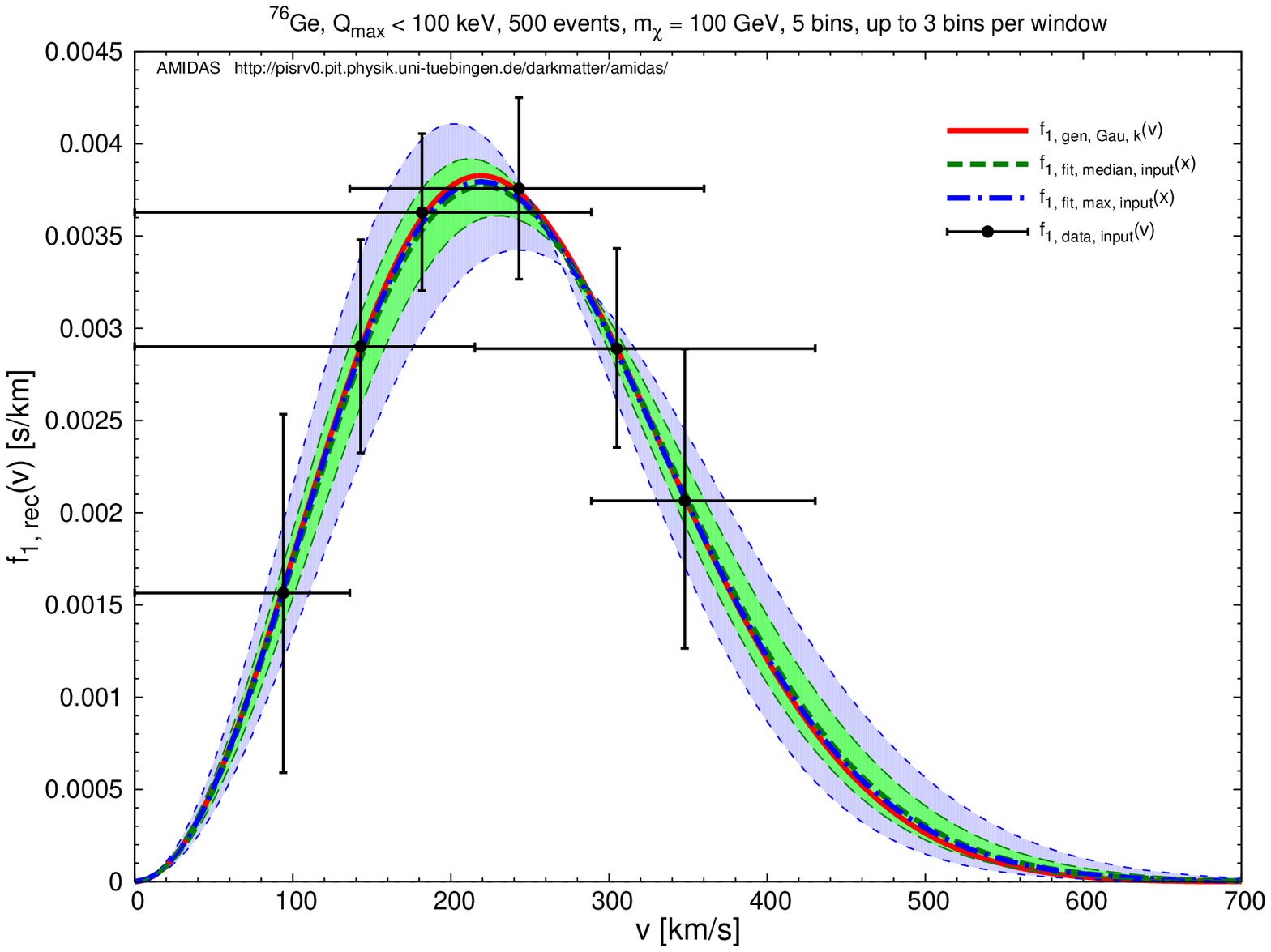}
\includegraphics[width=8.5cm]{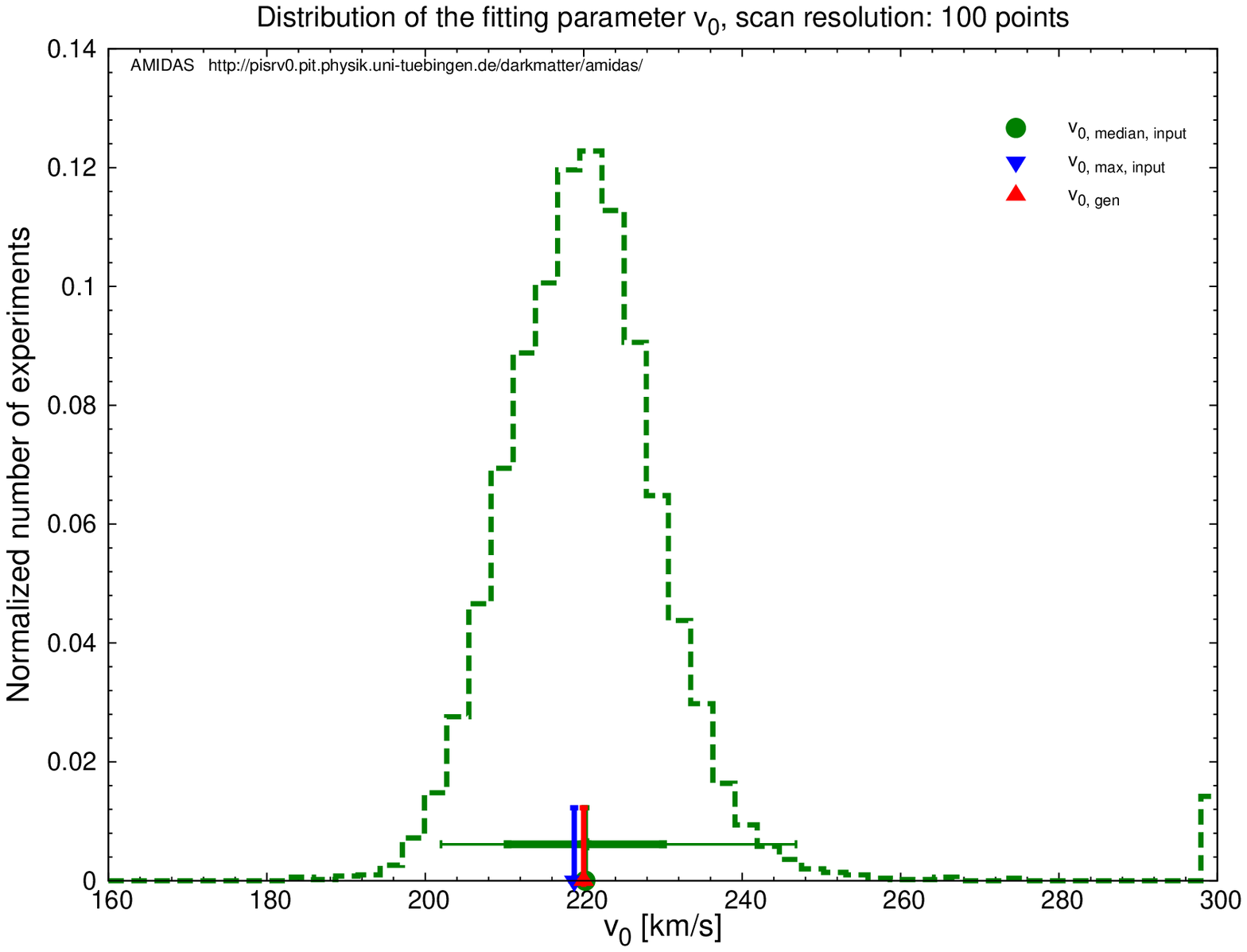}      \hspace*{-1.6cm} \par
\makebox[8.5cm]{(a)}\hspace{0.325cm}\makebox[8.175cm]{(b)}           \\ \vspace{0.5cm}
\hspace*{-1.6cm}
\includegraphics[width=8.5cm]{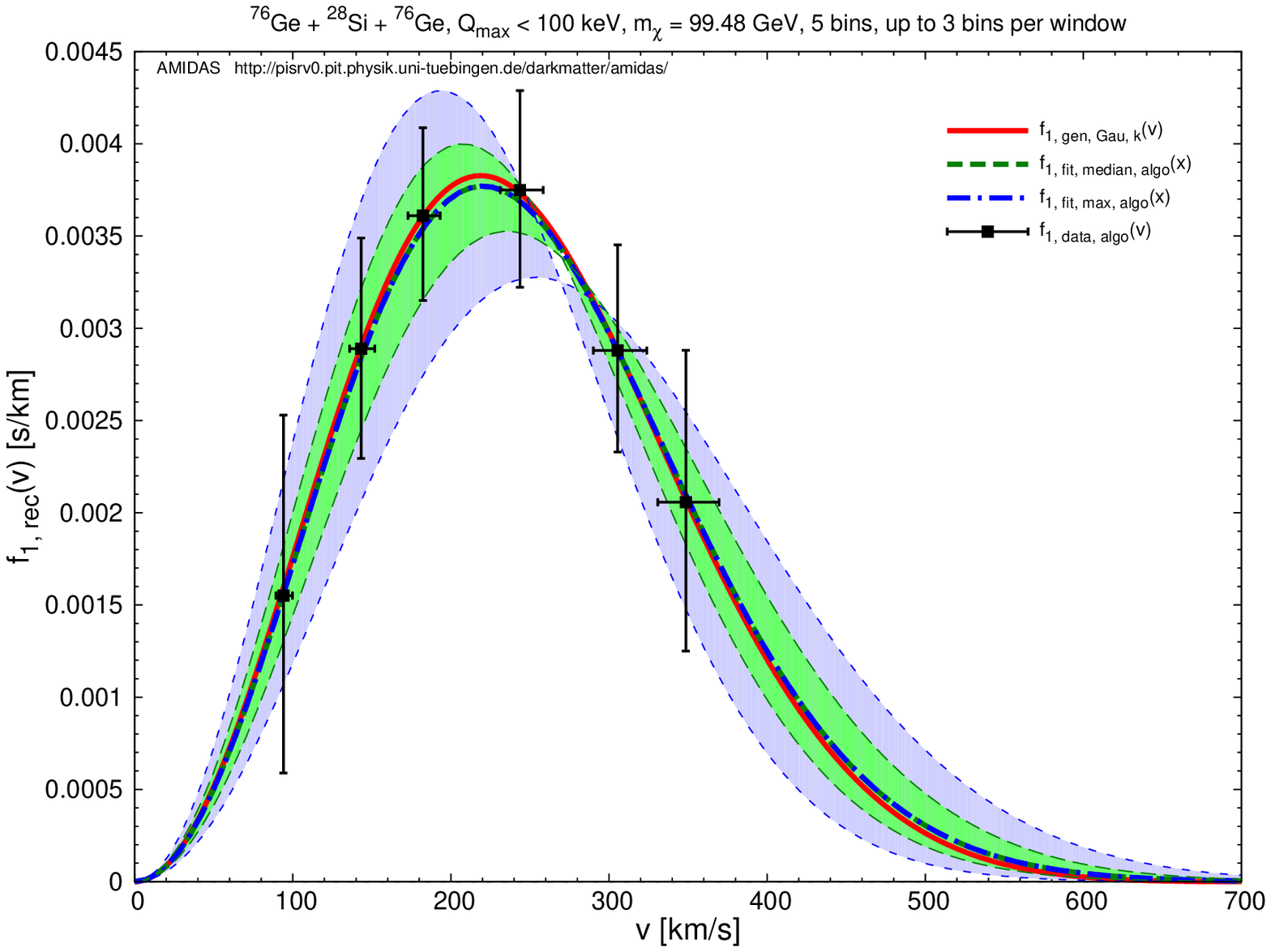}
\includegraphics[width=8.5cm]{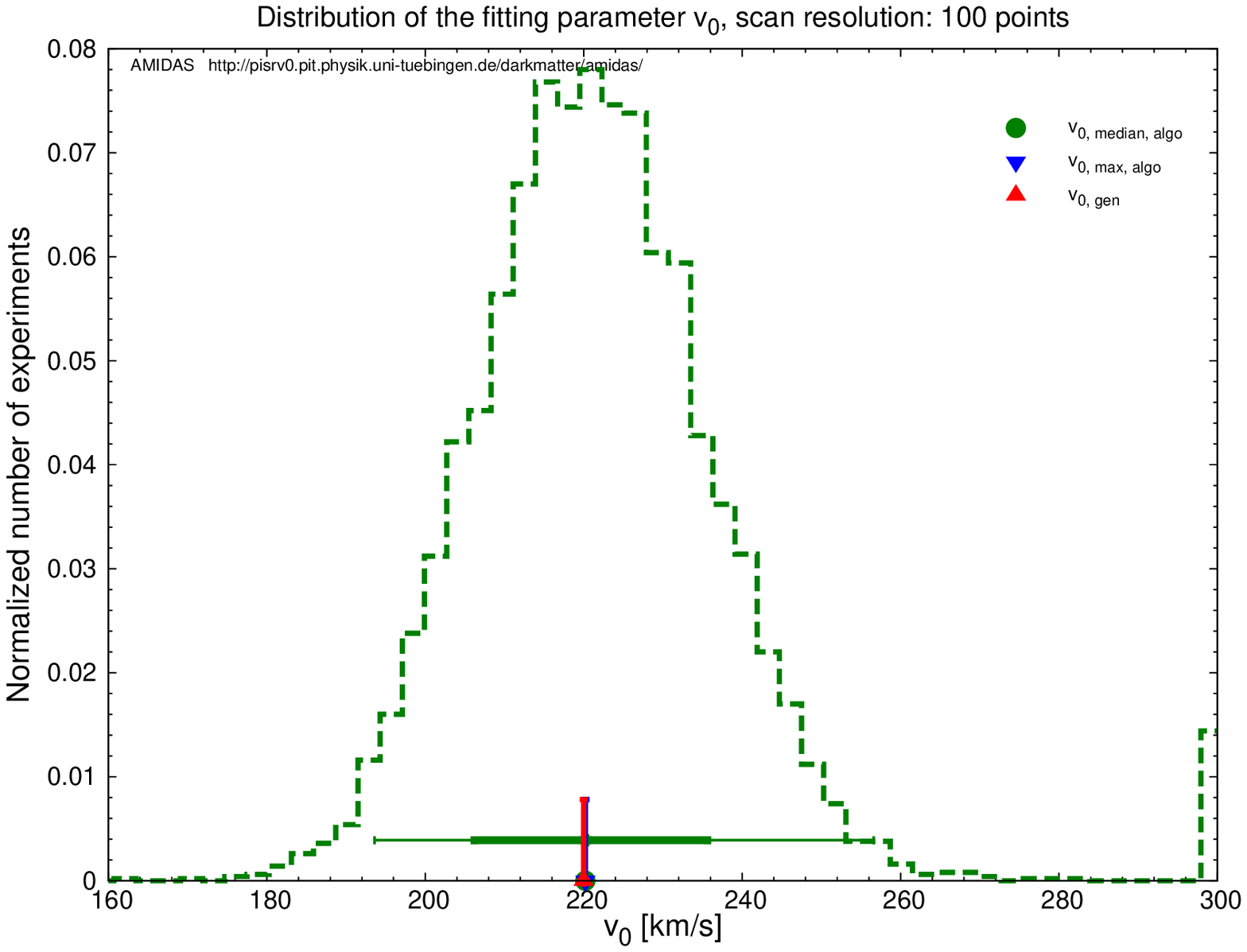} \hspace*{-1.6cm} \par
\makebox[8.5cm]{(c)}\hspace{0.325cm}\makebox[8.175cm]{(d)}%
}
\vspace{-0.35cm}
\end{center}
\caption{
 As in Figs.~\ref{fig:f1v-Ge-SiGe-100-0500-Gau_k-Gau-flat},
 except that
 the Gaussian probability distribution
 for $v_0$
 with an expectation value of \mbox{$v_0 = 230$ km/s}
 and a 1$\sigma$ uncertainty of \mbox{20 km/s}
 has been used.
}
\label{fig:f1v-Ge-SiGe-100-0500-Gau_k-Gau-Gau}
\end{figure}
}
\newcommand{\plotGeSiGeGaukshvflat}{
\begin{figure}[t!]
\begin{center}
\vspace{-0.25cm}
{
\hspace*{-1.6cm}
\includegraphics[width=8.5cm]{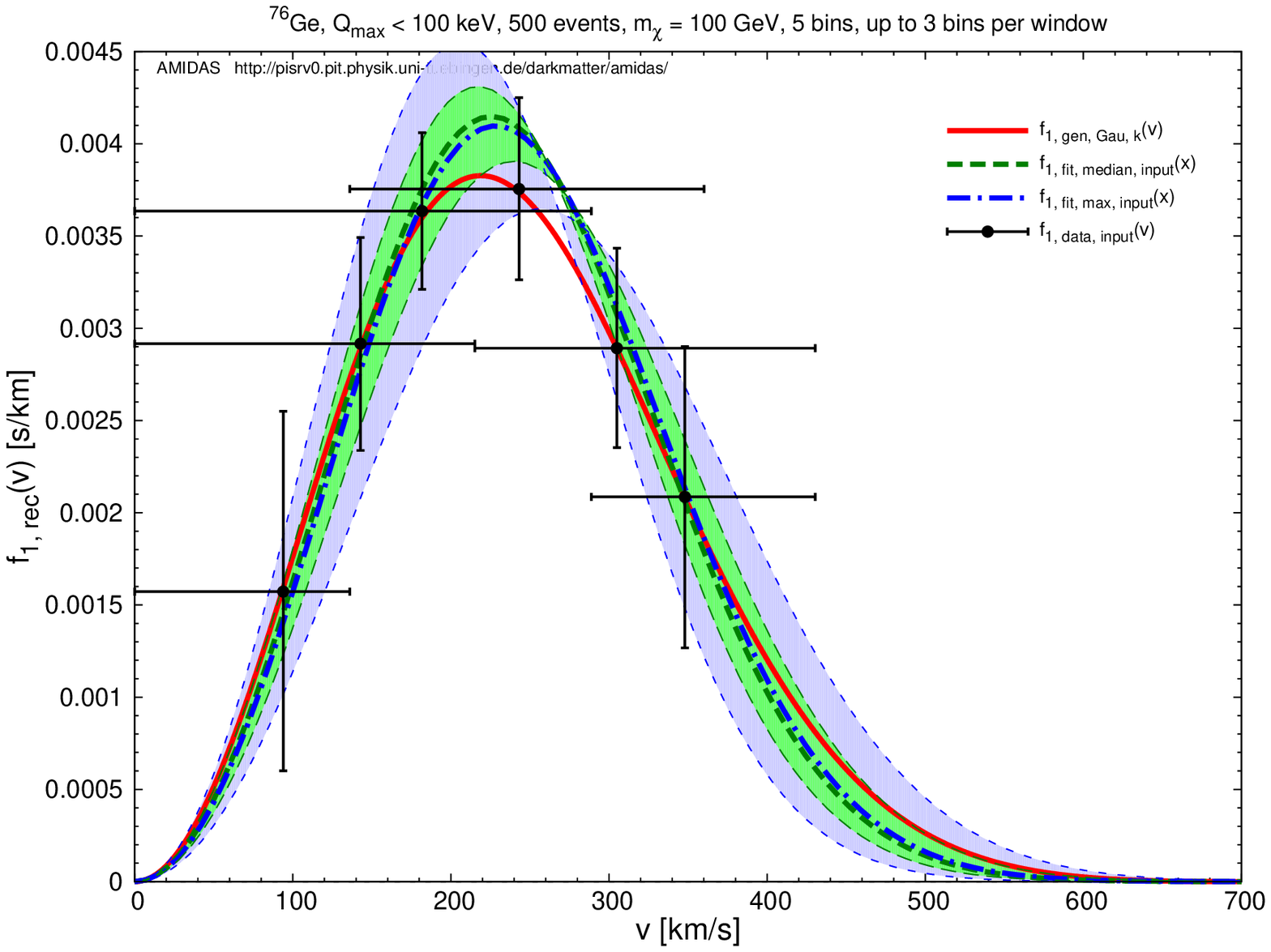}
\includegraphics[width=8.5cm]{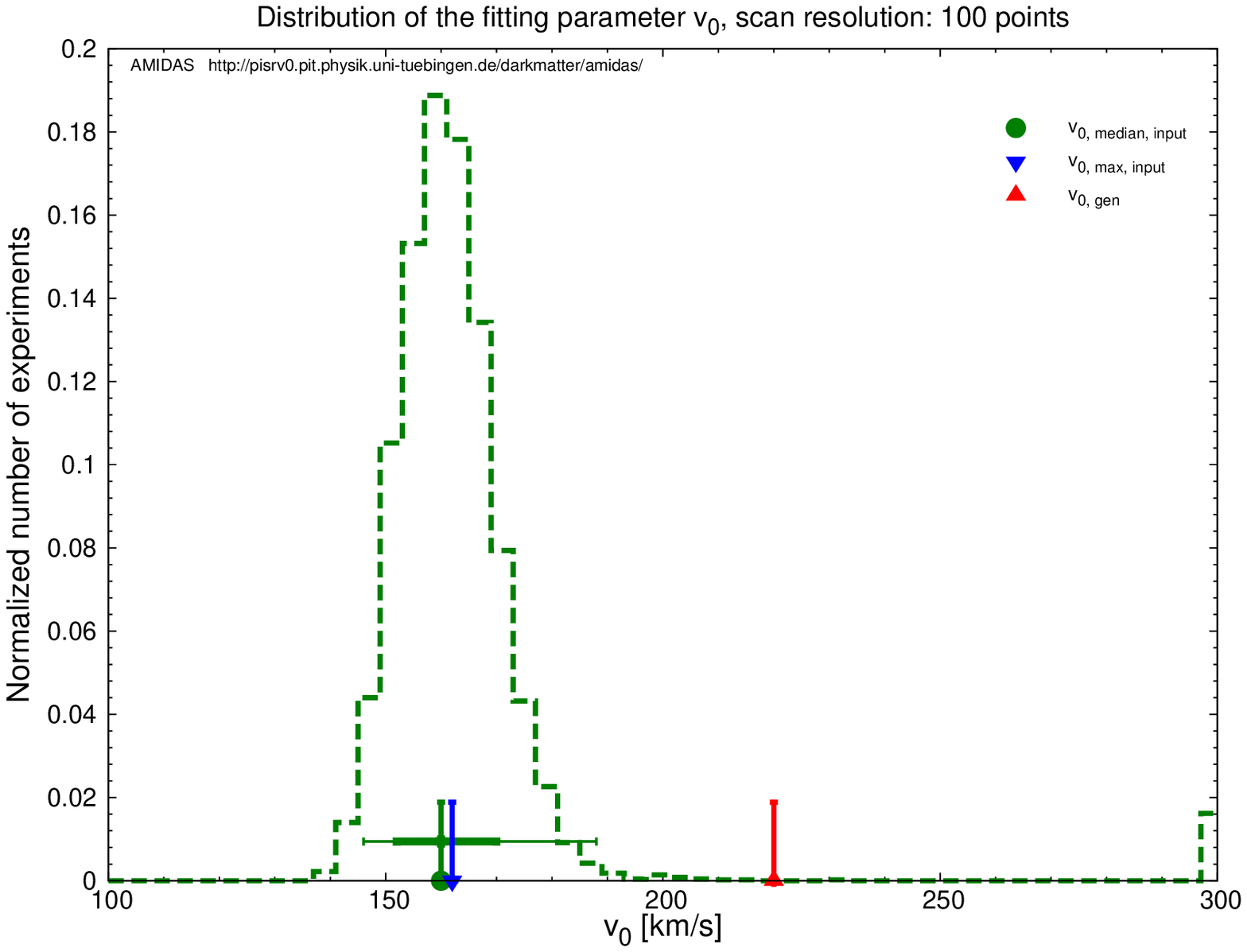}      \hspace*{-1.6cm} \par
\makebox[8.5cm]{(a)}\hspace{0.325cm}\makebox[8.175cm]{(b)}              \\ \vspace{0.5cm}
\hspace*{-1.6cm}
\includegraphics[width=8.5cm]{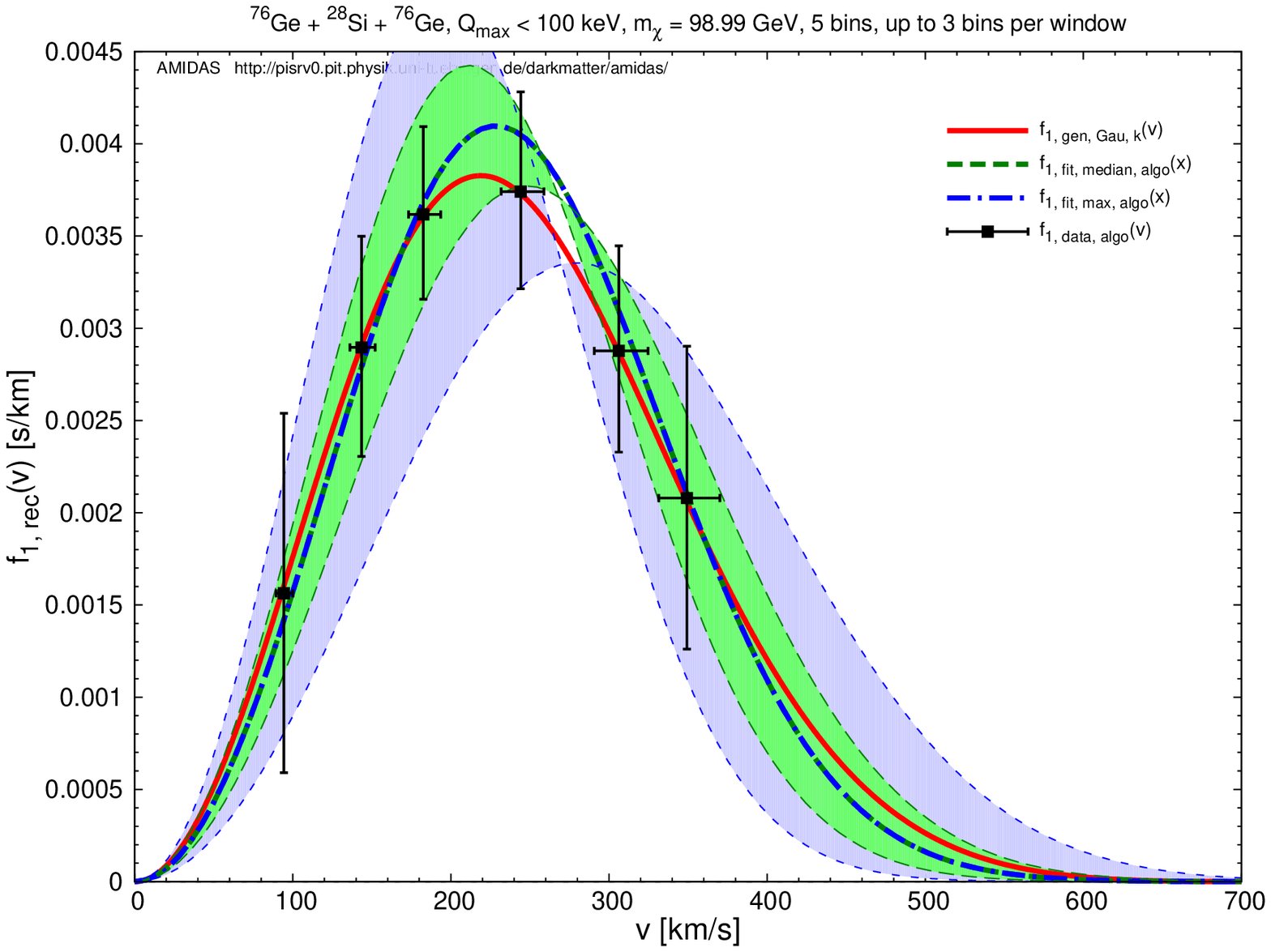}
\includegraphics[width=8.5cm]{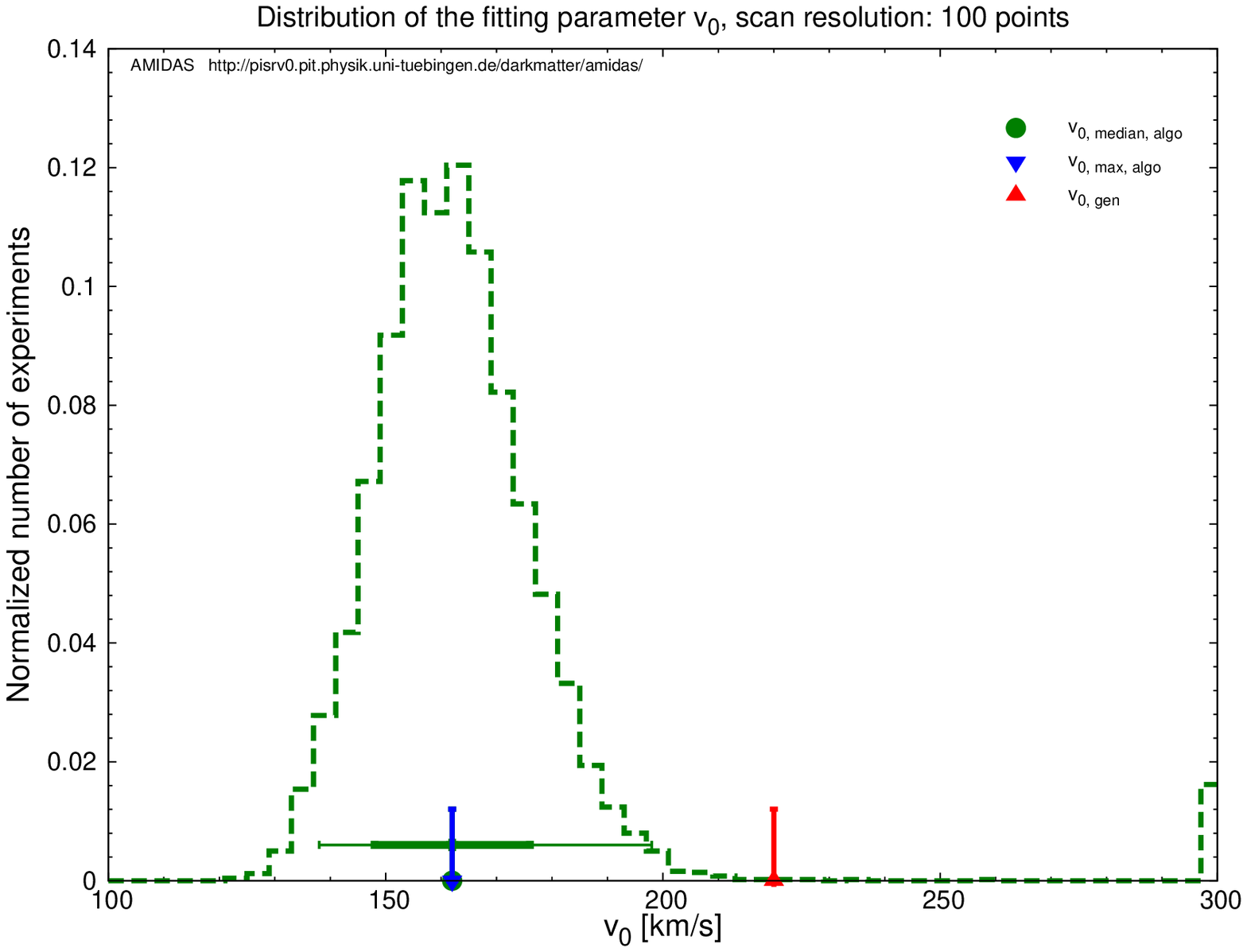} \hspace*{-1.6cm} \par
\makebox[8.5cm]{(c)}\hspace{0.325cm}\makebox[8.175cm]{(d)}%
}
\vspace{-0.35cm}
\end{center}
\caption{
 As in Figs.~\ref{fig:f1v-Ge-SiGe-100-0500-Gau_k-Gau-flat},
 except that
 the one--parameter shifted
 Maxwellian velocity distribution function $f_{1, \sh, v_0}(v)$
 with the unique fitting parameter $v_0$
 has been used
 as the fitting velocity distribution.
}
\label{fig:f1v-Ge-SiGe-100-0500-Gau_k-sh_v0-flat}
\end{figure}
}
\newcommand{\plotGeSiGeGaukshvGau}{
\begin{figure}[t!]
\begin{center}
\vspace{-0.25cm}
{
\hspace*{-1.6cm}
\includegraphics[width=8.5cm]{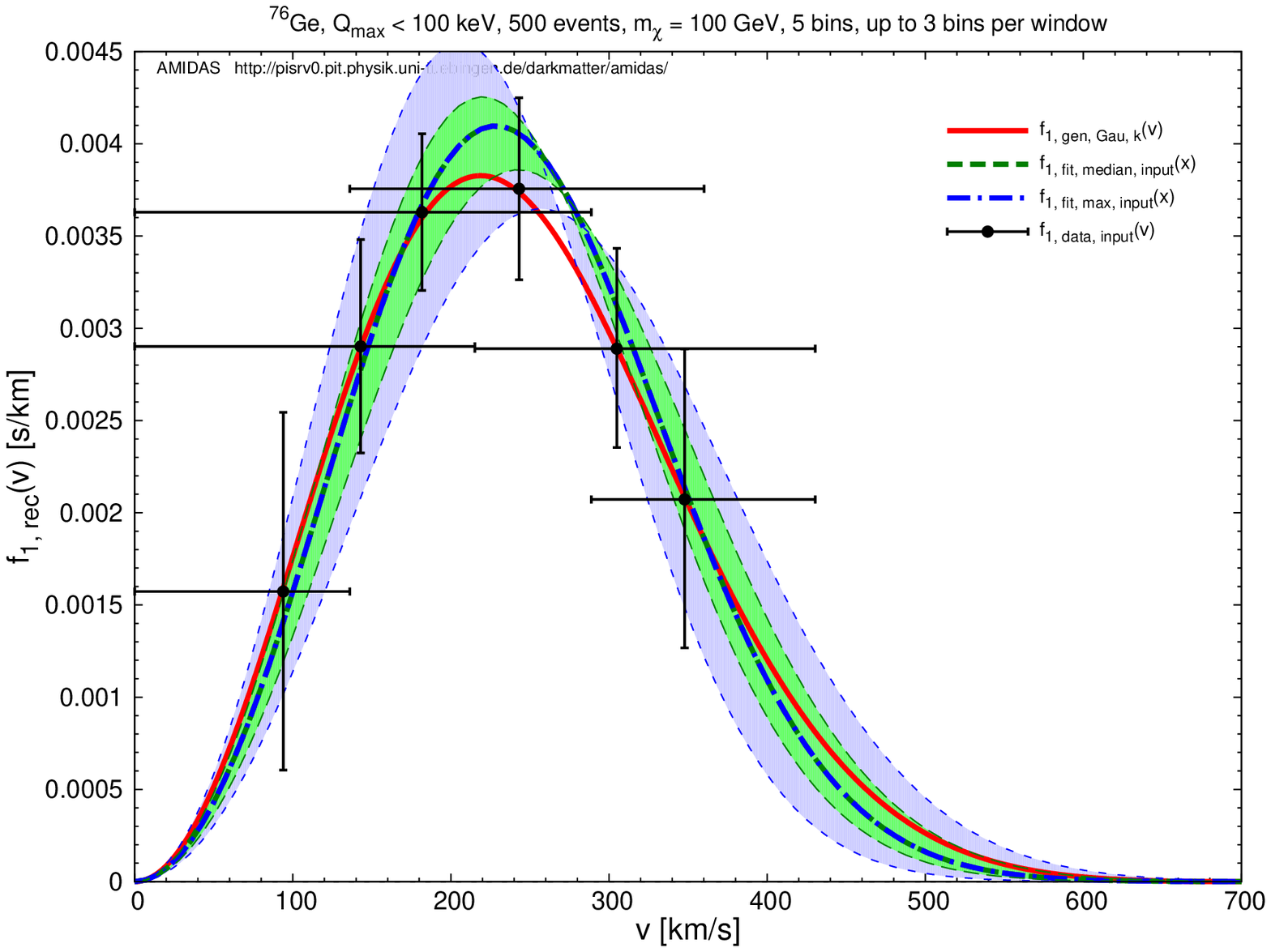}
\includegraphics[width=8.5cm]{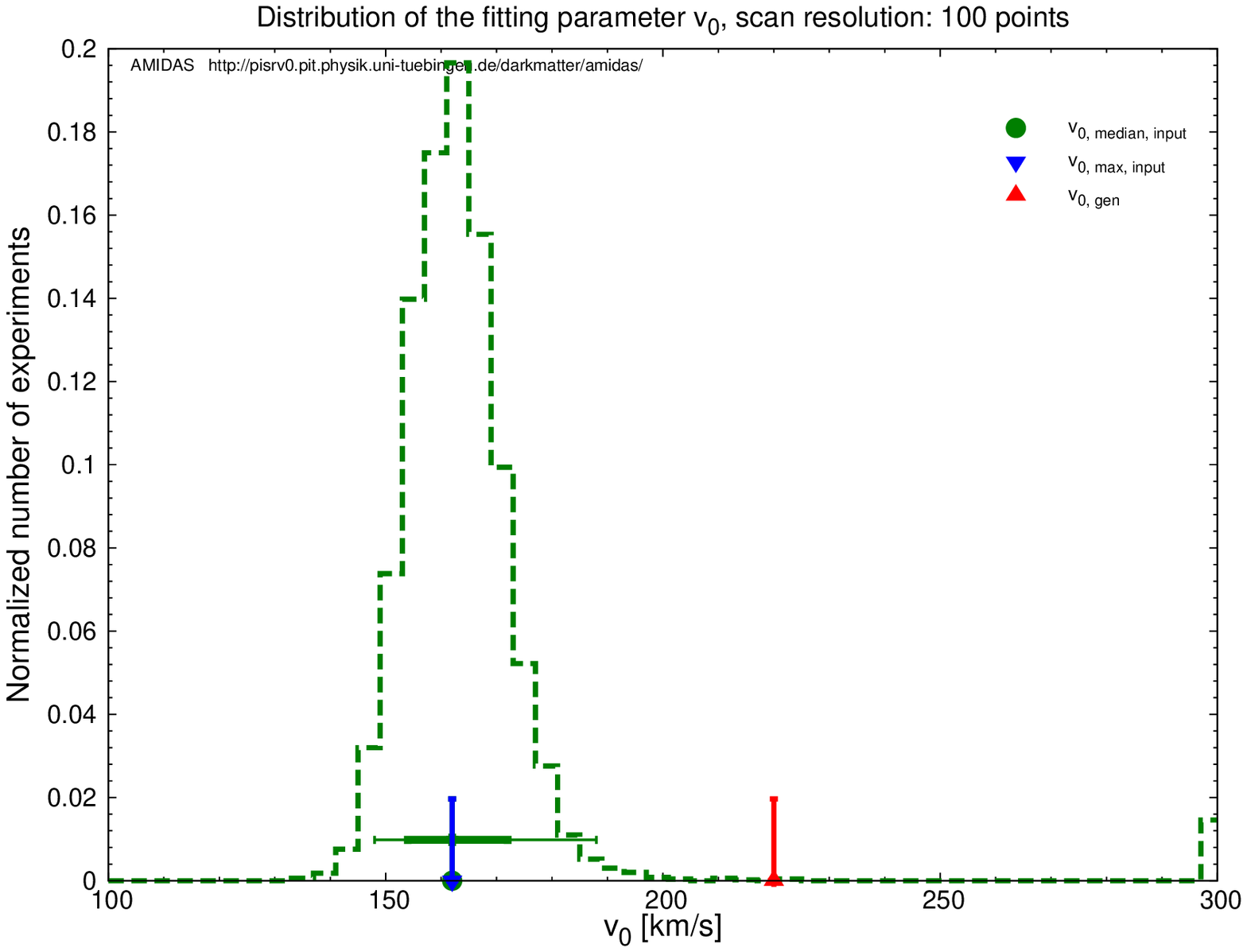}      \hspace*{-1.6cm} \par
\makebox[8.5cm]{(a)}\hspace{0.325cm}\makebox[8.175cm]{(b)}             \\ \vspace{0.5cm}
\hspace*{-1.6cm}
\includegraphics[width=8.5cm]{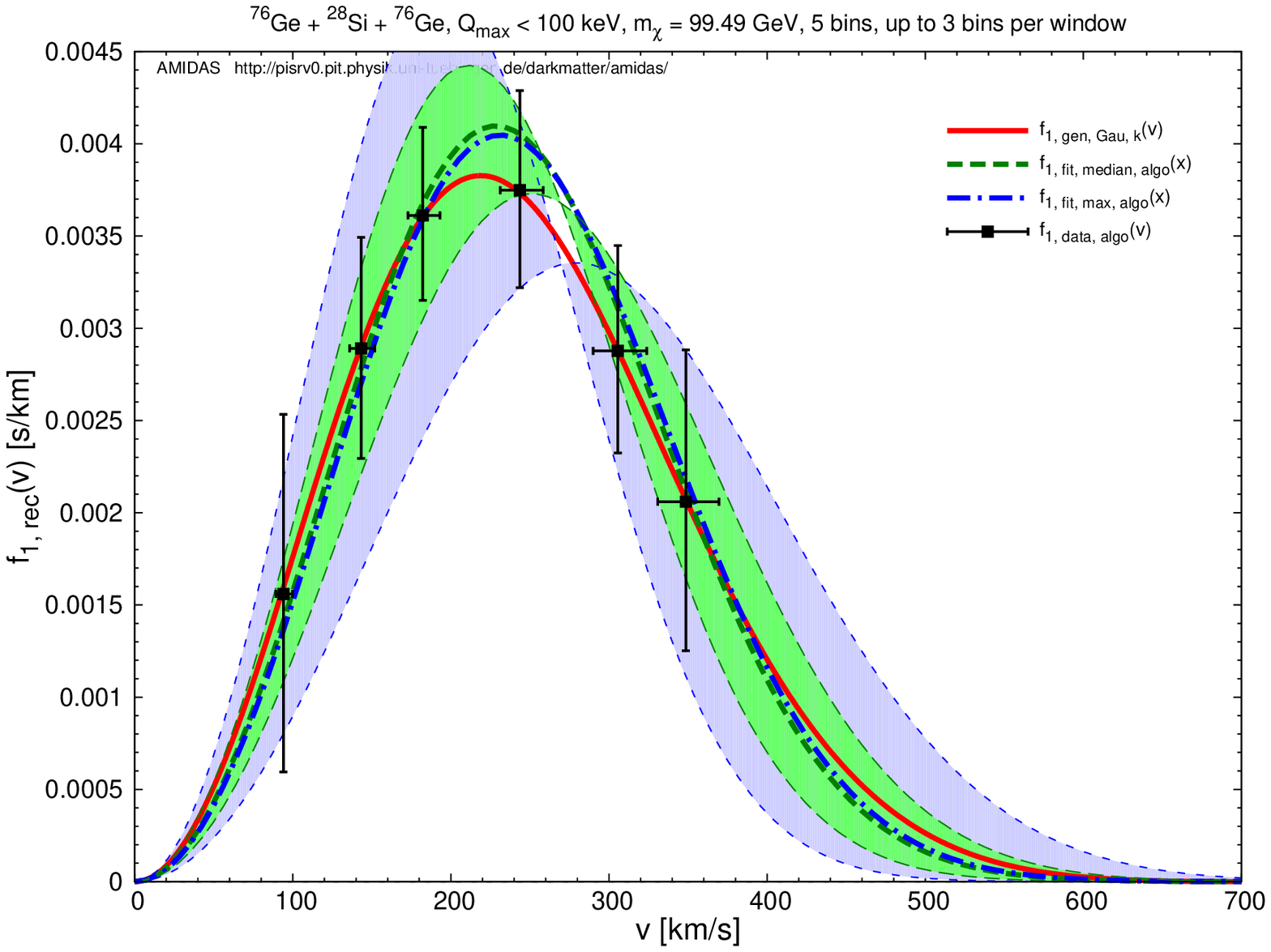}
\includegraphics[width=8.5cm]{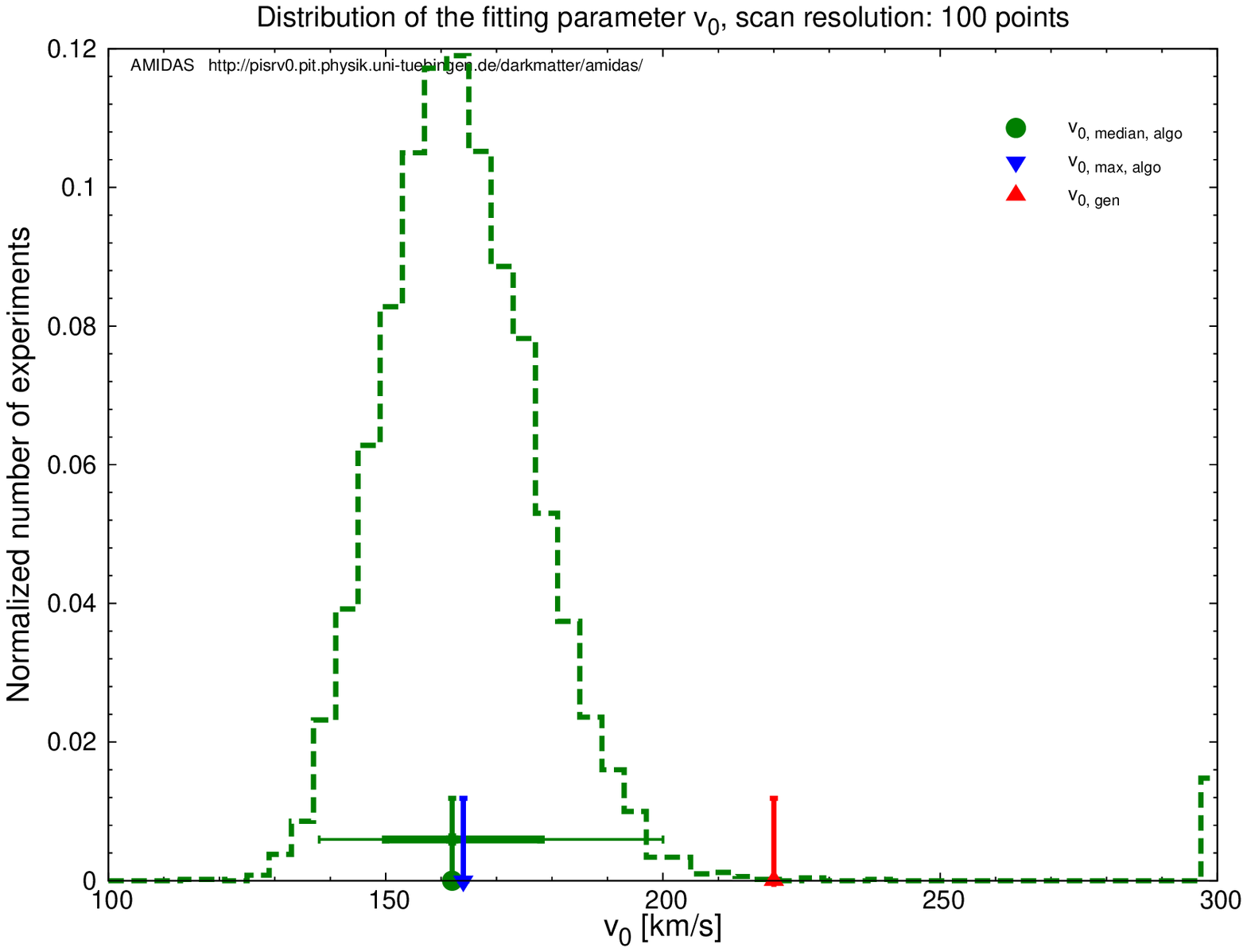} \hspace*{-1.6cm} \par
\makebox[8.5cm]{(c)}\hspace{0.325cm}\makebox[8.175cm]{(d)}%
}
\vspace{-0.35cm}
\end{center}
\caption{
 As in Figs.~\ref{fig:f1v-Ge-SiGe-100-0500-Gau_k-sh_v0-flat},
 except that
 the Gaussian probability distribution
 for $v_0$
 with an expectation value of \mbox{$v_0 = {\it 200}$ km/s}
 and a 1$\sigma$ uncertainty of \mbox{{\em 40} km/s}
 has been used.
}
\label{fig:f1v-Ge-SiGe-100-0500-Gau_k-sh_v0-Gau}
\end{figure}
}
\newcommand{\plotGeGaukshGau}{
\begin{figure}[t!]
\begin{center}
\vspace{-0.25cm}
{
\hspace*{-1.6cm}
\includegraphics[width=8.5cm]{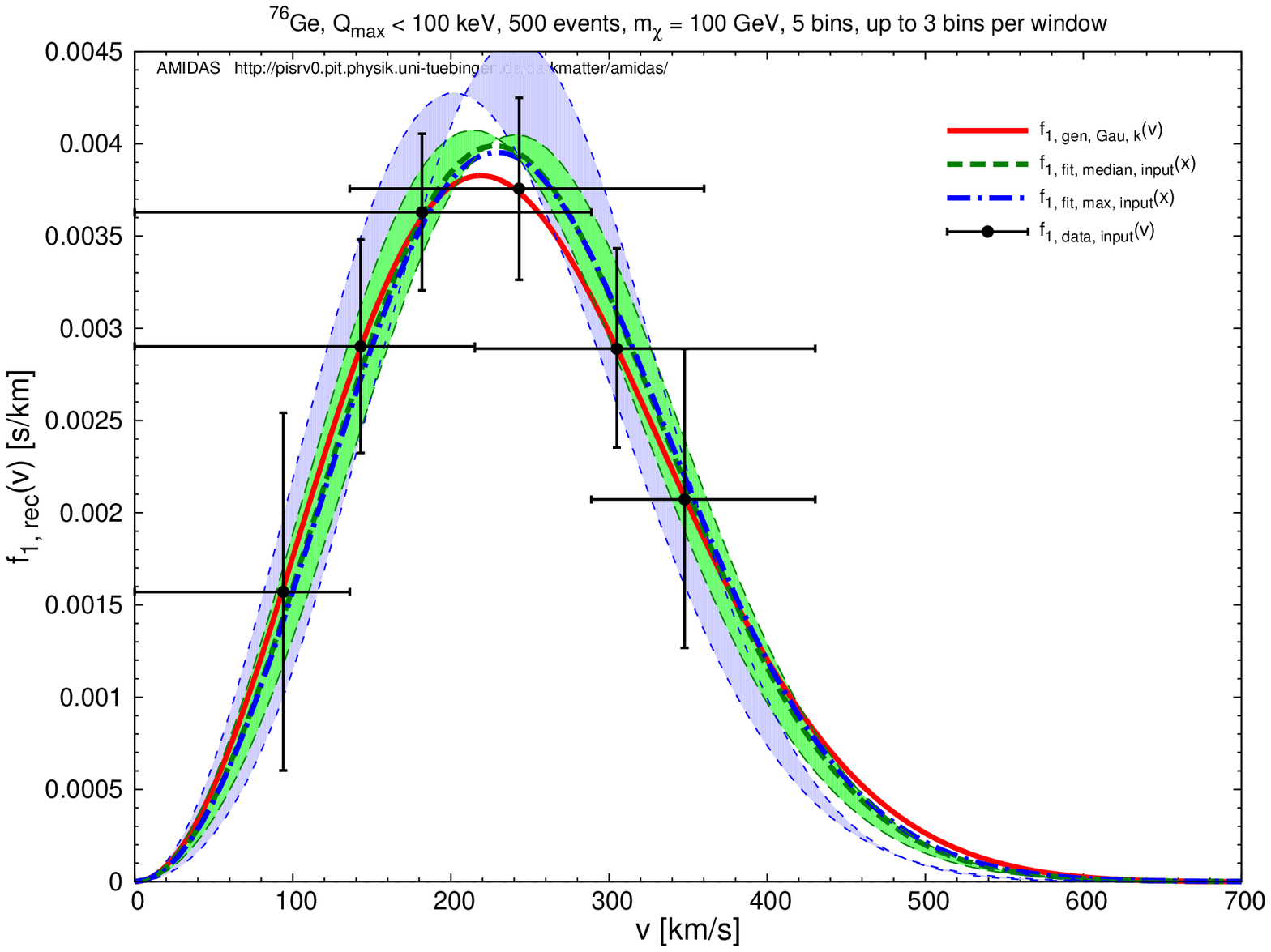}
\includegraphics[width=8.5cm]{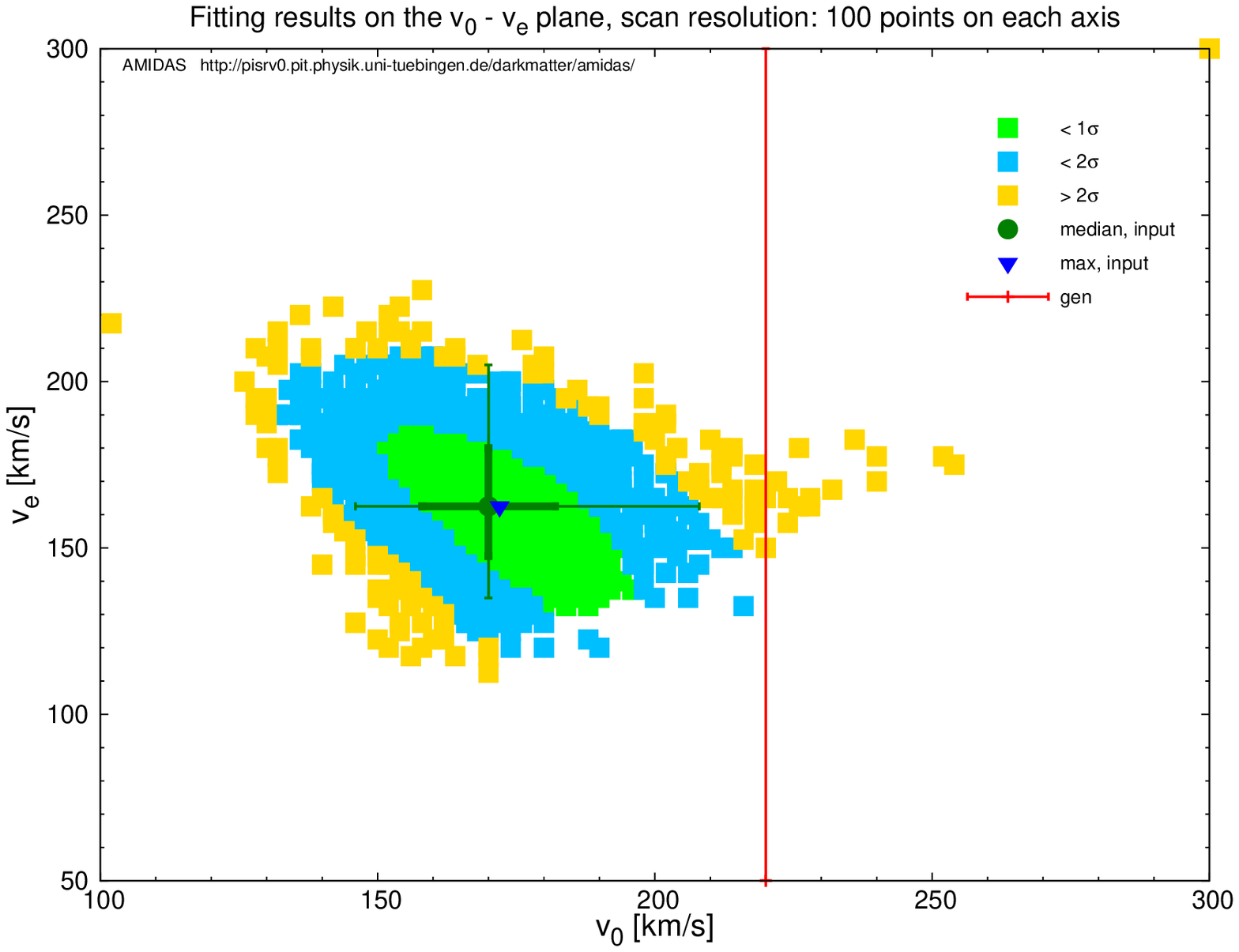} \hspace*{-1.6cm} \par
\makebox[8.5cm]{(a)}\hspace{0.325cm}\makebox[8.175cm]{(b)}        \\ \vspace{0.5cm}
\hspace*{-1.6cm}
\includegraphics[width=8.5cm]{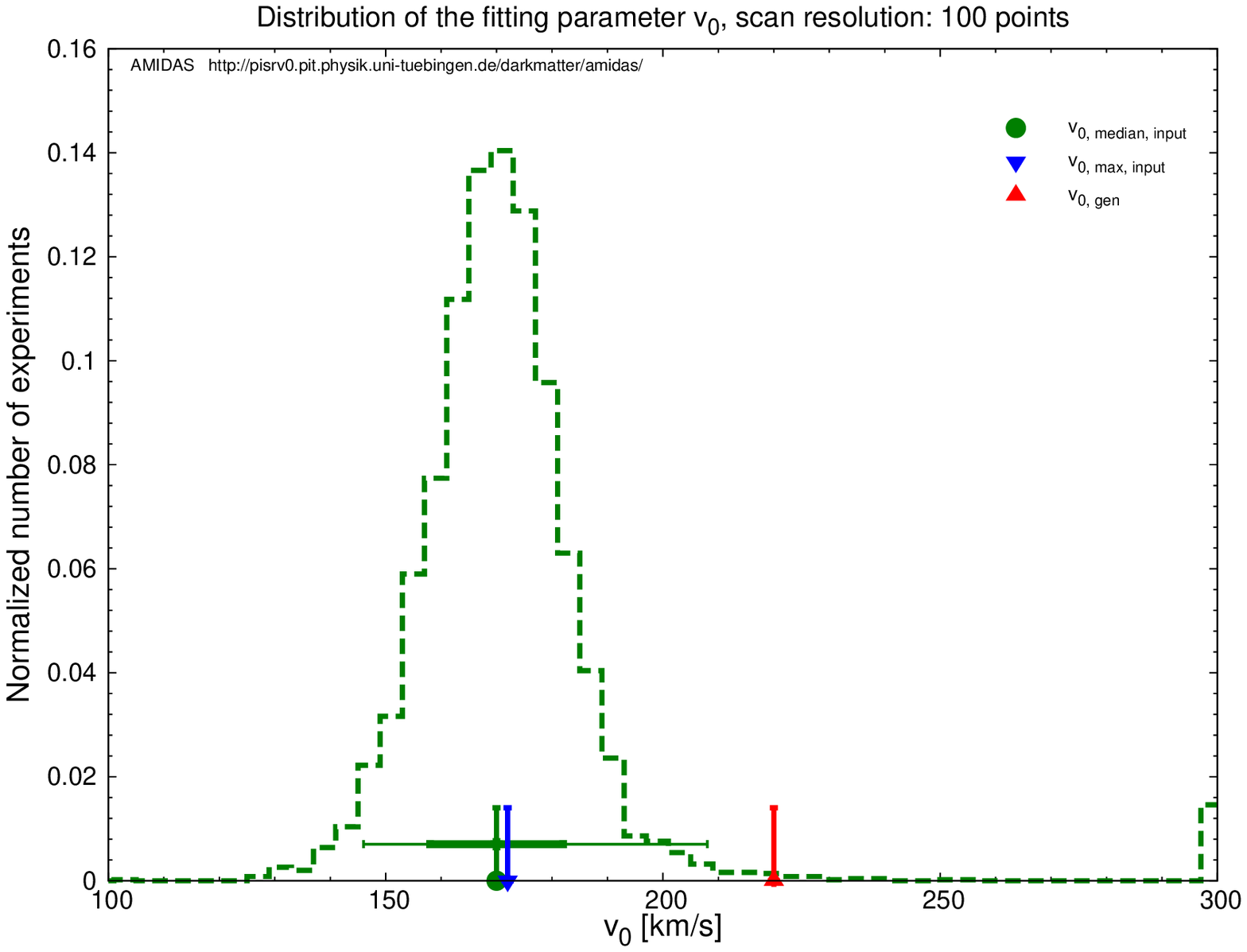}
\includegraphics[width=8.5cm]{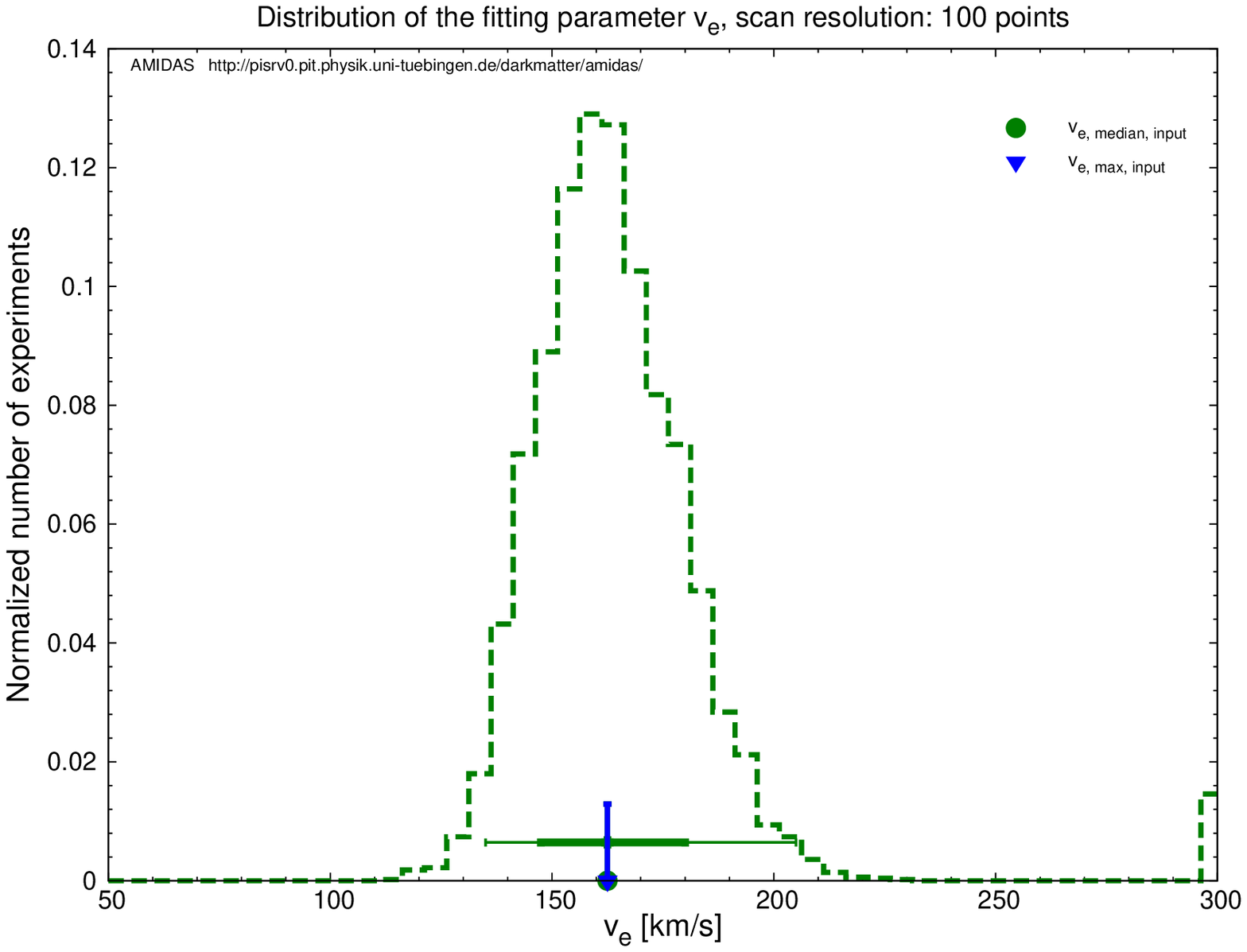}    \hspace*{-1.6cm} \par
\makebox[8.5cm]{(c)}\hspace{0.325cm}\makebox[8.175cm]{(d)}        \\
}
\vspace{-0.35cm}
\end{center}
\caption{
 As in Figs.~\ref{fig:f1v-Ge-100-0500-sh-sh-Gau},
 except that
 the modified Maxwellian velocity distribution function
 given in Eq.~(\ref{eqn:f1v_Gau_k})
 has been used
 for generating WIMP signals.
 Note that
 the solid red vertical line shown in (b)
 indicates the input value of the parameter $v_0$
 (since no input value for $\ve$).
 The Gaussian probability distribution
 for both fitting parameters
 with a common expectation value of \mbox{$v_0 = \ve = {\it 200}$ km/s}
 and a common 1$\sigma$ uncertainty of \mbox{{\em 40} km/s}
 has been used.
}
\label{fig:f1v-Ge-100-0500-Gau_k-sh-Gau}
\end{figure}
}
\newcommand{\plotGeSiGeGaukshGau}{
\begin{figure}[t!]
\begin{center}
\vspace{-0.25cm}
{
\hspace*{-1.6cm}
\includegraphics[width=8.5cm]{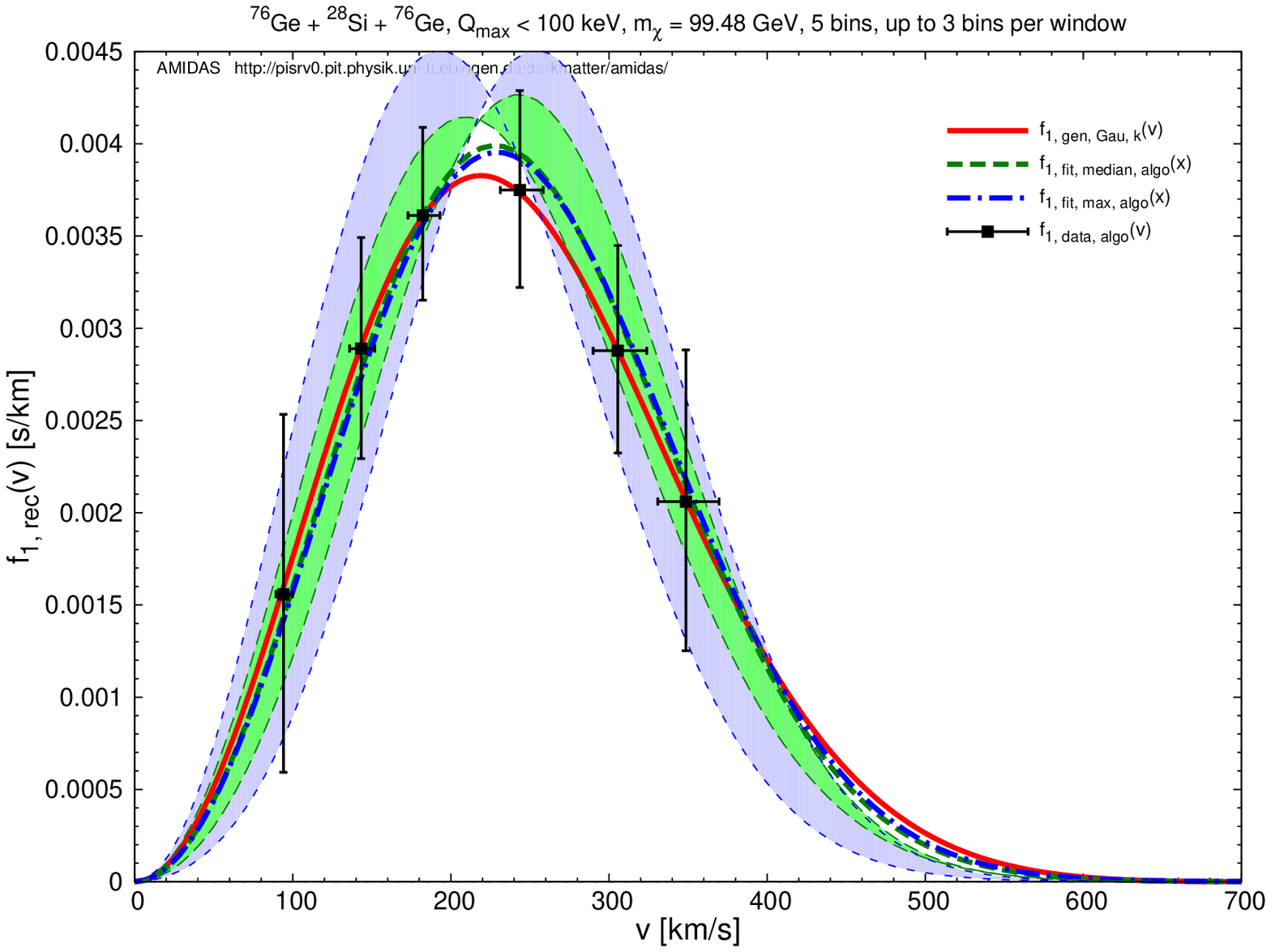}
\includegraphics[width=8.5cm]{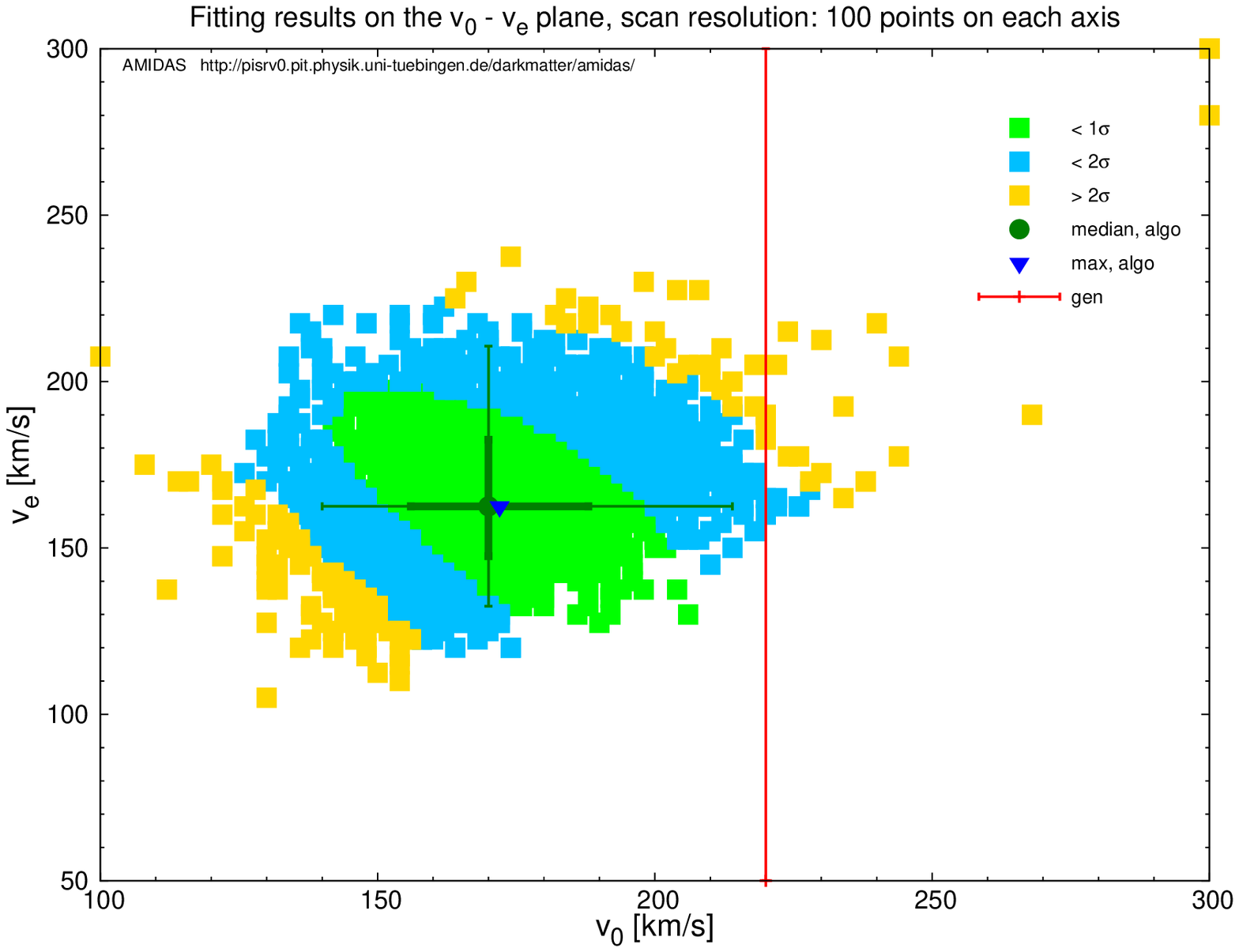} \hspace*{-1.6cm} \par
\makebox[8.5cm]{(a)}\hspace{0.325cm}\makebox[8.175cm]{(b)}             \\ \vspace{0.5cm}
\hspace*{-1.6cm}
\includegraphics[width=8.5cm]{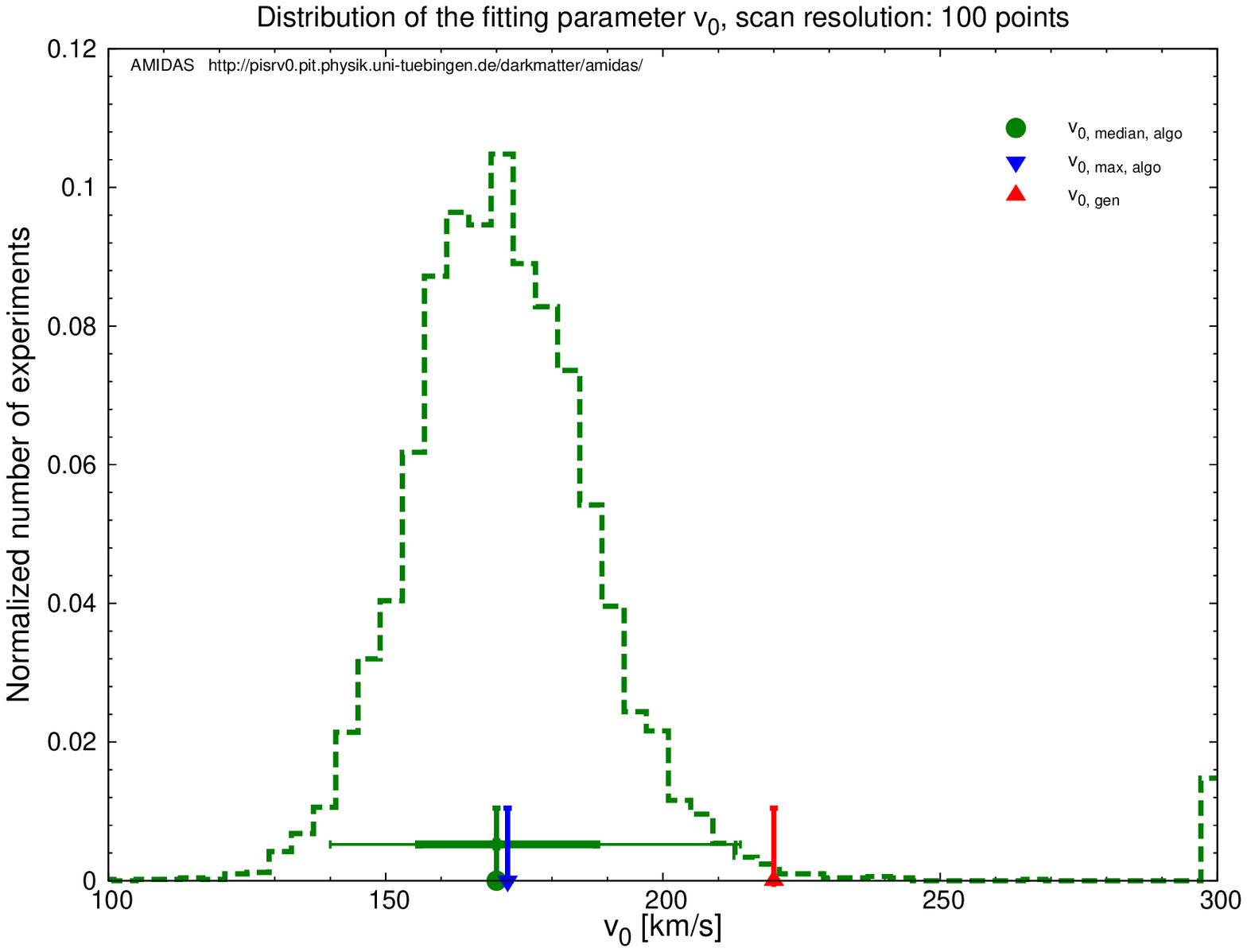}
\includegraphics[width=8.5cm]{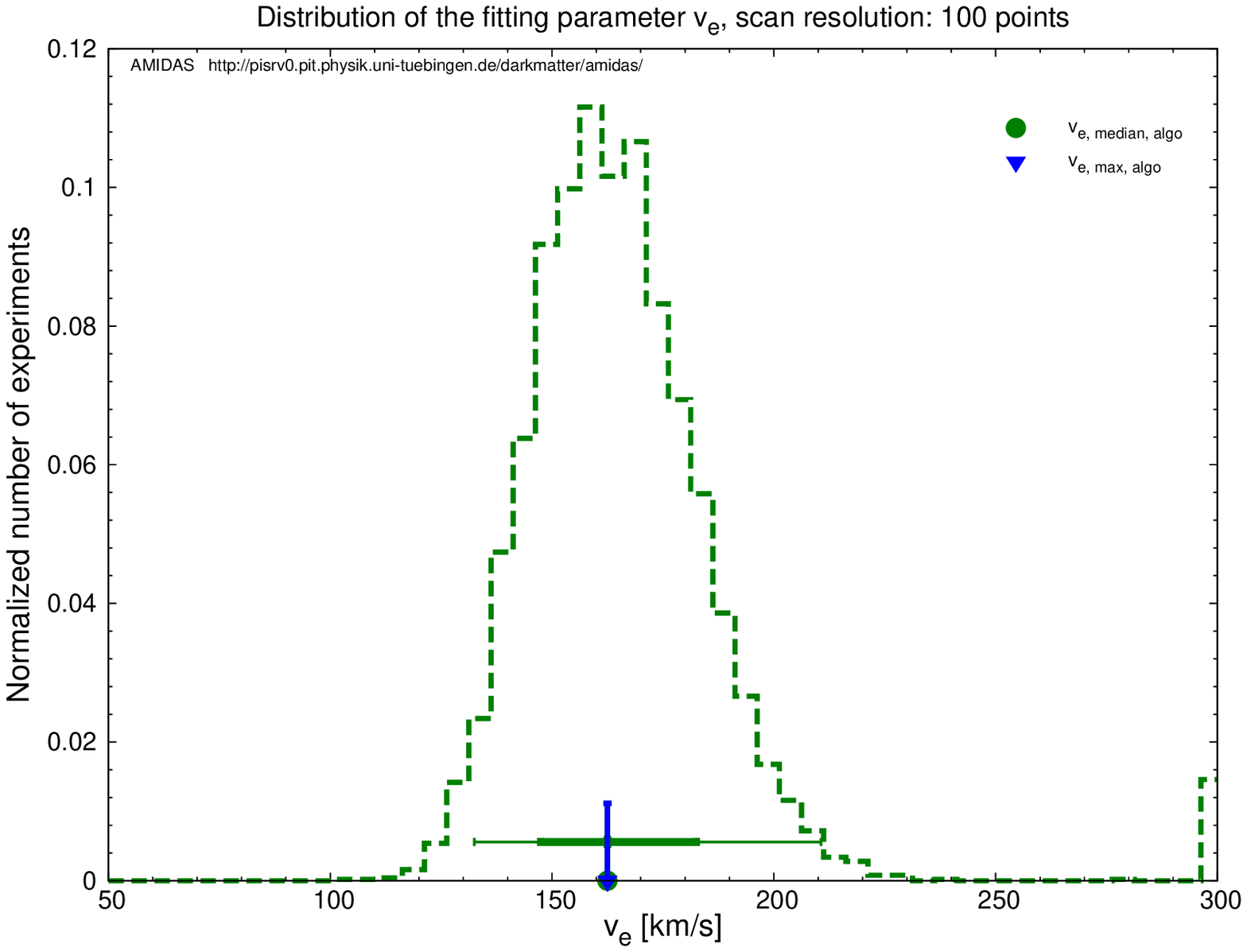}    \hspace*{-1.6cm} \par
\makebox[8.5cm]{(c)}\hspace{0.325cm}\makebox[8.175cm]{(d)}             \\
}
\vspace{-0.35cm}
\end{center}
\caption{
 As in Figs.~\ref{fig:f1v-Ge-100-0500-Gau_k-sh-Gau},
 except that
 the reconstructed WIMP mass
 has been
 used.
}
\label{fig:f1v-Ge-SiGe-100-0500-Gau_k-sh-Gau}
\end{figure}
}
\newcommand{\plotGeGaukshDvGau}{
\begin{figure}[t!]
\begin{center}
\vspace{-0.25cm}
{
\hspace*{-1.6cm}
\includegraphics[width=8.5cm]{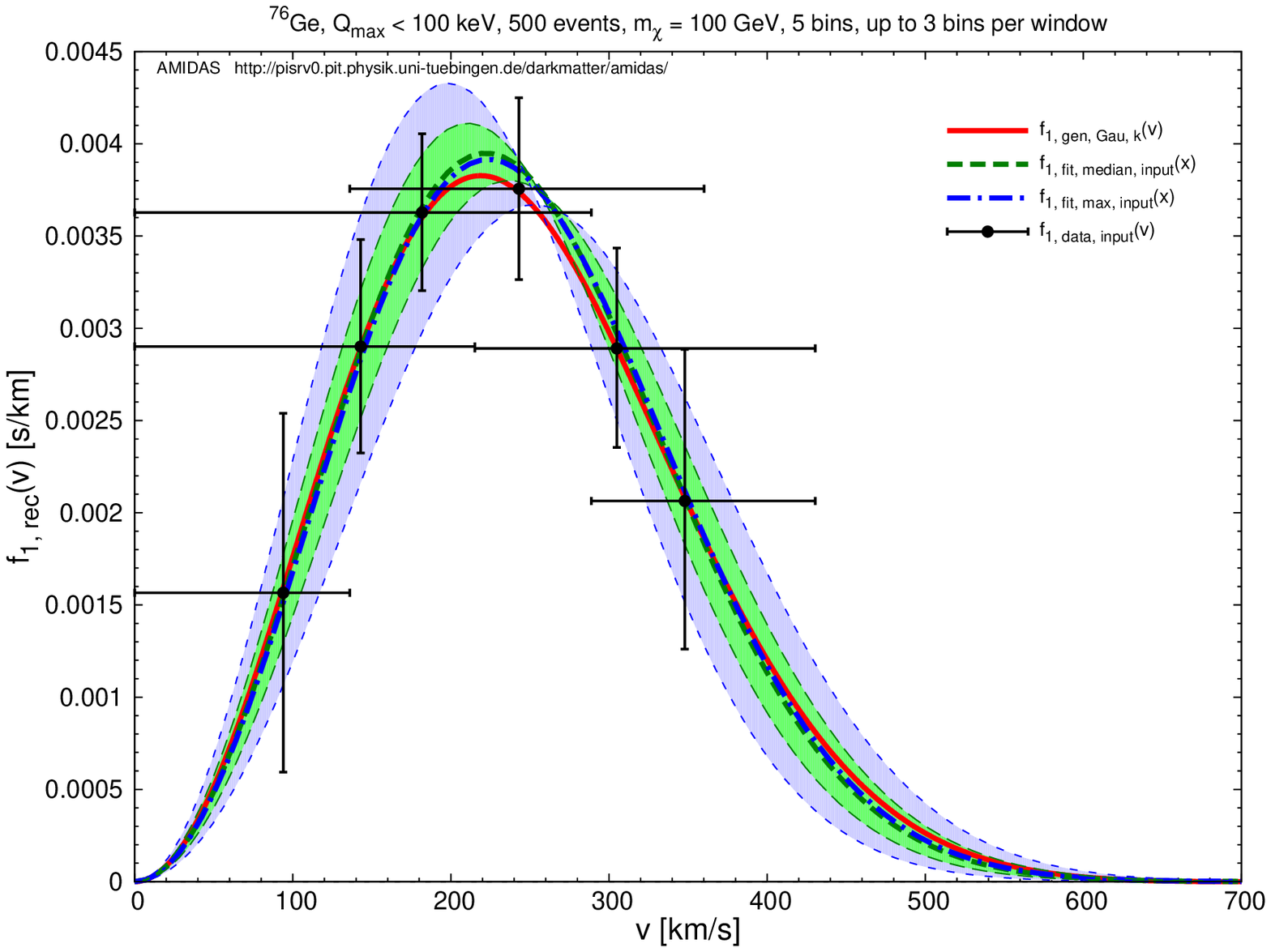}
\includegraphics[width=8.5cm]{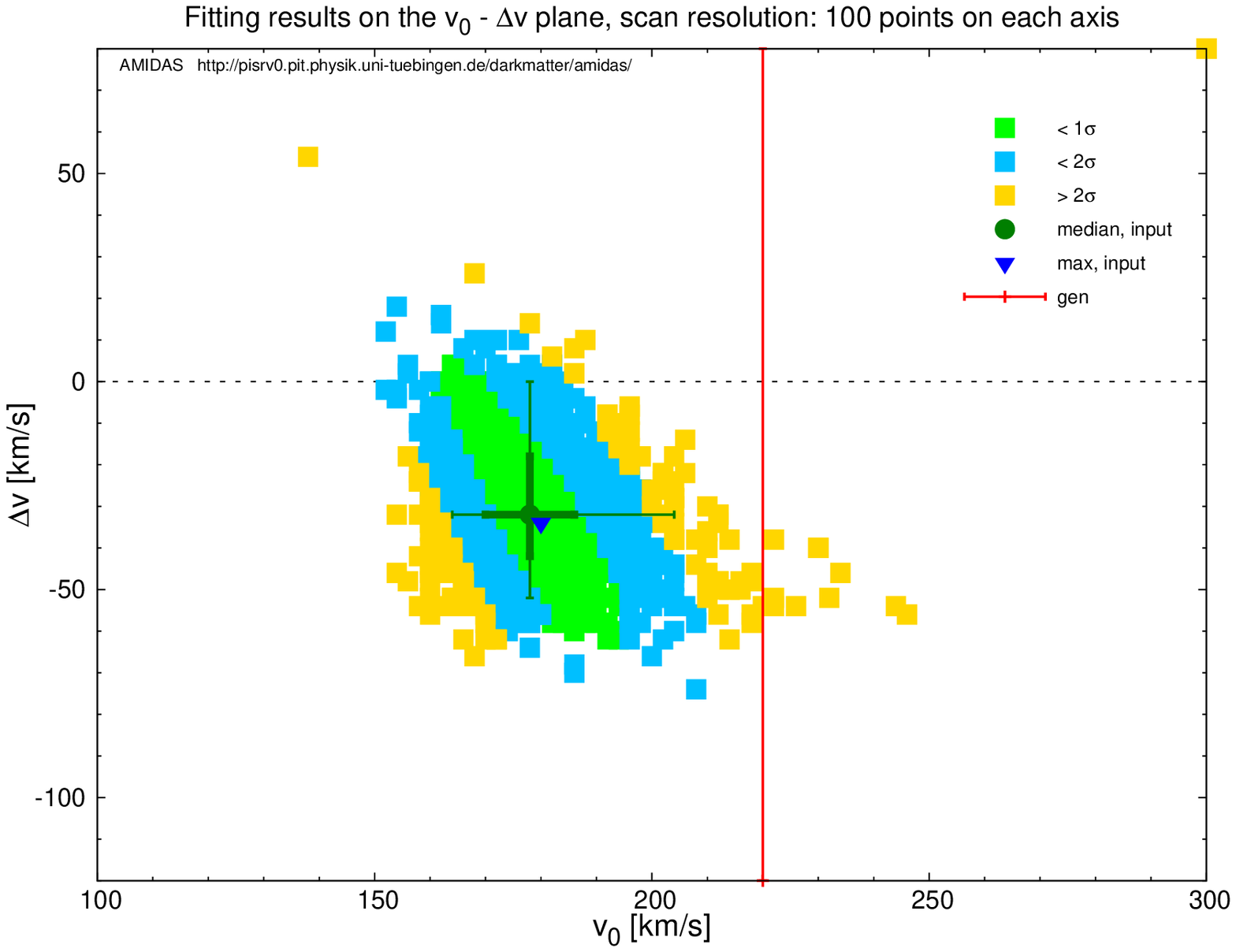} \hspace*{-1.6cm} \par
\makebox[8.5cm]{(a)}\hspace{0.325cm}\makebox[8.175cm]{(b)}           \\ \vspace{0.5cm}
\hspace*{-1.6cm}
\includegraphics[width=8.5cm]{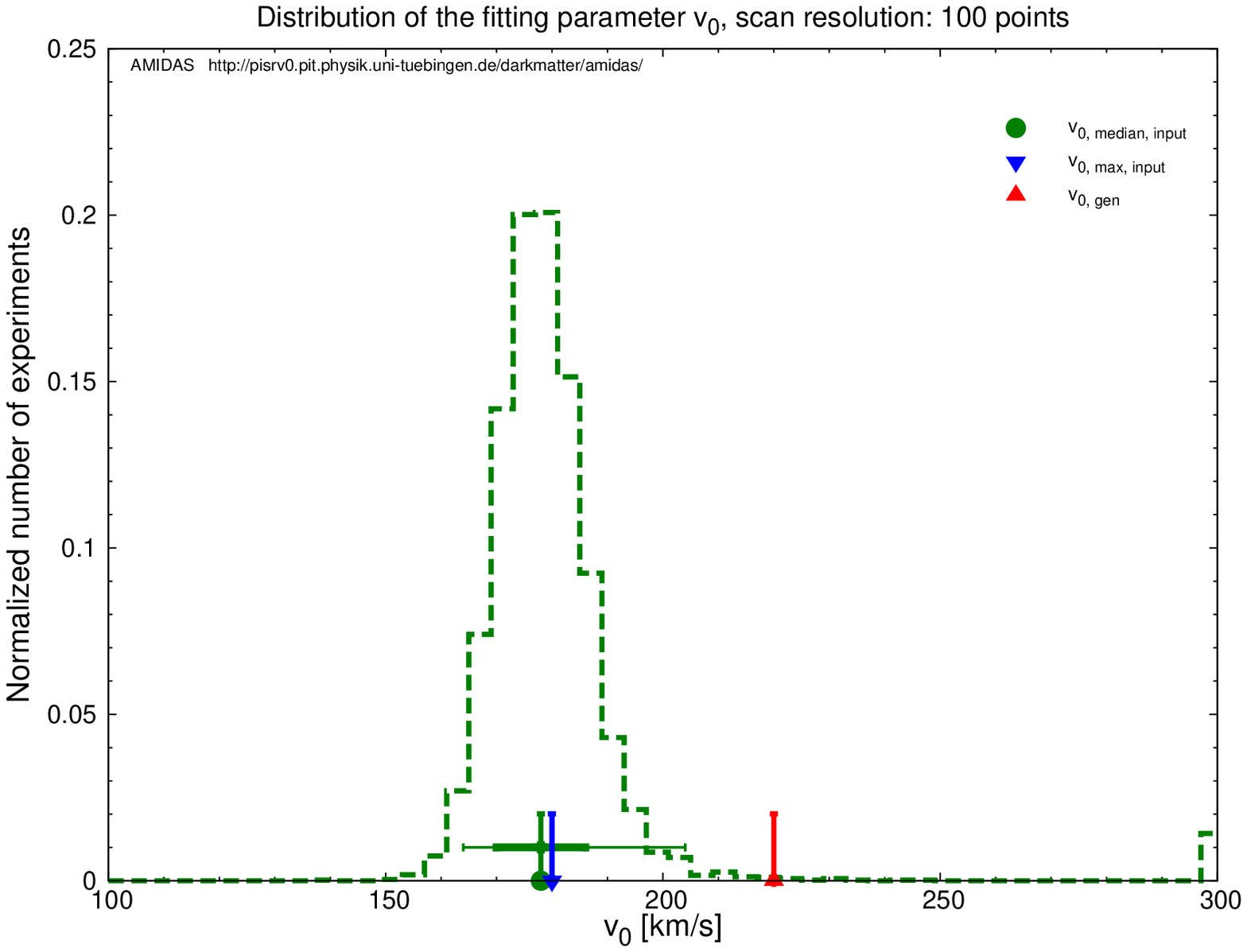}
\includegraphics[width=8.5cm]{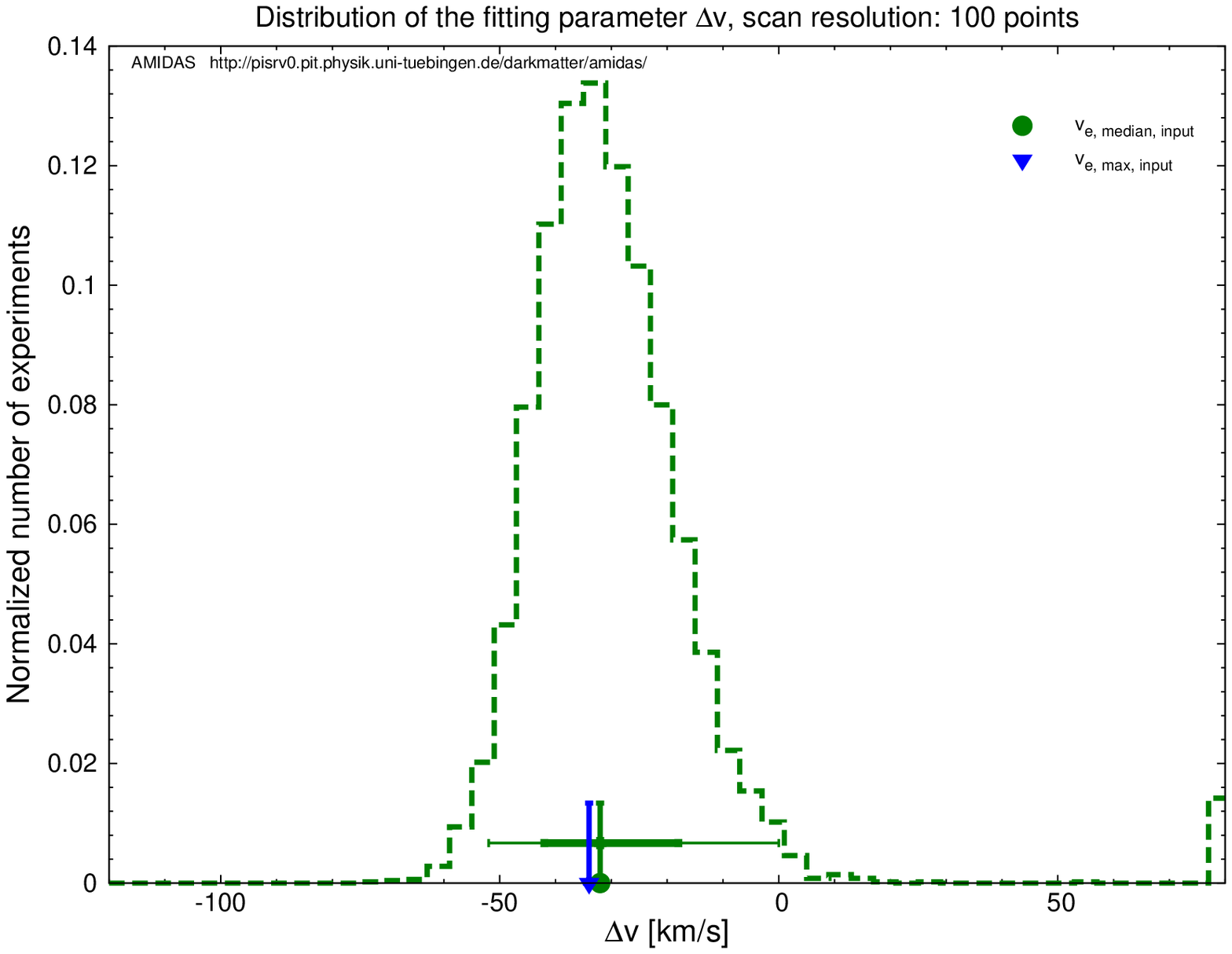}    \hspace*{-1.6cm} \par
\makebox[8.5cm]{(c)}\hspace{0.325cm}\makebox[8.175cm]{(d)}           \\
}
\vspace{-0.35cm}
\end{center}
\caption{
 As in Figs.~\ref{fig:f1v-Ge-100-0500-Gau_k-sh-Gau},
 except that
 the variated shifted Maxwellian velocity distribution function
 given in Eq.~(\ref{eqn:f1v_sh_Dv})
 with two fitting parameters $v_0$ and $\Delta v$
 is used.
 The Gaussian probability distribution
 for both fitting parameters
 with expectation values of \mbox{$v_0 = {\it 200}$ km/s}
 and \mbox{$\Delta v = {\it -20}$ km/s}
 and a common 1$\sigma$ uncertainty of \mbox{{\em 40} km/s}
 has been used.
}
\label{fig:f1v-Ge-100-0500-Gau_k-sh_Dv-Gau}
\end{figure}
}
\newcommand{\plotGeSiGeGaukshDvGau}{
\begin{figure}[t!]
\begin{center}
\vspace{-0.25cm}
{
\hspace*{-1.6cm}
\includegraphics[width=8.5cm]{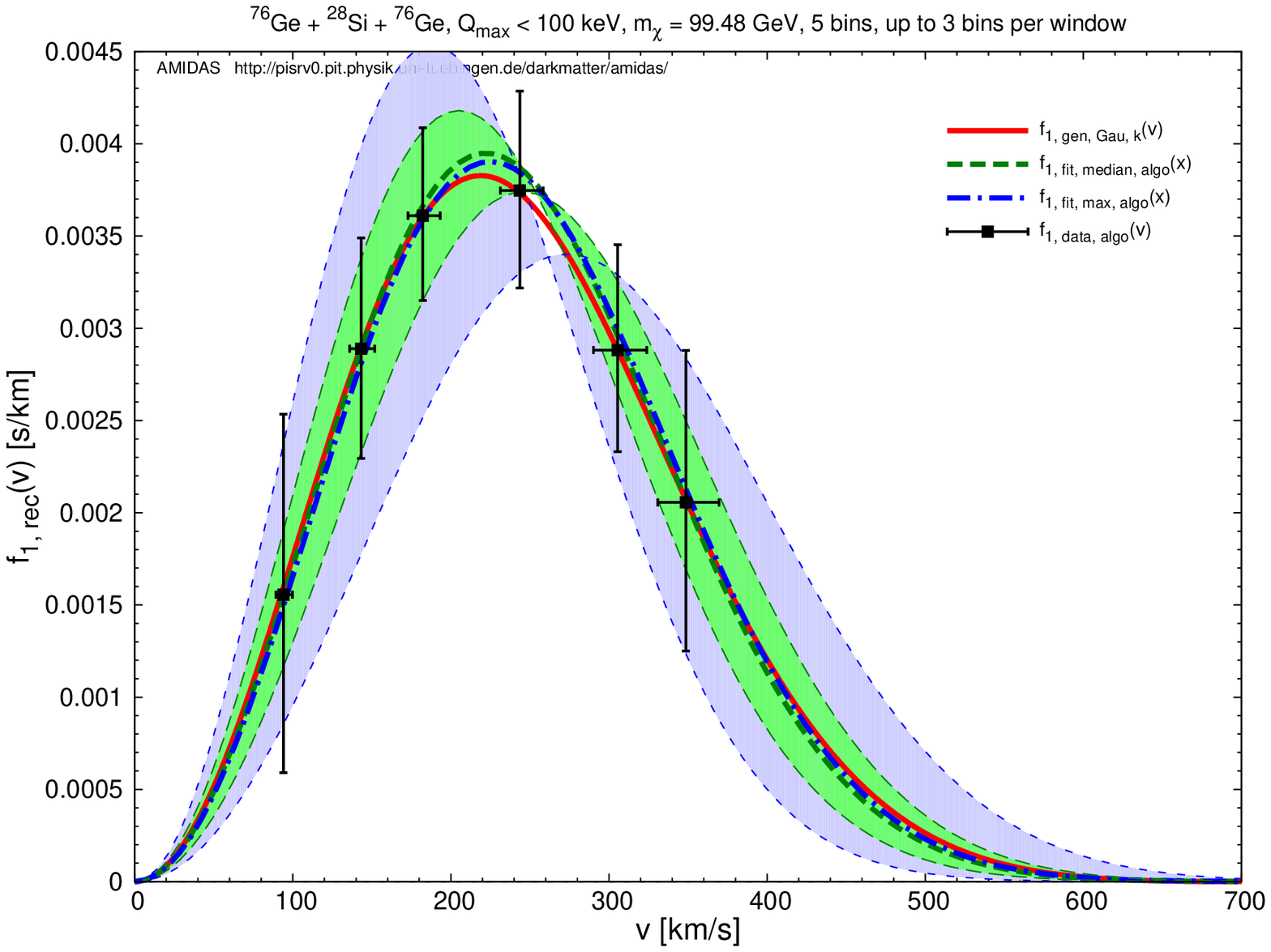}
\includegraphics[width=8.5cm]{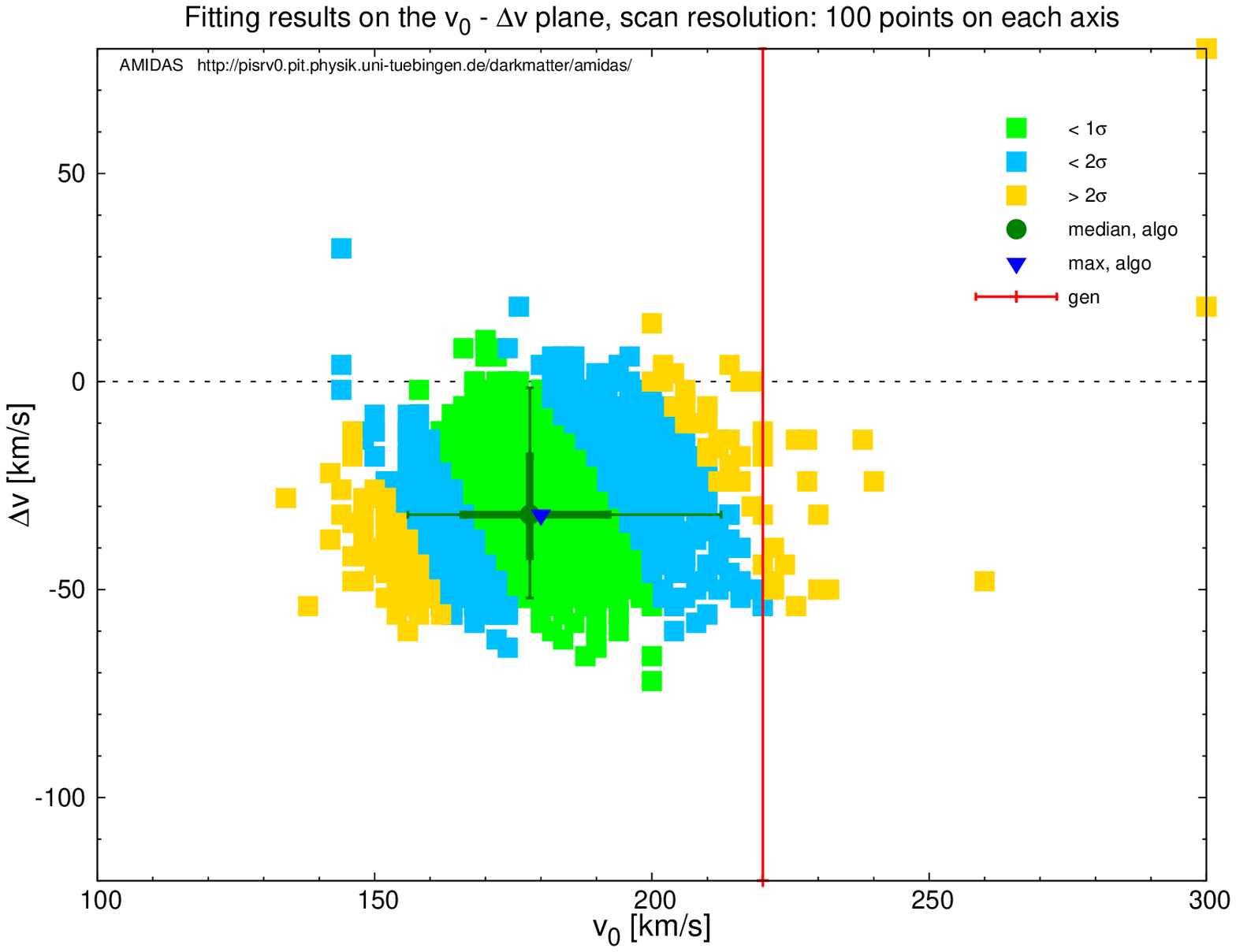} \hspace*{-1.6cm} \par
\makebox[8.5cm]{(a)}\hspace{0.325cm}\makebox[8.175cm]{(b)}                \\ \vspace{0.5cm}
\hspace*{-1.6cm}
\includegraphics[width=8.5cm]{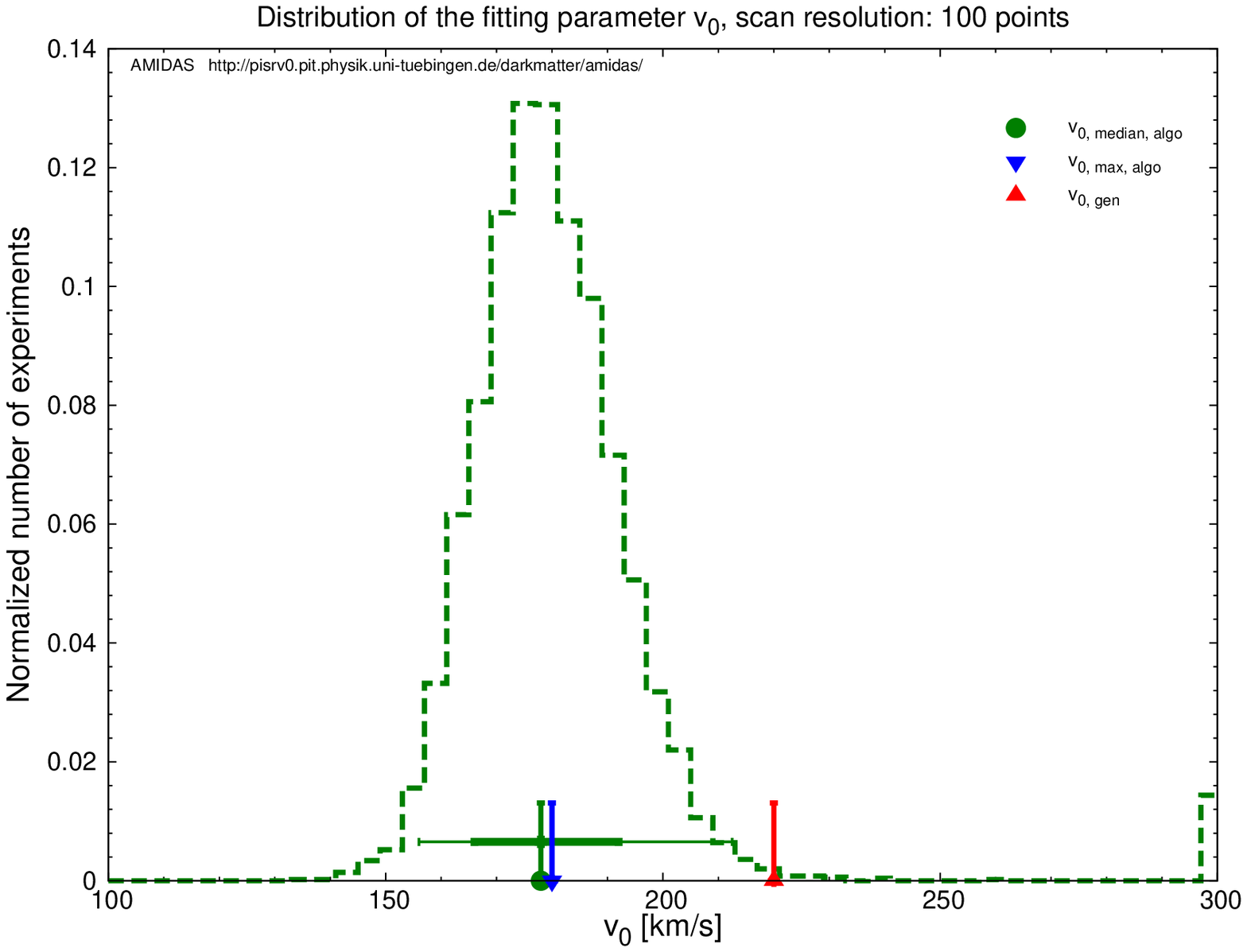}
\includegraphics[width=8.5cm]{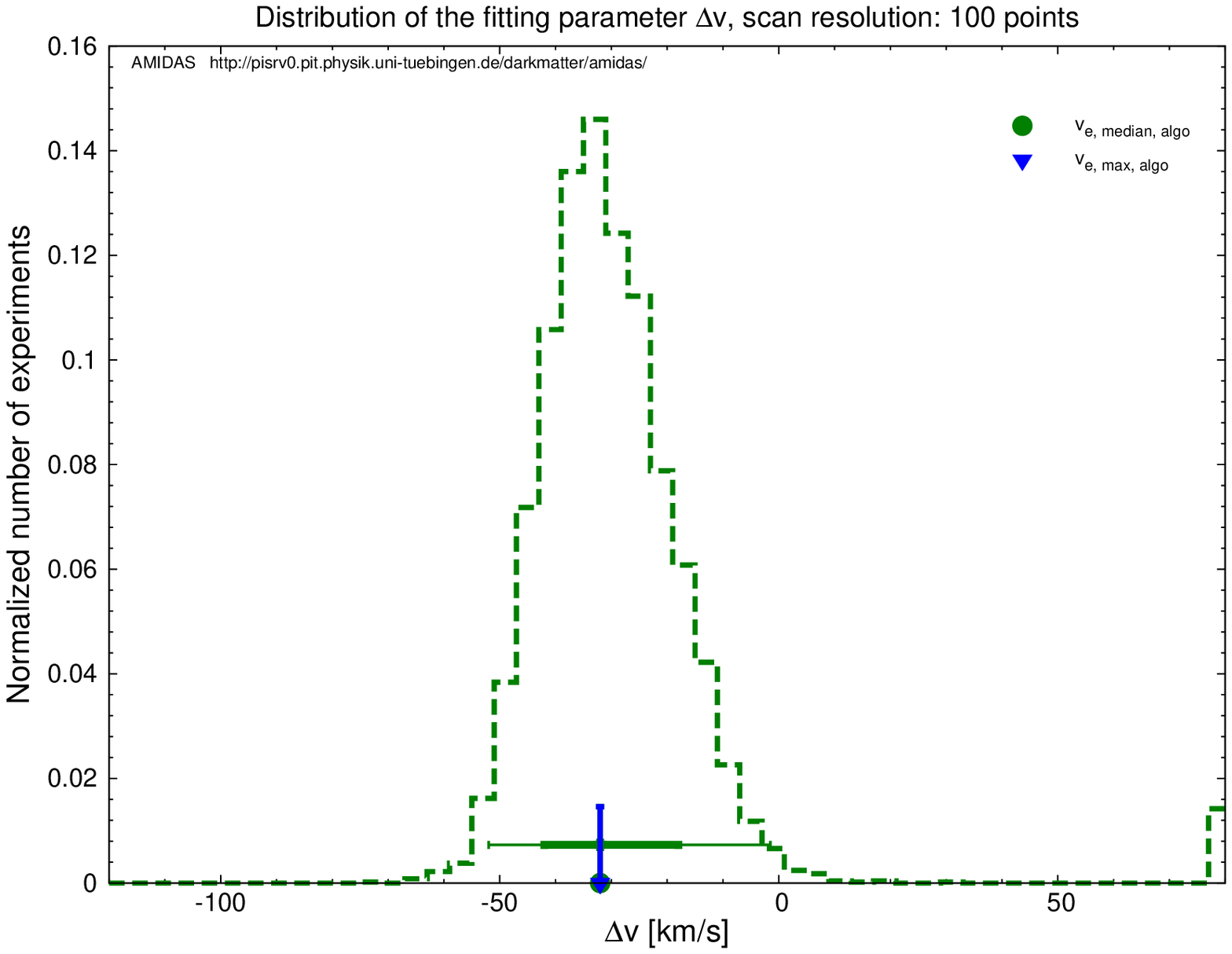}    \hspace*{-1.6cm} \par
\makebox[8.5cm]{(c)}\hspace{0.325cm}\makebox[8.175cm]{(d)}                \\
}
\vspace{-0.35cm}
\end{center}
\caption{
 As in Figs.~\ref{fig:f1v-Ge-100-0500-Gau_k-sh_Dv-Gau},
 except that
 the reconstructed WIMP mass
 has been
 used.
}
\label{fig:f1v-Ge-SiGe-100-0500-Gau_k-sh_Dv-Gau}
\end{figure}
}
\newcommand{\plotGeGaukGaukGauflat}{
\begin{figure}[t!]
\begin{center}
\vspace{-0.25cm}
{
\hspace*{-1.6cm}
\includegraphics[width=8.5cm]{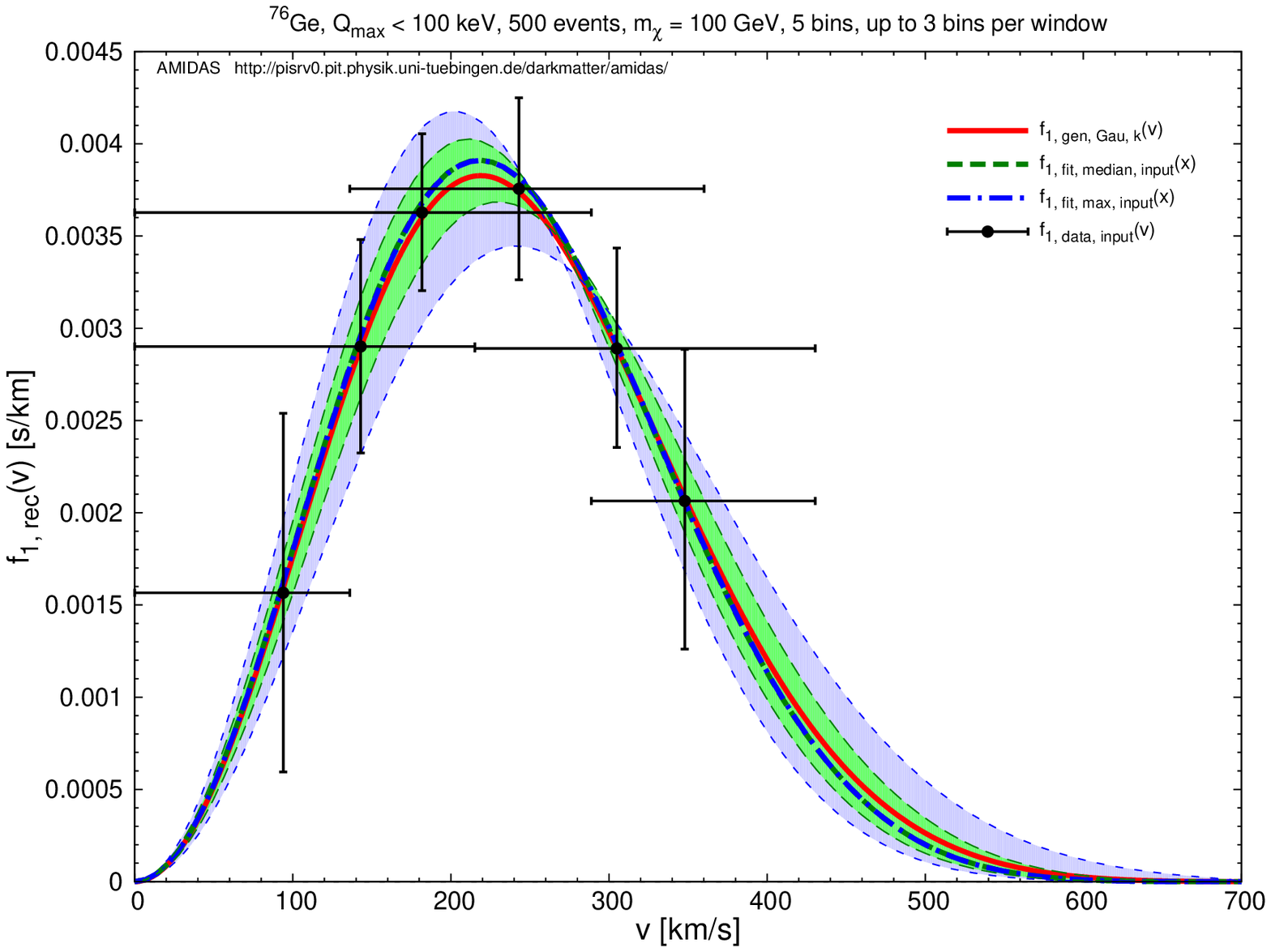}
\includegraphics[width=8.5cm]{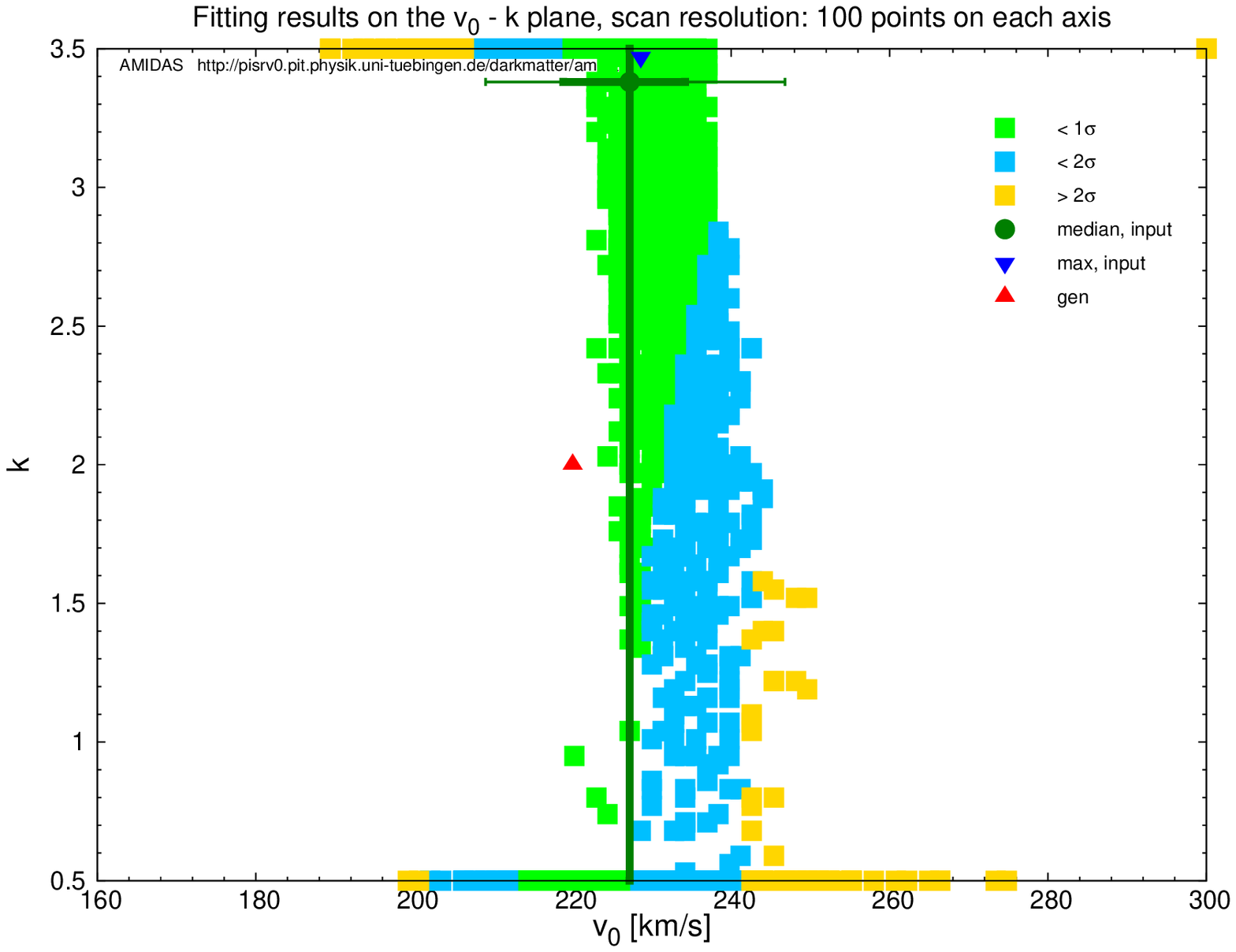} \hspace*{-1.6cm} \par
\makebox[8.5cm]{(a)}\hspace{0.325cm}\makebox[8.175cm]{(b)}               \\ \vspace{0.5cm}
\hspace*{-1.6cm}
\includegraphics[width=8.5cm]{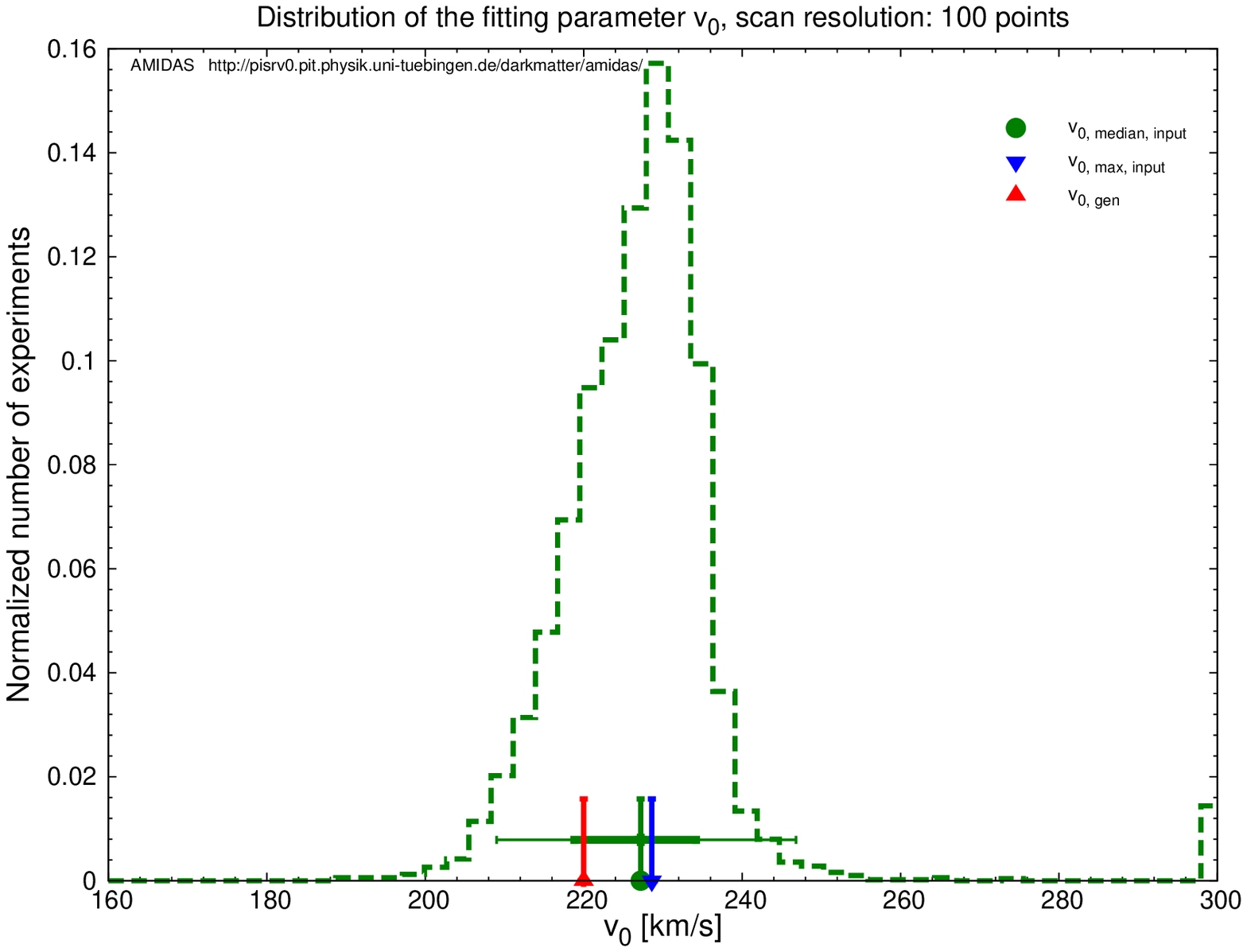}
\includegraphics[width=8.5cm]{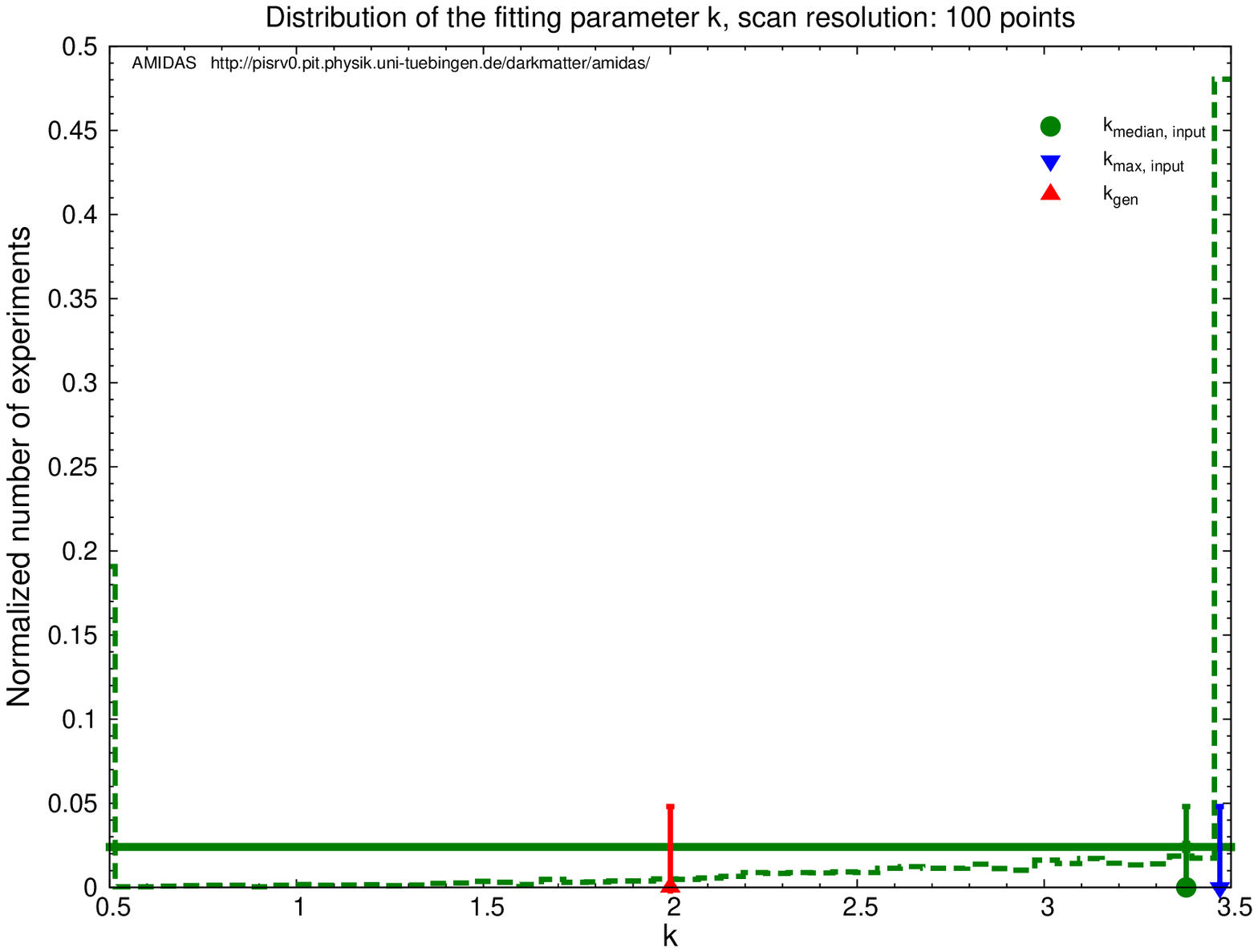}    \hspace*{-1.6cm} \par
\makebox[8.5cm]{(c)}\hspace{0.325cm}\makebox[8.175cm]{(d)}               \\
}
\vspace{-0.35cm}
\end{center}
\caption{
 As in Figs.~\ref{fig:f1v-Ge-100-0500-Gau_k-sh-Gau},
 except that
 the modified Maxwellian velocity distribution function
 given in Eq.~(\ref{eqn:f1v_Gau_k})
 with two fitting parameters $v_0$ and $k$
 is used.
 The Gaussian probability distribution
 for $v_0$
 with an expectation value of \mbox{$v_0 = 230$ km/s}
 and a 1$\sigma$ uncertainty of \mbox{20 km/s}
 but
 the flat distribution for $k$
 have been used.
 Remind that
 the bins at $k = 0.5$ and $k = 3.5$ are ``overflow'' bins,
 which contain also the experiments
 with the best--fit $k$ value of
 either $k < 0.5$ or $k > 3.5$.
}
\label{fig:f1v-Ge-100-0500-Gau_k-Gau_k-Gau-flat}
\end{figure}
}
\newcommand{\plotGeSiGeGaukGaukGauflat}{
\begin{figure}[t!]
\begin{center}
\vspace{-0.25cm}
{
\hspace*{-1.6cm}
\includegraphics[width=8.5cm]{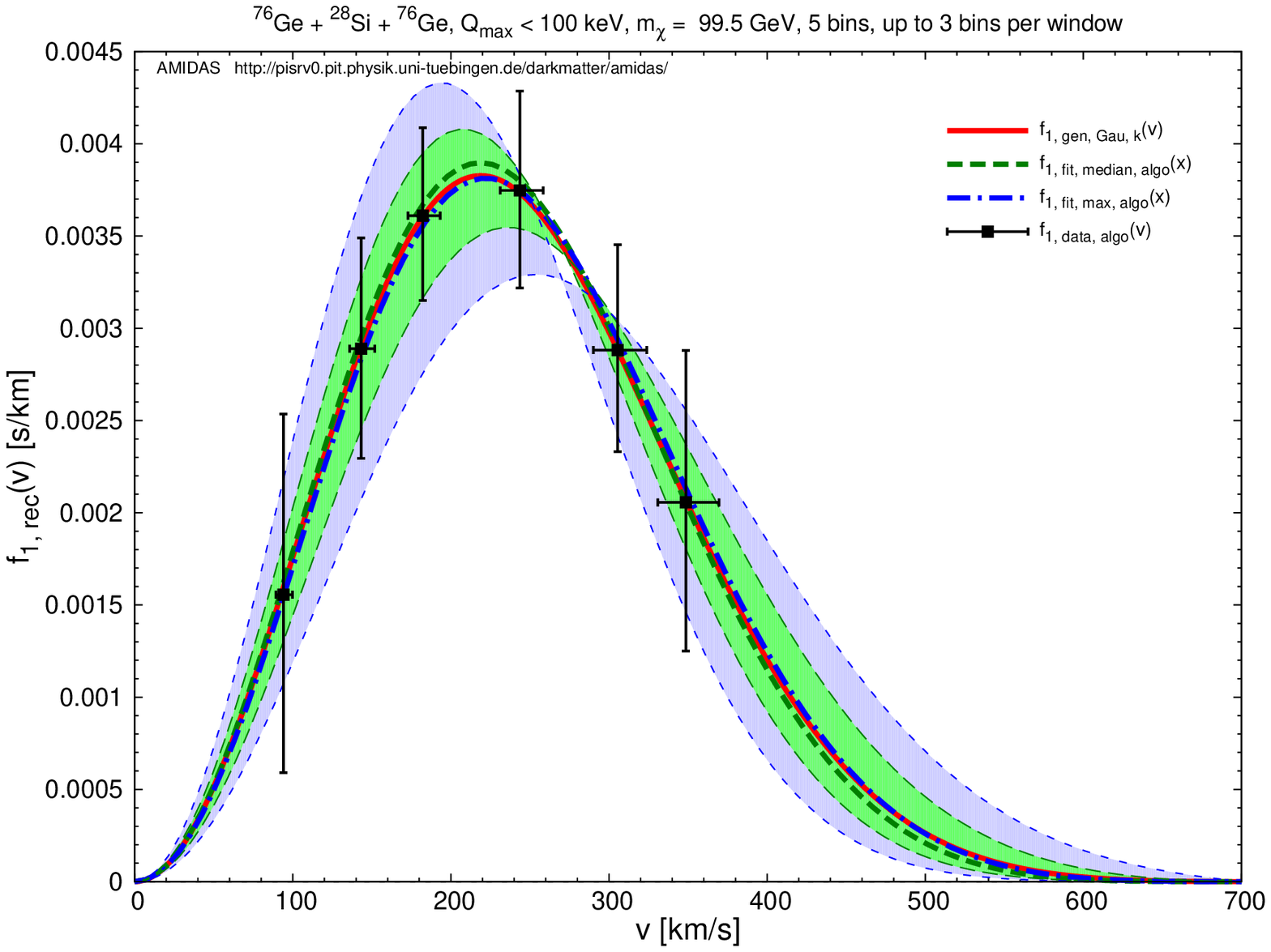}
\includegraphics[width=8.5cm]{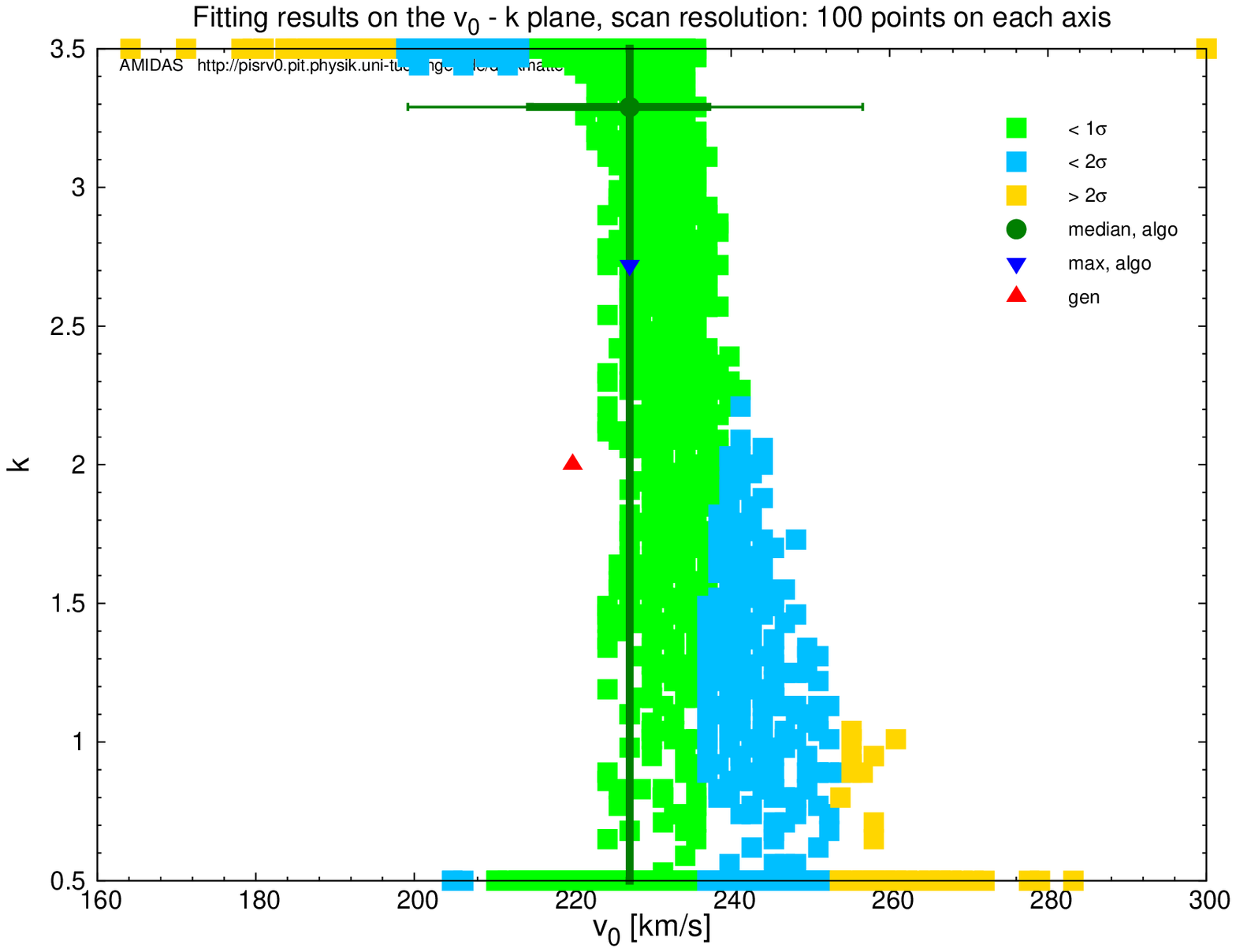} \hspace*{-1.6cm} \par
\makebox[8.5cm]{(a)}\hspace{0.325cm}\makebox[8.175cm]{(b)}                    \\ \vspace{0.5cm}
\hspace*{-1.6cm}
\includegraphics[width=8.5cm]{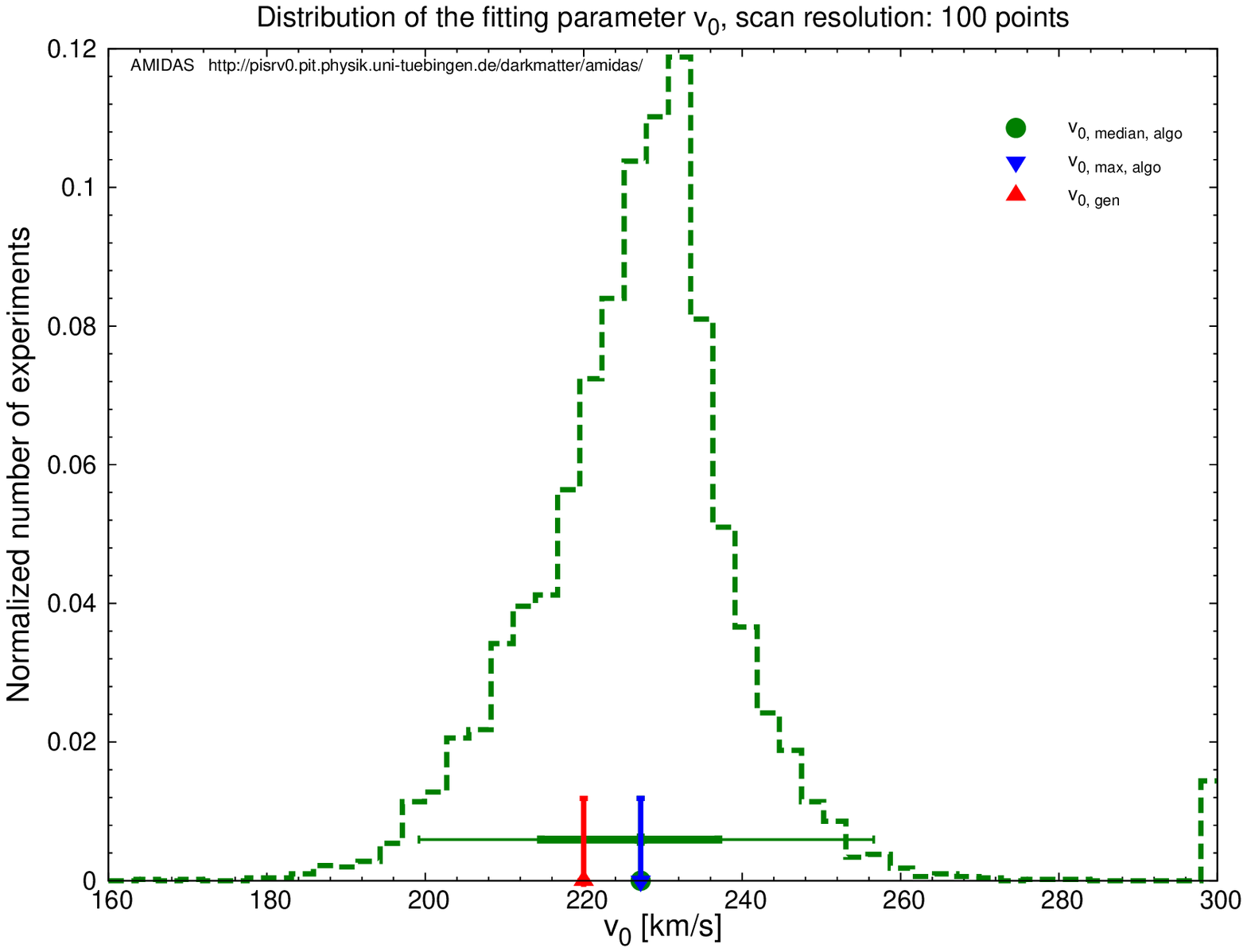}
\includegraphics[width=8.5cm]{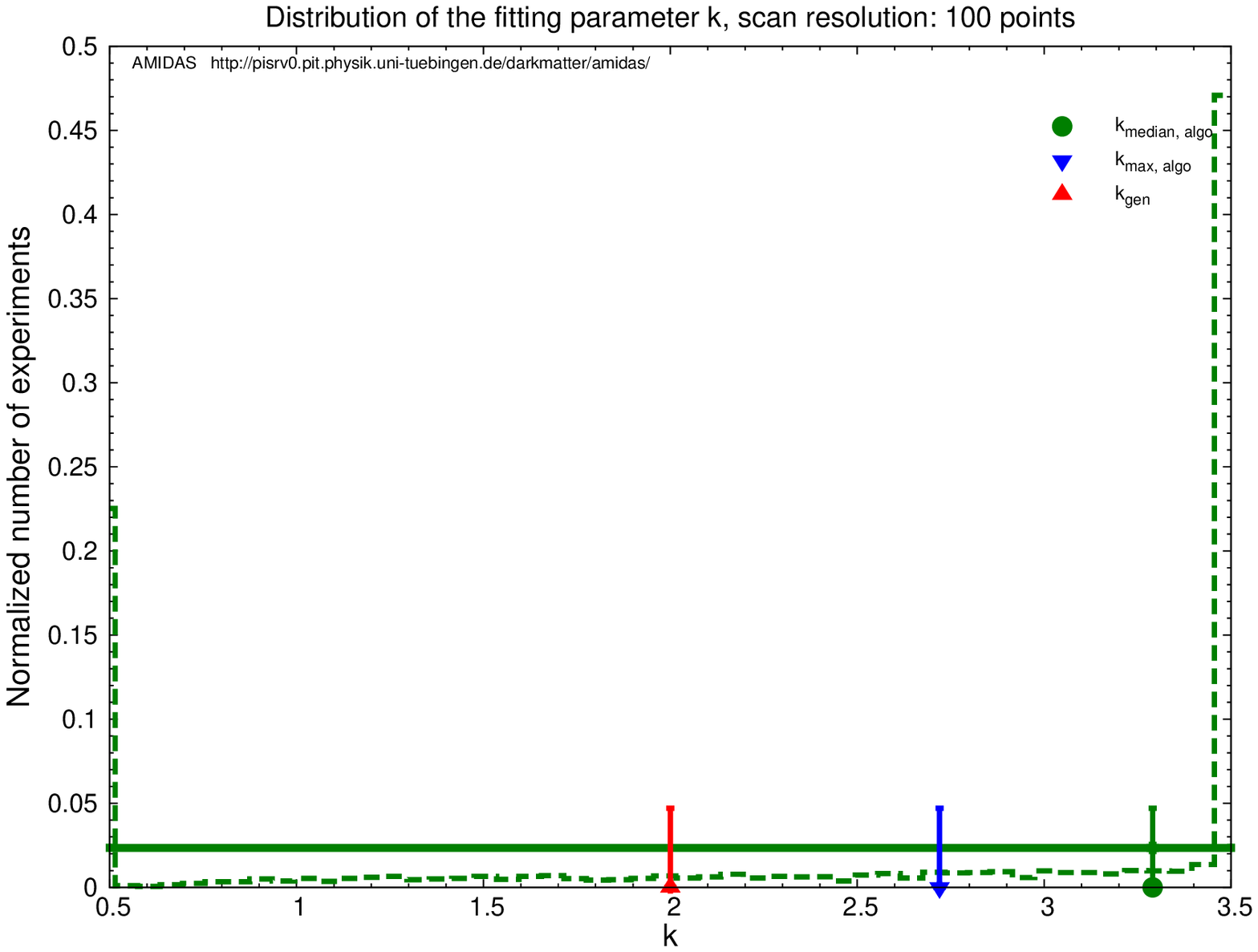}    \hspace*{-1.6cm} \par
\makebox[8.5cm]{(c)}\hspace{0.325cm}\makebox[8.175cm]{(d)}                    \\
}
\vspace{-0.35cm}
\end{center}
\caption{
 As in Figs.~\ref{fig:f1v-Ge-100-0500-Gau_k-Gau_k-Gau-flat},
 except that
 the reconstructed WIMP mass
 has been
 used.
}
\label{fig:f1v-Ge-SiGe-100-0500-Gau_k-Gau_k-Gau-flat}
\end{figure}
}
\newcommand{\plotGeSiGeshGauflatL}{
\begin{figure}[t!]
\begin{center}
\vspace{-0.25cm}
{
\hspace*{-1.6cm}
\includegraphics[width=8.5cm]{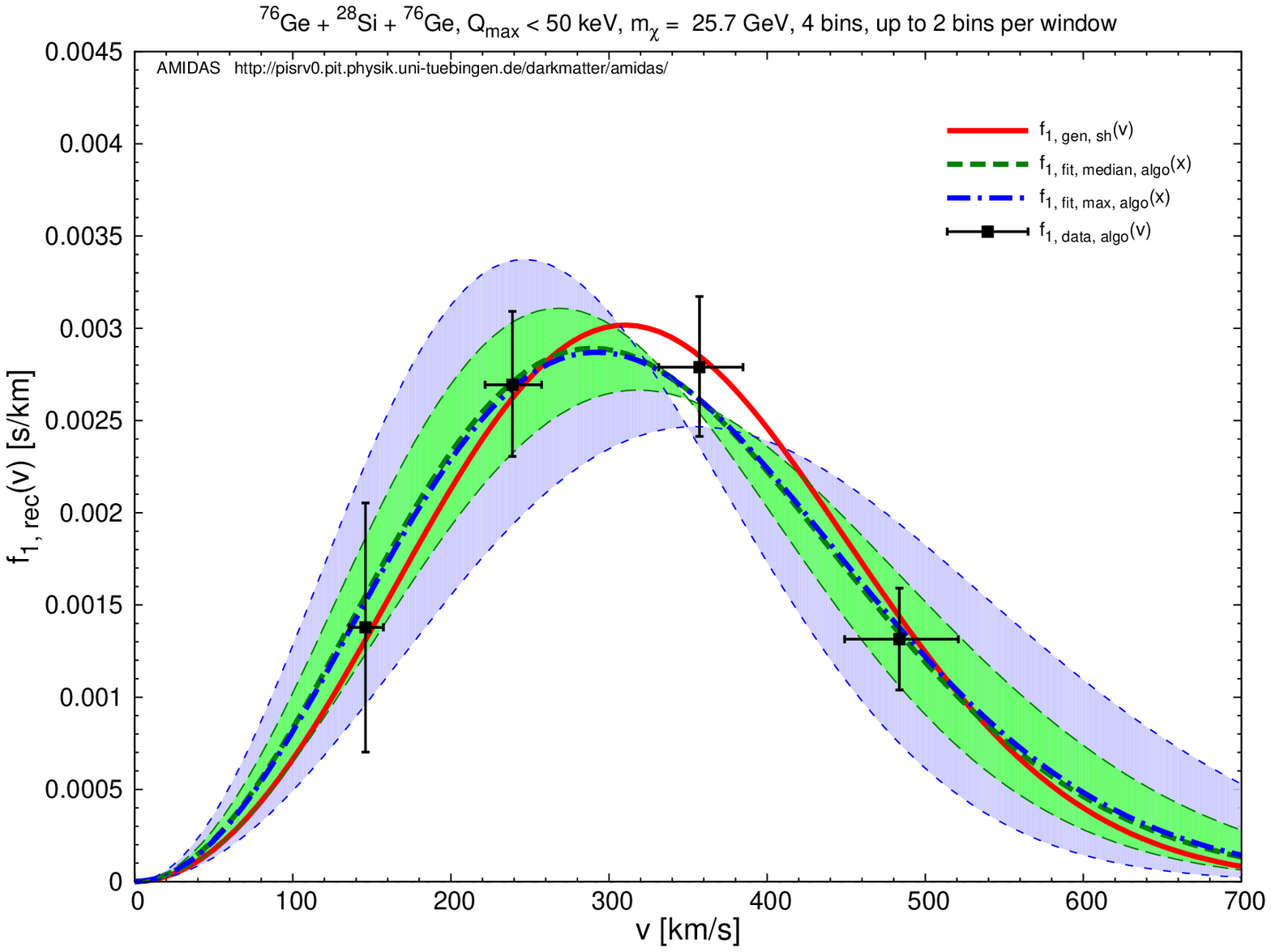}
\includegraphics[width=8.5cm]{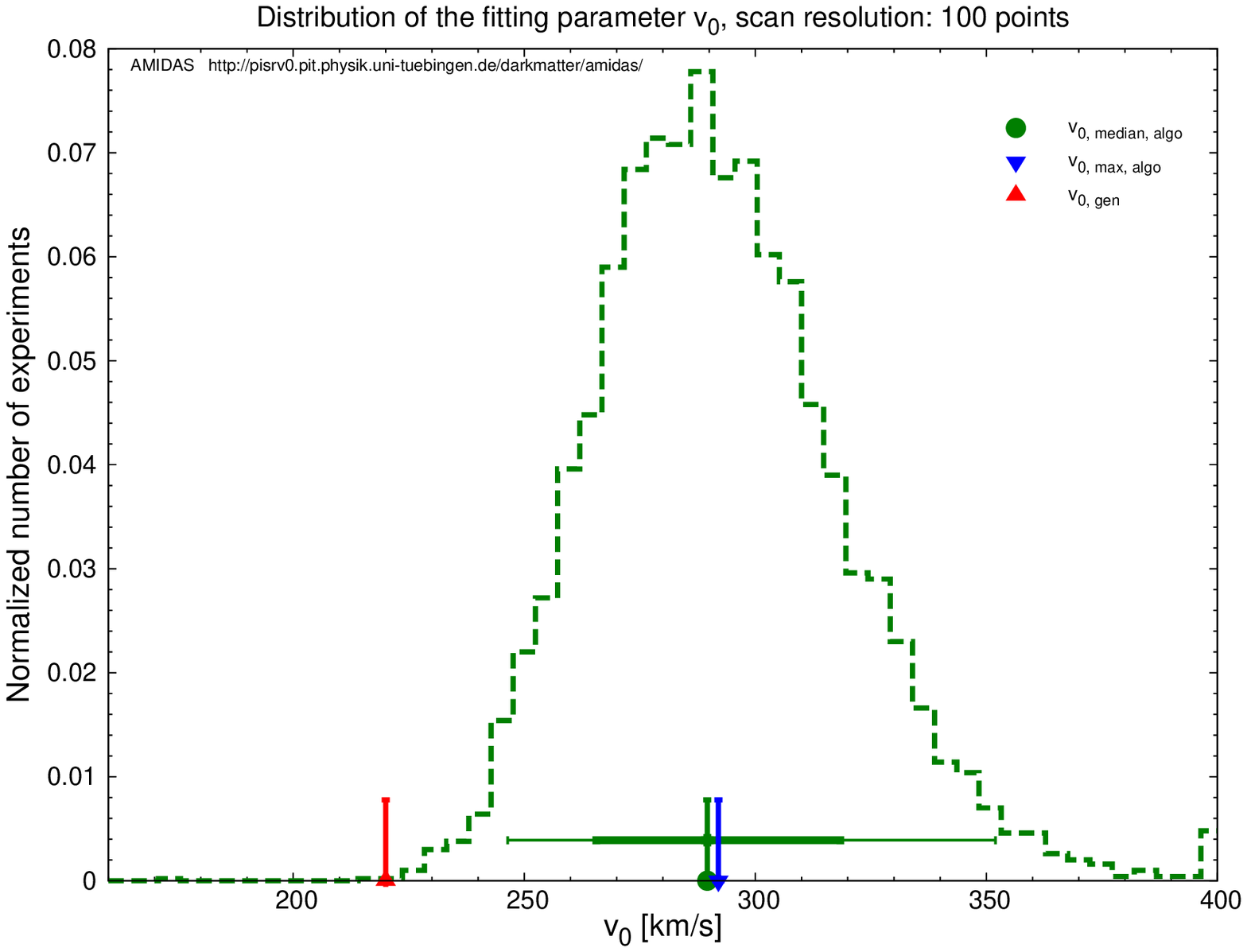} \hspace*{-1.6cm} \par
\makebox[8.5cm]{(a)}\hspace{0.325cm}\makebox[8.175cm]{(b)}%
}
\vspace{-0.35cm}
\end{center}
\caption{
 As in Figs.~\ref{fig:f1v-Ge-SiGe-100-0500-sh-Gau-flat}(c) and (d):
 shifted and simple Maxwellian velocity distributions
 have been used for generating WIMP events
 and as the fitting function,
 respectively;
 the flat probability distribution
 for $v_0$
 and
 the reconstructed WIMP mass
 has been used,
 except that
 the input WIMP mass has been set as \mbox{$\mchi = 25$ GeV}.
 See the text for further details
 about the simulation setup.
}
\label{fig:f1v-Ge-SiGe-025-0500-sh-Gau-flat}
\end{figure}
}
\newcommand{\plotGeSiGeshshGauL}{
\begin{figure}[t!]
\begin{center}
\vspace{-0.25cm}
{
\hspace*{-1.6cm}
\includegraphics[width=8.5cm]{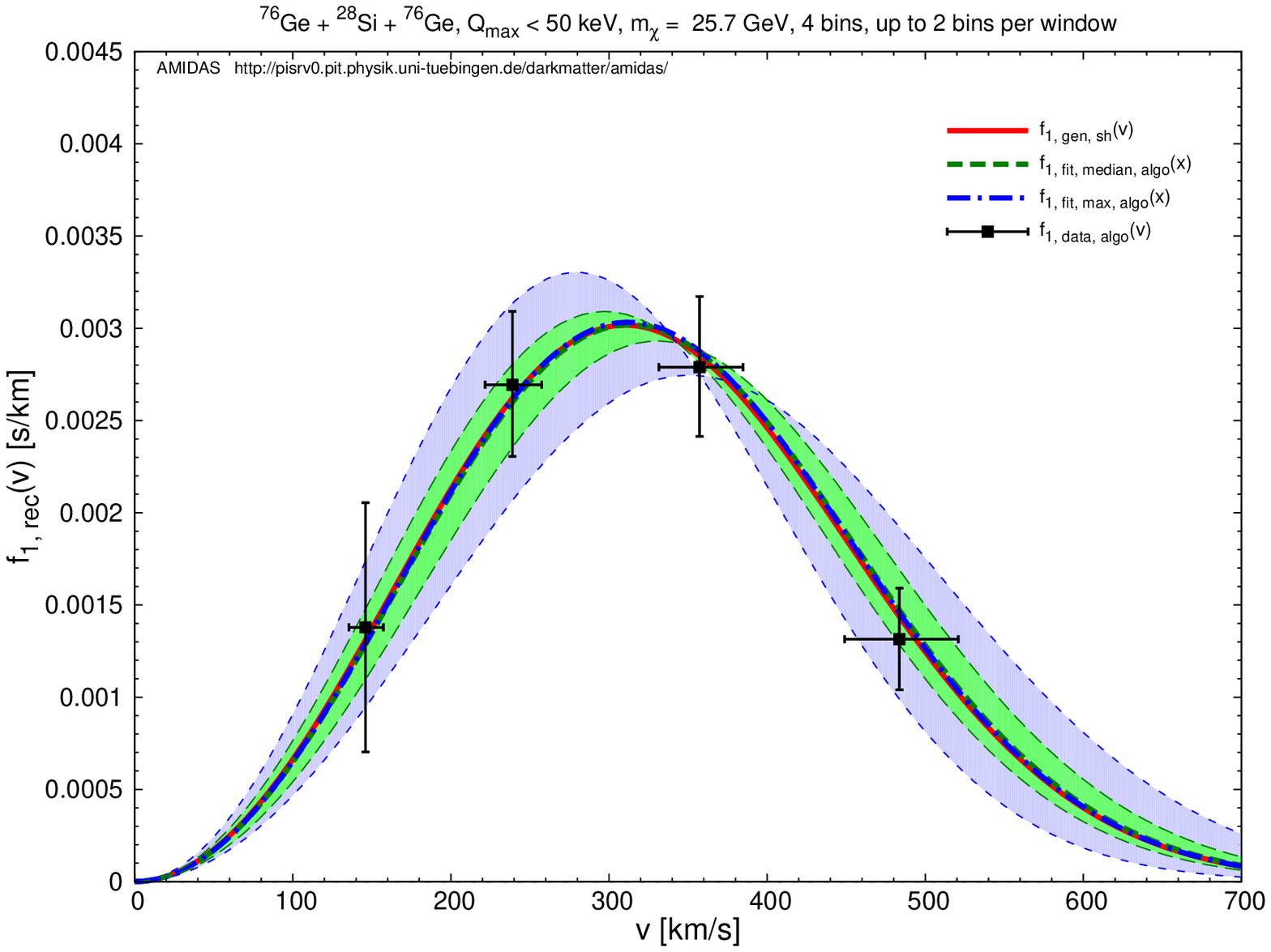}
\includegraphics[width=8.5cm]{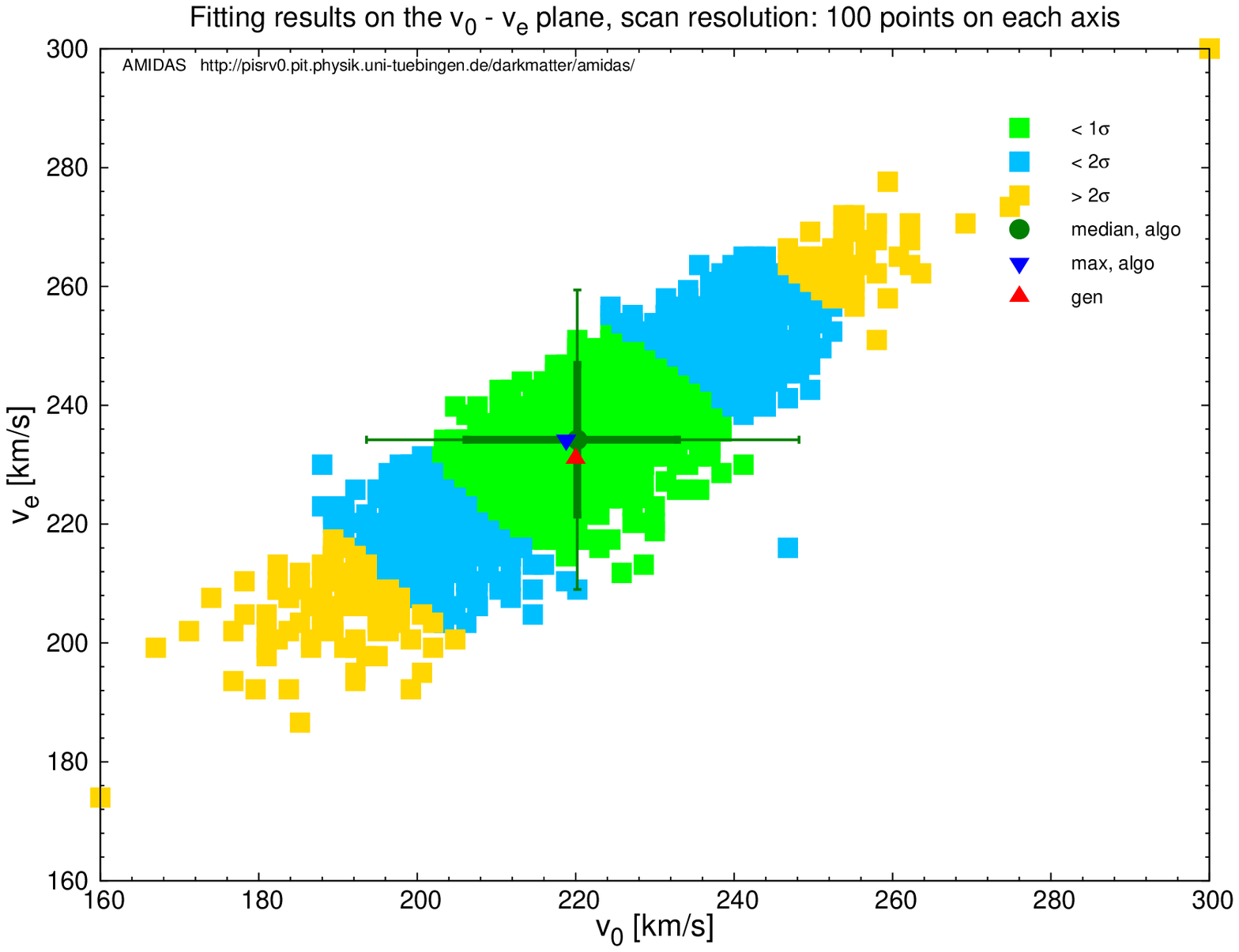} \hspace*{-1.6cm} \par
\makebox[8.5cm]{(a)}\hspace{0.325cm}\makebox[8.175cm]{(b)}          \\ \vspace{0.5cm}
\hspace*{-1.6cm}
\includegraphics[width=8.5cm]{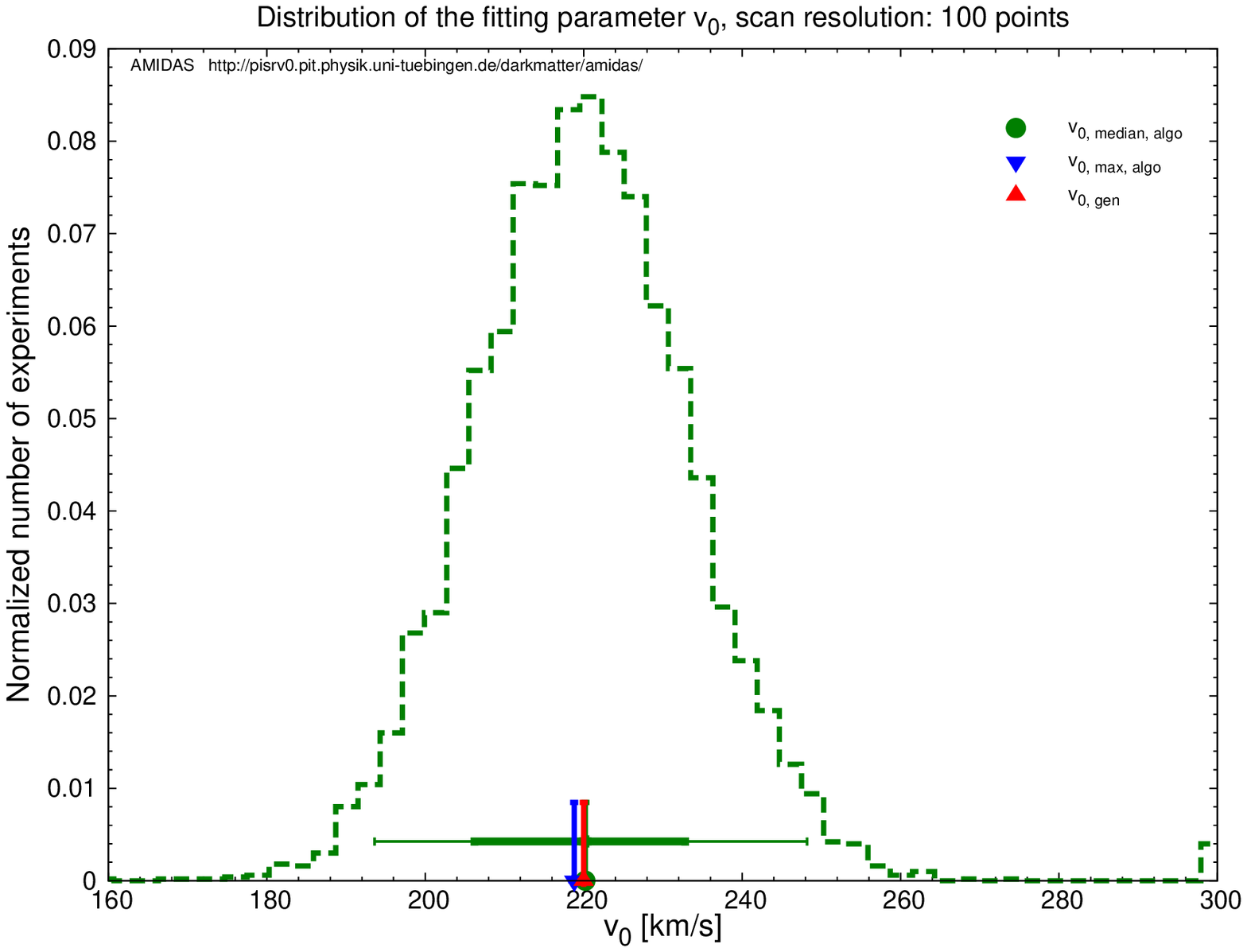}
\includegraphics[width=8.5cm]{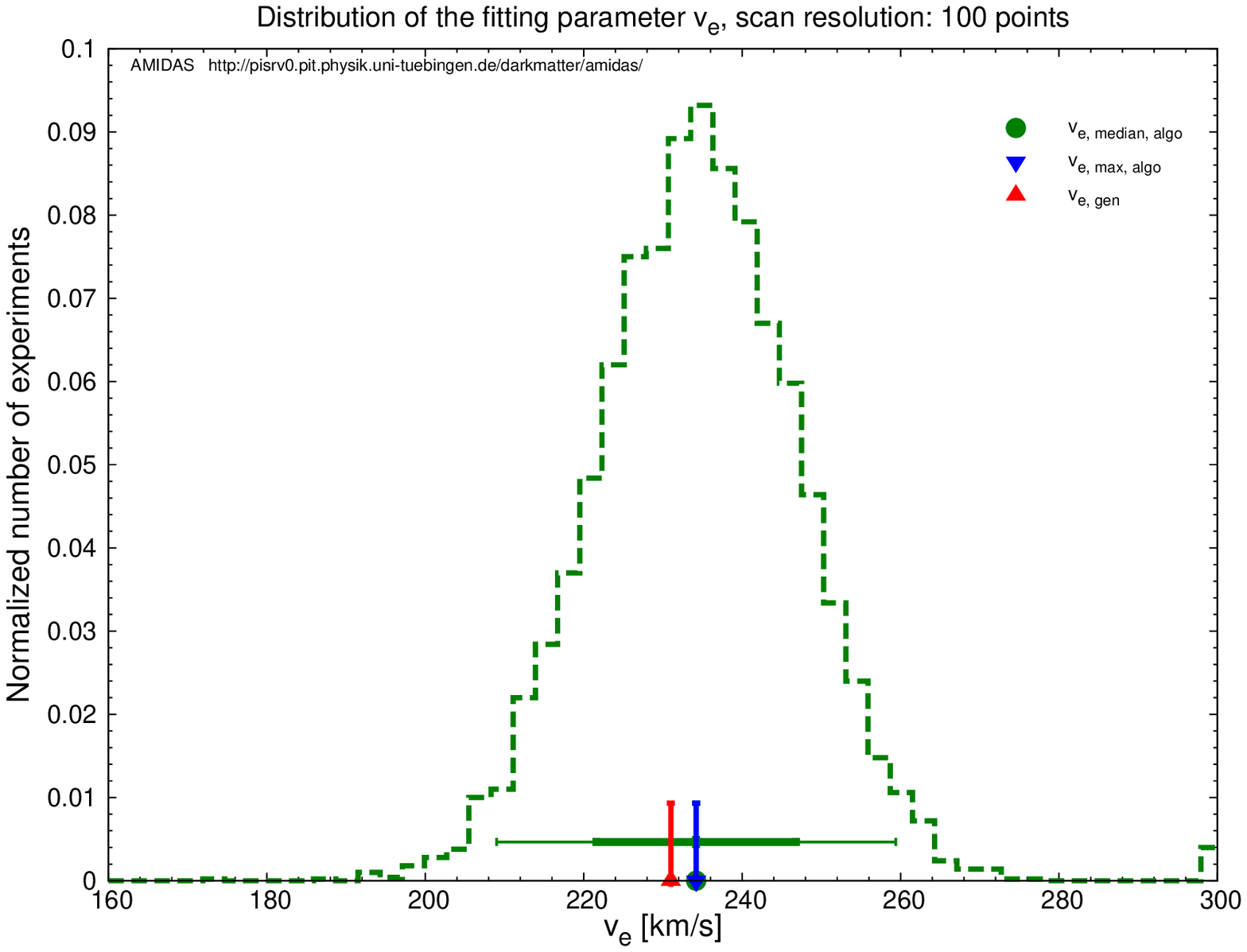}    \hspace*{-1.6cm} \par
\makebox[8.5cm]{(c)}\hspace{0.325cm}\makebox[8.175cm]{(d)}          \\
}
\vspace{-0.35cm}
\end{center}
\caption{
 As in Figs.~\ref{fig:f1v-Ge-SiGe-100-0500-sh-sh-Gau}:
 the shifted Maxwellian velocity distribution function
 and
 the Gaussian probability distribution
 for both fitting parameters
 with expectation values of \mbox{$v_0 = 230$ km/s}
 and \mbox{$\ve = 245$ km/s}
 and a common 1$\sigma$ uncertainty of \mbox{20 km/s}
 as well as 
 the reconstructed WIMP mass
 have been used,
 except that
 the input WIMP mass has been set as \mbox{$\mchi = 25$ GeV}.
}
\label{fig:f1v-Ge-SiGe-025-0500-sh-sh-Gau}
\end{figure}
}
\newcommand{\plotGeSiGeshshDvGauL}{
\begin{figure}[t!]
\begin{center}
\vspace{-0.25cm}
{
\hspace*{-1.6cm}
\includegraphics[width=8.5cm]{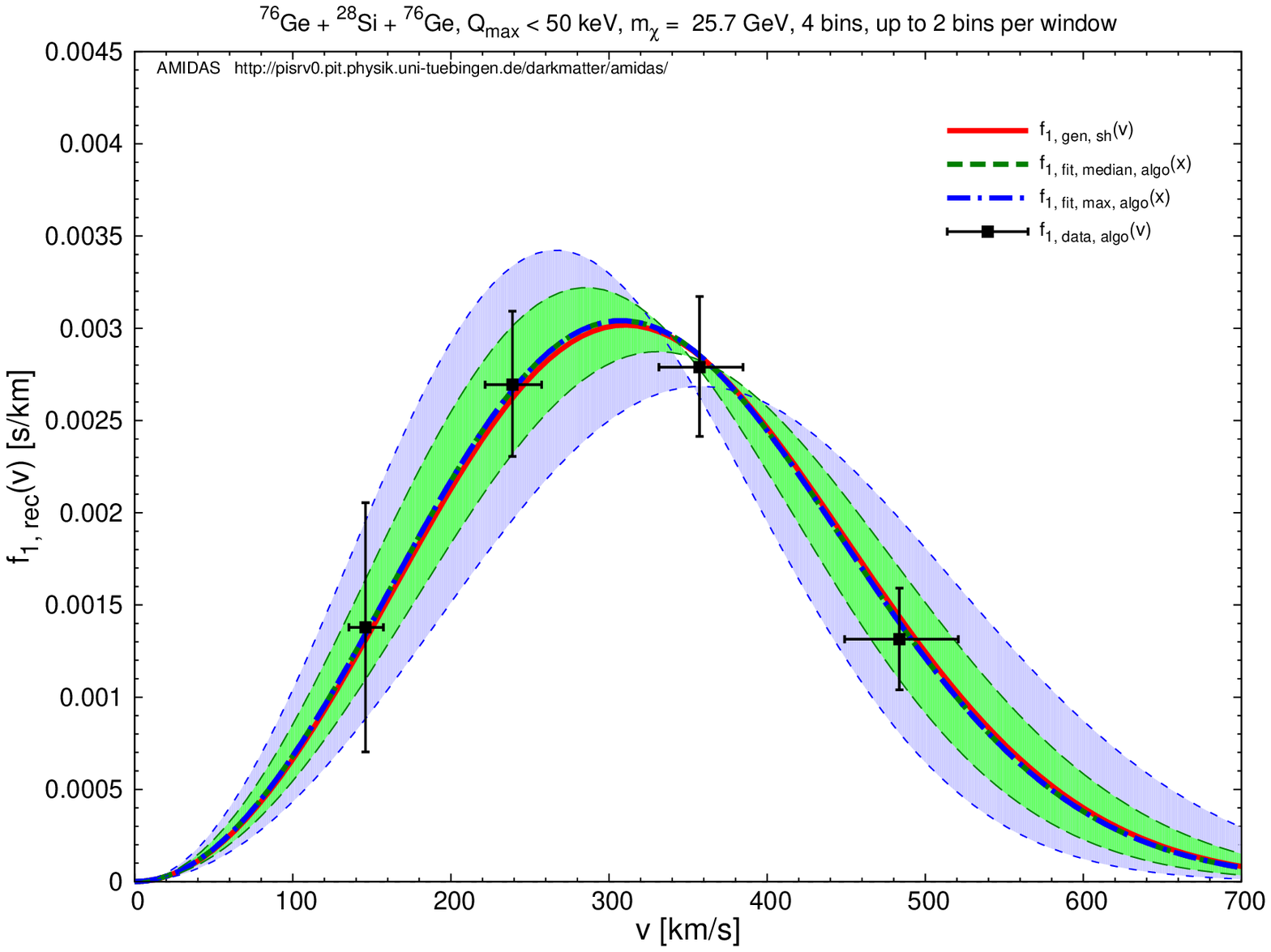}
\includegraphics[width=8.5cm]{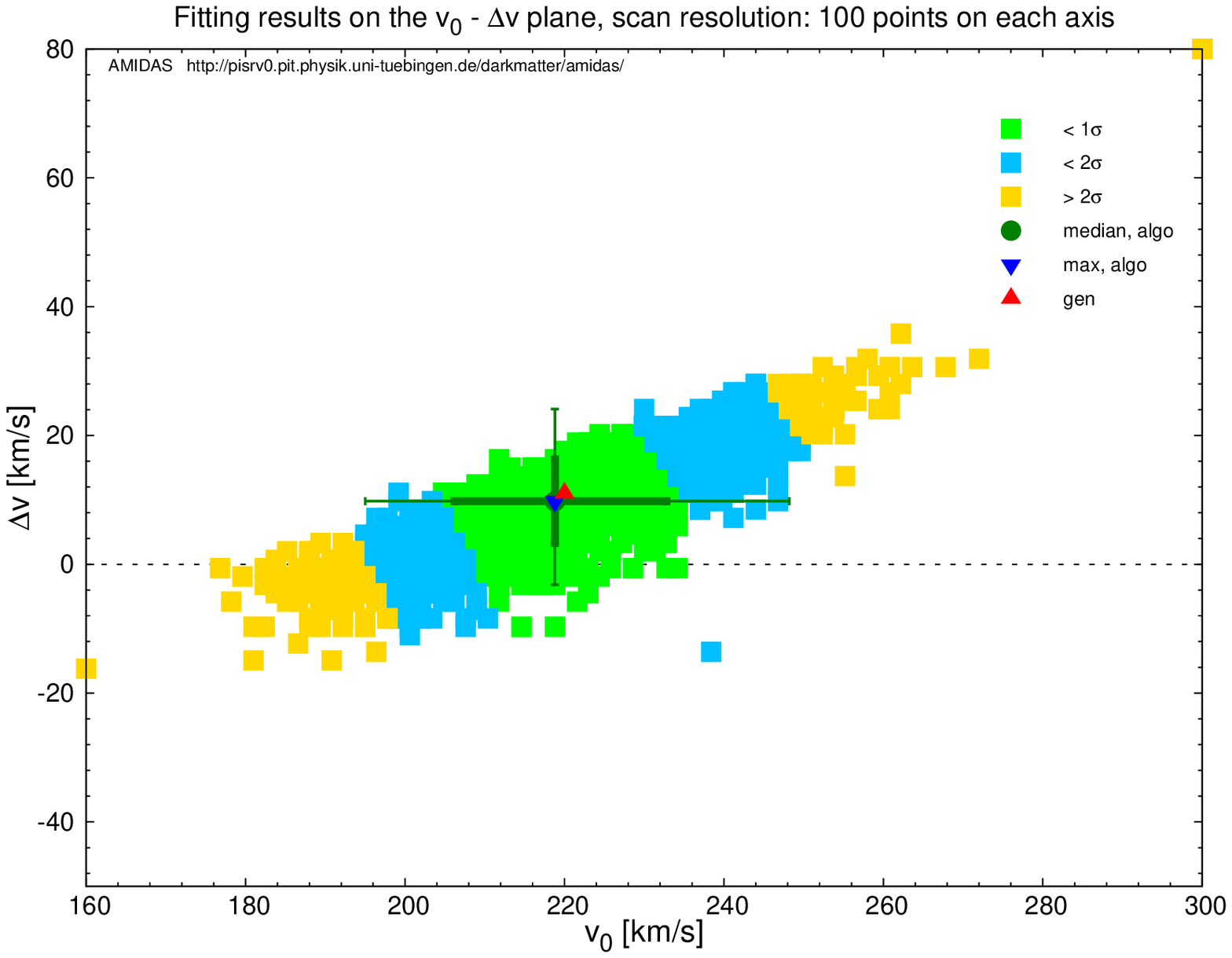} \hspace*{-1.6cm} \par
\makebox[8.5cm]{(a)}\hspace{0.325cm}\makebox[8.175cm]{(b)}             \\ \vspace{0.5cm}
\hspace*{-1.6cm}
\includegraphics[width=8.5cm]{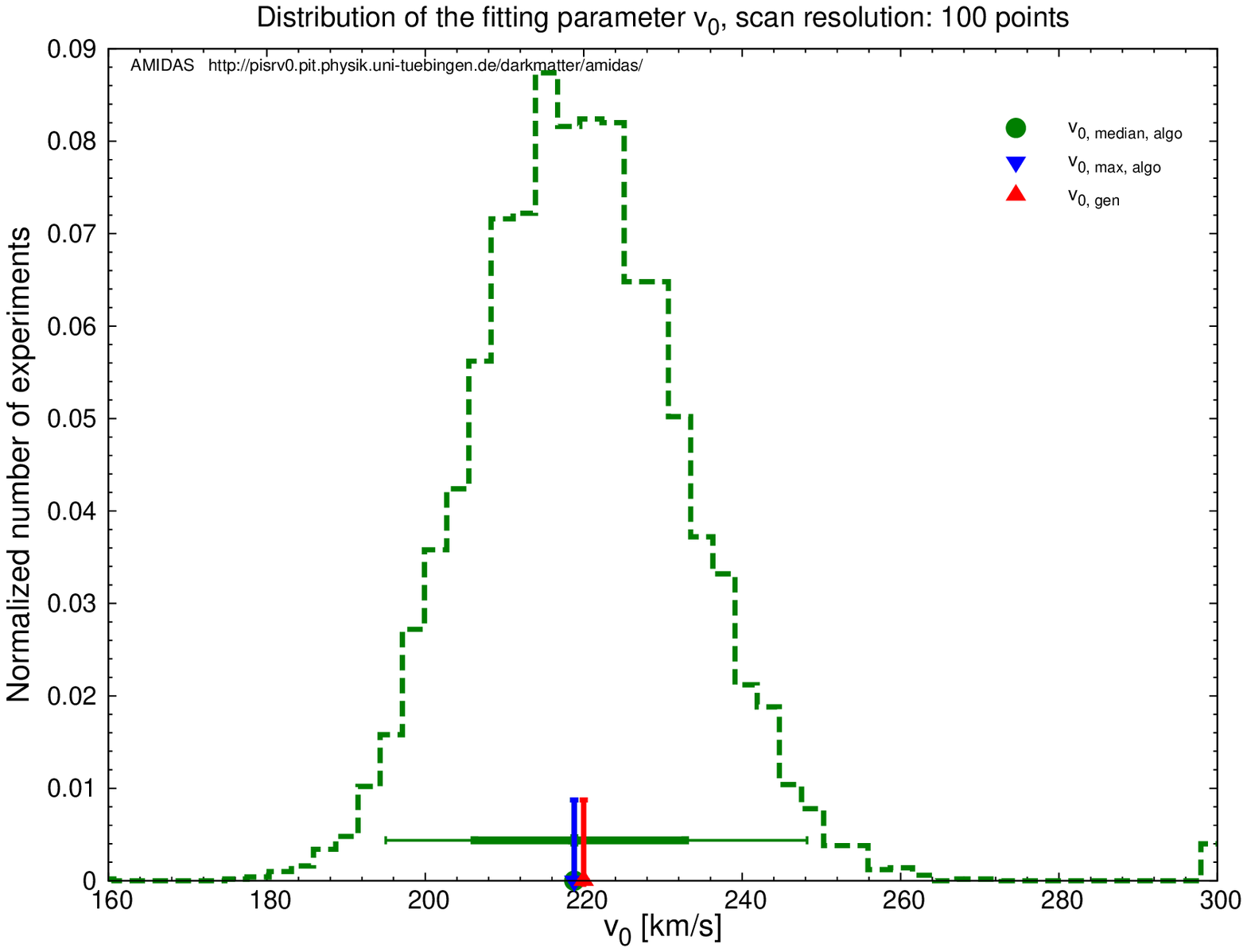}
\includegraphics[width=8.5cm]{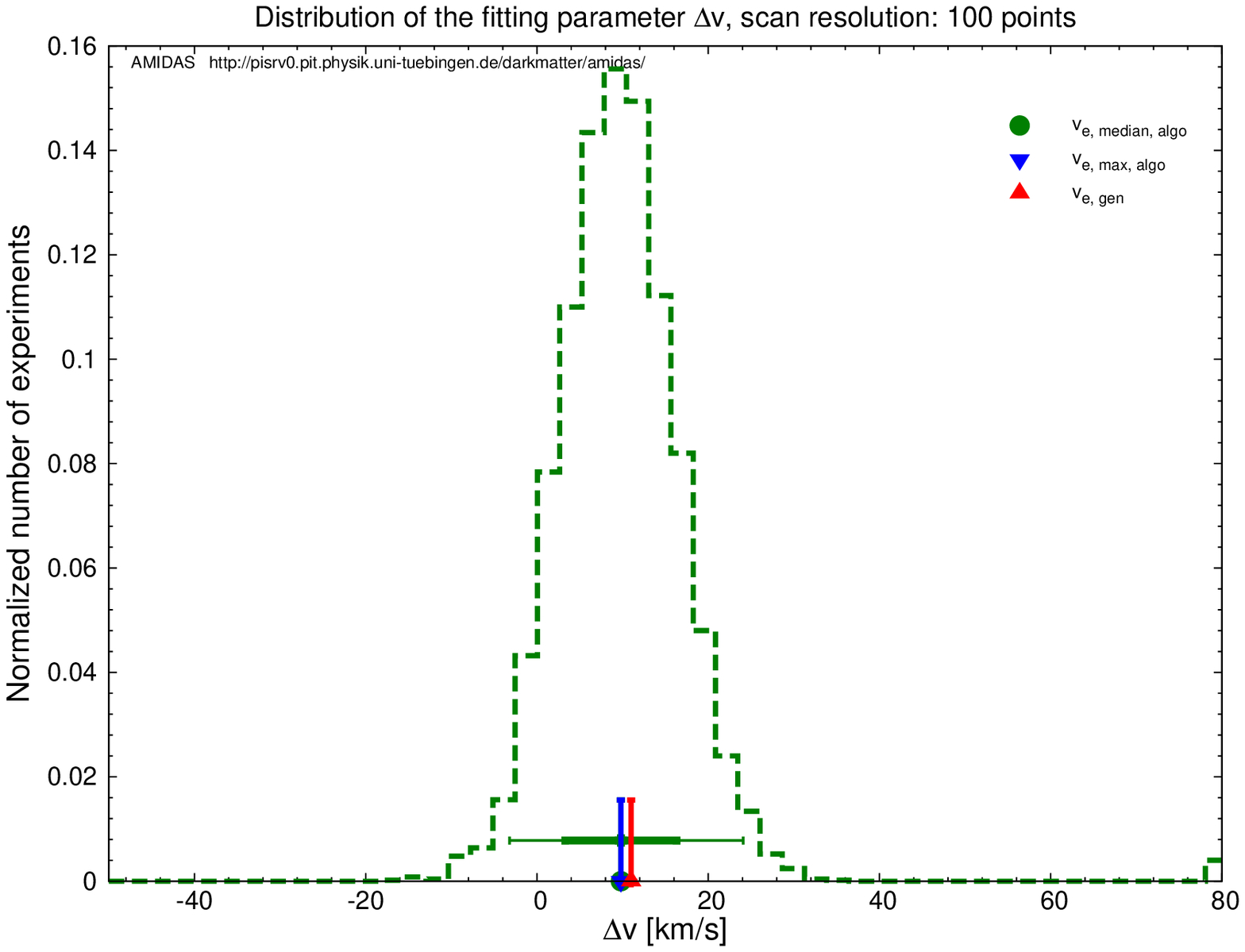}    \hspace*{-1.6cm} \par
\makebox[8.5cm]{(c)}\hspace{0.325cm}\makebox[8.175cm]{(d)}             \\
}
\vspace{-0.35cm}
\end{center}
\caption{
 As in Figs.~\ref{fig:f1v-Ge-SiGe-025-0500-sh-sh-Gau},
 except that
 the variated shifted Maxwellian velocity distribution function
 and
 the Gaussian probability distribution
 for both fitting parameters
 with expectation values of \mbox{$v_0 = 230$ km/s}
 and \mbox{$\Delta v = 15$ km/s}
 and a common 1$\sigma$ uncertainty of \mbox{20 km/s}
 has been used
 (cf.~also Figs.~\ref{fig:f1v-Ge-SiGe-100-0500-sh-sh_Dv-Gau}).
}
\label{fig:f1v-Ge-SiGe-025-0500-sh-sh_Dv-Gau}
\end{figure}
}
\newcommand{\plotGeSiGeshGauflatH}{
\begin{figure}[t!]
\begin{center}
\vspace{-0.25cm}
{
\hspace*{-1.6cm}
\includegraphics[width=8.5cm]{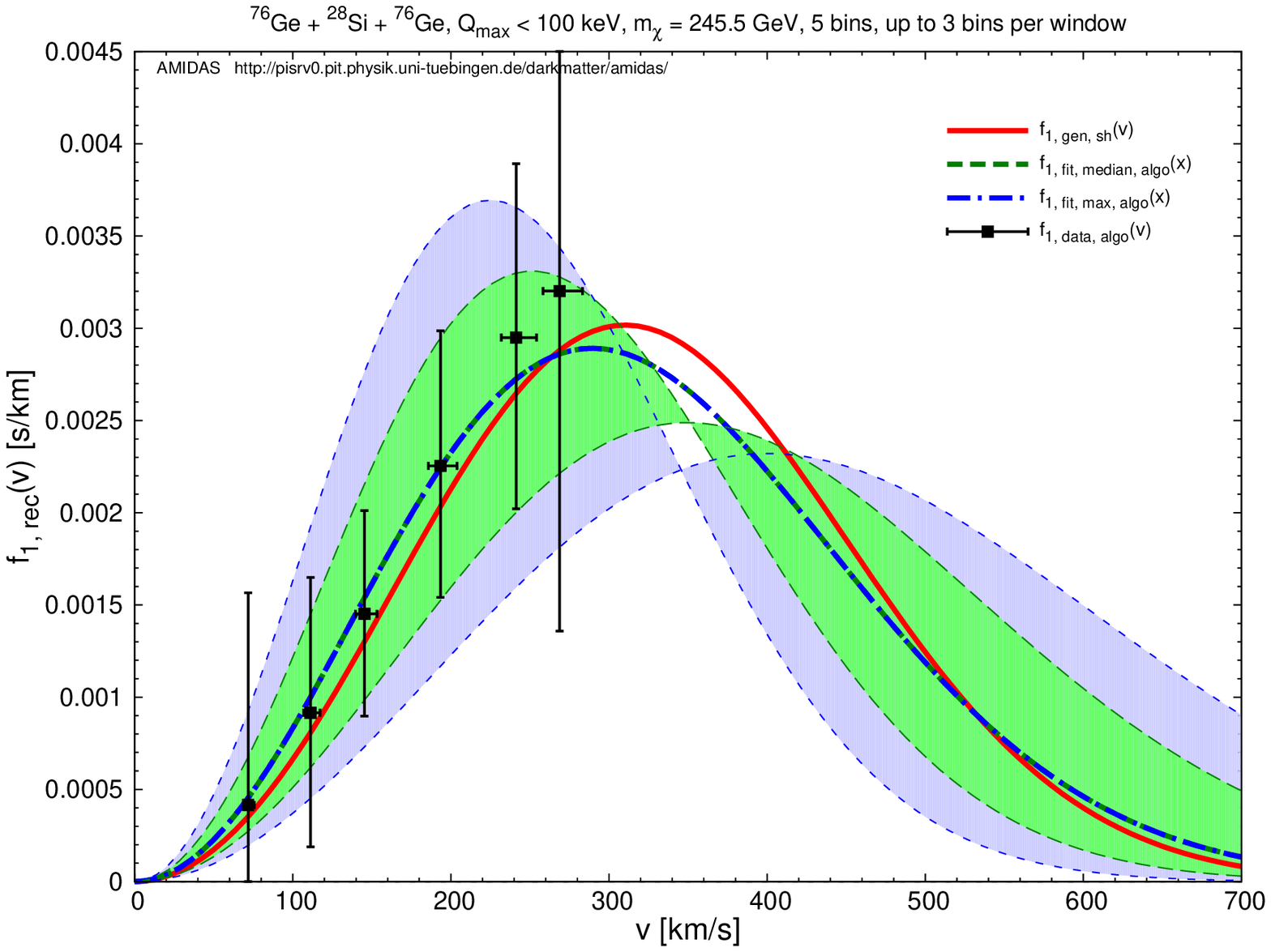}
\includegraphics[width=8.5cm]{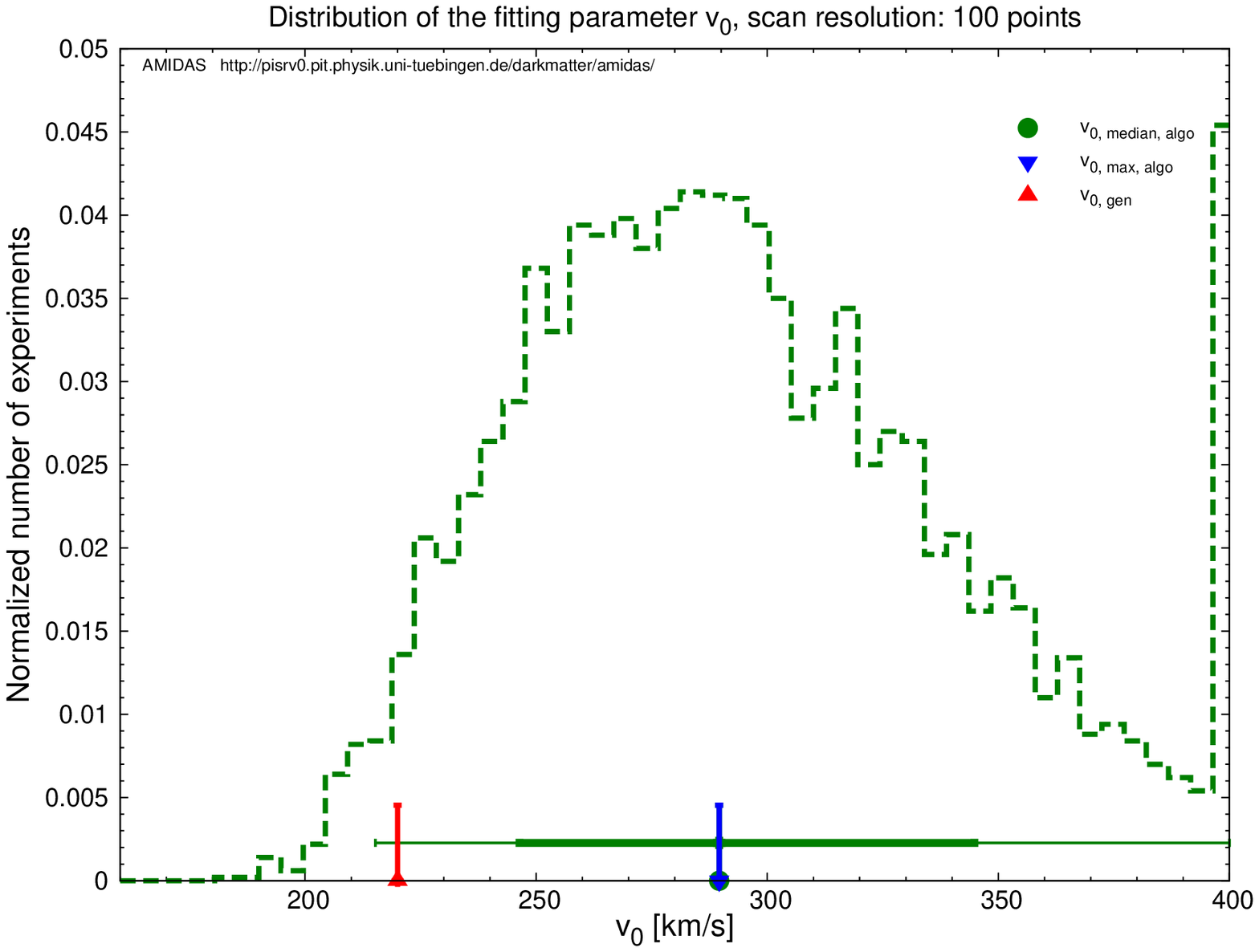} \hspace*{-1.6cm} \par
\makebox[8.5cm]{(a)}\hspace{0.325cm}\makebox[8.175cm]{(b)}%
}
\vspace{-0.35cm}
\end{center}
\caption{
 As in Figs.~\ref{fig:f1v-Ge-SiGe-025-0500-sh-Gau-flat}:
 the flat probability distribution
 for $v_0$
 and
 the reconstructed WIMP mass
 have been used,
 except that
 the input WIMP mass has been set as \mbox{$\mchi = 250$ GeV}.
 Remind that
 the bin at \mbox{$v_0 = 400$ km/s}
 is an ``overflow'' bin,
 which contains also the experiments
 with the best--fit $v_0$ value of \mbox{$> 400$ km/s}.
}
\label{fig:f1v-Ge-SiGe-250-0500-sh-Gau-flat}
\end{figure}
}
\newcommand{\plotGeSiGeshshGauH}{
\begin{figure}[t!]
\begin{center}
\vspace{-0.25cm}
{
\hspace*{-1.6cm}
\includegraphics[width=8.5cm]{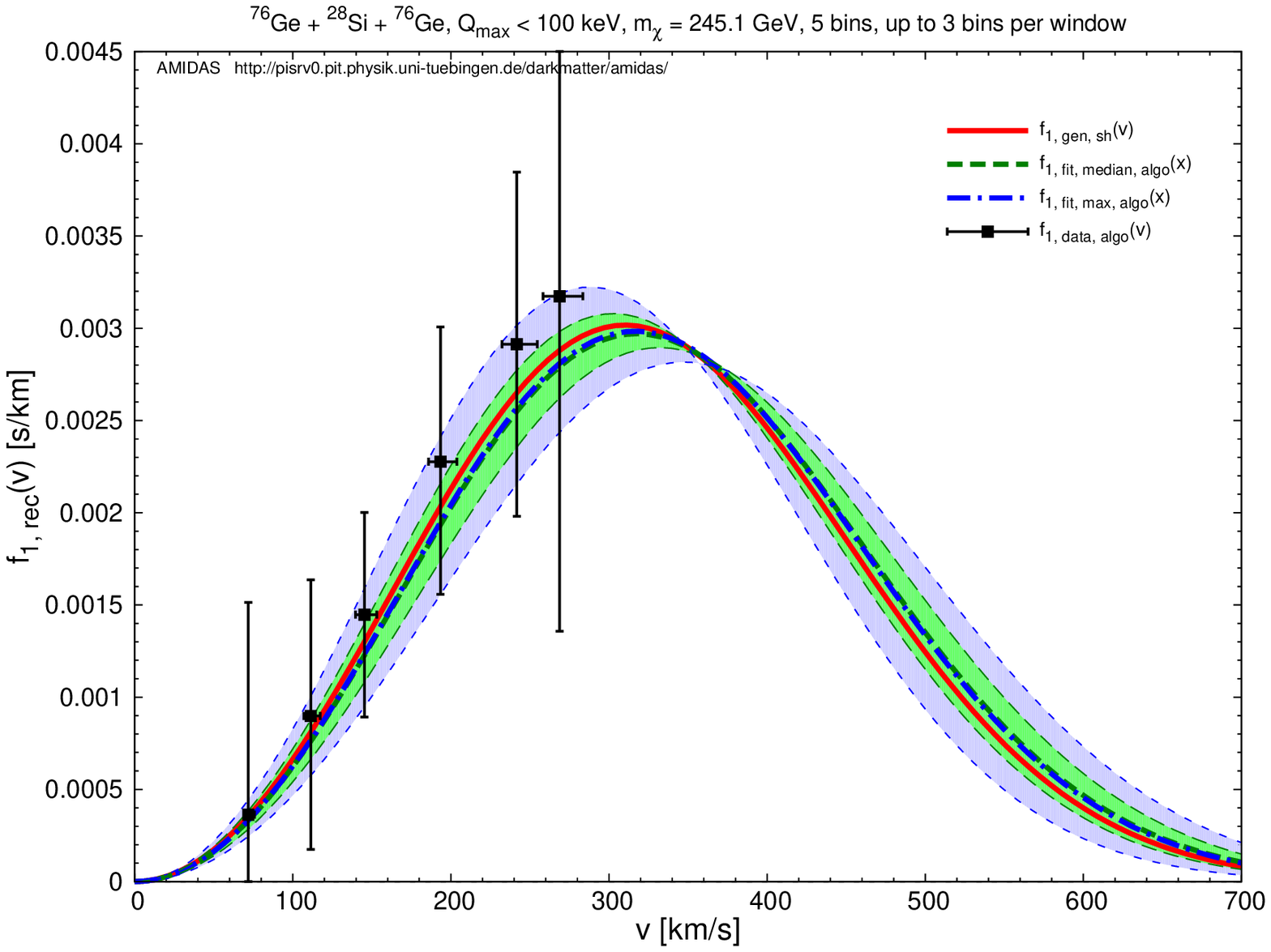}
\includegraphics[width=8.5cm]{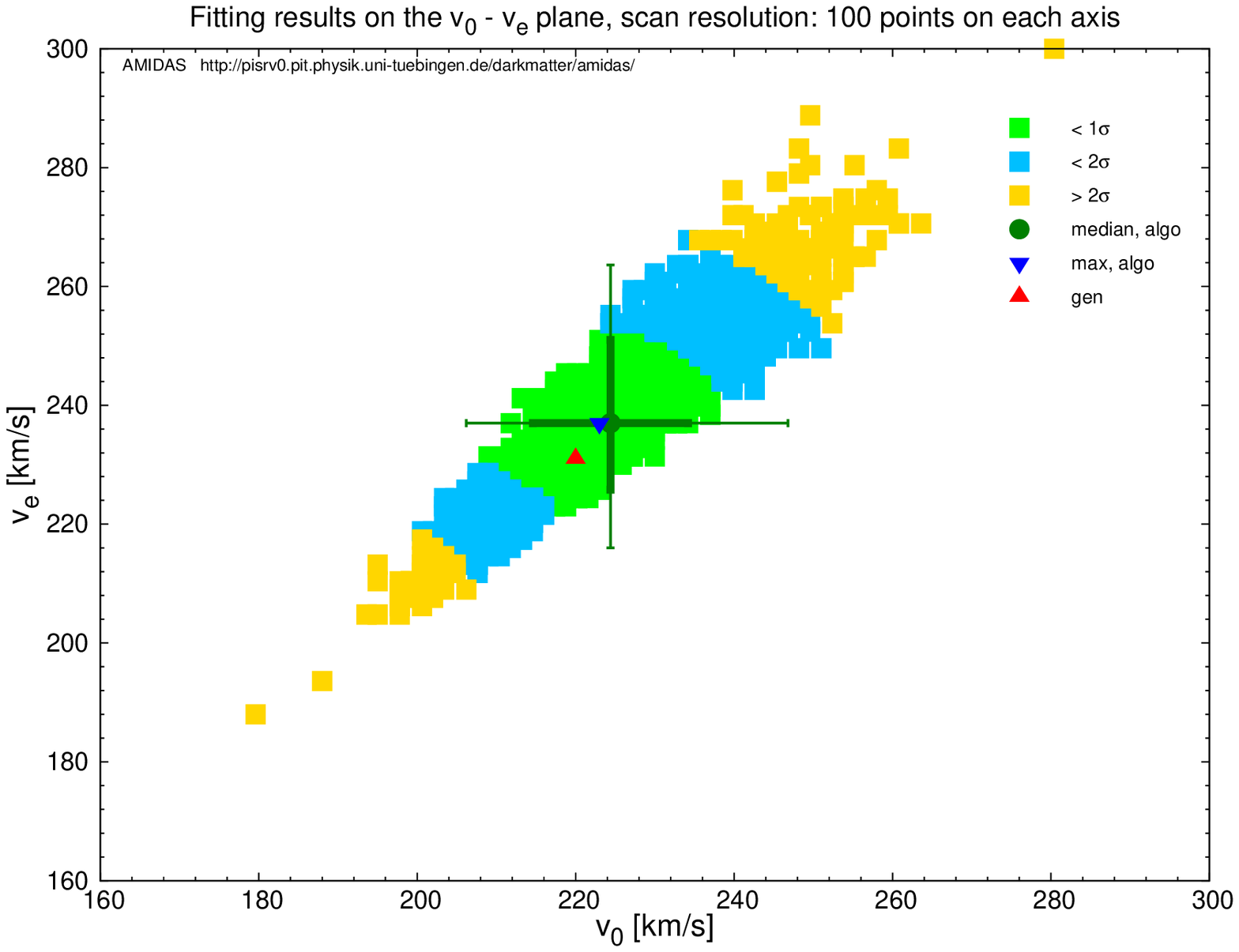} \hspace*{-1.6cm} \par
\makebox[8.5cm]{(a)}\hspace{0.325cm}\makebox[8.175cm]{(b)}          \\ \vspace{0.5cm}
\hspace*{-1.6cm}
\includegraphics[width=8.5cm]{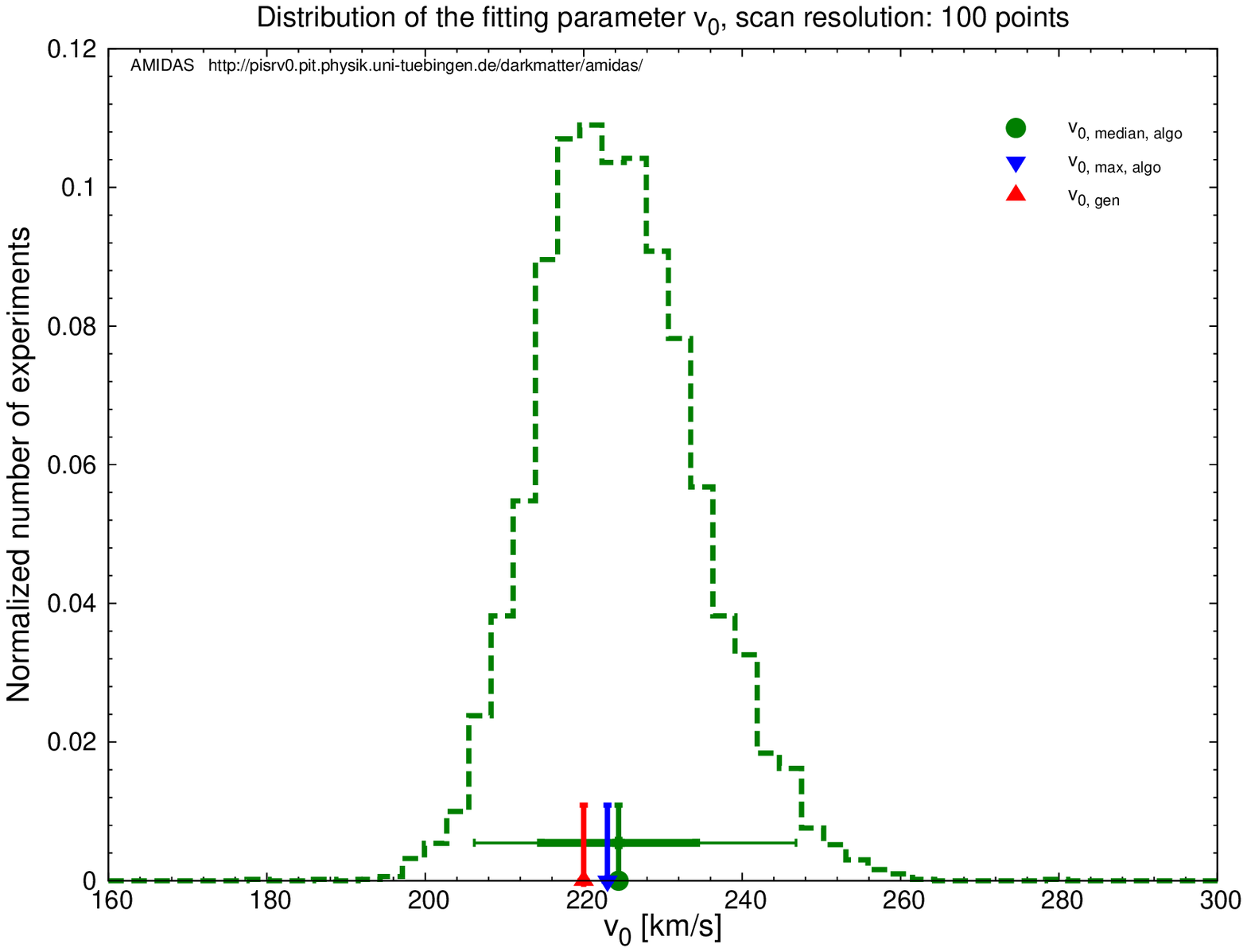}
\includegraphics[width=8.5cm]{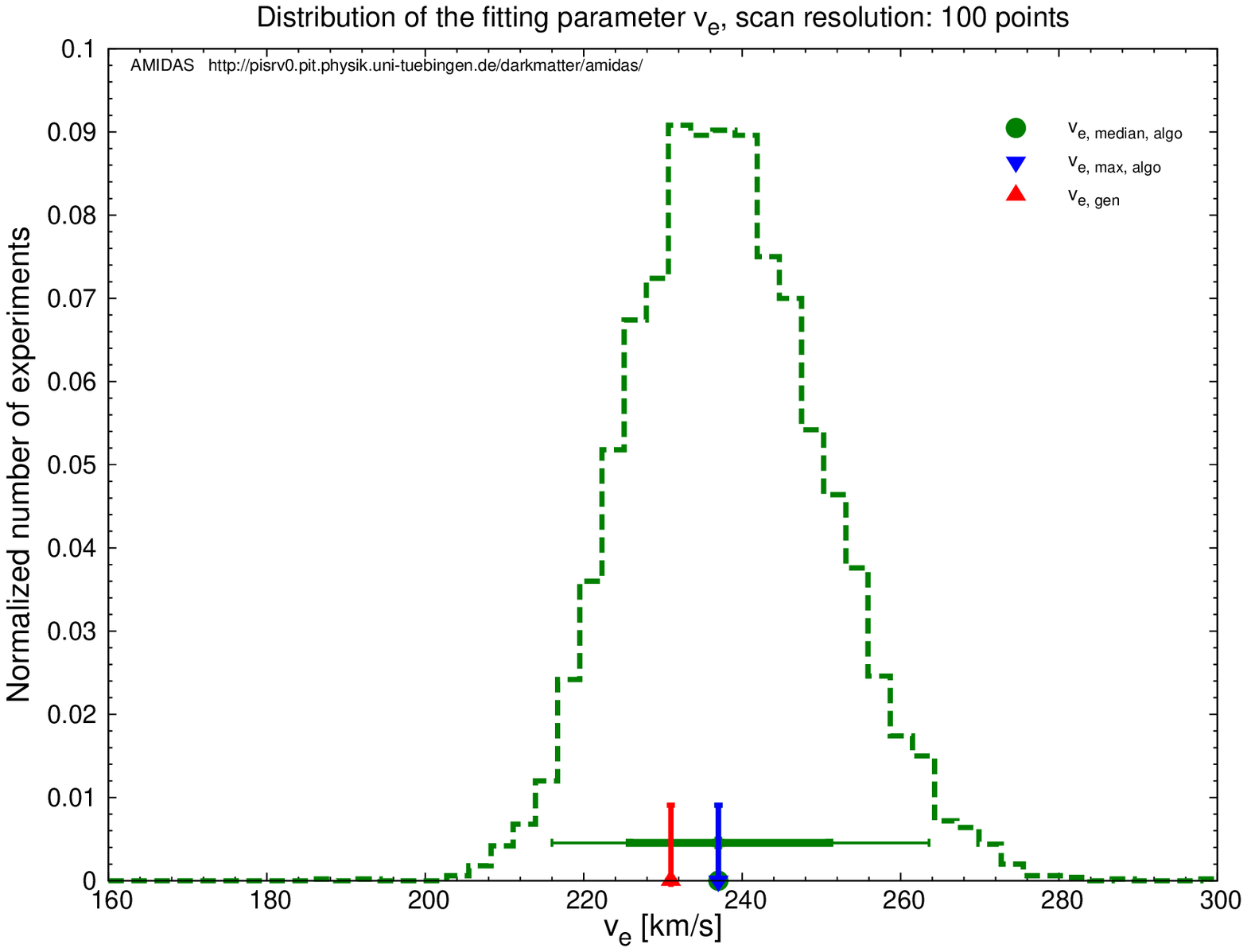}    \hspace*{-1.6cm} \par
\makebox[8.5cm]{(c)}\hspace{0.325cm}\makebox[8.175cm]{(d)}          \\
}
\vspace{-0.35cm}
\end{center}
\caption{
 As in Figs.~\ref{fig:f1v-Ge-SiGe-025-0500-sh-sh-Gau}:
 the Gaussian probability distribution
 for both fitting parameters $v_0$ and $\ve$
 as well as 
 the reconstructed WIMP mass
 have been used,
 except that
 the input WIMP mass has been set as \mbox{$\mchi = 250$ GeV}.
}
\label{fig:f1v-Ge-SiGe-250-0500-sh-sh-Gau}
\end{figure}
}
\newcommand{\plotGeSiGeshshDvGauH}{
\begin{figure}[t!]
\begin{center}
\vspace{-0.25cm}
{
\hspace*{-1.6cm}
\includegraphics[width=8.5cm]{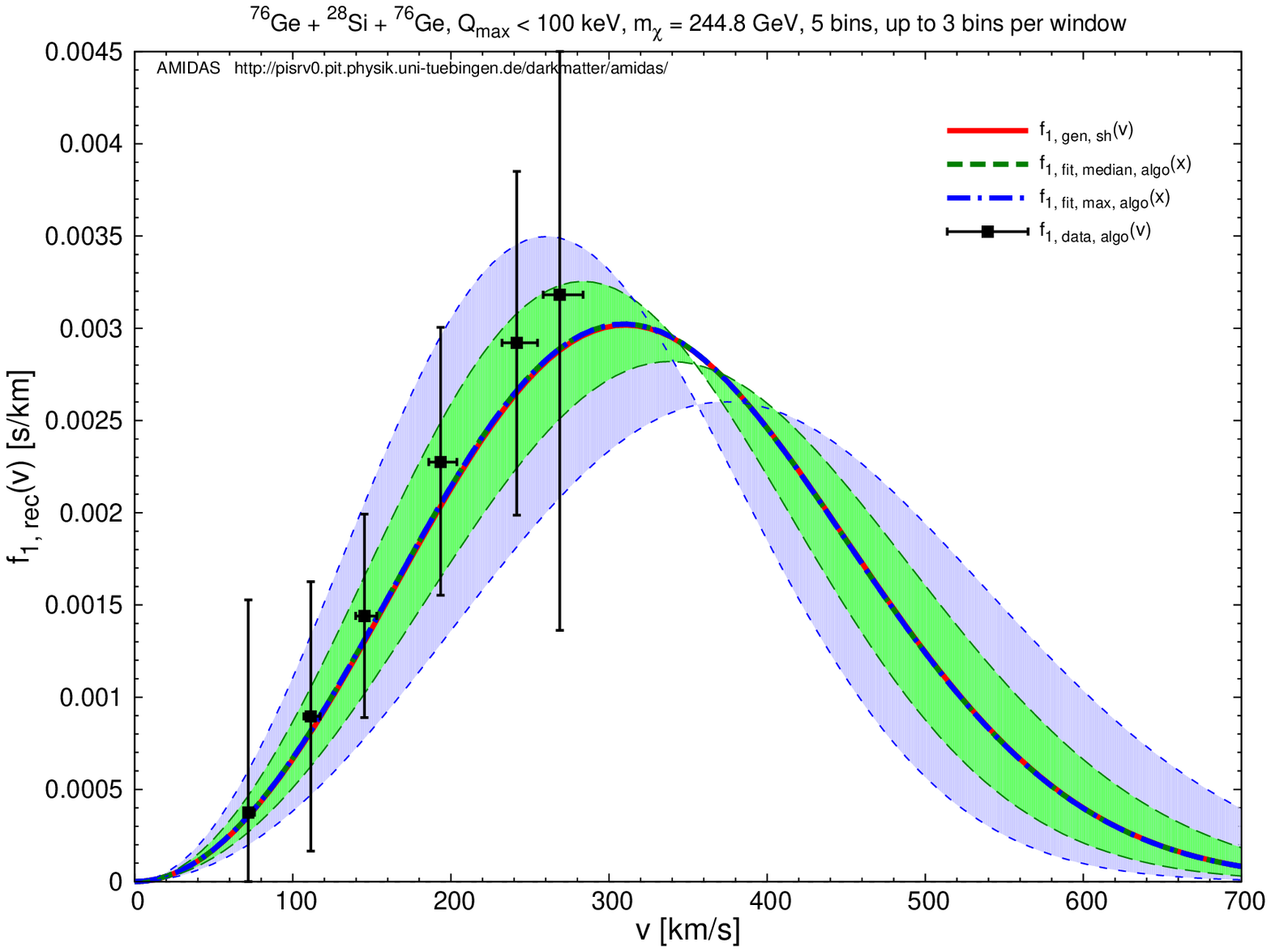}
\includegraphics[width=8.5cm]{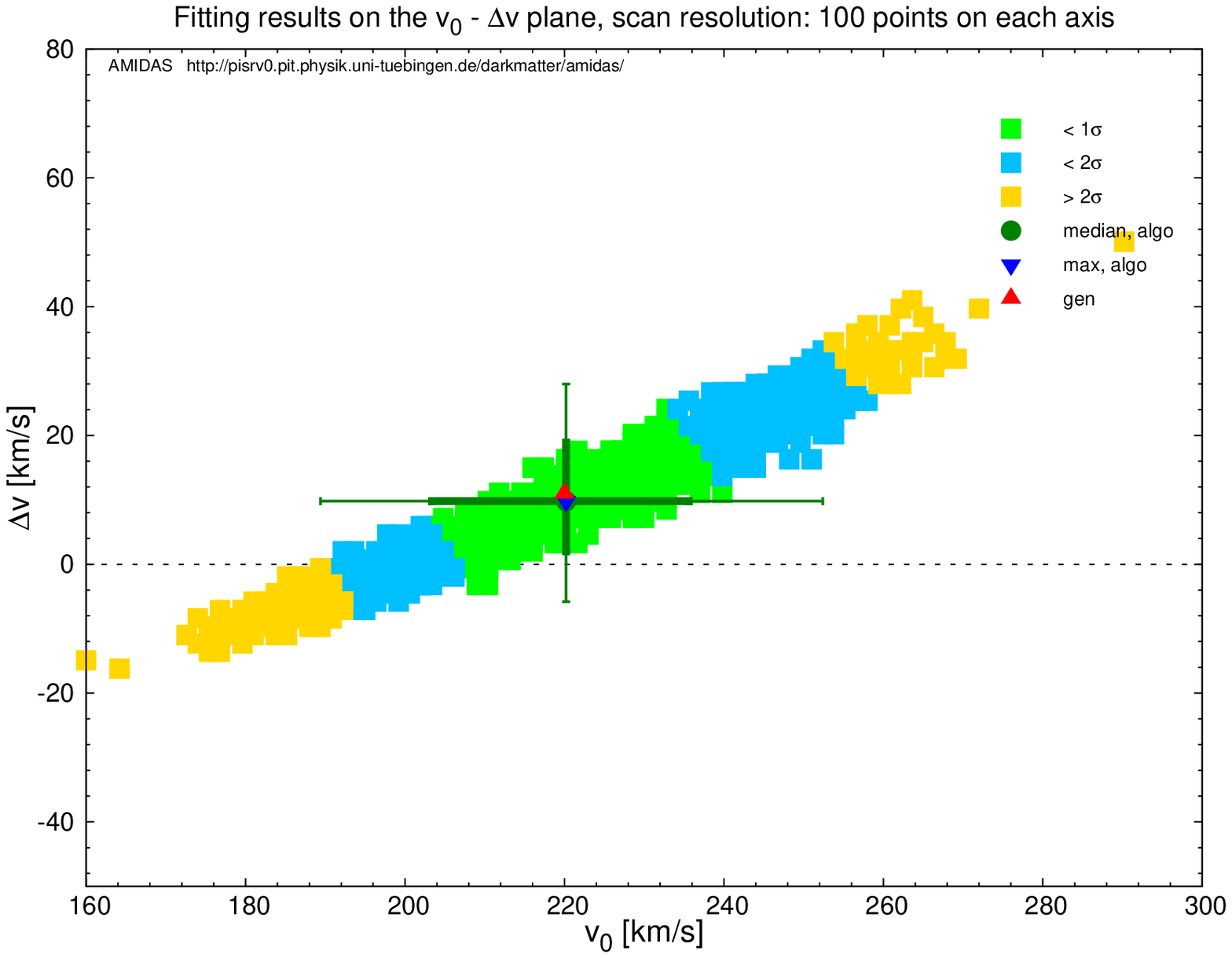} \hspace*{-1.6cm} \par
\makebox[8.5cm]{(a)}\hspace{0.325cm}\makebox[8.175cm]{(b)}             \\ \vspace{0.5cm}
\hspace*{-1.6cm}
\includegraphics[width=8.5cm]{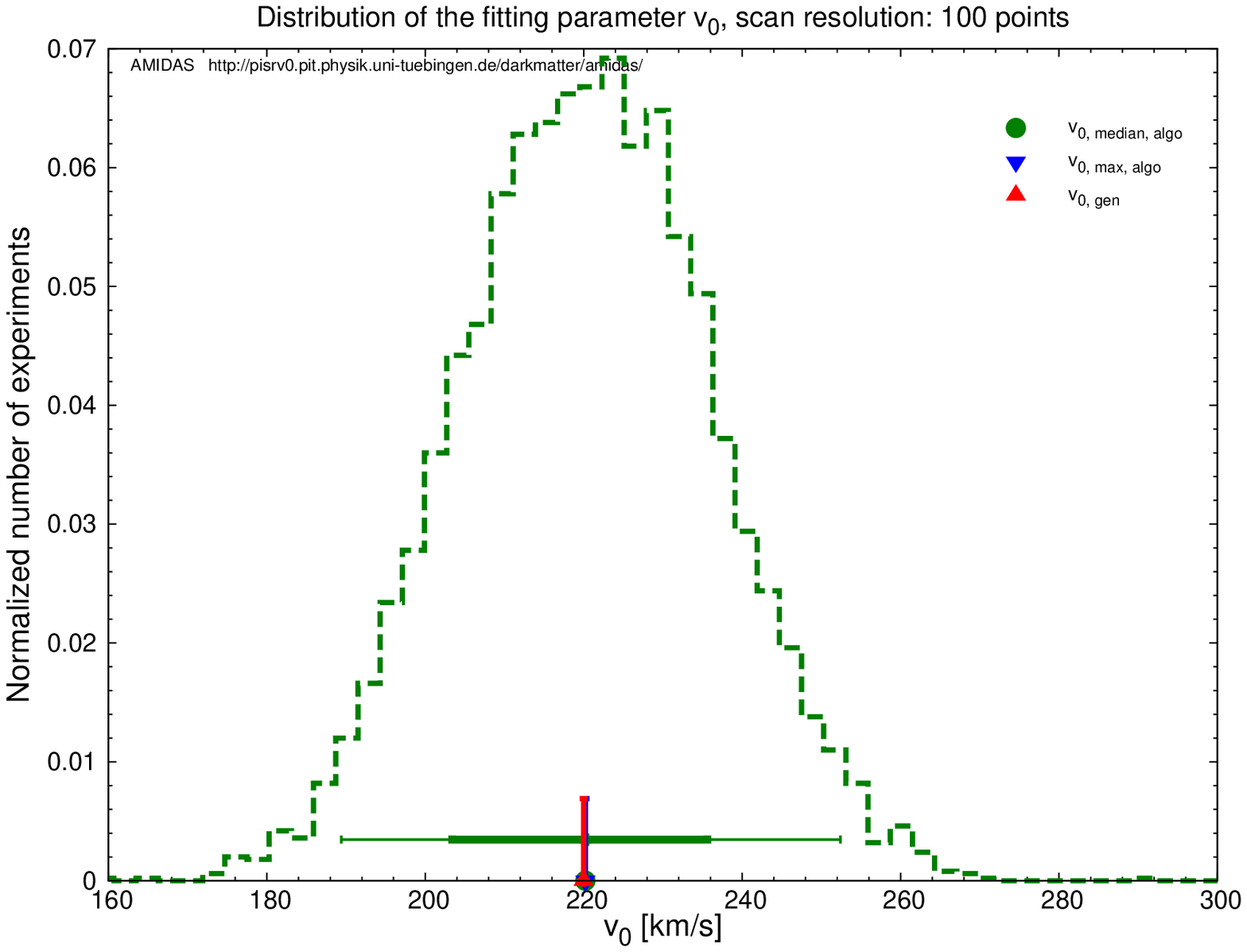}
\includegraphics[width=8.5cm]{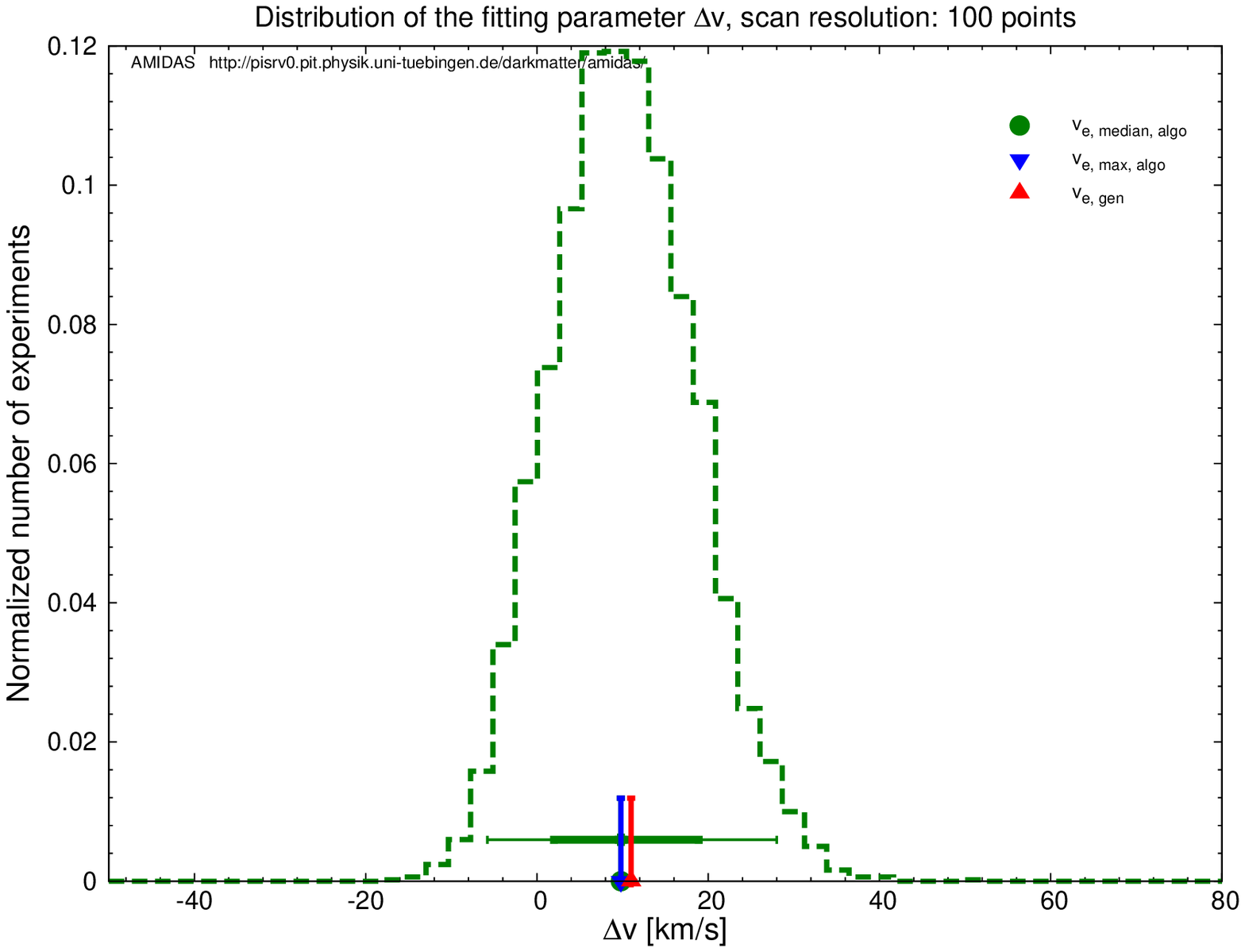}    \hspace*{-1.6cm} \par
\makebox[8.5cm]{(c)}\hspace{0.325cm}\makebox[8.175cm]{(d)}             \\
}
\vspace{-0.35cm}
\end{center}
\caption{
 As in Figs.~\ref{fig:f1v-Ge-SiGe-025-0500-sh-sh_Dv-Gau},
 the Gaussian probability distribution
 for both fitting parameters $v_0$ and $\Delta v$
 as well as 
 the reconstructed WIMP mass
 have been used,
 except that
 the input WIMP mass has been set as \mbox{$\mchi = 250$ GeV}.
}
\label{fig:f1v-Ge-SiGe-250-0500-sh-sh_Dv-Gau}
\end{figure}
}
\newcommand{\plotGeSiGeshshvGaubg}{
\begin{figure}[t!]
\begin{center}
\vspace{-0.25cm}
{
\hspace*{-1.6cm}
\includegraphics[width=8.5cm]{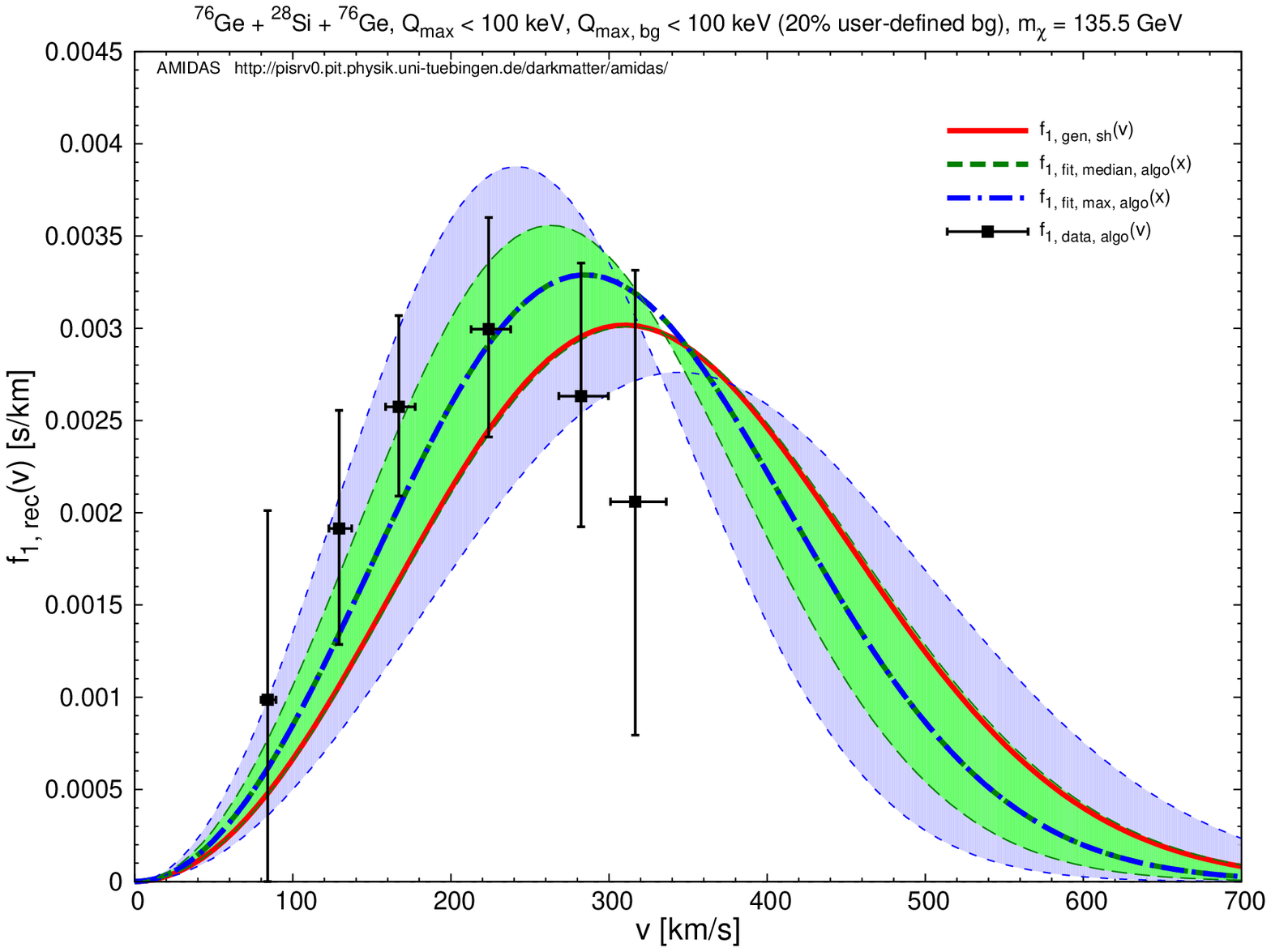}
\includegraphics[width=8.5cm]{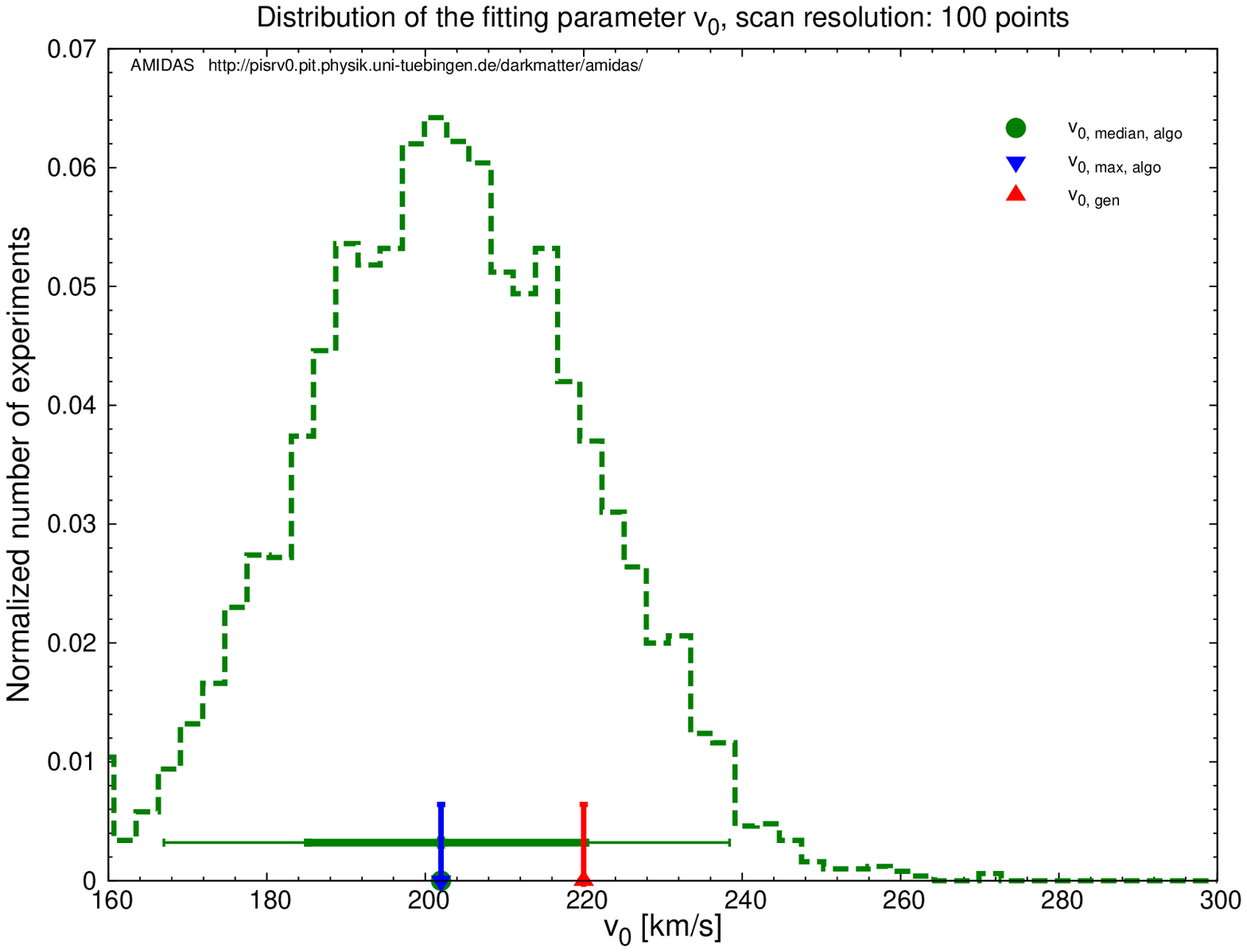} \hspace*{-1.6cm} \par
\makebox[8.5cm]{(a)}\hspace{0.325cm}\makebox[8.175cm]{(b)}%
}
\vspace{-0.35cm}
\end{center}
\caption{
 As in Figs.~\ref{fig:f1v-Ge-SiGe-100-0500-sh-sh_v0-Gau}(c) and (d):
 the one--parameter shifted Maxwellian velocity distribution
 and the Gaussian probability distribution
 for the unique fitting parameter $v_0$
 as well as
 the reconstructed WIMP mass
 have been used,
 except that
 a fraction of 20\% background events
 generated by the spectrum given in Eq.~(\ref{eqn:dRdQ_bg})
 has been taken into account.
}
\label{fig:f1v-Ge-SiGe-100-0500-sh-sh_v0-Gau-bg}
\end{figure}
}
\newcommand{\plotGeSiGeshshGaubg}{
\begin{figure}[t!]
\begin{center}
\vspace{-0.25cm}
{
\hspace*{-1.6cm}
\includegraphics[width=8.5cm]{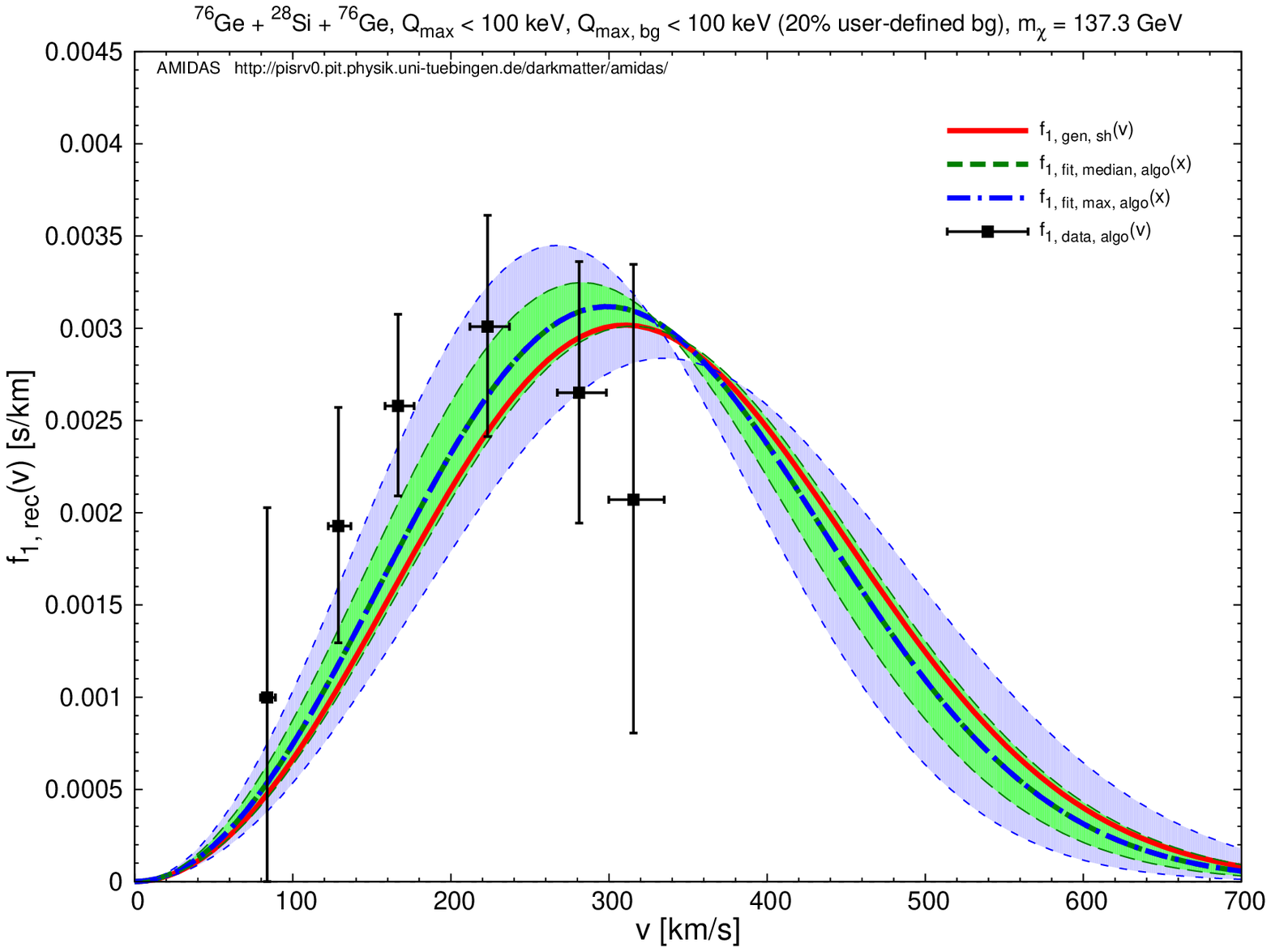}
\includegraphics[width=8.5cm]{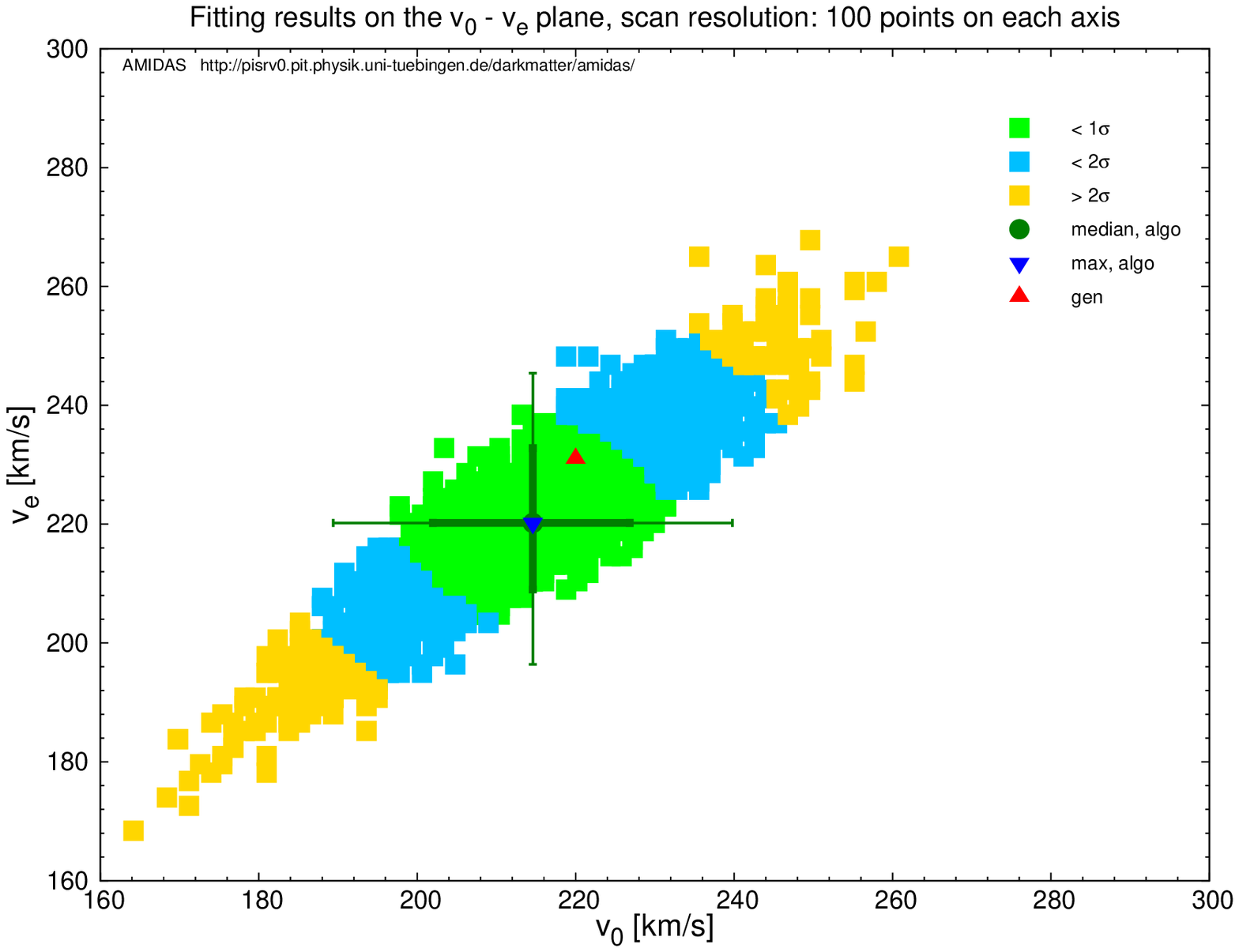} \hspace*{-1.6cm} \par
\makebox[8.5cm]{(a)}\hspace{0.325cm}\makebox[8.175cm]{(b)}             \\ \vspace{0.5cm}
\hspace*{-1.6cm}
\includegraphics[width=8.5cm]{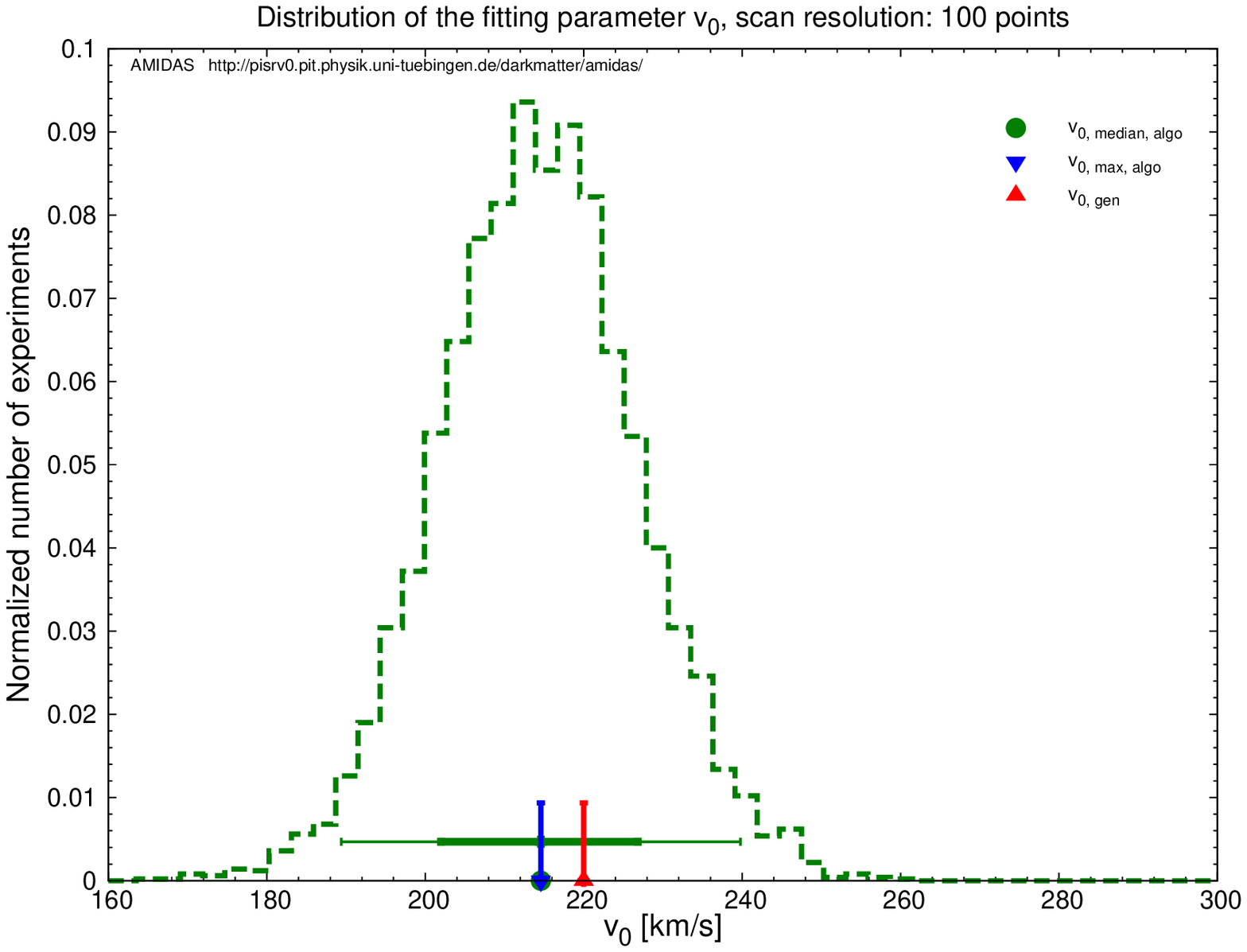}
\includegraphics[width=8.5cm]{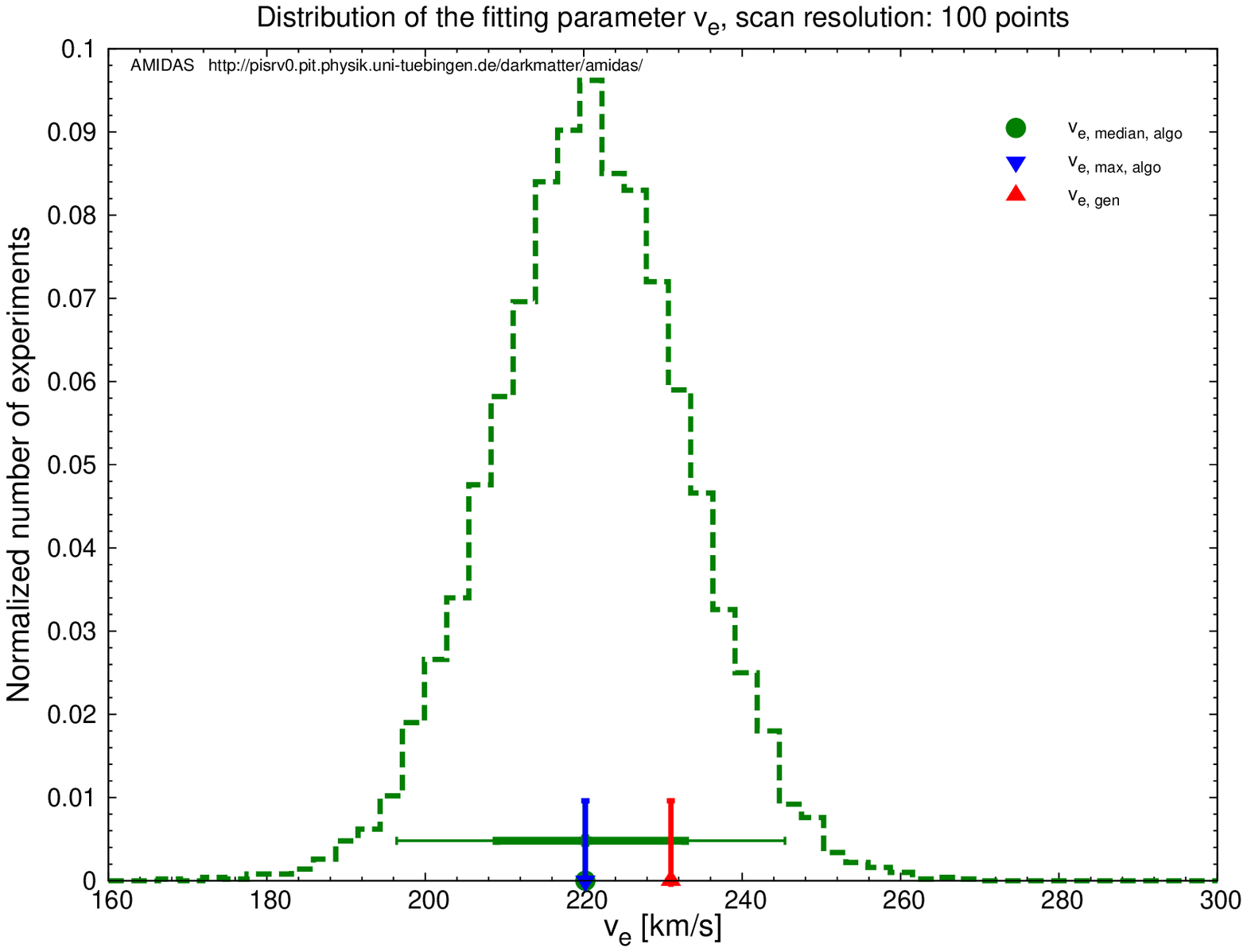}    \hspace*{-1.6cm} \par
\makebox[8.5cm]{(c)}\hspace{0.325cm}\makebox[8.175cm]{(d)}             \\
}
\vspace{-0.35cm}
\end{center}
\caption{
 As in Figs.~\ref{fig:f1v-Ge-SiGe-100-0500-sh-sh-Gau}:
 the shifted Maxwellian velocity distribution function
 and
 the Gaussian probability distribution
 for both fitting parameters $v_0$ and $\ve$
 as well as 
 the reconstructed WIMP mass
 have been used,
 except that
 a fraction of 20\% background events
 has been taken into account.
}
\label{fig:f1v-Ge-SiGe-100-0500-sh-sh-Gau-bg}
\end{figure}
}
\newcommand{\plotGeSiGeshshvGaubgL}{
\begin{figure}[p!]
\begin{center}
\vspace{-0.25cm}
{
\hspace*{-1.6cm}
\includegraphics[width=8.5cm]{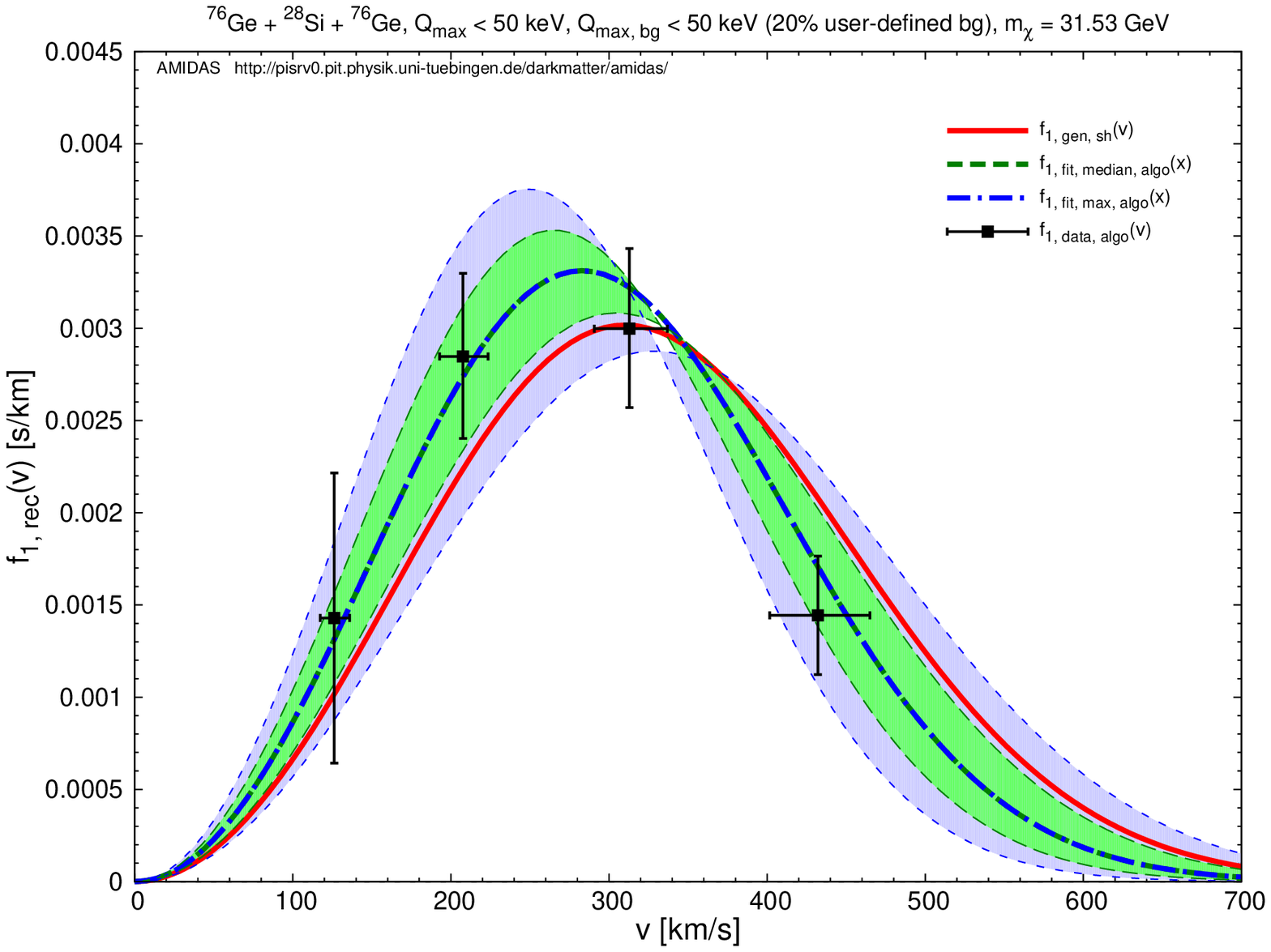}
\includegraphics[width=8.5cm]{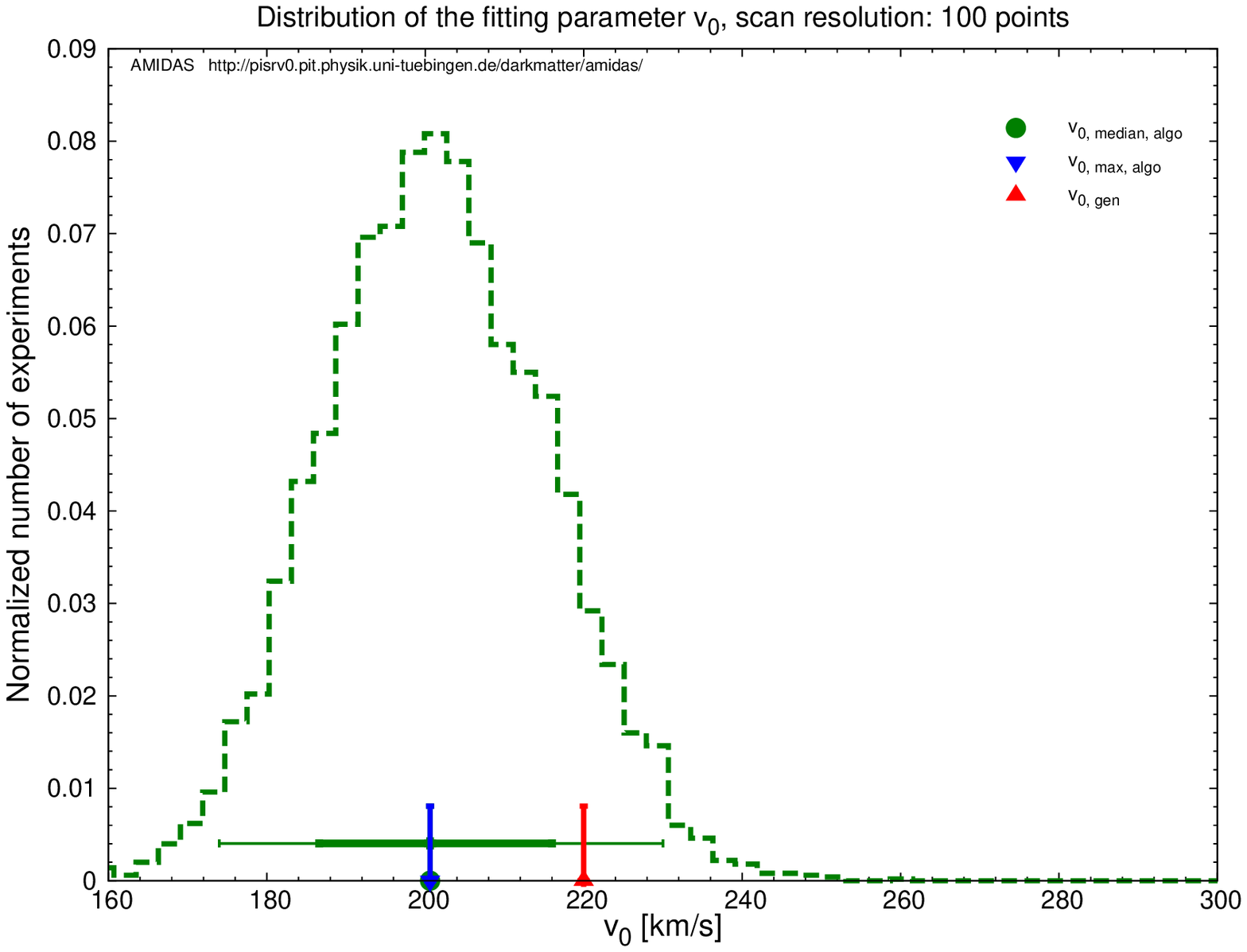} \hspace*{-1.6cm} \par
\makebox[8.5cm]{(a)}\hspace{0.325cm}\makebox[8.175cm]{(b)}%
}
\vspace{-0.35cm}
\end{center}
\caption{
 As in Figs.~\ref{fig:f1v-Ge-SiGe-100-0500-sh-sh_v0-Gau-bg},
 except that
 the input WIMP mass is set as \mbox{$\mchi = 25$ GeV}
 (simulation setup as in Sec.~3.4.1).
}
\label{fig:f1v-Ge-SiGe-025-0500-sh-sh_v0-Gau-bg}
\end{figure}
}
\newcommand{\plotGeSiGeshshGaubgL}{
\begin{figure}[p!]
\begin{center}
\vspace{0.25cm}
{
\hspace*{-1.6cm}
\includegraphics[width=8.5cm]{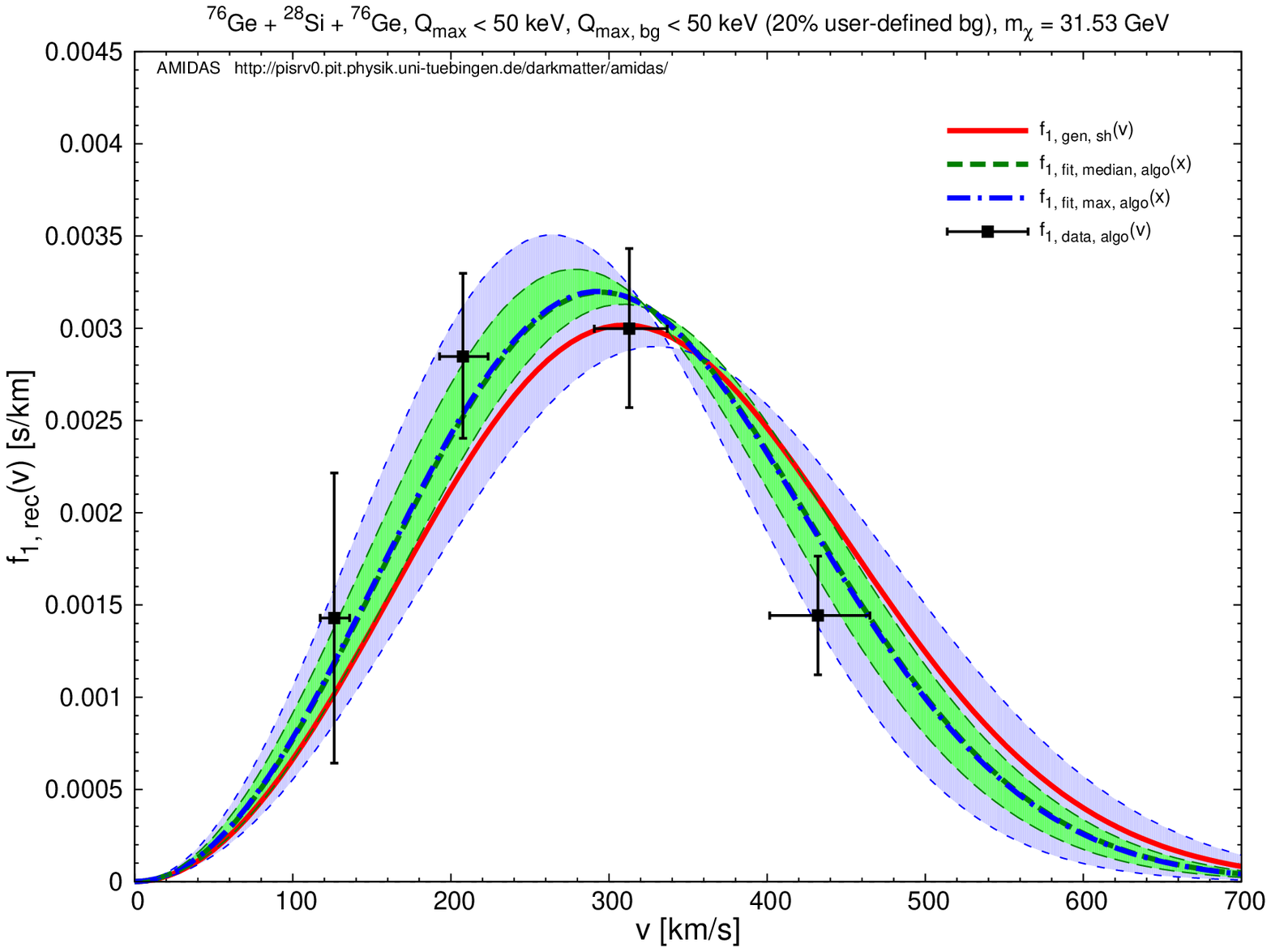}
\includegraphics[width=8.5cm]{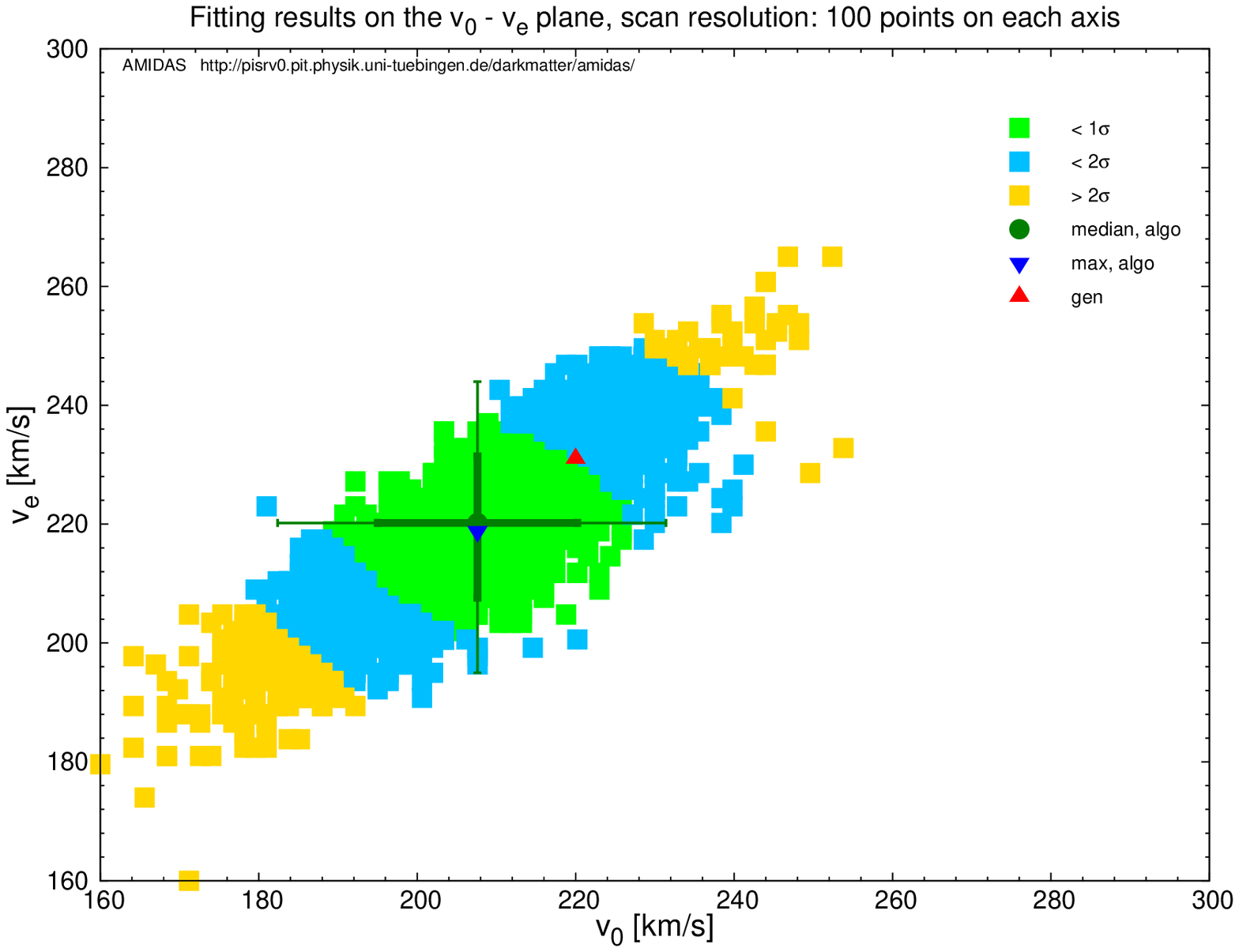} \hspace*{-1.6cm} \par
\makebox[8.5cm]{(a)}\hspace{0.325cm}\makebox[8.175cm]{(b)}             \\ \vspace{0.5cm}
\hspace*{-1.6cm}
\includegraphics[width=8.5cm]{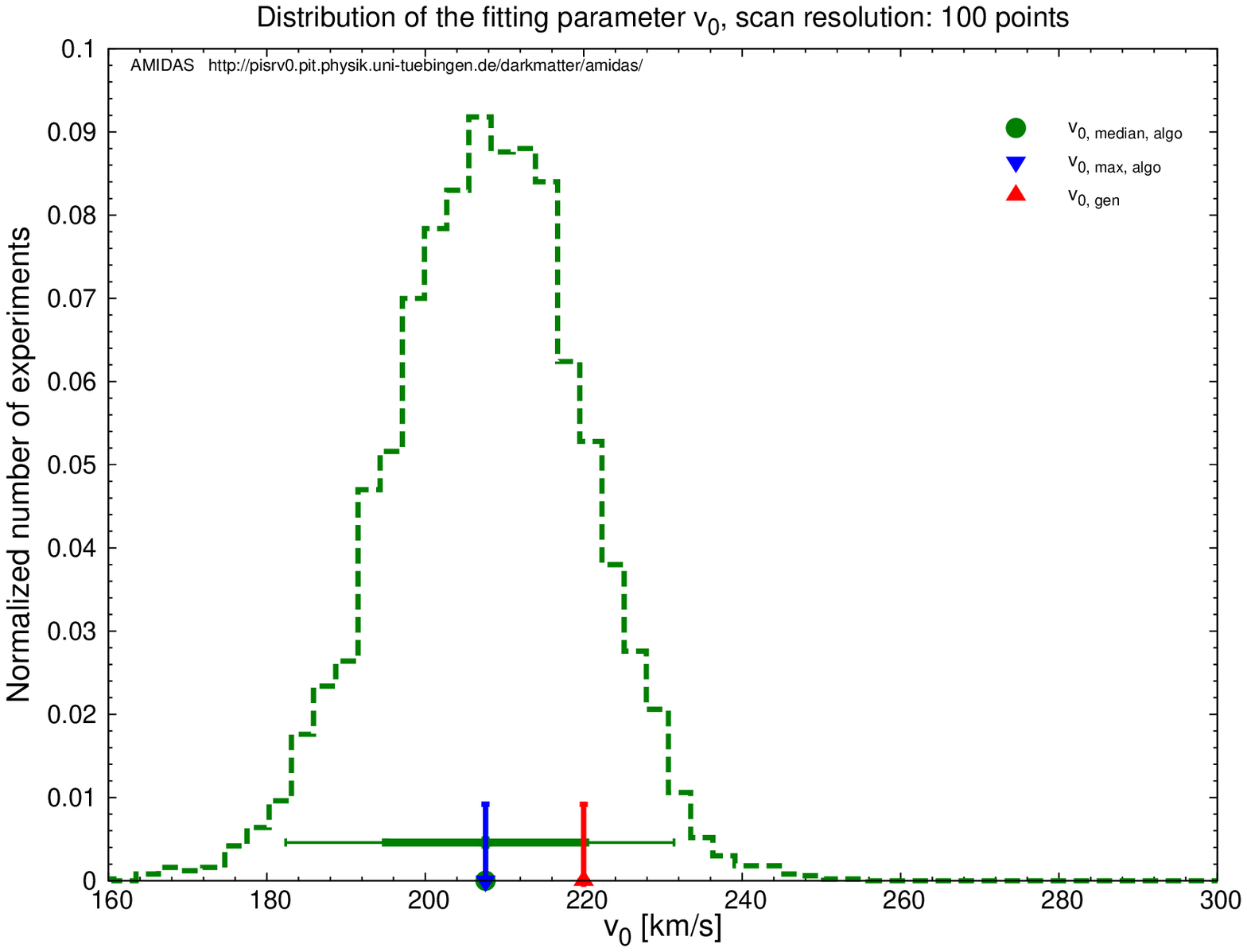}
\includegraphics[width=8.5cm]{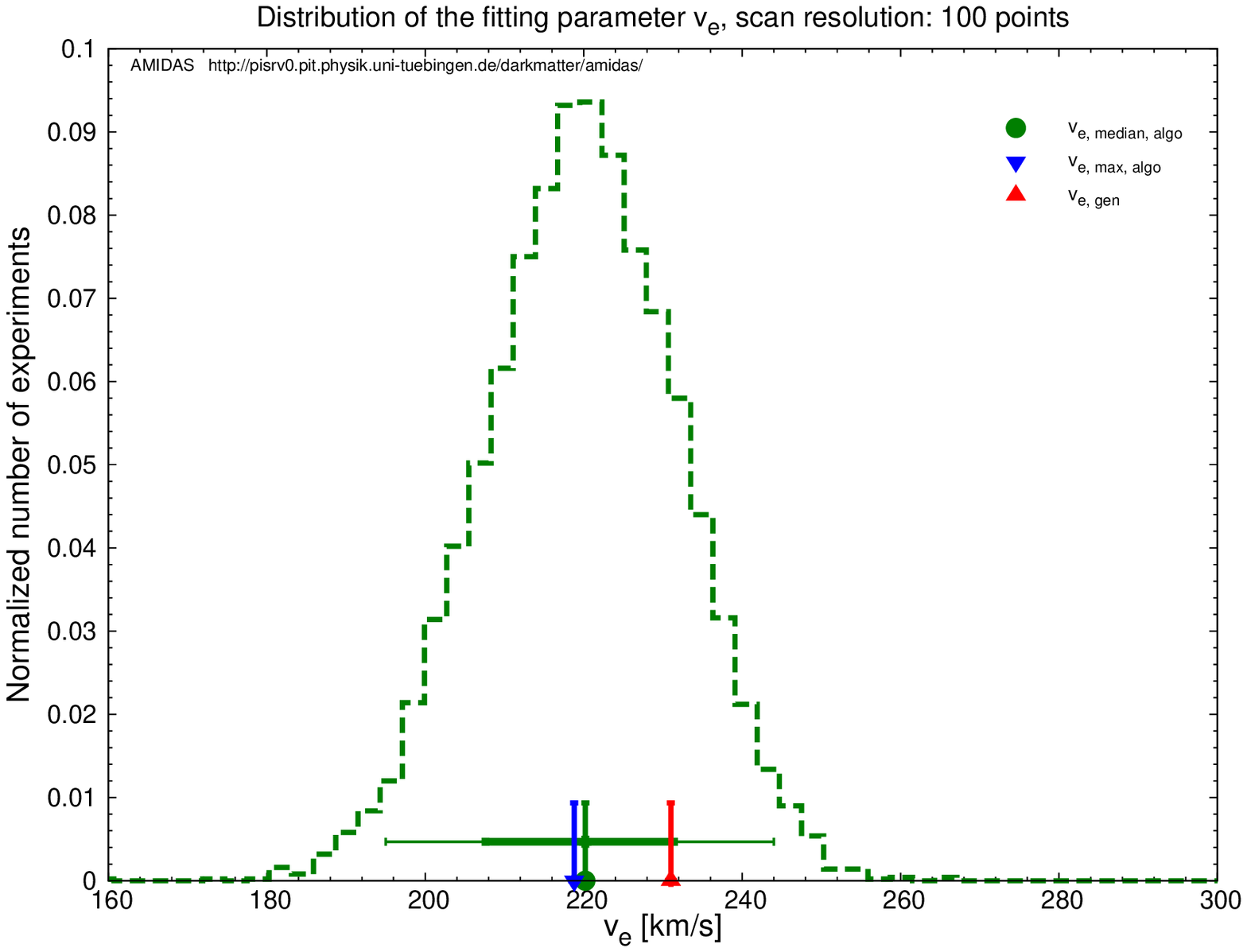}    \hspace*{-1.6cm} \par
\makebox[8.5cm]{(c)}\hspace{0.325cm}\makebox[8.175cm]{(d)}             \\
}
\vspace{-0.35cm}
\end{center}
\caption{
 As in Figs.~\ref{fig:f1v-Ge-SiGe-100-0500-sh-sh-Gau-bg},
 except that
 the input WIMP mass is set as \mbox{$\mchi = 25$ GeV}
 (simulation setup as in Sec.~3.4.1).
}
\label{fig:f1v-Ge-SiGe-025-0500-sh-sh-Gau-bg}
\end{figure}
}
\newcommand{\plotGeSiGeshshvGaubgH}{
\begin{figure}[p!]
\begin{center}
\vspace{-0.25cm}
{
\hspace*{-1.6cm}
\includegraphics[width=8.5cm]{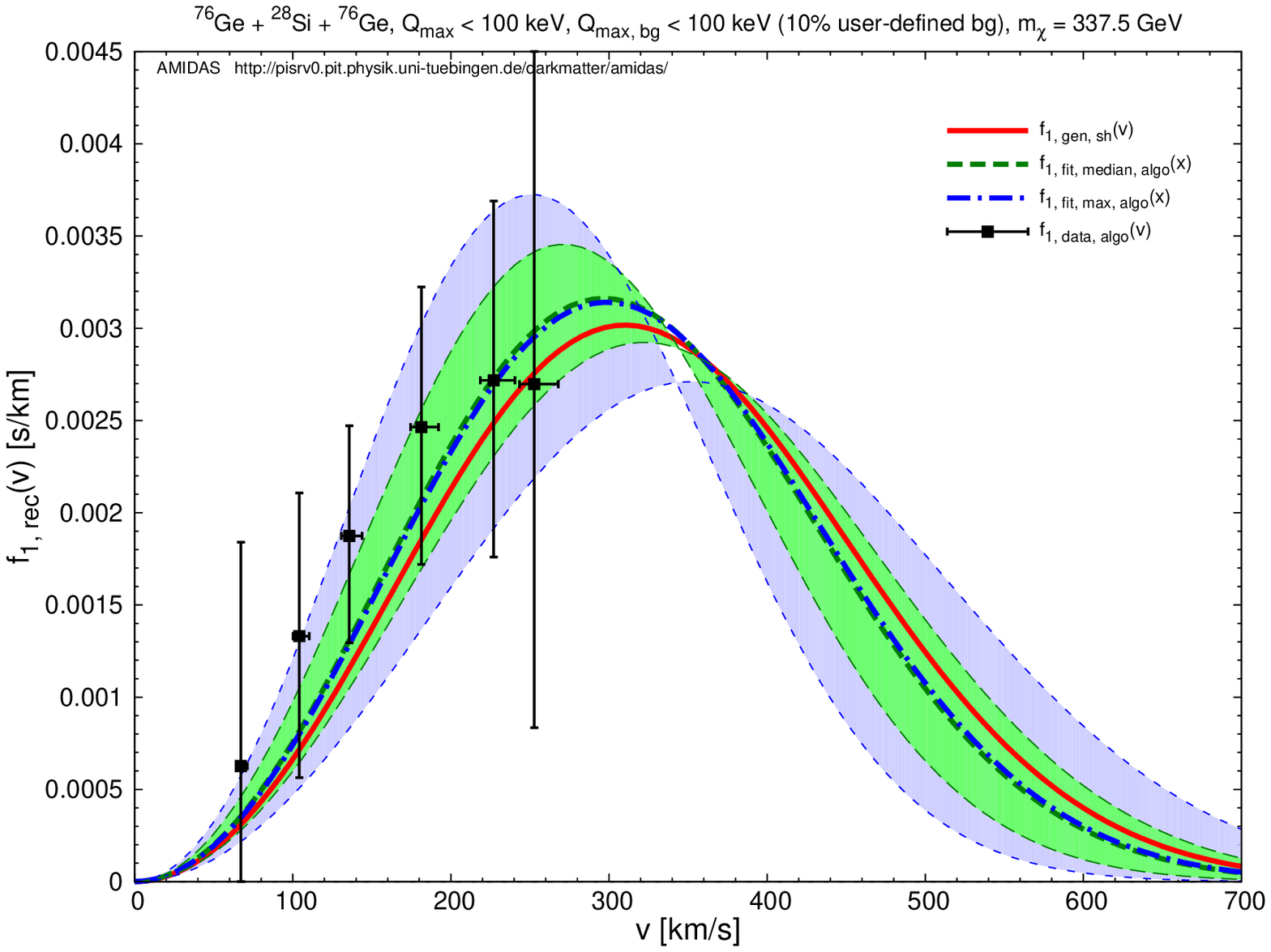}
\includegraphics[width=8.5cm]{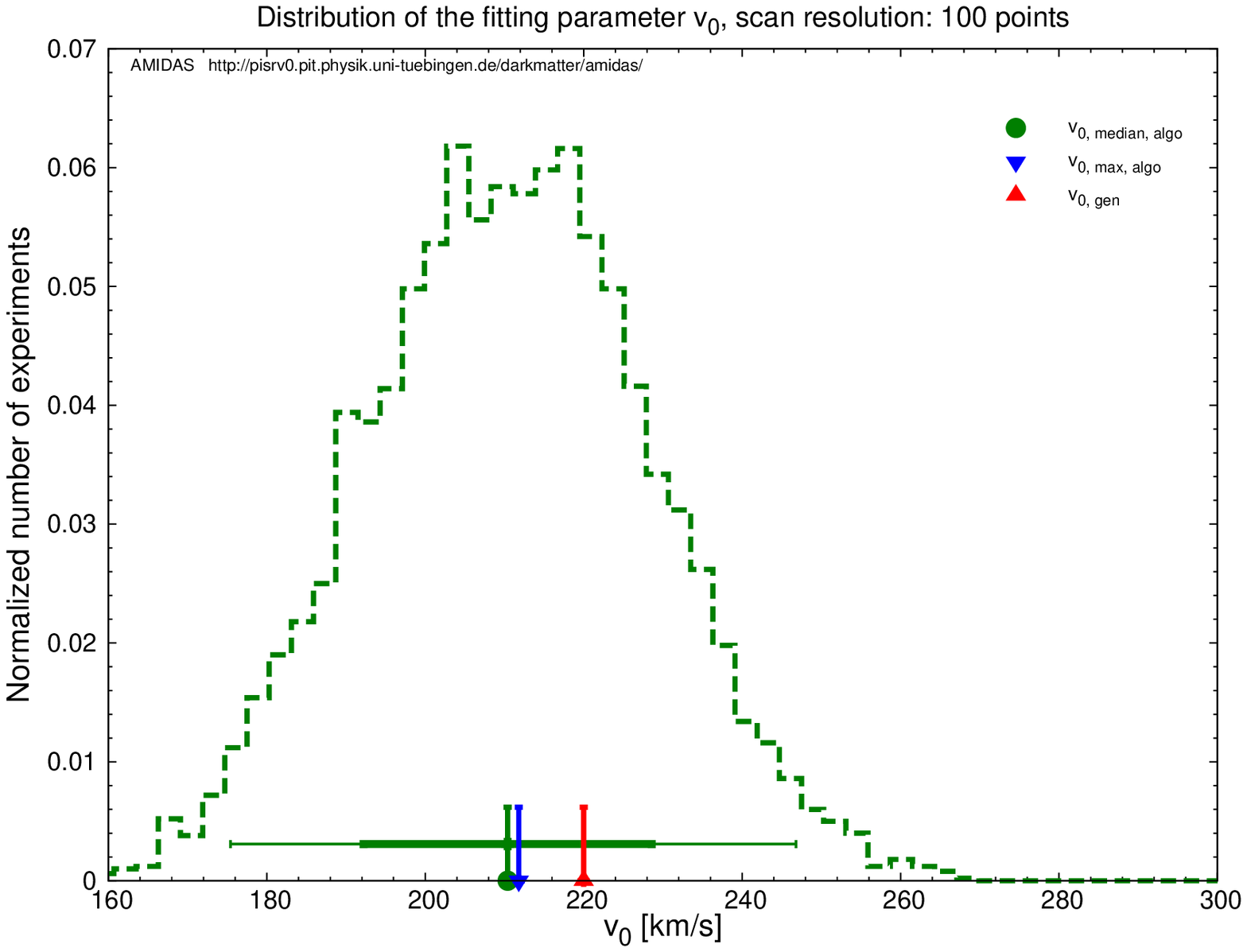} \hspace*{-1.6cm} \par
\makebox[8.5cm]{(a)}\hspace{0.325cm}\makebox[8.175cm]{(b)}%
}
\vspace{-0.35cm}
\end{center}
\caption{
 As in Figs.~\ref{fig:f1v-Ge-SiGe-100-0500-sh-sh_v0-Gau-bg},
 except that
 the input WIMP mass is set as \mbox{$\mchi = 250$ GeV}
 (simulation setup as in Sec.~3.4.2)
 and
 a fraction of {\em 10\%} background events
 has been taken into account.
}
\label{fig:f1v-Ge-SiGe-250-0500-sh-sh_v0-Gau-bg}
\end{figure}
}
\newcommand{\plotGeSiGeshshGaubgH}{
\begin{figure}[p!]
\begin{center}
\vspace{0.25cm}
{
\hspace*{-1.6cm}
\includegraphics[width=8.5cm]{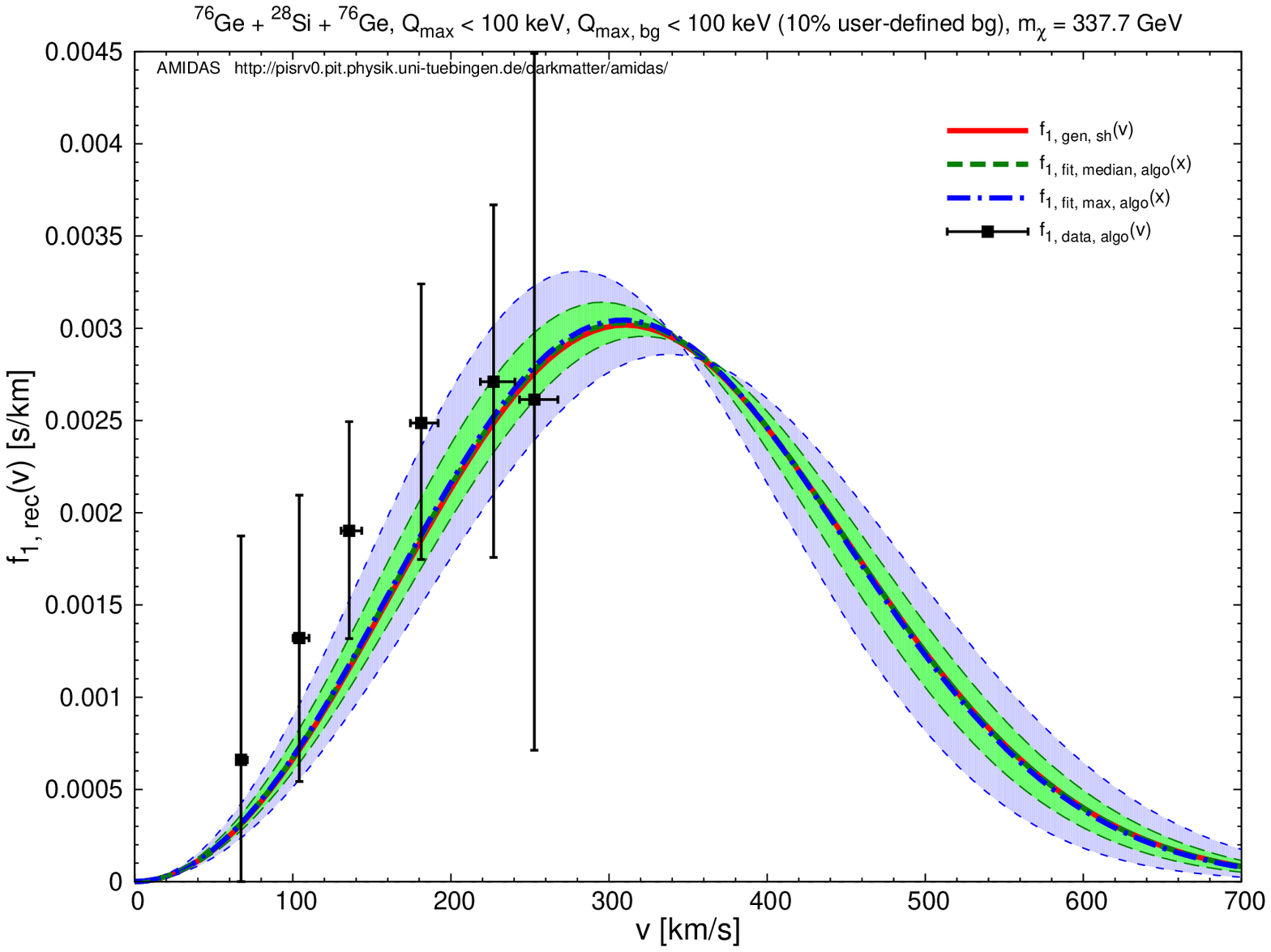}
\includegraphics[width=8.5cm]{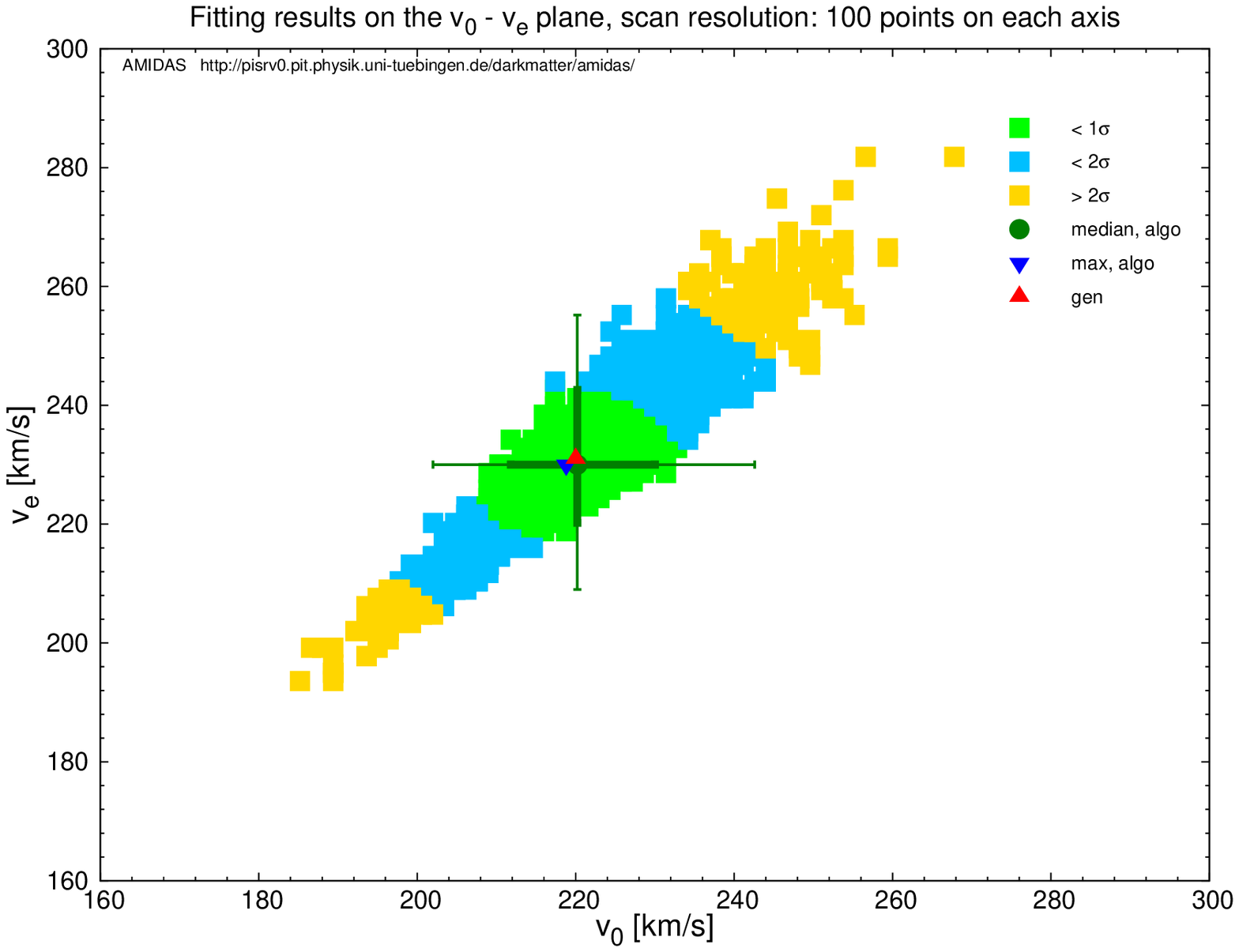} \hspace*{-1.6cm} \par
\makebox[8.5cm]{(a)}\hspace{0.325cm}\makebox[8.175cm]{(b)}             \\ \vspace{0.5cm}
\hspace*{-1.6cm}
\includegraphics[width=8.5cm]{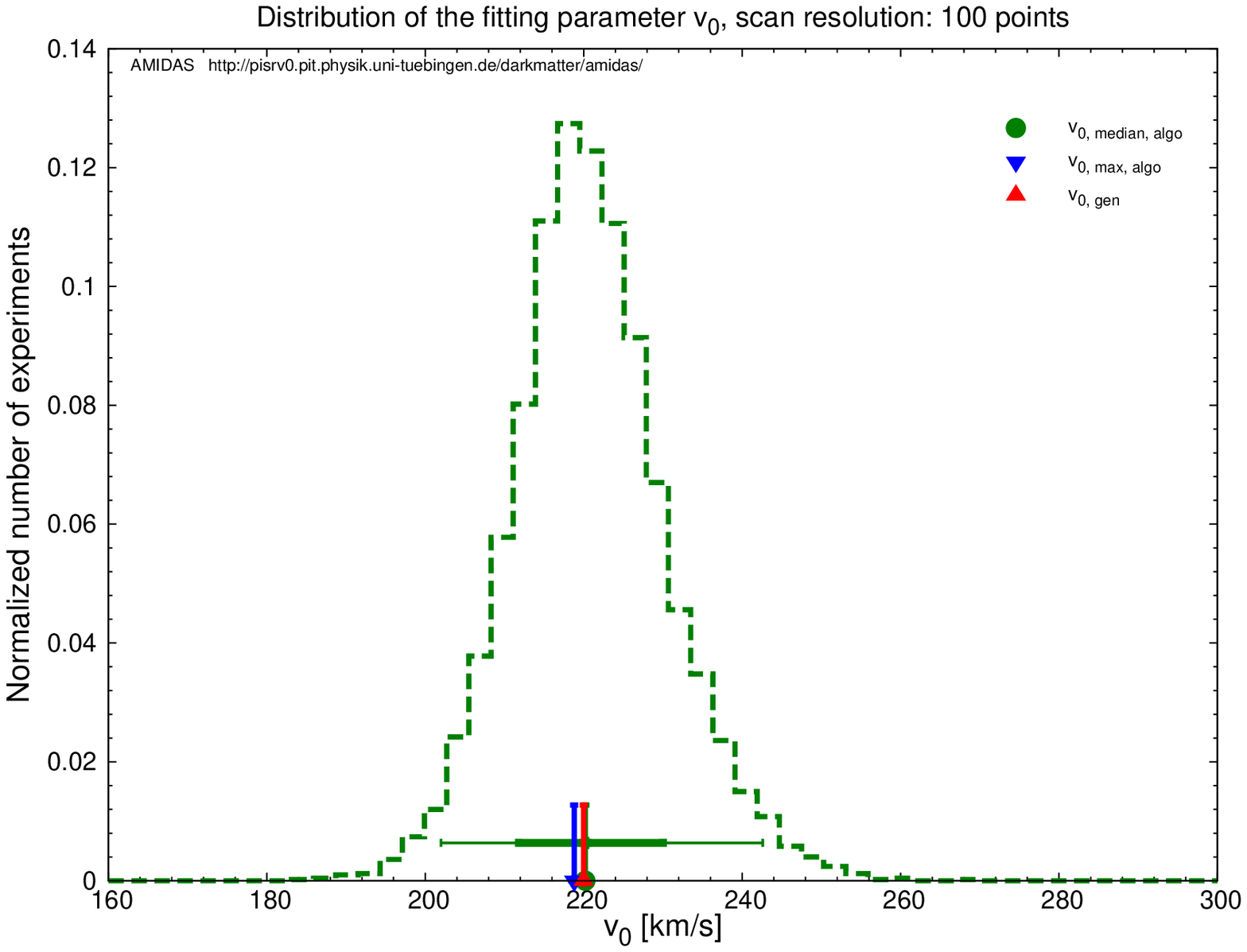}
\includegraphics[width=8.5cm]{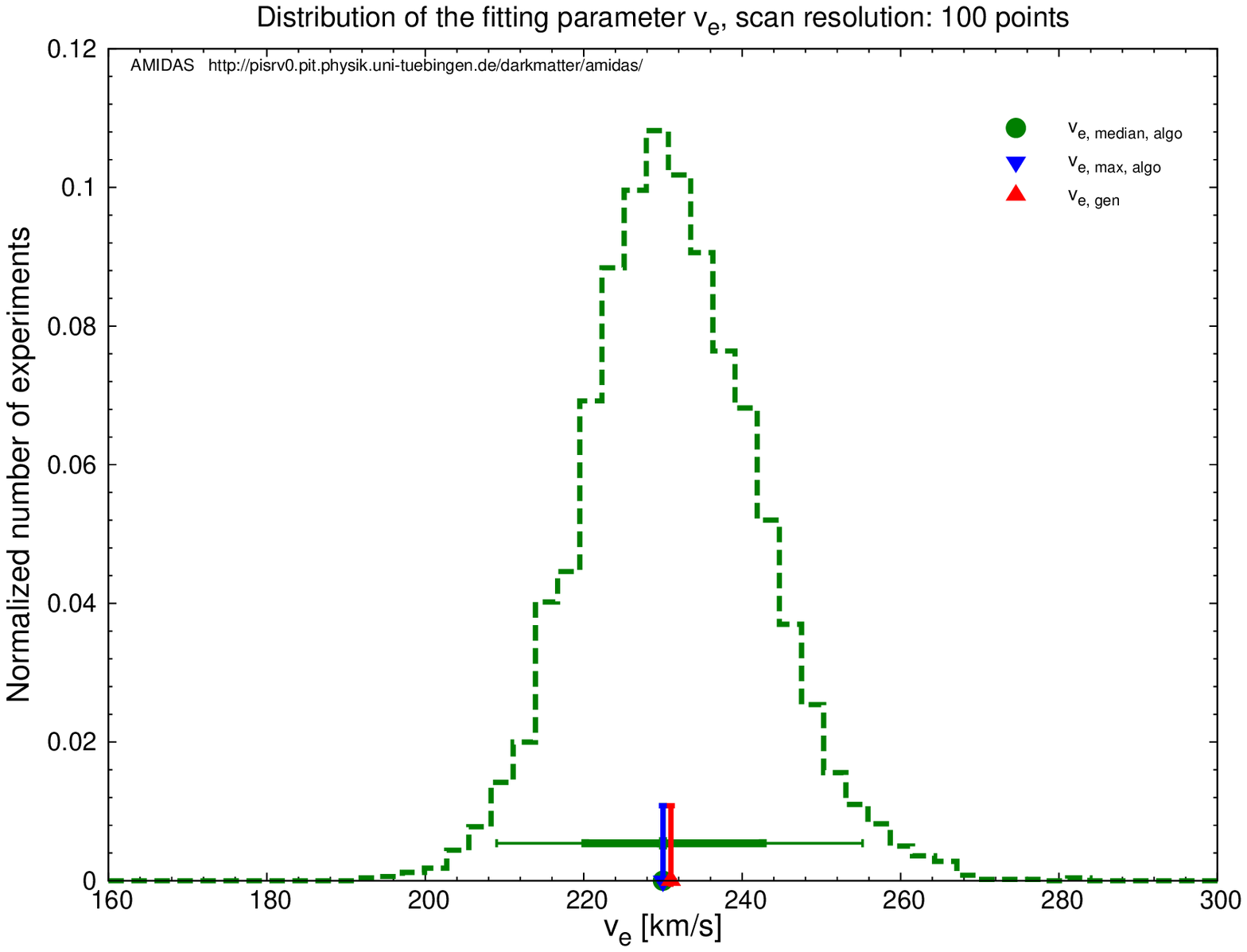}    \hspace*{-1.6cm} \par
\makebox[8.5cm]{(c)}\hspace{0.325cm}\makebox[8.175cm]{(d)}             \\
}
\vspace{-0.35cm}
\end{center}
\caption{
 As in Figs.~\ref{fig:f1v-Ge-SiGe-100-0500-sh-sh-Gau-bg},
 except that
 the input WIMP mass is set as \mbox{$\mchi = 250$ GeV}
 (simulation setup as in Sec.~3.4.2)
 and
 a fraction of {\em 10\%} background events
 has been taken into account.
}
\label{fig:f1v-Ge-SiGe-250-0500-sh-sh-Gau-bg}
\end{figure}
}